\documentclass[12pt]{article}

 \newread\testifexists
 \def\GetIfExists #1 {\immediate\openin\testifexists=#1
     \ifeof\testifexists\immediate\closein\testifexists\else
     \immediate\closein\testifexists\input #1\fi}

 \usepackage{gthstyle}\usepackage{amsfonts}
 \usepackage{amssymb}
 \usepackage{graphicx} \usepackage{epstopdf}
 \immediate\openin\testifexists=e
     \ifeof\testifexists\immediate\closein\testifexists\else
     \immediate\closein\testifexists\input e\fipsf

 \def\Bbb#1{\setbox0=\hbox{$\tt #1$}  \copy0\kern-\wd0\kern .1em\copy0}

 \def\bbf#1{\setbox0=\hbox{$#1$} \kern-.025em\copy0\kern-\wd0
         \kern.05em\copy0\kern-\wd0 \kern-.025em\raise.0433em\box0}

 \immediate\openin\testifexists=a
     \ifeof\testifexists\immediate\closein\testifexists\else
     \immediate\closein\testifexists\input a\fimssym.def  

 \newcommand{\tl}[1]{\tilde{#1}}        \renewcommand{\^}[1]{\hat{#1}}      \newcommand{\Tr}{{\mbox{Tr}}\,}
              \newcommand{\cl}{\centerline}       \newcommand{\fn}{\footnote}
 \newcommand{\nn}{\nonumber\\[2pt]}             \newcommand{\nm}{\nonumber}
 \newcommand{\be}{\begin{eqnarray}}             \newcommand{\ee}{\end{eqnarray}}
 \newcommand{\bi}[1]{\begin{itemize}\item[#1]}         \newcommand{\itm}[1]{\item[#1]}  \newcommand{\ei}{\end{itemize}}
 \newcommand{\eqn}[1]{(\ref{#1})}

 \newcommand{\crlb}[1]{\label{#1}\\[2pt]}
 \newcommand{\eela}[1]{\,\hbox{\scriptsize{#1}\qquad}\label{#1}\end{eqnarray}}
 \newcommand{\eelb}[1]{\label{#1}\end{eqnarray}}
 
 \newcommand{\newsecb}[2]{\section{#1}\label{#2}\setcounter{equation}{0}}
 	 	
	 \newcommand{\subsecb}[2]{\subsection{#1}\label{#2}}
	 	 	
	 	 \newcommand{\subsubsecb}[2]{\subsubsection{#1}\label{#2}}

 \newcommand{\nolabels} {\def\eel{\eelb} \def\crl{\crlb} \def\newsecl{\newsecb}\def\subsecl{\subsecb}
 	\def\subsubsecl{\subsubsecb}\def\bibiteml{\bibitem}\def\citel{\cite}\def\labell{\label}}

\newcommand\publishversion{\nolabels\setlength{\textheight}{9in}\setlength{\oddsidemargin}{0in}
    \setlength{\textwidth}{6.3in}\setlength{\topmargin}{-.7in}}

 \def\a{\alpha}      \def\b{\beta}   \def\g{\gamma}      \def\G{\Gamma}
 \def\d{\delta}      \def\D{\Delta}  \def\ep{\epsilon}  \def\e{\varepsilon} 
 \def\k{\kappa}      \def\l{\lambda} \def\L{\Lambda}     \def\m{\mu}
 \def\f{\phi}        \def\F{\Phi}    \def\vv{\varphi}    \def\n{\nu}
 \def\j{\psi}        \def\J{\Psi}    \def\r{\varrho}     \def\s{\sigma}  \def\SS{\Sigma}
 \def\t{\tau}        \def\tht{\theta}  
               
 \def\w{\omega}      \def\W{\Omega}  \def\z{\zeta}

 \def\HH{{\mathcal H}}   \def\LL{{\mathcal L}} \def\OO{{\mathcal O}} \def\DD{{\mathcal D}}  \def\BBB{{\mathcal B}}   \def\CCC{{\mathcal C}}   \def\SSS{{\mathcal S}} 
 \def\pa{\partial} \def\ra{\rightarrow}
  
 \def\dd{{\rm d}}  \def\bra{\langle}   \def\ket{\rangle}

 \def\qu{\ {\buildrel {\displaystyle ?} \over =}\ }
 \def\deff{\ {\buildrel{\rm def}\over{=}}\ }
 \def\iss{\ =\ }
 
 \newcommand{\rem}[1]{}	

 \def\fract#1#2{{\textstyle{#1\over#2}}}
 \def\ffract#1#2{\raise .2 em\hbox{$\scriptstyle#1$}\kern-.35em/
                 \kern-.2em\lower .15 em \hbox{$\scriptstyle#2$}}
 \def\fractje#1#2{{\scriptstyle{#1\over#2}}}
 \def\half{\fract12} \def\quart{\fract14} \def\halff{\ffract1{\!2}}
 \def\halfje{\fractje12}
 \def\ex#1{e^{\textstyle#1}} \def\dsp{\displaystyle} \def\low#1{\raise-.3em\hbox{$\scriptstyle{#1}$}} 
 	\def\lowtje#1{\raise-.2em\hbox{$\!\scriptscriptstyle{#1}\!$}}
  
\def\bmatrix{\begin{matrix}} \def\ematrix{\end{matrix}} \def\bpmatrix{\begin{pmatrix}}\def\epmatrix{\end{pmatrix}}
\def\bcenter{\begin{center}} \def\ecenter{\end{center}}

\def\lowerheightfig#1#2#3{\(\raise-#1\hbox{\includegraphics[height=#2]{#3}}\)}
\def\lowerwidthfig#1#2#3{\(\raise-#1\hbox{\includegraphics[width=#2]{#3}}\)}
\def\qqquad{\qquad\qquad}

\def\glt{\hbox{\,\raise .35em\hbox{$>$}\kern-.8em\raise-.15em\hbox{$<$}\,}}  \def\op{\mathrm{op}}
\def\ont{\mathrm{ont}}

\publishversion 
\begin{document} \begin{titlepage}

\title{\normalsize    
\vskip 5mm \LARGE\bf The Cellular Automaton Interpretation\\
of Quantum Mechanics\\[15pt]
}\author{Gerard 't~Hooft}
\date{\normalsize Institute for Theoretical Physics \\
Utrecht University  \\ 
Postbox 80.195 \\ 3508 TD Utrecht, the Netherlands \smallskip \\
e-mail:{ g.thooft@uu.nl} \\ internet: {
http://www.staff.science.uu.nl/\~{}hooft101/}}

\maketitle

{\small When investigating theories at the tiniest conceivable scales in nature, almost all researchers today revert to the quantum language,
accepting the verdict from the Copenhagen doctrine that the only way to describe what is going on will always involve states in Hilbert space,
controlled by operator equations. Returning to classical, that is, non quantum mechanical, descriptions will be forever impossible, unless one accepts
some extremely contrived theoretical constructions that may or may not reproduce the quantum mechanical phenomena observed in experiments.

Dissatisfied, this author investigated how one can look at things differently. 
This book is an overview of older material, but also contains many new observations and calculations. 
Quantum mechanics is looked upon as a tool, not as a theory. Examples are displayed of models that are classical in essence, but can be analysed by the use of quantum techniques, and we argue that even the Standard Model, together with gravitational interactions, might be viewed as a quantum mechanical approach to analyse a system that could be classical at its core. We explain how such thoughts can conceivably be reconciled with Bell's theorem, and how the usual objections voiced against the notion of `superdeterminism' can be overcome, at least in principle.
Our proposal would eradicate the collapse problem and the measurement problem. Even the existence of an ``arrow of time" can perhaps be explained in a more elegant way than usual.}\\

\vfill \flushleft{Version December, 2015 \ (extensively modified)}\\

\end{titlepage}


 \setcounter{page}{2}
\tableofcontents 
\listoffigures
	 \def\th{{\mathrm{\:\! th}}}  \def\E{\hbox{\scriptsize{E}}}
\hbox{ }\vskip 30pt \pagebreak[3]
\noindent      
 {\Large \textbf{ Preface}}  \\[5pt]

This book is not in any way  intended to serve as a replacement for the standard theory of quantum mechanics. A reader not yet thoroughly familiar with the basic concepts of quantum mechanics is advised first to learn this theory from one of the recommended text books\,\cite{Merzbacher-1961}\cite{Davidov-1963}\cite{Das-1986}, and only then pick up this book to find out that the doctrine called `quantum mechanics' can be viewed as part of a marvellous mathematical machinery that places physical phenomena in a greater context, and only in the second place as a theory of nature.

The present version, \#\,3, has been thoroughly modified. Some novelties, such as an unconventional view of \emph{the arrow of time}, have been added, and other arguments were further refined.
The book is now split in two. Part I deals with the many conceptual issues, without demanding excessive calculations. Part II adds to this our calculation techniques, occasionally returning to conceptual issues. Inevitably, the text in both parts will frequently refer to discussions in the other part, but they can be studied separately.
	
This book is not a novel that has to be read from beginning to end, but rather a collection of descriptions and derivations, to be used as a reference. Different parts can be read in random order. Some arguments are repeated several times, but each time in a different context.

\vskip30pt	 \pagebreak[3]
	 \part[\textbf{The Cellular Automaton Interpretation as a general\\  \hbox{\ }\quad\,doctrine}]   {\Large \textbf {The Cellular Automaton Interpretation as a general doctrine}}
\bigskip\bigskip
	
\newsecl{Motivation for this work}{motive}

This book is about a theory, and about an interpretation. The theory, as it stands, is highly speculative. It is born out of dissatisfaction with the existing explanations of a well-established fact.  The fact is that our universe appears to be controlled by the laws of quantum mechanics. Quantum mechanics looks weird, but nevertheless it provides for a very solid basis for doing calculations of all sorts that explain the peculiarities of the atomic and sub-atomic world. The theory developed in this book starts from assumptions that, at first sight, seem to be natural and straightforward, and we think they can be very well defended. 

Regardless whether the theory is completely right, partly right, or dead wrong, one may be inspired by the way it looks at quantum mechanics. We are assuming the existence of a definite `reality' underlying quantum mechanical descriptions. The assumption that this reality exists leads to a rather down-to-earth \emph{interpretation} of what quantum mechanical calculations are telling us. The interpretation works beautifully and seems to remove several of the difficulties encountered in other descriptions of how one might interpret the measurements and their findings. We propose this interpretation that, in our eyes, is superior to other existing dogmas.

However, numerous extensive investigations have provided very strong evidence that the assumptions that went into our theory cannot be completely right. The earliest arguments came from von Neumann\,\cite{Neumann-1932}, but these were later hotly debated \,\cite{g.hermann-1935},\cite{Bell-1964},\cite{Bub-2010}. 
The most convincing arguments came from John S.~Bell's theorem, phrased in terms of inequalities that are supposed to hold for any classical interpretation of quantum mechanics, but are \emph{strongly} violated by quantum mechanics. Later, many other variations were found of Bell's basic idea, some even more powerful. We will discuss these repeatedly, and at length, in this work. Basically, they all seemed to point in the same direction: from these theorems,  it was concluded by most researchers that the laws of nature cannot possibly be deterministic. So why this book?

There are various reasons why the author decided to hold on to his assumptions anyway. The first reason is that they fit very well with the quantum equations of various very simple models. It looks as if nature is telling us: ``wait, this approach is not so bad at all!". The second reason is that one could regard our approach simply as a \emph{first attempt} at a description of nature that is more realistic than other existing approaches. We can always later decide to add some twists that introduce indeterminism, in a way more in line with the afore mentioned theorems; these twists could be very \emph{different} from what is expected by many experts, but anyway, in that case, we could all emerge out of this fight victorious. Perhaps there is a subtle form of non-locality in the cellular automata, perhaps there is some quantum twist in the boundary conditions, or you name it. Why should Bell's inequalities forbid me to investigate this alley? I happen to find it an interesting one.

But there is a third reason. This is the strong suspicion that all those ``hidden variable models" that were compared with thought experiments as well as real experiments, are terribly naive.\fn{Indeed, in their eagerness to exclude local, realistic, and/or deterministic theories, authors rarely go into the trouble to carefully define what these theories are.} \emph{Real} deterministic theories have not yet been excluded. If a theory is deterministic \emph{all the way}, it implies that not only all observed phenomena, but also the observers themselves are controlled by deterministic laws. They certainly have no `free will', their actions all have roots in the past, even the distant past. Allowing an observer to have free will, that is, to reset his observation apparatus at will without even infinitesimal disturbances of the surrounding universe, \emph{including modifications in the distant past}, is fundamentally impossible. The notion that, also the actions by experimenters and observers are controlled by deterministic laws, is called \emph{superdeterminism}. When discussing these issues with colleagues the author got the distinct impression that it is here that the `no-go' theorems they usually come up with, can be put in doubt. 

We hasten to add that this is not the first time that this remark was made\,\cite{superdet-2011}. Bell noticed that superdeterminism could provide for a loophole around his theorem, but as most researchers also today, he was quick to dismiss it as ``absurd". As we hope to be able to demonstrate, however, superdeterminism may not quite be as absurd as it seems.\fn{We do find some ``absurd" correlation functions, see e.g. subsection~\ref{hiddeninfo}.}

In any case, realising these facts sheds an interesting new light on our questions, and the author was strongly motivated just to carry on. 

Having said all this, I do admit that what we have is still only a theory. It can and will be criticised and attacked, as it already was. I know that some readers will not be convinced. If, in the mind of some others, I succeed to generate some sympathy, even enthusiasm for these ideas, then  my goal has been reached. In a somewhat worse scenario, my ideas will be just used as an anvil, against which other investigators will sharpen their own, superior views.

In the mean time, we are developing mathematical notions that seem to be coherent and beautiful. Not very surprisingly, we do encounter some problems in the formalism as well, which we try to phrase as accurately as possible. They do indicate that the problem of generating quantum phenomena out of classical equations is actually quite complex. The difficulty we bounce into is that, although \emph{all} classical models allow for a reformulation in terms of some `quantum' system, the resulting quantum system will often not have a Hamiltonian that is local and properly bounded from below. It may well be that models that do produce acceptable Hamiltonians will demand inclusion of non-perturbative gravitational effects, which are indeed difficult and ill-understood at present. 

It is unlikely, in the mind of the author, that these complicated schemes can be wiped off the table in a few lines, as is asserted by some\fn{At various places in this book, we explain what is wrong with those `few lines'.}. Instead, they warrant intensive investigation. As stated, if we can make the theories more solid, they would provide for extremely elegant foundations that underpin the Cellular Automaton Interpretation of quantum mechanics. It will be shown in this book that we can arrive at Hamiltonians that are \emph{almost} both local and bounded from below. These models are like quantised field theories, which also suffer from mathematical imperfections, as is well-known.

Furthermore, one may question why we would have to require locality of the quantum model at all, as long as the underlying classical model is manifestly local by construction. What we exactly mean by all this will be explained, mostly in part II where we allow ourselves to perform detailed calculations.
 
\subsecl{Why an interpretation is needed}{why}

The discovery of quantum mechanics may well have been the most important scientific revolution of the \(20^\th\) century. Not only the world of atoms and subatomic particles appears to be completely controlled by the rules of quantum mechanics, but also the worlds of solid state physics, chemistry,  thermodynamics, and all radiation phenomena can only be understood by observing the laws of the quanta. The successes of quantum mechanics are phenomenal, and furthermore, the theory appears to be reigned by marvellous and impeccable internal mathematical logic.

Not very surprisingly, this great scientific achievement also caught the attention of scientists from other fields, and from philosophers, as well as the public in general. It is therefore perhaps somewhat curious that, even after nearly a full century, physicists still do not quite agree on what the theory tells us -- and what it does not tell us -- about \emph{reality}. 

The reason why quantum mechanics works so well is that, in practically all areas of its applications, exactly what reality means turns out to be immaterial. All that this theory\fn{Interchangeably, we use the word `theory' for quantum mechanics itself, and for models of particle interactions; therefore, it might be better to refer to quantum mechanics as a \emph{framework}, assisting us in devising theories for sub systems, but we expect that our use of the concept of `theory' should not generate any confusion.} says, and that needs to be said, is about the reality of the outcomes of an experiment. Quantum mechanics tells us exactly what one should expect, how these outcomes may be distributed statistically, and how these can be used to deduce details of its internal parameters. Elementary particles are one of the prime targets here. A theory\footnotemark[\value{footnote}] has been arrived at, the so-called Standard Model, that requires the specification of some 25 internal constants of nature, parameters that cannot be predicted using present knowledge. Most of these parameters could be determined from the experimental results, with varied accuracies. Quantum mechanics works flawlessly every time.

So, quantum mechanics, with all its peculiarities, is rightfully regarded as one of the most profound discoveries in the field of physics, revolutionising our understanding of many features of the atomic and sub-atomic world.

But physics is not finished. In spite of some over-enthusiastic proclamations just before the turn of the century, the \emph{Theory of Everything} has not yet been discovered, and there are other open questions reminding us that physicists have not yet done their job completely. Therefore, encouraged by the great achievements we witnessed in the past, scientists continue along the path that has been so successful. New experiments are being designed, and new theories are developed, each with ever increasing ingenuity and imagination. Of course, what we have learned to do is to incorporate every piece of knowledge gained in the past, in our new theories, and even in our wilder ideas.

But then, there is a question of strategy. Which roads should we follow if we wish to put the last pieces of our jig-saw puzzle in place? Or even more to the point: what do we expect those last jig-saw pieces to look like? And in particular: should we expect the ultimate future theory to be quantum mechanical?

It is at this point that opinions among researchers vary, which is how it should be in science, so we do not complain about this. On the contrary, we are inspired to search with utter concentration precisely at those spots where no-one else has taken the trouble to look before. The subject of this book is the `reality' behind quantum mechanics. Our suspicion is that it may be very different from what can be read in most text books. We actually advocate the notion that it might be \emph{simpler} than anything that can be read in the text books. If this is really so, this might greatly facilitate our quest for better theoretical understanding.

Many of the ideas expressed and worked out in this treatise are very basic. Clearly, we are not the first to advocate these ideas. The reason why one rarely hears about the obvious and simple observations that we will make, is that they have been made many times, in the recent and the more ancient past\,\cite{Neumann-1932}, and were subsequently categorically dismissed.

The primary reason why they have been dismissed is that they were unsuccessful; classical, deterministic models that produce the same results as quantum mechanics were devised, adapted and modified, but whatever was attempted ended up looking much uglier than the original theory, which was plain quantum mechanics with no further questions asked. The quantum mechanical theory describing relativistic, subatomic particles is called \emph{quantum field theory} (see part~II, chapter~\ref{QFT}), and it obeys fundamental conditions such as causality, locality and unitarity. Demanding all of these desirable properties was the core of the successes of quantum field theory, and that eventually gave us the Standard Model of the sub-atomic particles. If we try to reproduce the results of quantum field theory in terms of some deterministic underlying theory, it seems that one has to abandon at least one of these demands, which would remove much of the beauty of the generally accepted theory; it is much simpler not to do so, and therefore, as for the requirement of  \emph{the existence of a classical underlying theory}, one usually simply drops that.

Not only does it seem to be unnecessary to assume the existence of a classical world underlying quantum mechanics, it seems to be impossible also.  Not very surprisingly, researchers turn their heads in disdain, but just before doing so, there was one more thing  to do: if, invariably, deterministic models that were intended to reproduce typically quantum mechanical effects, appear to get stranded in contradictions, maybe one can \emph{prove} that such models are impossible. This may look like the more noble alley: close the door for good.

A way to do this was to address the famous Gedanken experiment designed by Einstein, Podolsky and Rosen\,\cite{EPR-1935}. This experiment suggested that quantum particles are associated with more than just a wave function; to make quantum mechanics describe `reality', some sort of `hidden variables' seemed to be needed. What could be done was to prove that such hidden variables are self-contradictory. We call this a `no-go theorem'. 
The most notorious, and most basic, example was \emph{Bell's theorem}\,\cite{Bell-1964}, as we already mentioned.  Bell studied the correlations between measurements of entangled particles, and found that, if the initial state for these particles is chosen to be sufficiently generic, the correlations found at the end of the experiment, as predicted by quantum mechanics, can \emph{never} be reproduced by information carriers that transport classical information. He expressed this in terms of the so-called Bell inequalities, later extended as CHSH inequality\,\cite{CHSH-1969}. They are obeyed by any classical system but strongly violated by quantum mechanics. It appeared to be inevitable to conclude that we have to give up producing classical, local, realistic theories. They don't exist.

So why the present treatise? Almost every day, we receive mail from amateur physicists telling us why established science is all wrong, and what they think a ``theory of everything" should look like. Now it may seem that I am treading in their foot steps. Am I suggesting that nearly one hundred years of investigations of quantum mechanics have been wasted?  Not at all.  I insist that the last century of research lead to magnificent results, and that the only thing missing so-far was a more radical description of what has been found. Not the equations were wrong, not the technology, but only the wording of what is often referred to as the Copenhagen Interpretation should be replaced. Up to today, the theory of quantum mechanics consisted of a set of very rigorous rules as to how amplitudes of wave functions refer to the probabilities for various different outcomes of an experiment. It was stated emphatically that they are not referring to `what is really happening'. One should not ask what is really happening, one should be content with the predictions concerning the experimental results. The idea that no such `reality' should exist at all sounds mysterious. 
It is my intention to remove every single bit of mysticism from quantum theory, and we intend to deduce facts about reality anyway.

Quantum mechanics is one of the most brilliant results of one century of science, and it is not my intention to replace it by some mutilated version, no matter how slight the mutilation would be. Most of the text books on quantum mechanics will not need the slightest revision anywhere, except perhaps when they state that questions about reality are forbidden. All practical calculations on the numerous stupefying quantum phenomena can be kept as they are. It is indeed in quite a few competing theories about the interpretation of quantum mechanics where authors are led to introduce non-linearities in the Schr\"odinger equation or violations of the Born rule that will be impermissible in this work. 

As for `entangled particles', since it is known how to produce such states in practice, their odd-looking behaviour must be completely taken care of in our approach.

The `collapse of the wave function' is a typical topic of discussion, where several researchers believe a modification of Schr\"odinger's equation is required. Not so in this work, as we shall explain. We also find surprisingly natural answers to questions concerning `Schr\"odinger's cat', and the `arrow of time'.

And as of `no-go theorems', this author has seen several of them, standing in the way of further progress. One always has to take the assumptions into consideration, just as the small print in a contract. 

\subsecl{Outline of the ideas exposed in part I}{outline}

Our starting point will be extremely simple and straightforward, in fact so much so that some readers may simply conclude that I am losing my mind. However, with questions of the sort I will be asking, it is inevitable to start at the very basic beginning. We start with just \emph{any} classical system that vaguely looks like our universe, with the intention to refine it whenever we find this to be appropriate. Will we need non-local interactions? Will we need information loss? Must we include some version of a gravitational force? Or will the whole project run astray? We won't know unless we try.

The price we do pay seems to be a modest one, but it needs to be mentioned: we have to select a very special set of mutually orthogonal states in Hilbert space that are endowed with the status of being `real'. This set consists of the states the universe can `really' be in. At all times, the universe chooses one of these states to be in, with probability 1, while all others carry probability 0. We call these states  \emph{ontological states}, and they form a special basis for Hilbert space, the \emph{ontological basis}. One could say that this is just wording, so this price we pay is affordable, but we will assume this very special basis to have special properties. What this does imply is that the quantum theories we end up with all form a very special subset of all quantum theories. This then, could lead to new physics, which is why we believe our approach will warrant attention: eventually, our aim is not just a reinterpretation of quantum mechanics, but the discovery of new tools for model building.

One might expect that our approach, having such a precarious relationship with both standard quantum mechanics and other insights concerning the interpretation of quantum mechanics, should quickly strand in contradictions. This is perhaps the more remarkable observation one then makes: it works quite well! Several models can be constructed that reproduce quantum mechanics \emph{without the slightest modification}, as will be shown in much more detail in part II. All our simple models are quite straightforward. The numerous responses I received, saying that the models I produce ``somehow aren't real quantum mechanics" are simply mistaken. They are really quantum mechanical. However, I will be the first to remark that one can nonetheless criticise our results: the models are either too simple, which means they do not describe interesting, interacting particles, or they seem to exhibit more subtle defects. In particular, reproducing realistic quantum models for locally interacting quantum particles along the lines proposed, has as yet shown to be beyond what we can do. As an excuse I can only plead that this would require not only the reproduction of a complete, renormalizable quantum field theoretical model, but in addition it may well demand the incorporation of a perfectly quantised version of the gravitational force, so indeed  it should not surprise anyone that this is hard. 

Numerous earlier attempts have been made to find holes in the arguments initiated by Bell, and corroborated by others. Most of these falsification arguments have been rightfully dismissed. But now it is our turn. Knowing what the locality structure is expected to be in our models, and why we nevertheless think they reproduce quantum mechanics, we can now attempt to locate the cause of this apparent disagreement. Is the fault in our models or in the arguments of Bell c.s.? What could be the cause of this  discrepancy? If we take one of our classical models, what goes wrong in a Bell experiment with entangled particles? Were assumptions made that do not hold? Do particles in our models perhaps refuse to get entangled? This way, we hope to contribute to an ongoing discussion.

The aim of the present study is to work out some fundamental physical principles. Some of them are nearly as general as the fundamental, canonical theory of classical mechanics. The way we deviate from standard methods is that, more frequently than usual, we introduce \emph{discrete} kinetic variables. We demonstrate that such models not only \emph{appear} to have much in common with quantum mechanics. In many cases, they \emph{are} quantum mechanical, but also classical at the same time. Some of our models occupy a domain in between classical and quantum mechanics, a domain often thought to be empty.

Will this lead to a revolutionary alternative view on what quantum mechanics is? The difficulties with the sign of the energy and the locality of the effective Hamiltonians in our theories have not yet been settled. In the real world there is a lower bound for the total energy, so that there is a \emph{vacuum state}. The subtleties associated with that are postponed to part~II, since they require detailed calculations. In summary: we suspect that there will be several ways to overcome this difficulty, or better still, that it can be used to explain some of the apparent contradictions in quantum mechanics. 
 
 The complete and unquestionable answers to many questions are not given in this treatise, but we are homing in to some important observations. As has happened in other examples of ``no-go theorems", Bell and his followers did make assumptions, and in their case also, the assumptions appeared to be utterly reasonable.  Nevertheless we now suspect that some of the premises made by Bell may have to be relaxed. Our theory is not yet complete, and a reader strongly opposed to what we are trying to do here, may well be able to find a stick that seems suitable  to destroy it. Others, I hope, will be inspired to continue along this path. 

We invite the reader to draw his or her own conclusions. We do intend to achieve that questions concerning the deeper meanings of quantum mechanics are illuminated from a new perspective.  This we do by setting up models and by doing calculations in these models. Now this has been done before, but most models I have seen appear to be too contrived, either requiring the existence of infinitely many universes all interfering with one another, or modifying the equations of quantum mechanics, while the original equations seem to be beautifully coherent and functional.

Our models suggest that Einstein may have been right, when he objected against the conclusions drawn by Bohr and Heisenberg. It may well be that, at its most basic level, there is no randomness in nature, no fundamentally statistical aspect to the laws of evolution. Everything, up to the most minute detail, is controlled by invariable laws. Every significant event in our universe takes place for a reason, it was caused by the action of physical law, not just by chance. This is the general picture conveyed by this book. We know that it looks as if Bell's inequalities have refuted this possibility, so yes, they raise interesting and important questions that we shall address at various levels.

 It may seem that I am employing rather long arguments to make my point\fn{A wise lesson to be drawn from one's life experiences is, that long arguments are often much more dubious than short ones.}. The most essential elements of our reasoning will show to be short and simple, but
just because I want chapters of this book to be self-sustained, well readable and understandable, there will be some repetitions of arguments here and there, for which I apologise. I also apologise for the fact that some parts of the calculations are at a very basic level; the hope is that this will also make this work accessible for a larger class of scientists and students.

The most elegant way to handle quantum mechanics in all its generality is Dirac's \emph{bra-ket} formalism (section~\ref{Dirac1}). We stress that Hilbert space is a central tool for physics, not only for quantum mechanics. It can be applied in much more general systems than the standard quantum models such as the hydrogen atom, and it will be used also in completely deterministic models (we can even use it in Newton's description of the planetary system, see subsection~\ref{EarthMars}). 

In any description of a model, one first chooses a \emph{basis} in Hilbert space. Then, what is needed is a Hamiltonian, in order to describe dynamics. A very special feature of Hilbert space is that one can use any basis one likes. The transformation from one basis to another is a unitary transformation, and we shall frequently make use of such transformations. Everything written about this in sections~\ref{Dirac1}, \ref{Copenhagen} and  \ref{Dirac2} is completely standard.

In part I of the book, we describe the philosophy of the Cellular Automaton Interpretation (CAI) without too many technical calculations. 
After the Introduction, we first demonstrate the most basic prototype of a model, the Cogwheel Model, in chapter~\ref{determm}.

In chapters~\ref{intpr} and \ref{detqu}, we begin to deal with the real subject of this research: the question of the interpretation of quantum mechanics. The standard approach, referred to as the Copenhagen Interpretation, is dealt with very briefly, emphasising those points where we have something to say, in particular the Bell and the CHSH inequalities. 

Subsequently, we formulate as clearly as possible what we mean with  \emph{deterministic quantum mechanics}. 
The \emph{Cellular Automaton Interpretation} of quantum mechanics (chapters~\ref{detqu} and \ref{CAshort}) must sound as a blasphemy to some quantum physicists, but this is because we do not go along with some of the assumptions usually made. Most notably, it is the assumption that space-like correlations in the \emph{beables} of this world cannot possibly generate the `conspiracy' that seems to be required to violate Bell's inequality. We derive the existence of such correlations.

We end chapter~\ref{intpr} with one of the more important fundamental ideas of the CAI: our hidden variables do contain `hidden information' about the future, notably the settings that will be chosen by Alice an Bob, but it is fundamentally non-local information, impossible to harvest even in principle (subsection~\ref{hiddeninfo}). This should not be seen as a violation of causality.

Even if it is still unclear whether or not the results of these correlations have a conspiratory nature, one can base a useful and functional interpretation doctrine from the assumption that the only conspiracy the equations perform is to fool some of today's physicists, while they act in complete harmony with credible sets of physical laws. The \emph{measurement process} and the \emph{collapse of the wave function} are two riddles that are completely resolved by this assumption, as will be indicated.

We hope to inspire more physicists to investigate these possibilities, to consider seriously the possibility that quantum mechanics as we know it is not a fundamental, mysterious, impenetrable feature of our physical world, but rather an instrument to  statistically  describe a world where the physical laws, at their most basic roots, are not quantum mechanical at all. Sure, we do not know how to formulate the most basic laws at present, but we are collecting indications that a classical world underlying quantum mechanics does exist. 

Our models show how to put quantum mechanics on hold when we are constructing models such as string theory and ``quantum" gravity, and this may lead to much improved understanding of our world at the Planck scale.
Many chapters are reasonably self sustained; one may choose to go directly to the parts where the basic features of the Cellular Automaton Interpretation (CAI) are exposed, chapters~\ref{intpr} -- \ref{conc1}, or look at the explicit calculations done in part II. 

In chapter~\ref{catcai}, we display the rules of the game. Readers might want to jump to this chapter directly, but might then be mystified by some of our assertions if one has not yet been exposed to the general working philosophy developed in the previous chapters. Well, you don't have to take everything for granted; there are still problems unsolved, and further alleys to be investigated. They are in chapter~\ref{miss}, where it can be seen how the various issues show up in calculations.

Part II of this book is not intended to impress the reader or to scare him or her away. The explicit calculations carried out there are displayed in order to develop and demonstrate our calculation tools; only few of these results are used in the more general discussions in the first part. Just skip them if you don't like them.

\subsection[A ${19^{\,\mathrm{th}}}$ century philosophy]{A $\mathbf {19^{\,\mathbf{th}}}$ century philosophy\labell{philo}}

Let us go back to the \(19^\th\) century. Imagine that mathematics were at a very advanced level, but nothing of the \(20^\th\) century physics was known. Suppose someone had phrased a detailed hypothesis about his world being a \emph{cellular automaton}\fn{One such person is E.~Fredkin, an expert in numerical computation techniques, with whom we had lengthy discussions. The idea itself was of course much older\,\cite{Zuse-1969}\cite{Wolfram-2002}.}. The cellular automaton will be precisely defined in section~\ref{CAgeneral} and in part II; for now, it suffices to characterise it by the requirement that the states Nature can be in are given by sequences of integers. The evolution law is a classical algorithm that tells unambiguously how these integers evolve in time. Quantum mechanics does not enter; it is unheard of.
The evolution law is sufficiently non-trivial to make our cellular automaton behave as a \emph{universal computer}\,\cite{Fredkin-1982}. This means that, at its tiniest time and distance scale, initial states could be chosen such that any mathematical equation can be solved with it. This means that it will be impossible to derive exactly how the automaton will behave at large time intervals; it will be far too complex.

Mathematicians will realise that one should not even try to deduce exactly what the large-time and large-distance properties of this theory will be, but they may decide to try something else. Can one, perhaps, make some \emph{statistical} statements about the large scale behaviour?

In first approximation, just white noise may be seen to emerge, but upon closer inspection, 
the system may develop non-trivial correlations in its series of integers; some of the correlation functions may be calculable, just the way these may be calculated in a Van der Waals gas. We cannot rigorously compute the trajectories of individual molecules in this gas, but we can derive free energy and pressure of the gas as a function of density and temperature, we can derive its viscosity and other bulk properties. Clearly, this is what our \(19^\th\) century mathematicians should do with their cellular automaton model of their world. In this book we will indicate how physicists and mathematicians of the \(20^\th\) and \(21^{\mathrm{st}}\) centuries can do even more: they have a tool called \emph{quantum mechanics} to derive and understand even more sophisticated details, but even they will have to admit that exact calculations are impossible. The only effective, large scale laws that they can ever expect to derive are statistical ones. The average outcomes of experiments can be predicted, but not the outcomes of individual experiments; for doing that, the evolution equations are far too difficult to handle.

In short, our imaginary \(19^\th\) century world will seem to be controlled by effective laws with a large stochastic element in them. This means that, in addition to an effective deterministic law, random number generators may seem to be at work that are fundamentally unpredictable. On the face of it, these effective laws together may look quite a bit like the quantum mechanical laws we have today for the sub-atomic particles.

The above metaphor is of course not perfect. The Van der Waals gas does obey general equations of state, one can understand how sound waves behave in such a gas, but it is not quantum mechanical. One could suspect that this is because the microscopic laws assumed to be at the basis of a Van der Waals gas are very different from a cellular automaton, but it is not known whether this might be sufficient to explain why the Van der Waals gas is clearly not quantum mechanical. 

What we do wish to deduce from this reasoning is that one feature of our world is not mysterious: the fact that we have effective laws that require a stochastic element in the form of an apparently perfect random number generator, is something we should not be surprised about. Our \(19^\th\) century physicists would be happy with what their mathematicians give them, and they would have been totally prepared for the findings of \(20^\th\) century physicists, which implied that indeed the effective laws controlling hydrogen atoms contain a stochastic element, for instance to determine at what moment exactly an excited atom decides to emit a photon.

This may be the deeper philosophical reason why we have quantum mechanics: not all features of the cellular automaton at the basis of our world allow to be extrapolated to large scales. Clearly, the exposition of this chapter is entirely non-technical and it may be a bad representation of all the subtleties of the theory we call quantum mechanics today. Yet we think it already captures some of the elements of the story we want to tell. If they had access to the mathematics known today, we may be led to the conclusion that our \(19^\th\) century physicists could have been able to derive an effective quantum theory for their automaton
model of the world. Would the \(19^\th\) century physicists be able to do experiments with entangled photons? This question we postpone to section~\ref{Bell} and onwards.

Philosophising about the different turns the course of history could have chosen, imagine the following. In the \(19^\th\) century, the theory of \emph{atoms} already existed. They could have been regarded as physicists' first successful step to discretise the world: atoms are the quanta of matter. Yet energy, momenta, and angular momenta were still assumed to be continuous. Would it not have been natural to suspect these to be discrete as well? In our world, this insight came with the discovery of quantum mechanics. But even today, space and time themselves are still strictly continuous entities. When will we discover that \emph{everything} in the physical world will eventually be discrete? This would be the discrete, deterministic world underlying our present theories. In this scenario, quantum mechanics as we know it, is the imperfect logic resulting from an incomplete discretisation\fn{As I write this, I expect numerous letters by amateurs, but beware, as it would be easy to propose some completely discretized concoction, but it is very hard to find the \emph{right} theory, one that helps us to understand the world as it is using rigorous mathematics.}.
	
\subsecl{Notation}{Dirac1}
In most parts of this book, quantum mechanics will be used as a tool kit, not a theory. Our theory may be anything; one of our tools will be Hilbert space and the mathematical manipulations that can be done in that space. Although we do assume the reader to be familiar with these concepts, we briefly recapitulate what a Hilbert space is.

Hilbert space \(\HH\) is a complex\fn{Some critical readers were wondering where the complex numbers in quantum mechanics should come from, given the fact that we start off from classical theories. The answer is simple: complex numbers are nothing but man-made inventions, just as \emph{real numbers} are. In Hilbert space, they are useful tools whenever we discuss something that is conserved in time (such as baryon number), and when we want to diagonalise a Hamiltonian. Note that quantum mechanics can be formulated without complex numbers, if we accept that the Hamiltonian is an \emph{anti}\,symmetric matrix. But then, its eigen values are imaginary. We emphasise that imaginary numbers are primarily used to do mathematics, and for that reason they are indispensable for physics.}
vector space, whose number of dimensions is usually infinite, but sometimes we allow that to be a finite number. Its elements are called states, denoted as \(|\j\ket,\ |\vv\ket\), or any other ``ket".

 We have \emph{linearity}: whenever \(|\j_1\ket\) and \(\j_2\ket\) are states in our Hilbert space, then
	\be |\vv\ket\iss\l|\j_1\ket+\m|\j_2\ket\ ,\eel{linHilb}
where \(\l\) and \(\m\) are complex numbers, is also a state in this Hilbert space. For every ket-state \(|\j\ket\) we have 
a `conjugated bra-state', \(\bra\j|\), spanning a  conjugated vector space, \(\bra\j|,\ \bra\vv|\). This means that, if Eq.~\eqn{linHilb} holds, then
	\be \bra\vv|\iss\l^*\bra\j_1|+\m^*\bra\j_2|\ . \eel{antilin}
Furthermore, we have an inner product, or inproduct: if we have a bra, \(\bra\chi|\), and a ket, \(|\j\ket\), then a complex number is defined, the inner product denoted by \(\bra\chi|\j\ket\), obeying 
	\be \bra\chi|(\l|\j_1\ket+\m|\j_2\ket)=\l\bra\chi|\j_1\ket+\m\bra\chi|\j_2\ket\ ;\qquad \bra\chi|\j\ket=\bra\j|\chi\ket^*\ . \eel{inprodprop}

The inner product of a ket state  \(|\j\ket\) with its own bra is real and positive:
	\be\|\j\|^2\equiv\bra\j|\j\ket = \hbox{ real and }\ge 0\ ,\eel{inprod1}
	\be\hbox{while }\qquad\bra\j|\j\ket=0\ \leftrightarrow\ |\j\ket=0\ .\eel{inprod2}
Therefore, the inner product can be used to define a norm. 
A state \(|\j\ket\) is called a physical state, or normalised state, if
	\be\|\j\|^2=\bra\j|\j\ket =1\ . \eel{norm}
Later, we shall use the word \emph{template} to denote such state (the word `physical state' would be confusing and is better to be avoided). The full power of Dirac's notation is exploited further in part II.

Variables will sometimes be just numbers, and sometimes operators in Hilbert space. If the distinction should be made, or if clarity may demand it, operators will be denoted as such. We decided to do this simply by adding a super- or subscript ``{\small{op}}" to the symbol in question.\fn{Doing this absolutely everywhere for all operators was a bit too much to ask. When an operator just amounts to multiplication by a function we often omit the super- or subscript ``{\scriptsize{op}}", and in some other places we just mention clearly the fact that we are discussing an operator.} 

The \emph{Pauli matrices}, \(\vec\s=(\s_x,\,\s_y,\,\s_z)\) are defined to be the \(2\times 2\) matrices
	\be\s^\op_x=\pmatrix{0&1\cr 1&0}\ ,\quad\s^\op_y=\pmatrix{0&-i\cr i&0}\,\quad\s^\op_z=\pmatrix{1&0\cr 0&-1}\ . \eel{pauli}

\newsecl{Deterministic models in quantum notation}	{determm}

  \subsecl{The basic structure of deterministic models}{basics}
	For deterministic models, we will be using the same Dirac notation. A physical  state \(|A\ket\), where \(A\) may stand for any array of numbers, not necessarily  integers or real numbers,  is called an \emph{ontological state} if it is a state our deterministic system can be in. These states themselves do not form a Hilbert space, since in a deterministic theory we have no superpositions, but we can declare that they form a \emph{basis} for  a Hilbert space that we may herewith define \,\cite{GtH-detqu-1988}\cite{GtH-I-K-1992}, by deciding, once and for all, that all ontological states form an orthonormal set:
		\be\bra A|B\ket\equiv\d_{AB}\ . \eel{ontoortho}
We can allow this set to generate a Hilbert space if we declare what we mean when we talk about superpositions. In Hilbert space, we now introduce the \emph{quantum  states} \(|\j\ket\), as being more general than the ontological states:
	\be|\j\ket=\sum_A\,\l_A|A\ket\ ,\qquad\sum_A|\l_A|^2\equiv 1\ . \eel{superposstate}
A quantum state can be used as a \emph{template} for doing physics. With this we mean the following:
	\begin{quotation}\noindent\emph{A template  is a quantum state of the form \eqn{superposstate}  describing a situation where the probability to find our system to be in the ontological state \(|A\ket\) is \(|\l_A|^2\).}\end{quotation}
Note, that \(\l_A\) is allowed to be a complex or negative number, whereas the phase of \(\l_A\) plays no role whatsoever. In spite of this,  complex numbers will turn out to be quite useful here, as we shall see. Using the \emph{square} in Eq.~\eqn{superposstate} and in our definition above, is a fairly arbitrary choice; in principle, we could have used a different power. Here, we use the squares because this is by far the most useful choice; different powers would not affect the physics, but would result in unnecessary mathematical complications. The squares ensure that probability conservation amounts to a proper normalisation of the template states, and enable the use of unitary matrices in our transformations.

Occasionally, we may allow the indicators \(A,\ B,\ \cdots\) to represent continuous variables, a straightforward generalisation. In that case, we have a continuous deterministic system; the Kronecker delta in Eq.~\eqn{ontoortho} is then replaced by a Dirac delta, and the sums in Eq.~\eqn{superposstate} will be replaced by integrals. For now, to be explicit, we stick to a discrete notation.

We emphasise that the template states are not ontological. Hence we have \emph{no} direct interpretation, as yet, for the inner products \(\bra\j_1|\j_2\ket\) if both \(|\j_1\ket\) and \(|\j_2\ket\) are template  states. Only the absolute squares of \(\bra A|\j\ket\), where\(\bra A|\) is the conjugate of an ontological state, denote the probabilities \(|\l_A|^2\). We briefly return to this in subsection \ref{inprod}.

The time evolution of a deterministic model can now be written in operator form:
	\be |A(t)\ket=|P_\op^{(t)}A(0)\ket\ , \eel{perm}
where \(P_\op^{(t)}\) is a permutation operator. We can write \(P_\op^{(t)}\) as a matrix \(P^{(t)}_{AB}\) containing only ones and zeros. Then, Eq.~\eqn{perm} is written as a matrix equation,
	\be|A(t\ket)=U_{AB}(t)|B(0)\ket\ , \qquad U(t)_{AB}=P^{(t)}_{AB}\ . \eel{permmatrix}
By definition therefore, the matrix elements of the operator \(U(t)\) in this bases can only be 0 or 1. 

It is very important, at this stage, that we choose \(P_\op^{(t)}\) to be a genuine permutator, that is, it should be invertible.\fn{One can imagine deterministic models where \(P_\op^{(t)}\) does not have an inverse, which means that two different ontological states might both evolve into the same state later. We will consider this possibility later, see chapter~\ref{infoloss}.}
If the evolution law is time-independent, we have
	\be P_\op^{(t)}=\bigg(P_\op^{(\d t)}\bigg)^{t/\d t}\ , \qquad U_\op(t)=\bigg(U_\op(\d t)\bigg)^{t/\d t}\ , \eel{timeevolv}
where the permutator \(P_\op^{(\d t)}\), and its associated matrix \(U_\op(\d t)\) describe the evolution over the shortest possible time step, \(\d t\).

Note, that no harm is done if some of the entries in the matrix \(U(\d t)_{ab}\), instead of 1, are chosen to be unimodular complex numbers. Usually, however, we see no reason to do so, since a simple rotation of an ontological state in the complex plane has no physical meaning, but it could be useful for doing mathematics (for example, in section~\ref{fermions} of part II, we use the entries \(\pm 1\) and \(0\) in our evolution operators).

We can now state our first important mathematical observation:\\
\emph{The quantum-, or template-,  states \(|\j\ket\) all obey the same evolution equation:}
	\be |\j(t)\ket=U_\op(t)|\j(0)\ket\ . \eel{psievolv}
It is easy to observe that, indeed, the probabilities \(|\l_A|^2\) evolve as expected.\fn{At this stage of the theory, one may still define probabilities to be given as different functions of \(\l_A\), in line with the observation just made after Eq.~\eqn{superposstate}.}

Much of the work described in this book will be about writing the evolution operators \(U_\op(t)\) as exponentials: \emph{Find a hermitean operator \(H_\op\) such that}
	\be U_\op(\d t)=e^{-iH_\op\,\d t}\ ,\qquad\hbox{so that}\qquad U_\op(t)=e^{-iH_\op t}\ . \eel{Hamilton}
This elevates the time variable \(t\) to be continuous, if it originally could only be a multiple of \(\d t\). Finding an example of such an operator is actually easy. If, for simplicity, we restrict ourselves to template states \(|\j\ket\) that are orthogonal to the eigenstate of \(U_\op\) with eigenvalue 1,  then
	\be H_\op\,\d t=\pi-i\sum_{n=1}^\infty{1\over n}\left(U_\op(n\,\d t)-U_\op(-n\,\d t)\right)\eel{firstH}
is a solution of Eq.~\eqn{Hamilton}. This equation can be checked by Fourier analysis, see part II, section~\ref{infinitediscr}, Eqs~\eqn{omegasumsin} -- \eqn{HfrUn}.
	
Note that a correction is needed: the lowest eigenstate \(|\emptyset\ket\) of \(H\), the ground state, has \(U_\op|\emptyset\ket=|\emptyset\ket\) and \(H_\op|\emptyset\ket=0\), so that Eq.~\eqn{firstH} is invalid for that state, but here this is a minor detail\fn{In part II, we shall see the importance of having one state for which our identities fail, the so-called \emph{edge state.}} (it is the \emph{only} state for which Eq.~\eqn{firstH} fails).  If we have a periodic automaton, the equation can be replaced by a finite sum, also valid for the lowest energy state, see subsection~\ref{cogwheelN}.

There is one more reason why this is \emph{not} always the Hamiltonian we want: its eigenvalues will always be between \(-\pi/\d t\) and \(\pi/\d t\), while sometimes we may want expressions for the energy that take larger values (see for instance section~\ref{CAgeneral}).  
 
 We do conclude that there is always a Hamiltonian.  We repeat that the ontological states, as well as all other template  states \eqn{superposstate} obey the Schr\" odinger equation,
 	\be {\dd\over\dd t}|\j(t)\ket=-iH_\op|\j(t)\ket\ , \eel{Schro}
which reproduces the discrete evolution law~\eqn{Hamilton} at all times \(t\) that are integer multiples of \(\d t\). Therefore, 
we always reproduce \emph{some kind} of ``quantum" theory!  \def\tot{\mathrm{tot}}\def\openstr{\mathrm{open}}\def\closedstr{\mathrm{closed}}

 \subsubsection{Operators: Beables, Changeables and Superimposables \labell{operators}}
 
We plan to distinguish three types of operators:
\bi{(I)}  \emph{beables}: 
these denote a property of the ontological states, so that beables are diagonal in the ontological basis \(\{|A\ket,\,|B\ket,\cdots\}\) of Hilbert space:
 	\be\OO_\op|A\ket = \OO(A)|A\ket\ ,\qquad\hbox{(beable)}\ . \eel{beable} 
\itm{(II)} \emph{changeables}: operators that replace  an ontological state by another ontological state, such as a permutation operator:
	\be\OO_\op|A\ket=|B\ket\ ,\ \qquad\hbox{(changeable)}\ ; \eel{changeable} 
	These operators act as pure permutations.
\itm{(III)} \emph{superimposables}: these map ontological states onto superpositions of ontological states:
	\be\OO_\op|A\ket=\l_1|A\ket+\l_2|B\ket+\cdots\ . \eel{superimposable}
 \ei
 Now, we will construct a number of examples.  In part II, we shall see more examples of constructions of beable operators (e.g. section \ref{nu}).
 
\subsecl{The Cogwheel Model}{Cogwheelm} 

		\begin{figure}[htb!] \setcounter{figure}{0} \begin{quotation}\begin{center}
	$a)$\lowerwidthfig{0pt}{22mm}{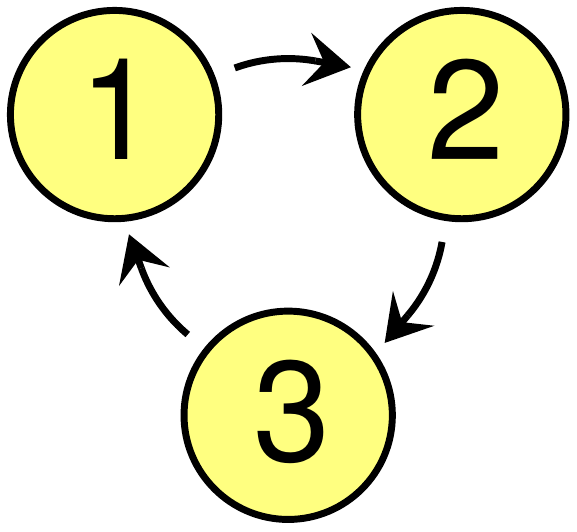}\qqquad$b)$\lowerwidthfig{0pt}{20mm}{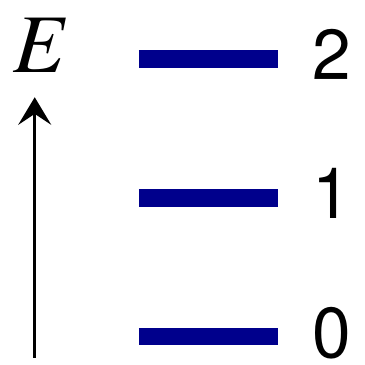}
 \caption{\small $a)$ Cogwheel model with three states.\quad $b)$ Its three energy levels.	 \labell{threestate.fig}  }
\end{center} \end{quotation}
		\end{figure}
	
	One of the simplest deterministic models is a system that can be in just 3 states, called (1), (2), and (3). The time evolution law is that, at the beat of a clock, (1) evolves into (2), (2) evolves into (3), and state (3) evolves into (1), see Fig.~\ref{threestate.fig}$a$. Let the clock beat with time intervals \(\d t\). As was explained in the previous section, we associate Dirac kets to these states, so we have the states \(|1\ket,\ |2\ket,\) and \(|3\ket\). The evolution operator \(U_\op(\d t)\) is then the matrix
	\be U_\op(\d t)=\pmatrix{0&0&1\cr 1&0&0\cr 0&1&0}\ . \eel{clockU}
It is now useful to diagonalise this matrix. Its eigenstates are \(|0\ket_H\,,\ |1\ket_H\,,\) and \(|2\ket_H\), defined as
	\be |0\ket_H&=&\fract 1{\sqrt 3}\bigg(|1\ket+|2\ket+|3\ket\bigg)\ ,\nn
		|1\ket_H&=&\fract 1{\sqrt 3}\bigg(|1\ket+e^{2\pi i/3}|2\ket+e^{-2\pi i/3}|3\ket\bigg)\ ,\nn
		|2\ket_H&=&\fract 1{\sqrt 3}\bigg(|1\ket+e^{-2\pi i/3}|2\ket+e^{2\pi i/3}|3\ket\bigg)\ , \eel{clockeigen}
for which we have
	\be U_\op(\d t)\pmatrix{|0\ket_H\cr|1\ket_H\cr|2\ket_H}=
	\pmatrix{\qquad{|0\ket_H}_{\vphantom{\|}}\cr e^{-2\pi i/3}|1\ket_H\cr e^{-4\pi i/3}\,|2\ket_H}\ . \eel{diagclock}	
In this basis, we can write this as
	\be U_\op=e^{-iH_\op\d t}\ ,\qquad\hbox{with}\qquad H_\op=\fract{2\pi}{ 3\,\d t}\hbox{ diag}(0,\ 1,\ 2)\ . \eel{clockhamdiag}
At times \(t\) that are integer multiples of \(\d t\), we have, in this basis,
	\be U_\op(t)=	e^{-iH_\op\,t}\ , \eel{evolveH}
but of course, this equation holds in every basis. In terms of the \emph{ontological basis} of the original states \(|1\ket,\ |2\ket,\) and \(|3\ket\), the Hamiltonian \eqn{clockhamdiag} reads
	\be H_\op=\fract{2\pi}{ 3\,\d t}\pmatrix{1&\k&\k^*\cr\k^*&1&\k\cr\k&\k^*&1}\ ,\qquad\hbox{with }\ \k=-\half+\fract {i\sqrt 3}6\ ,\quad 
	\k^*=-\half-\fract{i\sqrt 3}6\ . \eel{clockham}
Thus, we conclude that a template  state \(|\j\ket=\l(t)|1\ket+\m(t)|2\ket+\n(t)|3\ket\) that obeys the Schr\"odinger equation
	\be{\dd\over\dd t}|\j\ket=-iH_\op|\j\ket\ , \eel{clockschro}
with the Hamiltonian \eqn{clockham}, will be in the state described by the cogwheel model at all times \(t\) that are an integral multiple of \(\d t\). This is enough reason to claim that the ``quantum" model obeying this Schr\"odinger equation is \emph{mathematically equivalent} to our deterministic cogwheel model. 

The fact that the equivalence only holds at integer multiples of \(\d t\) is not a restriction. Imagine \(\d t\) to be as small as the Planck time, \(10^{-43}\) seconds (see chapter~\ref{grav}), then, if any observable changes take place only on much larger time scales, deviations from the ontological model will be unobservable. The fact the the ontological and the quantum model coincide at all integer multiples of the time \(d t\), is physically important. Note, that the original ontological model was not at all defined at non-integer time; we could simply define it to be described by the quantum model at non-integer times.

The eigenvalues of the Hamiltonian are \(\fract{2\pi}{ 3\,\d t}\,n\), with \(n=0,\ 1,\ 2\), see Fig.~\ref{threestate.fig}$b$. This is reminiscent of an atom with spin one that undergoes a Zeeman splitting due to a homogeneous magnetic field. One may conclude that such an atom is actually a deterministic system with three states, or, a cogwheel, but only if the `proper' basis has been identified.

The reader may remark that this is only true if, somehow, observations faster than the time scale \(\d t\) are excluded. We can also rephrase this. To be precise, a Zeeman atom is a system that needs only 3 (or some other integer \(N\)) states to characterise it. These are the states it is in at three (or \(N\)) equally spaced moments in time. It returns to itself after the period \(T=N\d t\).

	\subsubsection{Generalisations of the cogwheel model:  cogwheels with \textit{N} teeth\labell{cogwheelN}}
	The first generalisation of the cogwheel model (section~\ref{Cogwheelm}) is the system that permutes \(N\) `ontological' states \(|n\ket_\ont\), with \(n=0,\cdots N-1\,,\) and \(N\) some positive integer \(>1\). Assume that the evolution law is that, at the beat of the clock,  \def\mod{{\mathrm{\ mod\ }}}
	\be |n\ket_\ont\ra|n+1\mod N\ket_\ont\ . \eel{Nevolv}
This model can be regarded as the universal description of any system that is periodic with a period of \(N\) steps in time. The states in this evolution equation are regarded as `ontological' states. The model does not say anything about ontological states in between the integer time steps. We call this the \emph{simple periodic cogwheel model} with period \(N\).

As a generalisation of what was done in the previous section, we perform a discrete Fourier transformation on these states:
 	\be\hskip-10pt |k\ket_H&\deff&\fract 1{\sqrt N}\sum_{n=0}^{N-1}e^{2\pi ikn/N}|n\ket_\ont\ ,\qquad k=0,\cdots N-1\ ;\crl{Hstates}
	|n\ket_\ont&=&\fract 1{\sqrt N}\sum_{k=0}^{N-1}e^{-2\pi i kn/N}|k\ket_H\ .	 \eel{ontstates}
Normalising the time step \(\d t\) to one, we have
	\be U_\op(1)\,|k\ket_H=\fract 1{\sqrt N}\sum_{n=0}^{N-1}e^{2\pi ikn/N}|n+1\mod N\ket_\ont=e^{-2\pi i k/N}|k\ket_H\ , \eel{Hstatesevolve}
and we can conclude
	\be U_\op(1)=e^{-iH_\op}\ ;\qquad H_\op|k\ket_H=\fract{2\pi k}N|k\ket_H\ . \eel{Hamiltonperiodic}	
This Hamiltonian is restricted to have eigenvalues in the interval \([0,\,2\pi)\). where the notation means that 0 is included while \(2\pi\) is excluded. Actually, its definition implies that the Hamiltonian is periodic with period \(2\pi\), but in most cases we will treat it as being defined to be restricted to within the interval. The most interesting physical cases will be those where the time interval is very small, for instance close to the Planck time, so that the highest eigenvalues of the Hamiltonian will be considered unimportant in practice.
	
	In the original, ontological basis, the matrix elements of the Hamiltonian are
	\be {}_\ont\bra m|H_\op|n\ket_\ont=\fract{2\pi}{N^2}\sum_{k=1}^{N-1}k\,e^{2\pi ik(m-n)/N}\ . \eel{hammatrixelements}
This sum can we worked out further to yield	
	\be H_\op=\pi\left(1-{1\over N}\right)-{\pi\over N}\sum_{n=1}^{N-1}\left({i\over\tan(\pi n/N)}+1\right) U_\op(n)\ . \eel{NHam}
Note that,  unlike Eq.~\eqn{firstH}, this equation includes the corrections needed for the ground state. For the other energy eigenstates,  one can check that Eq.~\eqn{NHam} agrees with Eq.~\eqn{firstH}.
	
	For later use, Eqs.~\eqn{NHam} and \eqn{firstH}, without the ground state correction when \(U(t)|\j\ket =|\j\ket\),  can be generalised to the form
	\be H_\op=C-{\pi i\over T}\sum_{t_n>0}^{t_n<T}{U_\op(t_n)\over \tan(\pi t_n/T)}\  \stackrel{T\ra\infty}{\longrightarrow}\ C-i\sum_{t_n\ne 0}{U_\op(t_n)\over t_n}\ , \eel{genH}
where \(C\) is a (large) constant, \(T\) is the period, and \(t_n=n\,\d t\) is the set of times where the operator \(U(t_n)\) is required to have some definite value. We note that this is a sum, not an integral, so when the time values are very dense, the Hamiltonian tends to become very large. There seems to be no simple continuum limit. Nevertheless, in part~II, we will attempt to construct a continuum limit, and see what happens
(section~\ref{harmosc}).

Again, if we impose the Schr\" odinger equation \(\fract{\dd}{\dd t}|\j\ket_t=-iH_\op|\j\ket_t\) and the boundary condition \(|\j\ket_{t=0}=|n_0\ket_\ont\), then
this state obeys the deterministic evolution law \eqn{Nevolv} at integer times \(t\). If we take superpositions of the states \(|n\ket_\ont\) with the Born rule interpretation of the complex coefficients, then the {Schr\"odinger} equation still correctly describes the evolution of these Born probabilities.

	It is of interest to note that the energy spectrum \eqn{Hamiltonperiodic} is frequently encountered in physics: it is the spectrum of an atom with total angular momentum \(J=\half(N-1)\) and magnetic moment \(\m\) in a weak magnetic field: the Zeeman atom. We observe that, after the discrete Fourier transformation \eqn{Hstates}, a Zeeman atom may be regarded as the simplest deterministic system that hops from one state to the next in discrete time intervals, visiting \(N\) states in total.
	
	As in the Zeeman atom, we may consider the option of adding a finite, universal quantity \(\d E\) to the Hamiltonian. It has the effect of rotating all states with the complex amplitude \(e^{-i\,\d E}\) after each time step. For a simple cogwheel, this might seem to be an innocuous modification, with no effect on the physics, but below we shall see that the effect of such an added constant might become quite significant later.

	Note that, if we introduce any kind of perturbation on the Zeeman atom, causing the energy levels to be split in intervals that are no longer equal, it will no longer look like a cogwheel. Such systems will be a lot more difficult to describe in a deterministic theory; they must be seen as parts of a much more complex world.

\subsubsection{The most general deterministic, time reversible, finite model \labell{generalfinite}}
	
		\begin{figure}[htb!] \begin{quotation}
\begin{center} \lowerwidthfig{0pt} {100mm}{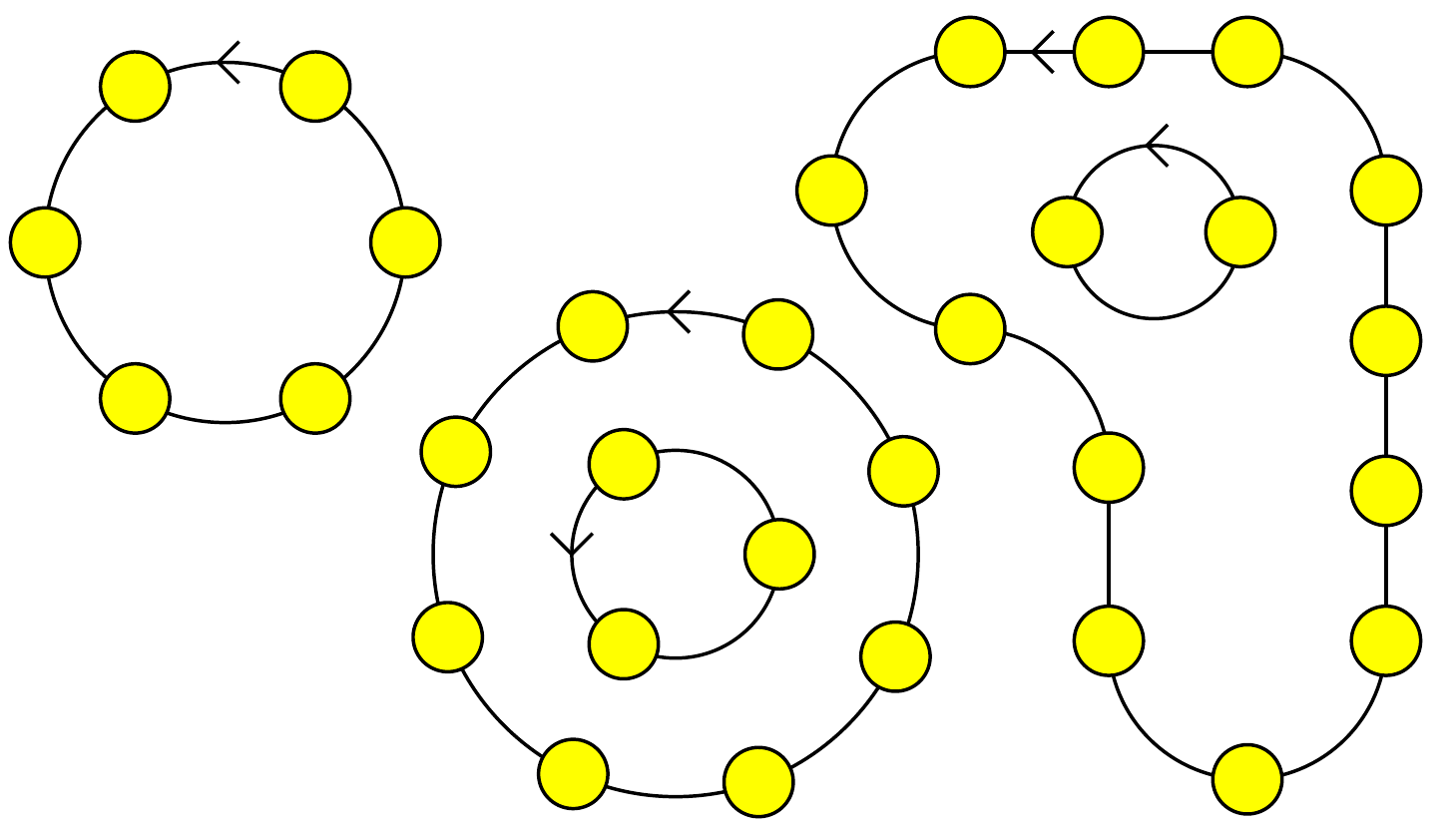}
\caption{\small Example of a more generic finite, deterministic, time reversible model\labell{genmod.fig}}\end{center}\end{quotation}
		\end{figure}

	Generalising the finite models discussed earlier in this chapter, consider now a model with a finite number of states, and an arbitrary time evolution law.  Start with any state \( |n_0\ket_\ont\), and follow how it evolves. After some finite number, say \(N_0\), of time steps, the system will be back at \(|n_0\ket_\ont\). However, not all states \(|n\ket_\ont\) may have been reached. So, if we start with any of the remaining states, say \(|n_1\ket_\ont\), then a new series of states will be reached, and the periodicity might be a different number, \(N_1\). Continue until all existing states of the model have been reached. We conclude that the most general model will be described as a set of 
simple periodic cogwheel models with varying periodicities, but all working with the same universal time step \(\d t\), which we could normalise to one;  see Fig.~\ref{genmod.fig}.

		\begin{figure}[htb!]
\begin{center}$a)$\hskip-5mm \lowerheightfig{0pt}{55mm}{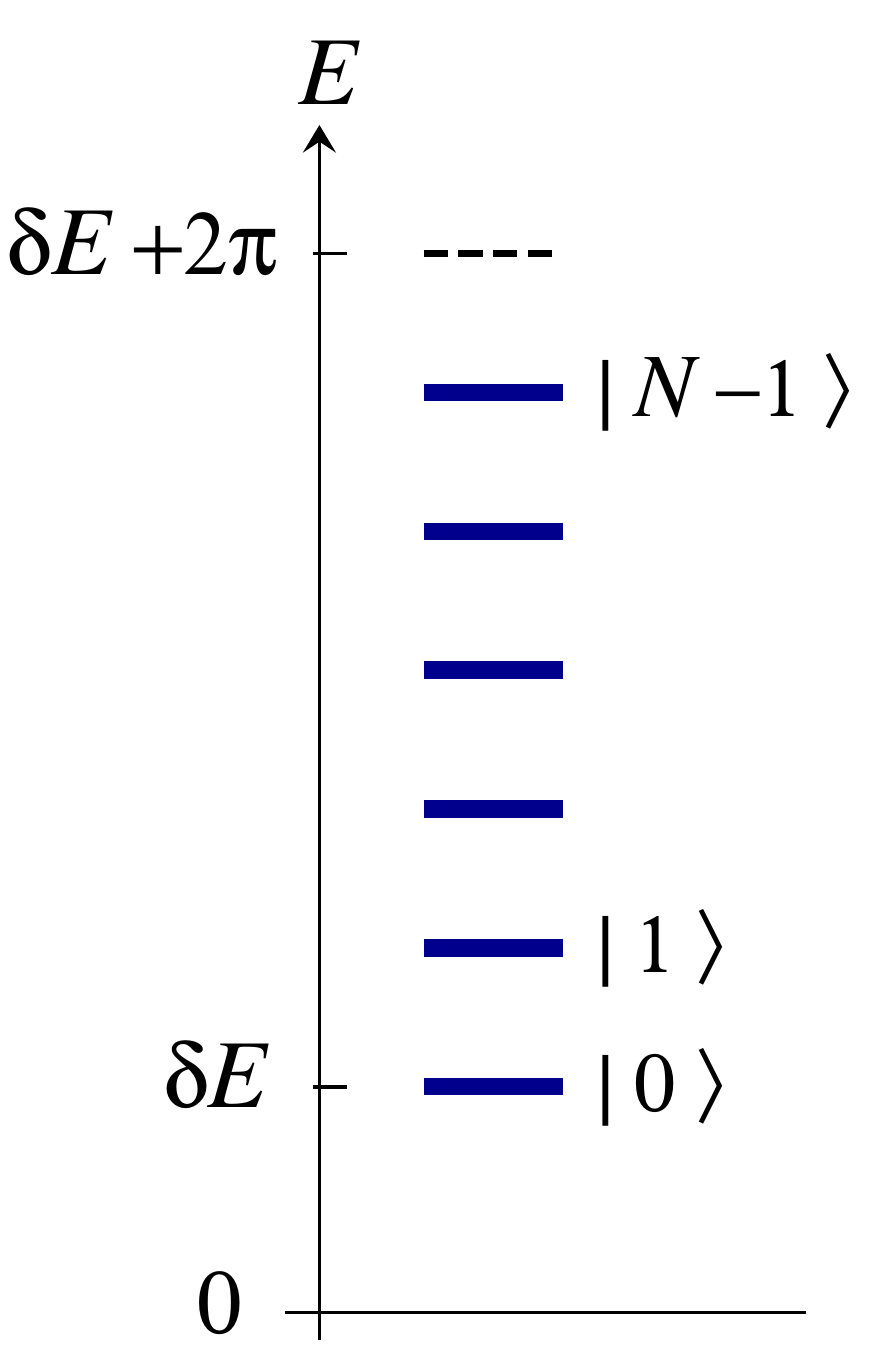}\qquad $b)$\lowerheightfig{0pt}{55mm}{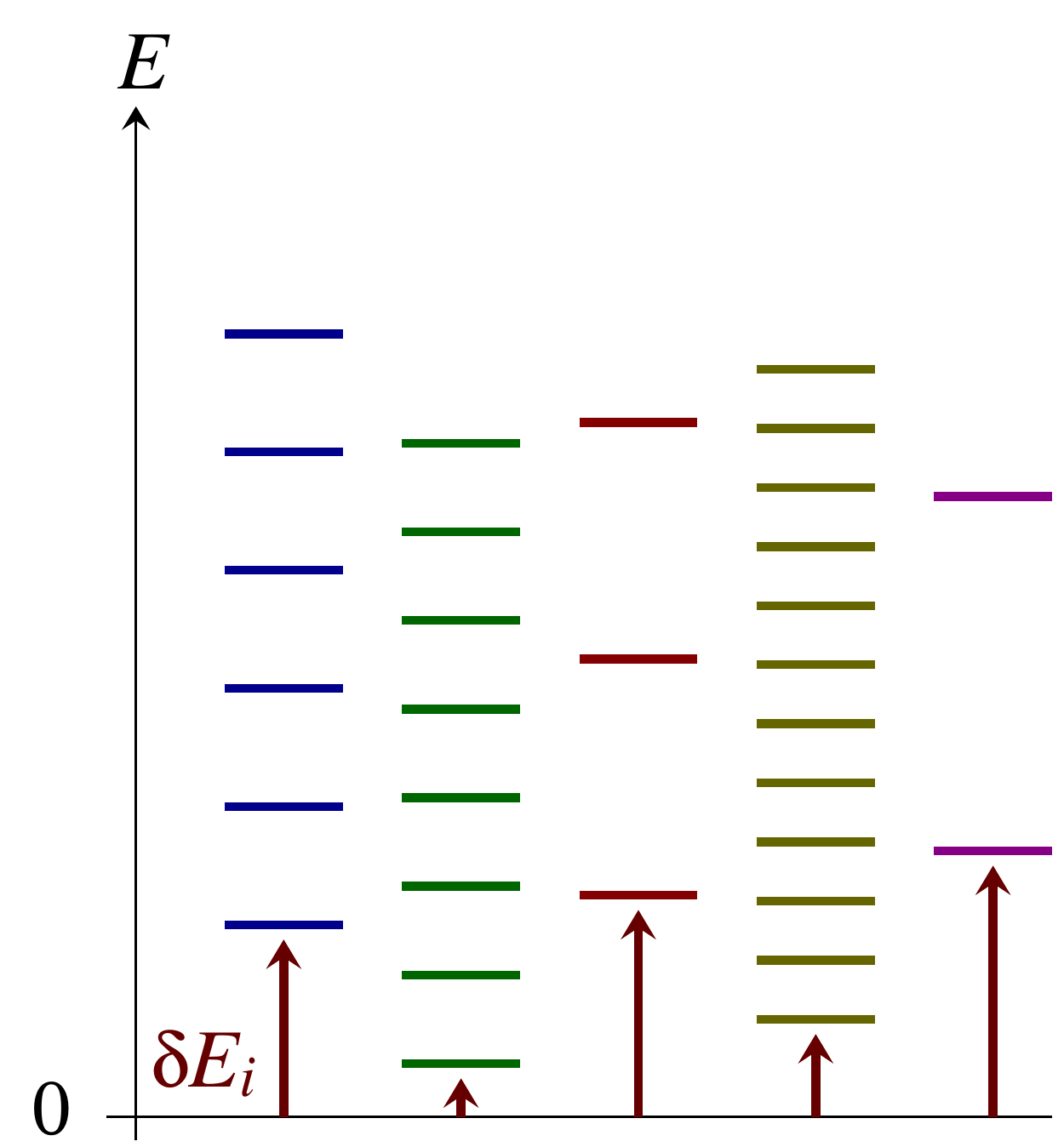}
\qquad$c)$\lowerheightfig{0pt}{55mm}{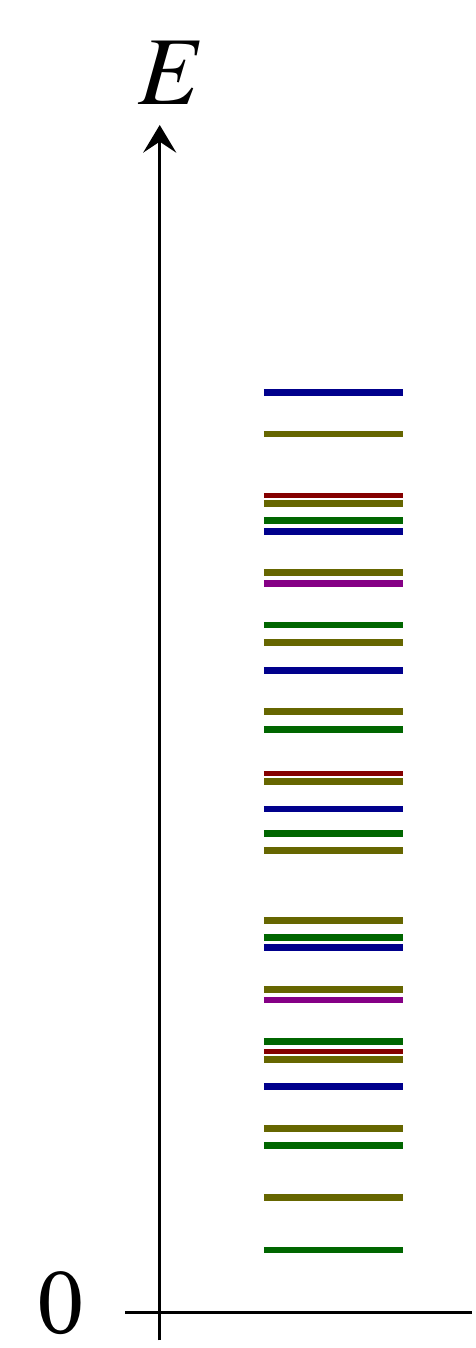} \end{center}\begin{quotation}
 \caption[\small $a)$ Energy spectrum of the simple periodic cogwheel model. $b)$ Energy spectrum of various cogwheels. $c)$ Energy spectrum of composite model of Fig.~\ref{genmod.fig}.]{\small  $a)$ Energy spectrum of the simple periodic cogwheel model.  $\d E$ is an arbitrary energy shift. $b)$ Energy spectrum of the model sketched in Fig.~\ref{genmod.fig}, where several simple cogwheel models are combined. Each individual cogwheel $i$ may be shifted by an arbitrary amount $\d E_i$. $c)$ Taking these energy levels together we get the spectrum of a generic finite model.\labell{generalfinite.fig}}\end{quotation}
		\end{figure}

	Fig.~\ref{generalfinite.fig} shows the energy levels of a simple periodic cogwheel model (left), a combination of simple periodic cogwheel models (middle), and the most general deterministic, time reversible, finite model (right).  Note that we now shifted the energy levels of all cogwheels by different amounts \(\d E_i\). This is allowed because the index \(i\), telling us which cogwheel we are in, is a conserved quantity; therefore these shifts have no physical effect. We do observe the drastic consequences however when we combine the spectra into one, see Fig.~\ref{generalfinite.fig}$c$.

 	Fig.~\ref{generalfinite.fig} clearly shows that the energy spectrum of a finite discrete deterministic model can quickly become quite complex\fn{It should be self-evident that the models displayed in the figures, and subsequently discussed, are just simple examples; the real universe will be infinitely more complicated than these. One critic of our work was confused: ``Why this model with 31 states? What's so special about the number 31?" Nothing, of course, it is just an example to illustrate how the math works.}. It raises the following question: given any kind of quantum system, whose energy spectrum can be computed. Would it be possible to identify a deterministic model that mimics the quantum model? To what extent would one have to sacrifice locality when doing this? Are there classes of deterministic theories that can be mapped on classes of quantum models? Which of these would be potentially interesting?

\newsecl{Interpreting quantum mechanics}{intpr}

	This book will not include an exhaustive discussion of all proposed interpretations of what quantum mechanics actually is. Existing approaches have been described in excessive detail in the literature\,\cite{Bohm-1952}\cite{Everett-1957}\cite{Bell-1964}\cite{Bell-1982}\cite{Adler-2004}, but we think they all contain weaknesses. The most conservative attitude is the so-called Copenhagen Interpretation. It is also a very pragmatic one, and some mainstream researchers insist that it contains all we need to know about quantum mechanics.	
	
	Yet it is the things that are not explained in the Copenhagen picture that often capture our attention. Below, we begin with indicating  how the cellular Automaton interpretation will address some of these questions.

\subsecl{The Copenhagen Doctrine}{Copenhagen} 
	It must have been a very exciting period of early modern science, when researchers began to understand how to handle quantum mechanics, in the late 1920s and subsequent years.\,\cite{PaisBohr-1991} The first coherent picture of how one should \emph{think} of quantum mechanics, is what we now shall call the Copenhagen Doctrine. In the early days, physicists were still struggling with the equations and the technical difficulties. Today, we know precisely how to handle all these, so that now we can rephrase the original starting points much more accurately.  Originally, quantum mechanics was formulated in terms of wave functions, with which one referred to the states electrons are in; ignoring spin for a moment, they were the functions \(\j(\vec x,\,t)=\bra\vec x\,|\j(t)\ket\).  Now, we may still use the words `wave function' when we really mean to talk of ket states in more general terms.  

Leaving aside who said exactly what in the 1920s, here are the main points of what one might call the \emph{Copenhagen Doctrine}. Somewhat anachronistically, we employ Dirac's notation:
	\begin{quote} A system is completely described by its `wave function' \(|\j(t)\ket\), which is an element of Hilbert space, and any basis in Hilbert space can be used for its description. This wave function obeys a linear first order differential equation in time, to be referred to as Schr\"odinger's equation, of which the exact form can be determined by repeated experiments.\\
	A measurement can be made using any observable \(\OO\) that one might want to choose (observables are hermitean operators in Hilbert space). The theory then predicts the average measured value of \(\OO\), after many repetitions of the experiment, to be
		\be\bra\OO\ket=\bra\j(t)|\OO|\j(t)\ket\ . \eel{expvalue}
As soon as the measurement is made, the wave function of the system collapses to a state in the subspace of Hilbert space that is an eigenstate of the observable \(\OO\), or a probabilistic distribution of eigenstates, according to Eq.~\eqn{expvalue}.\\
	When two observables \(\OO_1\) and \(\OO_2\) do not commute, they cannot both be measured accurately. The commutator \([\OO_1,\,\OO_2]\) indicates how large the product of the `uncertainties' \(\d\OO_1\) and \(\d\OO_2\) should be expected to be.\\
The measuring device itself must be regarded as a classical object, and
for large systems the quantum mechanical measurement approaches closely the classical description.
	\end{quote}
Implicitly included in Eq.~\eqn{expvalue} is the element of probability. If we expand the wave function \(|\j\ket\) into eigenstates \(|\vv\ket\) of an observable \(\OO\), then we find that the \emph{probability} that the experiment on \(|\j\ket\) actually gives as a result that the eigenvalue of the state \(|\vv\ket\) is found, will be given by \(P=|\bra\vv|\j\ket|^2\). This is referred to as Born's probability rule\,\cite{Born-1926}.

We note that the wave function may not be given any ontological significance. The existence of a `pilot wave' is not demanded; one cannot actually measure \(\bra\vv|\j\ket\) itself; only by repeated experiments, one can measure the probabilities, with intrinsic margins of error. We say that the wave function, or more precisely, the amplitudes, are \emph{psi-epistemic} rather than \emph{psi-ontic}.

An important element in the Copenhagen interpretation is that one may \emph{only ask what the outcome of an experiment will be}. In particular, it is forbidden to ask: \emph{what is it that is actually happening?} It is exactly the latter question that sparks endless discussions; the important point made by the Copenhagen group is that such questions are unnecessary. If one knows the Schr\"odinger equation, one knows everything needed to predict the outcomes of an experiment, no further questions should be asked.

This is a strong point of the Copenhagen doctrine, but it also yields severe limitations. If we know the Schr\"odinger equation, we know everything there is to be known.

But what if we do not yet know the Schr\"odinger equation? How does one arrive at the correct equation? In particular, how do we arrive at the correct Hamiltonian if the gravitational force is involved? 

Gravity has been a major focus point of the last 30 years and more, in elementary particle theory and the theory of space and time. Numerous wild guesses have been made. In particular, (super)string theory has made huge advances. Yet no convincing model that unifies gravity with the other forces has been constructed; models proposed so-far have not been able to explain, let alone predict, the values of the fundamental constants of Nature, including the masses of many fundamental particles, the fine structure constant, and the cosmological constant. And here it is, according to the author's opinion, where we \emph{do} have to ask: \emph{What is it, or what could it be, that is actually going on?}

One strong feature of the Copenhagen approach to quantum theory was that it was also clearly shown how a Schr\"odinger equation can be obtained if the classical limit is known:
	\begin{quote} If a classical system is described by the (continuous) Hamilton equations, this means that we have classical variables \(p_i\) and \(q_i\), for which one can define \emph{Poisson brackets}: any pair of observables \(A(\vec q,\vec p\,)\) and \(B(\vec q,\vec p\,)\), allow for the construction of an observable variable called \(\{A,B\}\), defined by
		\be \{A,B\}\equiv\sum_i\bigg({\pa A\over\pa q_i}{\pa B\over\pa p_i}-{\pa A\over\pa p_i}{\pa B\over\pa q_i}\bigg)\ , \eel{poisson}
in particular:
		\be \{q_i,p_j\}=-\{p_i,q_j\}=\d_{ij}\ ,\qquad\{q_i,q_j\}=\{p_i,p_j\}=0\ . \eel{poissonpq}
A quantum theory is obtained, if we replace the Poisson brackets by commutators, which involves a factor \(i\) and a new constant of Nature, \(\hbar\). Basically:
		\be [A,B]\ra i\hbar\{A,B\}\ . \eel{poissoncomm}
\end{quote}
In the limit \(\hbar\ra0\), this quantum theory reproduces the classical system. This very powerful procedure allows us, more often than not, to bypass the question ``what is going on?". We have a theory, we know where to search for the appropriate Schr\"odinger equation, and we know how to recover the classical limit.

Unfortunately, in the case of the gravitational force, this `trick' is not good enough to give us `quantum gravity'. The problem with gravity is not just that the gravitational force appears not to be renormalizable, or that it is difficult to define the quantum versions of space- and time coordinates, and the physical aspects of non-trivial space-time topologies; some authors attempt to address these problems as merely technical ones, which can be handled by using some tricks. The real problem is that space-time curvature runs out of control at the Planck scale. We will be forced to turn to a different book keeping system for Nature's physical degrees of freedom there.

A promising approach was to employ local conformal symmetry\,\cite{GtHconf-2010} as a more fundamental principle than usually thought; this could be a way to make distance and time scales relative, so that what was dubbed as `small distances' ceases to have an absolute meaning. The theory is recapitulated in Appendix~\ref{confgrav}. It does need further polishing, and it too could eventually require a Cellular Automaton interpretation of the quantum features that it will have to include.

\subsecl{The Einsteinian view}{Einstein}
	
This section is called ``The Einsteinian view', rather than `Einstein's view', because we do not want to go into a discussion of what it actually was that Einstein thought. It is well-known that Einstein was uncomfortable with the Copenhagen Doctrine. The notion that there might be ways to rephrase things such that all phenomena in the universe are controlled by equations that leave nothing to chance, will now be referred to as the Einsteinian view. We do ask further questions, such as \emph{Can quantum-mechanical description of physical reality be considered complete?}\,\cite{EPR-1935}, or, does the theory tell us everything we might want to know about what is going on? 

In the Einstein-Podolsky-Rosen discussion of a Gedanken experiment, two particles (photons, for instance), are created in a state 
	\be x_1-x_2=0\ , \qquad p_1+p_2=0\ . \eel{EPR}
Since \([x_1-x_2,\,p_1+p_2]=i-i=0\), both equations in \eqn{EPR} can be simultaneously sharply imposed. 

What bothered Einstein, Podolsky and Rosen was that, long after the two particles ceased to interact, an observer of particle \# 2 might decide either to measure its momentum \(p_2\), after which we know for sure the momentum \(p_1\) of particle \# 1,  or its position \(x_2\), after which we would know for sure the position \(x_1\) of particle \#1.  How can such a particle be described by a quantum mechanical wave function at all? Apparently, the measurement at particle \# 2 affected the state of particle \#1, but how could that have happened?

In modern quantum terminology, however, we would have said that the measurements proposed in this Gedanken experiment would have disturbed the wave function of the entangled particles. The measurements on particle \# 2 affects the probability distributions for particle \# 1, which in no way should be considered as the effect of a spooky signal from one system to the other.

In any case, even Einstein, Podolsky and Rosen had no difficulty in computing the quantum mechanical probabilities for the outcomes of the measurements, so that, in principle, quantum mechanics emerged unharmed out of this sequence of arguments.

It is much more difficult to describe the two EPR photons in a classical model. Such questions will be the topic of section~\ref{Bell}.

Einstein had difficulties with the relativistic invariance of quantum mechanics (``does the spooky information transmitted by these particles go faster than light?"). These, however, are now seen as technical difficulties that have been resolved. It may be considered part of Copenhagen's Doctrine, that the transmission of information over a distance can only take place, if we can identify operators \(A\) at space-time point  \(x_1\) and operators \(B\) at space-time point  \(x_2\) that do not commute: \([A,B]\ne 0\). We now understand that, in elementary particle theory, all space-like separated observables mutually commute, which precludes any signalling faster than light. It is a built-in feature of the Standard Model, to which it actually owes much of its success.

So, with the technical difficulties out of the way, we are left with the more essential Einsteinian objections against the Copenhagen doctrine for quantum mechanics: it is a probabilistic theory that does not tell us what actually is going on. It is sometimes even suggested that we have to put our ``classical" sense of logic on hold. Others deny that: ``Keep remembering what you should never ask, while reshaping your sense of logic, and everything will be fine."  According to the present author, the Einstein-Bohr debate is not over.
A theory must be found that does not force us to redefine \emph{any} aspect of classical, logical reasoning.

What Einstein and Bohr did seem to agree about is the importance of the role of an observer. Indeed, this was the important lesson learned in the \(20^{\mathrm{th}}\) century: \emph{if something cannot be observed, it may not be a well-defined concept -- it may even not exist at all. We have to limit ourselves to observable features of a theory.} It is an important ingredient of our present work that we propose to part from this doctrine, at least to some extent: Things that are not directly observable may still exist and as such play a decisive role in the observable properties of an object. They may also help us to construct realistic models of the world.

Indeed, there are big problems with the dictum that everything we talk about must be observable. While observing microscopic objects, an observer may disturb them, even in a classical theory; moreover, in gravity theories, observers may carry gravitational fields that disturb the system they are looking at, so we cannot afford to make an observer infinitely heavy (carrying large bags full of ``data", whose sheer weight  gravitationally disturbs the environment), but also not infinitely light (light particles do not transmit large amounts of data at all), while, if the mass of an observer would be ``somewhere in between", this could entail that our theory will be inaccurate from its very inception.

An interesting blow was given to the doctrine that observability should be central, when quark theory was proposed. Quarks cannot be isolated to be observed individually, and for that reason the idea that quarks would be physical particles was attacked. Fortunately, in this case  the theoretical coherence of the evidence in favour of the quarks became so overwhelming, and experimental methods for observing them, even while they are not entirely separated, improved so much, that all doubts evaporated.

In short, the Cellular Automaton Interpretation tells us to return to classical logic and build models. These models describe the evolution of large sets of data, which eventually may bring about classical phenomena that we can observe. The fact that these data themselves cannot be directly observed, and that our experiments will provide nothing but statistical information, including fluctuations and uncertainties, can be fully explained within the settings of the models; if the observer takes no longer part in the \emph{definition} of physical degrees of freedom and their values, then his or her limited abilities will no longer stand in the way of accurate formalisms.

We suspect that this view \emph{is} closer to Einstein's than it can be to Bohr, but, in a sense, neither of them would fully agree. We do not claim the wisdom that our view is obviously superior, but rather advocate that one should try to follow such paths, and learn from our successes and failures.

\subsecl{Notions not admitted in the C.A.I.}{notadmitted}

It is often attempted to attach a physical meaning to the wave function beyond what it is according to Copenhagen. Could it have an ontological significance as a `pilot wave function'\,\cite{Broglie-1927}\cite{Bohm-1952}? It should be clear from nearly every page of this book that we do not wish to attach any ontological meaning to the wave function, \emph{if we are using it as a template}. 

In an ontological description of our universe, in terms of its ontological basis, there are only two values a wave function can take: 1 and 0.  A state is actually realised when the wave function is 1, and it does not describe our world when the wave function is zero. It is only this `universal wave function', that for that reason may be called ontological. 

It is only for mathematical reasons that one might subsequently want to equip this wave function with a phase, \(e^{i\vv}\). \emph{In the ontological basis}, this phase \(\vv\) has no physical meaning at all, but as soon as one considers operators, including the time-evolution operator \(U(t)\), and therefore also the Hamiltonian, these phases have to be chosen. From a physical point of view, any phase is as good as any other, but for keeping the mathematical complexity under control, precise definitions of these phases is  crucial. One can then perform the unitary transformations to any of the basis choices usually employed in physics. The template states subsequently introduced, all come with precisely defined phases.

A semantic complication is caused as soon as we apply second quantisation. Where a single particle state is described by a wave function, the second-quantized version of the theory sometimes replaces this by an operator field. Its physical meaning is then completely different. Operator fields are usually not ontological since they are superimposables rather than beables (see subsection~\ref{operators}), but in principle they could be; wave functions, in contrast, are elements of Hilbert space and as such should not be confused with operators, let alone beable operators.

How exactly to phrase the so-called `Many World Interpretation'\,\cite{Everett-1957} of quantum mechanics, is not always agreed upon\,\cite{DeWitt-1970}. When doing ordinary physics with atoms and elementary particles, this interpretation may well fulfil the basic needs of a researcher, but from what has been learned in this book it should be obvious that our theory contrasts strongly with such ideas. There is only one single world that is being selected out in our theory as being `the real world', while all others simply are not realised.

The reader may have noticed that the topic in this book is being referred to alternately as a `theory' and as an `interpretation'. The \emph{theory} we describe consists not only of the assumption that an ontological basis exists, but also that it can be derived, so as  to provide an ontological description of our universe. It suggests pathways to pin down the nature of this ontological basis. When we talk of an \emph{interpretation}, this means that, even if we find it hard or impossible to identify the  ontological basis, the mere assumption that one might exist suffices to help us understand what the quantum mechanical expressions normally employed in physics, are actually standing for, and how a physical reality underlying them can be imagined.

\subsecl{The collapsing wave function and Schr\"odinger's cat}{cat}

The following ingredient in the Copenhagen interpretation, section~\ref{Copenhagen}, is often the subject of discussions: \begin{quote}
	\emph{As soon as an observable \(\OO\) is measured, the wave function of the system collapses to a state in the subspace of Hilbert space that is an eigenstate of the observable \(\OO\), or a probabilistic distribution of eigenstates.} \end{quote}
This is referred to as the ``collapse of the wave function". It appears as if the action of the measurement itself causes the wave function to attain its new form. The question then asked is what physical process is associated to that.

Again, the official reply according to the Copenhagen doctrine is that this question should not be asked. Do the calculation and check your result with the experiments. However, there appears to be a contradiction, and this is illustrated by Erwin Schr\"odinger's Gedanken experiment with a cat.\,\cite{cat-1935} The experiment is summarised as follows:
\begin{quote}
	In a sealed box, one performs a typical quantum experiment. It could be a Stern Gerlach experiment where a spin \(\half\) particle with spin up is sent through an inhomogeneous magnetic field that splits the wave function according to the values of the spin in the \(y\) direction, or it could be a radioactive atom that has probability \(\half\) to decay within a certain time. In any case, the wave function is well specified at \(t=0\), while at \(t=1\) it is in a superposition of two states, which are sent to a detector that determines which of the two states is realised. It is expected that the wave function `collapses' into one of the two possible final states.\\
	The box also contains a live cat (and air for the cat to breathe). Depending on the outcome of the measurement, a capsule with poison is broken, or kept intact. The cat dies when one state is found, and otherwise the cat stays alive. At the end of the experiment, we open the box and inspect the cat.\end{quote}
Clearly, the probability that we find a dead cat is about \(\half\), and otherwise we find a live cat. However, we could also regard the experiment from a microscopic point of view. The initial state was a pure quantum state. The final state is a superposition. Should the cat, together with the other remains of the experiment, upon opening the box, not be found in a superimposed state: dead \emph{and} alive?

The collapse axiom tells us that the state should be `dead cat' \emph{or} `live cat', whereas the first parts of our description of the quantum mechanical states of Hilbert space, clearly dictates that if two states, \(|\j_1\ket\) and \(|\j_2\ket\) are possible in a quantum system, then we can also have \(\a|\j_1\ket+\b|\j_2\ket\). According to Schr\"odinger's equation, this superposition of states always evolves into a superposition of the final states. The collapse seems to violate Schr\"odinger's equation. Something is not quite right.

An answer that several investigators have studied\,\cite{Pearle-1982}, is that, apparently, Schr\"odinger's equation is only an approximation, and that tiny non-linear `correction terms' bring about the \linebreak[3] collapse \,\cite{GRW-1986}\cite{BGh-2003}\cite{MS-2004}. One of the problems with this is that observations can be made at quite different scales of space, time, energy and mass. How big should the putative correction terms be? Secondly, how do the correction terms know in advance which measurements we are planning to perform?

This, we believe, is where the cellular automaton interpretation of quantum mechanics will come to the rescue.   It is formulated using no wave function at all, but there are ontological states instead.  It ends up with just one wave function, taking the value 1 if we have a state the universe is in, and 0 if that state is not realised. There are no other wave functions, no superposition.

How this explains the collapse phenomenon will be explained in chapter~\ref{detqu}. In summary: quantum mechanics is not the basic theory but a tool to solve the mathematical equations. This tool works just as well for superimposed states (the templates) as for the ontological states, but they are not the same thing. The dead cat is in an ontological state and so is the live one. The superimposed cat solves the equations mathematically in a perfectly acceptable way, but it does not describe a state that can occur in the real world. We postpone the precise explanation to chapter~\ref{detqu}. It will sound very odd to physicists who have grown up with standard quantum mechanics, but it does provide the logical solution to the Schr\"odinger cat paradox.\fn{Critical readers will say: Of course, this theory isn't quantum mechanics, so it doesn't share any of its problems. True, but our theory is supposed to \emph{generate} quantum mechanics, without generating its associated problems.}

One may ask what this  may imply when we have transformations between ontological states and template states. Our experiences tell us that all template states that are superpositions \(\a|\j_1\ket+\b|\j_2\ket\) of ontological states, may serve as suitable approximations describing probabilistic situations in the real world. 
How can it be that, sometimes, they \emph{do} seem to be ontological? The most likely response to that will be that the transformation does not always have to be entirely local, but in practice may involve many spectator states in the environment. What we can be sure of is that \emph{all} ontological states form an orthonormal set. So, whenever we use \(\a|\j_1\ket+\b|\j_2\ket\) to describe an ontological state, there must be other wave functions in the environment which must be chosen differently for any distinct pair \(\a\) and \(\b\), such that the entire set that we use to describe physical situations are always orthonormal.

This should be taken in mind in the next sections where we comment on the \emph{Alice and Bob} Gedanken experiments.

\subsecl{Decoherence and Born's probability axiom}{decoBorn} The cellular automaton interpretation does away with one somewhat murky ingredient of the more standard interpretation schemes: the role of `decoherence'. It is the argument often employed to explain why macroscopic systems are never seen in a quantum superposition. Let \(|\j_1\ket\) and \(|\j_2\ket\) be two states a classical system can be in, such as a cat being dead and a cat being alive. According to Copenhagen, in its pristine form, quantum mechanics would predict the possibility of a third state, \(|\j_3\ket = \a|\j_1\ket+\b|\j_2\ket\), where \(\a\) and \(\b\) can be any pair of complex numbers with \(|\a|^2+|\b|^2=1\). 

Indeed, it seems almost inevitable that a system that can evolve into state \(|\j_1\ket\) or into state \(|\j_2\ket\), should also allow for states that evolve into \(|\j_3\ket\). Why do we not observe such states? The only thing we do observe is a situation whose probability of being in \(|\j_1\ket\) might be \(|\a|^2\) and the probability to be in \(|\j_2\ket\) is \(|\b|^2\). But that is not the same as state \(|\j_3\ket\).

The argument is that, somehow, the state \(|\j_3\ket\) is unstable. According to Copenhagen, the probability of a state \(|\j\ket\) to be in state \(|\j_3\ket\) is
\be P_3=|\bra\j_3|\j\ket|^2=|\a|^2 |\bra \j_1|\j\ket|^2+|\b|^2|\bra \j_2|\j\ket|^2+2\mathrm{Re}\bigg(\a^*\b\bra\j|\j_1\ket\bra \j_2|\j\ket\bigg)\ . \ee
The last term here is the interference term. It distinguishes the real quantum theory from classical theories. Now it is said that, if \(|\j_1\ket\) and \(|\j_2\ket\) become classical, they cannot stay immune for interactions with the environment.
In the presence of such interactions, the energies of \(|\j_1\ket\) and \(|\j_2\ket\) will not exactly match, and consequently, the interference term, will oscillate badly. This term might then be seen to average out to zero. The first two terms are just the probabilities to have either \(|\j_1\ket\) or \(|\j_2\ket\), which would be the classical probabilities. 

If indeed the last term becomes unobservable, we say that the two states decohere\,\cite{Zurek-1991}\cite{MS-2004}, so that the interference term should be replaced by zero. The question is, if we include the environment in our description, the energies should still be exactly conserved, and there is no rapid oscillation. Is it legal to say nevertheless that the interference term will disappear? Note that its absolute value on average does remain large.

The CAI will give a much more direct answer: if states \(|\j_1\ket\) and \(|\j_2\ket\) are classical, then they are ontological states. State \(|\j_3\ket\) will then not be an ontological state, and the states of the real universe, describing what happens if an actual experiment is carried out, \emph{never} include state \(|\j_3\ket\). It is merely a template, useful for calculations, but not describing reality. What it \emph{may} describe is a situation where, from the very start, the coefficients \(\a\) and \(\b\) were declared to represent probabilities.

 Copenhagen quantum mechanics contains an apparently irreducible axiom: the \emph{probability} that a state \(|\j\ket\) is found to agree with the properties of another state \(|\vv\ket\), must be given by
	\be P=|\bra\vv|\j\ket|^2\ . \eel{bornprob2}
This is the famous Born rule\,\cite{Born-1926}. What is the physical origin of this axiom?

Note, that Born did not have much of a choice.  The completeness theorem of linear algebra implies that the eigenstates \(|\vv\ket\) of an hermitean operator span the entire Hilbert space, and therefore,
	\be\sum_\vv|\vv\ket\bra\vv|=\mathbb{I}\ ;\qquad \sum_\vv|\bra\vv|\j\ket|^2=\sum_\vv\bra\j|\vv\ket\bra\vv|\j\ket=\bra\j|\j\ket=1\  ,\ee
where \(\mathbb I\) stands for the identity operator. If Born would have chosen any other expression to represent probabilities, according to Gleason's theorem\,\cite{Gleason-1957}, they would not have added up to one. The expression~\eqn{bornprob2} turns out to be ideally suited to serve as a probability.

Yet this is a separate axiom, and the question why it works so well is not misplaced. In a hidden variable theory, probabilities may have a different origin. The most natural explanation as to why some states are more probable than others may be traced to their initial states much earlier in time. One can ask which initial states may have lead to a state seen at present, and how probable these may have been. There may be numerous answers to that question. One now could attempt to  estimate their combined probabilities. The relative probabilities of some given observed final states could then be related to the ratios of the numbers found. Our question then is, can we explain whether and how the expression \eqn{bornprob2} is related to these numbers?
This discussion is continued in subsection~\ref{mousedr} and in section~\ref{born.sec}.

\subsecl{Bell's theorem, Bell's inequalities and the CHSH  inequality.}{Bell}

One of the main reasons why `hidden variable' theories are usually dismissed, and emphatically so when the theory obeys local equations, is the apparent difficulty in such theories to represent entangled quantum states. Just because the De Broglie-Bohm theory (not further discussed here) is intrinsically non-local, it is generally concluded that all hidden variable theories are either non-local or unable to reproduce quantum features at all. When J.S.~Bell was investigating the possibility of hidden variable theories, he hit upon the same difficulties, upon which he attempted to prove that local hidden variable theories are impossible.

As before, we do not intend to follow precisely the historical development of Bell's theory\,\cite{Bell-1982}, but limit ourselves to a summary of the most modern formulation of the principles. Bell designed a Gedanken experiment, and at the heart of it is a pair of quantum-entangled particles. They could be spin-\(\half\) particles, which each can be in just two quantum states described by the Pauli matrices \eqn{pauli}, or alternatively spin 1 photons. There are a few subtle differences between these two cases, although these are not essential to the argument. The first is that the two orthonormal states for photons are the ones where they are polarised horizontally or vertically, while the two spin-\(\half\) states are polarised up or down. Indeed, quite generally when polarised particles are discussed, the angles for the photons are handled as being half the angles for spin-\(\half\) particles.

The second difference concerns the entangled state, which in both cases has total spin 0. For spin-\(\half\), this means that \((\vec\s_1+\vec\s_2)|\j\ket=0\), 
where \(\vec\s\) are the Pauli matrices, Eqs.~\eqn{pauli},
so that
	\be|\j\ket=\fract 1{\sqrt 2}(|\uparrow\downarrow\ket-|\downarrow\uparrow\ket)\ ,\eel{spinhalfentangled}
which means that the two electrons are polarised in \emph{opposite} directions.

For spin 1, this is different. Let us have these photons move in the \(\pm z\)-direction. Defining \(A_\pm=\fract 1{\sqrt{2}}(A_x\pm iA_y)\) as the operators that create or annihilate one unit of spin in the \(z\)-direction, and taking into account that photons are bosons, the 2 photon state with zero spin in the \(z\)-direction is
	\be|\j\ket=\fract 1{\sqrt 2}(A^{(1)}_+A^{(2)}_-+A^{(1)}_-A^{(2)}_+)\,|\,\ket=\fract 1{\sqrt 2}\,|z,\,-z\ket+\fract 1{\sqrt 2}|-z,\,z\ket\ ,\eel{spin1tot0}
and since helicity is spin in the direction of motion, while the photons go in opposite directions, we can rewrite this state as
	\be|\j\ket=\fract 1{\sqrt 2}(\,|++\ket+|--\ket\,)\ , \eel{helicitystates}
where \(\pm\) denote the helicities. Alternatively one can use the operators \(A_x\) and \(A_y\) to indicate the creators of linearly polarised photons, and then we have
	\be|\j\ket=\fract 1{\sqrt 2}(A_x^{(1)}A_x^{(2)}+A_y^{(1)}A_y^{(2)})\,|\,\ket=\fract 1{\sqrt 2}(\,|xx\ket+|yy\ket\,)\ . \eel{linearpolar}
Thus, the two photons are linearly polarised in the \emph{same} direction.

Since the experiment is mostly done with photons, we will henceforth describe the entangled photon state.
	\begin{figure}[h!] \begin{quote} \begin{center}
 \lowerwidthfig{0pt}  {130mm}{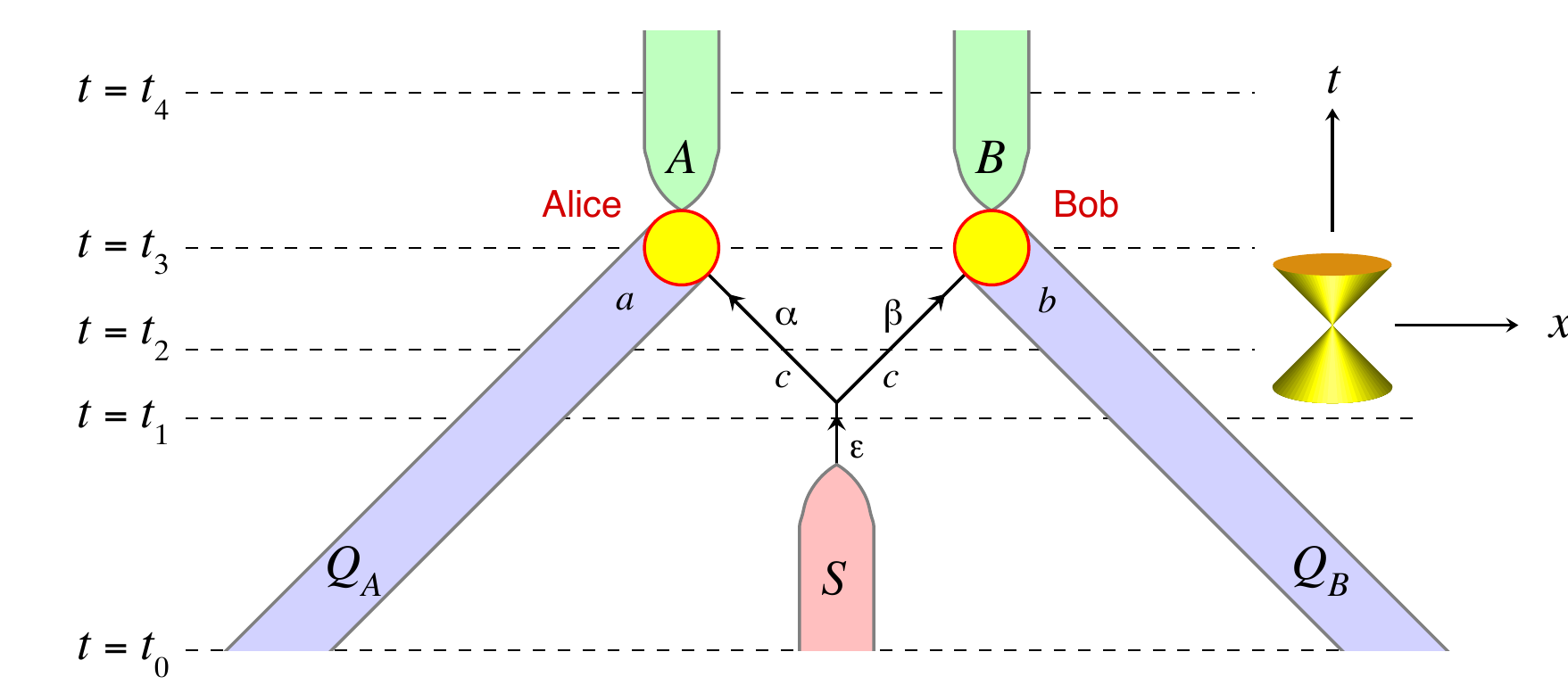} \end{center}
\caption[\small A bell-type experiment. Space runs horizontally, time vertically.]{\small A bell-type experiment. Space runs horizontally, time vertically. Single lines denote single bits or qubits travelling; widened lines denote classical information carried by millions of bits. Light travels along \(45^\circ\), as indicated by the light cone on the right. Meaning of the variables: see text. \labell{bell.fig}}\end{quote}
		\end{figure}

The Bell experiment is now illustrated in Fig.~\ref{bell.fig}. At the point \(S\), an atom \(\e\) is prepared to be in an unstable \(J=0\) state at \(t=t_1\), such that it can decay only into an other \(J=0\) state, by simultaneously emitting two photons such that \(\D J=0\), and the two photons, \(\a\) and \(\b\),  must therefore be in the entangled \(S_\tot=0\) state, at \(t=t_2\). 

After having travelled for a long time, at \(t=t_3\), photon \(\a\) is detected by observer \(A\) (Alice), and photon \(\b\) is detected by \(B\) (Bob). Ideally,  the observers use a polarisation filter oriented at an angle \(a\) (Alice) and \(b\) (Bob). If the photon is transmitted by the filter, it is polarised in the direction of the polarisation filter's angle, if it is reflected, its polarisation was orthogonal to this angle. At both sides of the filter there is a detector, so both Alice and Bob observe that one of their two detectors gives a signal. We call Alice's signal \(A=+1\) if the photon passed the filter, and \(A=-1\) if it is reflected. Bob's signal \(B\) is defined the same way.

According to quantum theory,  if \(A=1\), Alice's photon is polarised in the direction \(a\), so Bob's photon must also be polarised in that direction, and the intensity of the light going through Bob's filter will be \(\cos^2(a-b)\). Therefore, according to quantum theory, the probability that \(B=1\) is \(\cos^2(a-b)\). The probability that \(B=-1\) is then \(\sin^2(a-b)\), and the same reasoning can be set up if \(A=-1\). The expectation value of the product \(AB\) is thus found to be
	\be \bra\,A\,B\,\ket=\cos^2(a-b)-\sin^2(a-b)=\cos2(a-b)\ , \eel{ABquant}
according to quantum mechanics.

In fact, these correlation functions can now be checked experimentally. Beautiful experiments\,\cite{Aspect-1982} confirmed that correlations can come close to Eq.~\eqn{ABquant}.

The point made by Bell is that it seems to be impossible to reproduce this strong correlation between the findings \(A\) and \(B\)  in any theory where classical information is passed on from the atom \(\e\) to Alice (A) and Bob (B). All one needs to assume is that the atom emits a signal to Alice and one to Bob, regarding the polarisation of the photons emitted. It could be the information that  both photons \(\a\) and \(\b\) are polarised in direction \(c\). Since this information is emitted long before either Alice or Bob decided how to orient their polarisation filters, it is obvious that the signals in \(\a\) and \(\b\) should not depend on that.
Alice and Bob are free to choose their polarisers. 

The correlations then directly lead to a contradiction, regardless the nature of the classical signals used.
The contradiction is arrived at as follows. Consider two choices that Alice can make: the angles \(a\) and \(a\,'\). Similarly, Bob can choose between angles \(b\) and \(b\,'\). Whatever the signal is that the photons carry along, it should entail an expectation value for the four observations that can be made:  Alice observes \(A\) or \( A'\) in the two cases, and Bob observes \( B\) or \(B'\). If both Alice and Bob make large numbers of observations, every time using either one of their two options, they can compare notes afterwards, and measure the averages of \( AB\,,\  A'\,B\,,\  AB'\,,\) and \( A'\,B'\). They calculate the value of
	\be S=\bra AB\ket+\bra A'\,B\ket+\bra A\,B'\,\ket-\bra A'\,B'\,\ket\ ,\eel{CHSHcomb}
and see how it depends on the polarisation angles \(a\) and \(b\).

Now suppose we assume that, every time photons are emitted, they have well-defined values for \(A,\ A',\ B\), and \(B'\).
Whatever the signals are that are carried by the photons,  at each measurement these four terms will take the values \(\pm 1\), but they can never all contribute to the quantity \(S\) with the same sign (because of the one minus sign in \eqn{CHSHcomb}). Because of this, it is easy to see that \(S\) is always \(\pm 2\), and its average value will therefore obey:
	\be | \bra S\ket |\le 2\ ; \eel{CHSH}
this modern version of Bell's original observation is called the Clauser-Horne-Shimony-Holt (CHSH) inequality\,\cite{CHSH-1969}. However, if we choose the angles
	\be a=22.5^\circ\ ,\quad a'=-22.5^\circ\ ,\quad b=0^\circ\ ,\quad b'=45^\circ\ , \eel{bellangles}
then, according to Eq.~\eqn{ABquant}, quantum mechanics gives for the expectation value
	\be S=3\cos(45^\circ)-\cos 135^\circ=2\sqrt 2>2\ . \eel{CHSHquant}

How can this be? Apparently, quantum mechanics does not give explicit values \(\pm1\) for the measurements of \(A\) and \(A'\); it only gives a value to the quantity actually measured, which in each case is \emph{either} \(A\) \emph{or} \(A'\) and also either \(B\) or \(B'\).
	If Alice measures \(A\), she cannot also measure \(A'\) because the operator for \(A'\) does not commute with \(A\); the polarisers differ by an angle of \(45^\circ\), and a photon polarised along one of these angles is a quantum superposition of a photon polarised along the other angle and a photon polarised orthogonally to that. So the quantum outcome is completely in accordance with the Copenhagen prescriptions, but it seems that it cannot be realised in a local hidden variable theory.
	
We say that, if \(A\) is actually measured, the measurement of \(A'\) is \emph{counterfactual}, which means that we imagine measuring \(A'\) but we cannot actually do it, just as we are unable to measure position if we already found out what exactly the momentum of a particle is. If two observables do not commute, one can measure one of them, but the measurement of the other is counterfactual.

Indeed, in the arguments used, it was assumed that the hidden variable theory should allow an observer actually to carry out counterfactual measurements. This is called \emph{definiteness}. Local hidden variable theories that allow for counterfactual observations are said to have local definiteness. Quantum mechanics forbids local counterfactual definiteness.

However, to use the word `definiteness', or `realism', for the possibility to perform counterfactual observations is not very accurate. `realism' should mean that there is actually something happening, not a superposition of things; something is happening for sure, while something else is not happening. That is not the same thing as saying that both Alice and Bob at all times can choose to modify their polarisation angles, \emph{without allowing for any modification at other places}\,\cite{Vervoort-2013}.

It is here that the notion of ``free will" is introduced \,\cite{Newscientist-2006}\cite{CK-2008}, again, in an imprecise manner\fn{One can certainly attempt to define `free will' in terms of sound mathematical physics. See Part II, subsection~\ref{will}.}. The real extra assumption made by Bell, and more or less tacitly by many of his followers, is that both Alice and Bob should have the ``free will'' to modify their settings at any moment, without having to consult the settings at the system \(S\) that produces an unstable atom. If this were allowed to happen in a hidden variable theory, we would get local counterfactual definiteness, which was ruled out.

The essence of the argument that now follows has indeed been raised before. The formulation by C.H. Brans\,\cite{Brans-1987} is basically correct, but we add an extra twist, something to be called the `ontology conservation law', in order to indicate why violation of Bell's theorem does not require `absurd physics'.

How can we deny Alice and/or Bob their free will? Well, precisely in a deterministic hidden variable theory, Alice and Bob can only change their minds about the setting of their polarisers, if their brains follow different laws than they did before, and, like it or not, Alice's and Bob's actions \emph{are} determined by laws of physics\,\cite{GtH-FreeWill-2007}, even if these are only local laws. Their decisions, logically, have their roots in the distant past, going back all the way to the Big Bang. So why should we believe that they can do counterfactual observations?

The way this argument is usually countered is that the correlations between the photons  \(c\) from the decaying atom and the settings \(a\) and \(b\) chosen by Alice and Bob have to be amazingly strong. A gigantically complex algorithm could make Alice an Bob take their decisions, and yet the decaying atom, long before Alice and Bob applied this algorithm, knew about the result. This is called `conspiracy', and conspiracy is said to be ``disgusting". ``One could better stop doing physics than believe such a weird thing", is what several investigators quipped.

In subsections~\ref{hiddeninfo}, \ref{entangle} and \ref{correlinfo}, we go to the core of this issue.

\subsubsection{The mouse dropping function\labell{mousedr}}

To illustrate how crazy things can get, a polished version of Bell's experiment was proposed: both Alice and Bob carry along with them a mouse in a cage\fn{This version was brought forward in a blog discussion. Unfortunately, I do not remember who raised it.}, with food. Every time they want to set the angles of their polarisers, they count the number of mouse droppings. If it is even, they choose one angle, if it is odd, they choose the other. ``Now, the decaying atom has to know ahead of time how many droppings the mouse will produce. Isn't this disgusting?"

To see what is needed to obtain this ``disgusting" result, let us consider a simple model. We assume that there are correlations between the joint polarisations of the two entangled photons, called \(c\), and the settings \(a\) chosen by Alice and \(b\) chosen by Bob. All these angles are taken to be in the interval \([0,\,180^\circ]\). We define the conditional probability to have the photons polarised in the direction \(c\), given \(a\) and \(b\), to be given by a function \(W(c\,|\,a,b)\). Next, assume that Alice's outcome \(A=+1\) as soon as her `ontological' photon has \(|a-c|<45^\circ\) or \(>135^\circ\), otherwise \(A=-1\). For Bob's measurement, replacing \(a\leftrightarrow b\), we assume the same.

To see how the CHSH inequality gets violated, consider the probability \(W(c|a,b)\) of the angle \(c\) when \(a\) and \(b\) are given. We now compute this assuming the quantum mechanical result is right. Everything will be periodic in \(a,\ b,\) and \(c\) with period \(\pi\ (180^\circ).\)

 It is reasonable to expect that \(W\) only depends on the relative angles \(c-a\) and \(c-b\). 
 	\be W(c\,|\,a,b)=W(c-a,\,c-b)\ ;\qquad \int_0^\pi \dd c\,W(c-a,\,c-b)=1\ . \eel{condprobbellangles}

Introduce a sign function \(s_{(\vv)}\) as follows:\fn{We use the brackets in subscript so as to avoid confusion with simple multiplications in our subsequent calculations.} \def\sign{\mathrm{sign}}
	\be s_{(\vv)}\equiv\sign(\cos(2\vv))\ ;\qquad A=s_{(c-a)}\ ,\quad B=s_{(c-b)}\ . \eel{ABmodel}
The expectation value of the product \(AB\) is
	\be\bra A\,B\ket=\int\dd c\,W(c\,|\,a,b)\,s_{(c-a)}\,s_{(c-b)}\ , \eel{ABexpectation}
How should \(W\) be chosen so as to reproduce the quantum expression \eqn{ABquant}?

Introduce the new variables
	\be x=c-\half(a+b)\ ,\quad z=\half(b-a)\ ,\qquad W=W(x+z,\,x-z)\ . \eel{xzvariables}
Quantum mechanics demands that
	\be \int_0^\pi\dd x\, W(x+z,\,x-z)\,s_{(x+z)}\,s_{(x-z)}=\cos 4z\ .\eel{eq729}
Writing
	\be s_{(x+z)}\,s_{(x-z)}=\sign(\cos2(x+z)\,\cos2(x-z))=\sign(\cos 4x+\cos 4z)\ , \ee
we see that the equations are periodic with period \(\pi/2\), but looking more closely, we see that both sides of Eq.~\eqn{eq729} change sign if \(x\) and \(z\) are shifted by an amount \(\pi/4\). Therefore, one first tries solutions with periodicity \(\pi/4\). Furthermore, we have the symmetry \(x\leftrightarrow -x,\ z\leftrightarrow -z\).

Eq.~\eqn{eq729} contains more unknowns than equations, but  if we assume \(W\) only to depend on \(x\) but not on \(z\) then the equation can readily be solved. Differentiating Eq.~\eqn{eq729} to \(z\), one only gets Dirac delta functions inside the integral, all adding up to yield:
	\be 4\int_0^{\pi /4}W(x)\dd x	(-2 \d(x+z-\pi /4))= -4 \sin4z\ ,\quad\hbox{if }\ 0<z<\quart\pi \ee
(the four parts of the integral in Eq.~\eqn{eq729} each turn out to give the same contribution, hence the first factor 4).
Thus one arrives at
	\be W(c\,|\,a,b)\iss W(x+z,\,x-z)\iss\half|\sin 4x|\iss\half|\,\sin(4c-2a-2b)\,|\ . \eel{bellcorrel}
This also yields a normalised 3-point probability distribution,
	\be W(a,b,c)\iss \fract 1{2\pi^2}|\sin(4c-2a-2b)|\ . \eel{bellprobl}

By inspection, we find that this correlation function \(W\) indeed leads to the quantum expression \eqn{ABquant}. We could call this the `mouse dropping function' (see Fig.~\ref{mousedrop.fig}). If Alice wants to perform a counterfactual measurement, she modifies the angle \(a\), while \(b\) and \(c\) are kept untouched. She herewith chooses a configuration that is less probable, or more probable, than the configuration she had before. Taking in mind a possible interpretation of the Born probabilities, as expressed in sections~\ref{decoBorn} and \ref{born.sec}, this means that the configuration of initial states where Alice's mouse produced a different number of droppings, may be more probable or less probable than the state she had before. In quantum mechanics, we have learned that this is acceptable. If we say this in terms of an underlying deterministic theory, there seem to be problems with it.

\begin{figure}[h!] \begin{quote} 
\begin{center} \lowerwidthfig{0pt}{70mm}{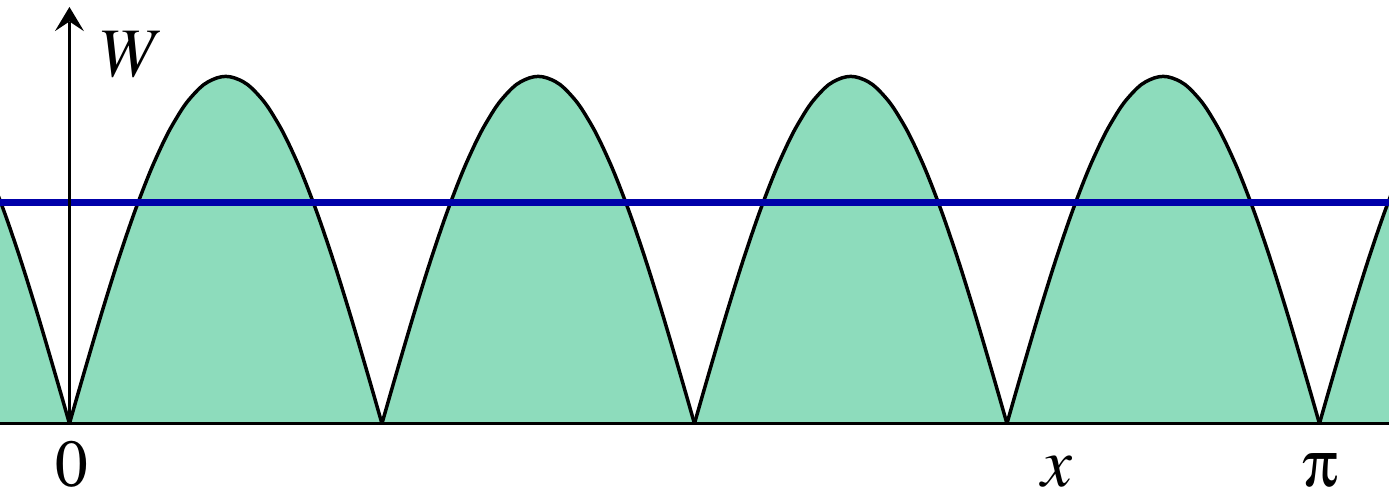} 
 \caption[\small The mouse dropping function, Eq.~\eqn{bellprobl}. ]{\small  The mouse dropping function, Eq.~\eqn{bellprobl}. Horizontally the variable \(x=c-\half(a+b)\). Averaging over any of the three variables \(a\), \(b\), or \(c\), gives the same result as the flat line; the correlations then disappear.\labell{mousedrop.fig}}
\end{center}  \end{quote}
\end{figure}

\subsubsection{Ontology conservation and hidden information\labell{hiddeninfo}}

In fact, all we have stated here is that, even in a deterministic theory obeying local equations, configurations of template  states may have non-trivial space-like correlations. It is known that this happens in many physical systems. A liquid close to its thermodynamical critical point shows a phenomenon called critical opalescence: large fluctuations in the local density. This means that the density correlation functions are non-trivial over relatively large space-like distances. This does not entail a violation of relativity theory or any other principle in physics such as causality; it is a normal phenomenon. A liquid does not have to be a quantum liquid to show critical opalescence.

It is still true that the effect of the mouse droppings seems to be mysterious, since, in a sense, we do deny that Alice, Bob, and their mice have ``free will". What exactly is `free will'? We prefer a mathematical definition rather than an emotional one (see subsection~\ref{will}), and we
 also return to this subject, in our final discussion in sections~\ref{counter} and \ref{conspire}. All we can bring forward now is that mice guts also obey energy, momentum and angular momentum conservation because these are general laws. In our `hidden variable' theory, a general law will have to be added to this: \emph{an ontological state evolves into an ontological state}; superpositions evolve into superpositions. If the mouse droppings are ontological in one description and counterfactual in another, the initial state from which they came was also ontological or counterfactual, accordingly. This should not be a mysterious statement.

There is a problematic element in this argument however, which is that, somehow, the entangled photons leaving the source, already carry information about the settings that will be used by Alice and Bob. They don't tell us the settings themselves, but they do carry information about the correlation function of these settings. Thus, \emph{non-local information} about the future is already present in the `hidden' ontological data of the photons, information that disappears when we rephrase what happens in terms of standard quantum mechanical descriptions. Thus, there is non-trivial information about the future in the ontological description of what happens. We claim that, as long as this information is subject to a rigorous conservation law -- the law of the conservation of ontology as described above -- there is no contradiction here, but we do suspect that this may shed at least an unusual light on the idea of superdeterminism.

Actually, there is an even stronger arrangement than the mouse droppings by which Alice and Bob can make their decisions. They could both monitor the light fluctuations caused by light coming from different quasars, at opposite points in the sky\,\cite{MIT-2014}, and use these to decide about the settings \(a\) and \(b\) of their filters. These quasars, called \(Q_A\) and \(Q_B\) in Fig.~\ref{bell.fig}, may have emitted their photons shortly after the Big Bang, at time \(t=t_0\) in the Figure, when they were at billions of light years separation from one another. The fluctuations of these quasars should also obey the mouse dropping formula \eqn{bellcorrel}. How can this be? The only possible explanation is the one offered by the inflation theory of the early universe: these two quasars, together with the decaying atom, do have a common past, and therefore their light is correlated.

Note that the correlation generated by the probability distribution \eqn{bellprobl} is a genuine three-body correlation. Integrating over any one of the three variables gives a flat distribution. The quasars are correlated only through the state the decaying atom is in, but not directly with one another. It clearly is a mysterious correlation, but it is not at odds with what we know about the laws of physics, see sections~\ref{vacfluct} and \ref{commsignals} in part II.

In fact, these space-like correlations must be huge: who knows where else in the universe some alien Alices and Bobs are doing experiments using the same quasars \dots

\newsecl{Deterministic quantum mechanics}{detqu}
	\subsecl{Introduction}{axioms}
	Most attempts at formulating hidden variable theories in the presently existing literature consist of some sort of modification of the real quantum theory, and of a replacement of ordinary classical physics by some sort of stochastic formalism. The idea behind this is always that quantum mechanics seems to be so different from the classical regime, that some deep modifications of standard procedures seem to be necessary.
	
	In the approach advocated here, what we call deterministic quantum mechanics is claimed to be much closer to standard procedures than usually thought to be possible.
\begin{quote} \emph{Deterministic quantum mechanics is neither a modification of standard quantum mechanics, nor a modification of classical theory.} It is a cross section of the two. \end{quote}
This cross section is claimed to be much larger and promising than usually thought. We can phrase the theory in two ways: starting from conventional quantum mechanics, or starting from a completely classical setting. We have seen already in previous parts of this work what this means; here we recapitulate.

Starting from conventional quantum mechanics, deterministic quantum mechanics is a small subset of all quantum theories: \emph{we postulate the existence of a very special basis in Hilbert space}: the ontological basis. Very likely, there will be many different choices for an ontological basis (often related by symmetry transformations), and it will be difficult to decide which of these is ``the real one". This is not so much our concern. Any of these choices for our ontological basis will serve our purpose. Finding quantum theories that have an ontological basis will be an important and difficult exercise. Our hope is that this exercise might lead to new theories that could help elementary particle physics and quantum gravity theories to further develop. Also it may help us find special theories of cosmology. 

An ontological basis is a basis in terms of which the Schr\"odinger equation sends basis elements into other basis elements at sufficiently dense moments in time.

This definition is deliberately a bit vague. We do not specify how dense the moments in time have to be, nor do we exactly specify how time is defined; in special relativity, we can choose different frames of reference where time means something different. In general relativity, one has to specify Cauchy surfaces that define time slices. What really matters is that an ontological basis allows a meaningful subset of observables to be defined as operators that are diagonal in this basis.  We postulate that they evolve into one another, and this implies that their eigenvalues remain sharply defined as time continues. Precise definitions of an ontological basis will be needed only if we have specific theories in mind; in the first simple examples that we shall discuss, it will always be clear what this basis is. In some cases, time is allowed to flow continuously, in others, notably when we have discrete operators, time is also limited to a discrete subset of a continuous time line.
	
Once such a basis has been identified, we may have in our hands a set of observables in terms of which the time evolution equations appear to be classical equations. This then links the system to a classical theory. But it was a quantum theory from where we started, and this quantum theory allows us without much ado to transform to any other basis we like. Fock space for the elementary particles is such a basis, and it still allows us to choose any orthonormal sets of wave functions we like for each particle type. In general, Fock space will not be an ontological basis. We might also wish to consider the basis spanned by the field operators \(\f(\vec x,t)\,,\ A_\m(\vec x,t)\,,\) and so on. This will also not be an ontological basis, in general.

Clearly, an ontological basis for the Standard Model has not yet been found, and it is very dubious whether anything resembling an ontological basis for the Standard Model exists. More likely, the model will first have to be extended to encompass gravity in some way. In the mean time, it might be a useful exercise to identify operators that stay diagonal longer than others, so they may be closer to the ontological variables of the theory than other operators. In general, this means that we have to investigate commutators of operators. Can we identify operators that, against all odds, accidentally commute? We shall see a simple example of such a class of operators when we study our ``neutrino" models (section~\ref{nu} in part II\,); ``neutrinos" in quotation marks, because these will be  idealised neutrino-like fermions. We shall also see that anything resembling a harmonic oscillator can be rephrased in an ontological basis to describe classical variables that evolve periodically, the period being that of the original oscillator (section~\ref{harmosc} in part II\,).

We shall also see in part II that some of the mappings we shall find are not at all fool-proof. Most of our examples cease to be linked to a classical system if interactions are turned on, unless one accepts that negative-energy states emerge (section~\ref{secondquant}). Furthermore, we have the so-called edge states. These are states that form a subset of Hilbert space with measure zero, but their contributions may still spoil the exact correspondence.

Rather than searching for an ontological basis in an existing quantum system, we can also imagine defining a theory for deterministic quantum mechanics by starting with some completely classical theory. Particles, fields, and what not, move around following classical laws. These classical laws could resemble the classical theories we are already familiar with: classical mechanics, classical field theories such as the Navier Stokes equations, and so on. There is however also a  very interesting class of models usually called ``cellular automata". A cellular automaton is a system with localised, classical, discrete degrees of freedom\fn{Besides this, one may also imagine \emph{quantum} cellular automata.  These would be defined by quantum operators (or \emph{qubits}) inside their cells. These are commonly used as `lattice quantum field theories', but would not, in general, allow for an ontological basis.}, typically arranged in a lattice, which obey evolution equations. The evolution equations take the shape of a computer program, and could be investigated in computers exactly, by running these programs in a model. A typical feature of a cellular automaton is that the evolution law for the data in every cell only depends on the data in the adjacent cells, and not on what happens at larger distances. This is a desirable form of locality, which indeed ensures that information cannot spread faster than some limiting speed, usually assumed to be the local speed of light.

In principle, these classical theories may be chosen to be much more general than the classical models most often used in physics. As we need a stabilisation mechanism, our classical model will usually be required to obey a Hamiltonian principle, which however, for discrete theories, takes a shape that differs substantially from the usual Hamiltonian system, see part II, chapter~\ref{Hamiltonform}. A very important limitation would then be the demand of time reversibility. If a classical model is not time reversible, it seems at first sight that our procedures will fail. For instance, we wish our evolution operator to be unitary, so that the quantum Hamiltonian will turn out to be a hermitean operator. But, as we shall see, it may be possible to relax even this condition. The Navier Stokes equations, for instance, although time reversible at short time scales, do seem to dissipate information away. When a Navier Stokes liquid comes to rest, due to the viscosity terms, it cannot be followed back in time anymore. Nevertheless, time non reversible systems may well be of interest for our theories anyway, as will be discussed in section~\ref{infoloss}.
	
Starting from any classical system, we consider its book keeping procedure, and identify a basis element for every state the system can be in. These basis elements are declared to be orthonormal. In this artificial Hilbert space, the states evolve, and it appears to be a standard exercise first to construct the evolution operator that describes the postulated evolution, and then to identify a quantum Hamiltonian that generates this evolution operator, by exponentiation.

As soon as we have our Hilbert space, we are free to perform any basis transformation we like. Then, in a basis where  quantum calculations can be done to cover long distances in space and time, we find that the states we originally called ``ontological" now indeed are quantum superpositions of the new basis elements, and as such, they can generate interference phenomena. The central idea behind deterministic quantum mechanics is, that at this stage our transformations might tend to become so complex that the original ontological states can no longer be distinguished from any other superposition of states, and this is why, in conventional quantum mechanics, we treat them all without distinction. We lost our ability to identify the ontological states in today's `effective' quantum theories.

\subsecl{The classical limit revisited}{classical}

	Now there are a number of interesting issues to be discussed. One is the act of measurement, and the resulting `collapse of the wave function'. What is a measurement?\,\cite{Zeh-1970}
	
	The answer to this question may be extremely interesting. A measurement allows a single bit of information at the quantum level, to evolve into something that can be recognised and identified at large scales. Information becomes classical if it can be magnified to arbitrary strength. Think of space ships that react on  the commands of a computer, which in turn may originate in just a few electrons in its memory chips. A single cosmic ray might affect these data. The space ship in turn might affect the course of large systems, eventually forcing planets to alter their orbits, first in tiny ways, but then these modifications might get magnified.
	
	Now we presented this picture for a reason: we define measurement as a process that turns a single bit of information into states where countless many bits and bytes react on it. Imagine a planet changing its course. Would this be observable in terms of the original, ontological variables, the beables? It would be very hard to imagine that it would \emph{not} be.  The interior of a planet may have its ontological observables arranged in a way that differs ever so slightly  from what happens in the vacuum state. Whatever these minute changes are, the planet itself is so large that the tiny differences can be added together statistically so that the classical orbit parameters of a planet will be recognisable in terms of the original ontological degrees of freedom. 
	
	In equations, consider a tiny fraction \(\d V\) of the volume \(V\) of a planet. Consider the ontological variables inside \(\d V\) and compare these with the ontological variables describing a similar volume \(\d V\) in empty space. Because of the `quantum' fluctuations, there may be some chance that these variables coincide, but it is hard to imagine that they will coincide completely. So let the probability \(P(\d V)\) that these coincide be somewhat less than 1, say:
	\be P(\d V)=1-\e\ , \eel{probplanet}
with a small value for \(\e>0\). Then the odds that the planet as a whole is indistinguishable from the vacuum will be
	\be P_\mathrm{tot}=(1-\e)^{V/\d V}\approx e^{-\e V/\d V}\ra 0\ ,\eel{probtotalplanet}
if the volume \(V\) of the planet is sufficiently large. This means that large planets must be well distinguishable from the vacuum state.
		
	This is a very important point, because it means that, at a large scale, all other classical observables of our world must also be diagonal in terms of the ontological basis: \emph{large scale observables}, such as the orbits of planets, and then of course also the classical data shown in a detector, \emph{are beables}. They commute with our microscopic beable operators. See also Figure~\ref{states.fig}.
	
	Let us now again address the nature of the wave functions, or states \(|\j\ket\), that represent real observed phenomena. In terms of the basis we would normally use in quantum mechanics, these states will be complicated quantum superpositions. In terms of the original, ontological basis, the beables will just describe the elementary basis elements. And now what we just argued is that they will also be elementary elements of the \emph{classical} observables at large scales! What this means is that the states \(|\j\ket\) that we actually produce in our laboratories, will \emph{automatically collapse} into states that are distinguishable classically. There will be no need to modify Schr\"odinger's equation to realise the collapse of the wave function; it will happen automatically.

This does away with Schr\"odinger's cat problem. The cat will definitely emerge either dead or alive, but never in a superposition. This is because all states \(|\j\ket \) that we can ever produce inside the cat-killing machine, are ontological. When we write them as superpositions, it is because the exact state, in terms of ontological basis elements, is not precisely known. 

In Schr\"odinger's Gedanken experiment, the state actually started from was an ontological state, and for that reason could only evolve into either a dead cat or a live cat. If we would have tried to put the superimposed state, \(\a|\mathrm{dead}\ket+\b|\mathrm{alive}\ket\) in our box, we would not have had an ontological state but just a template state. 
We \emph{can't} produce such a state! What we can do is repeat the experiment; in our simplified description of it, using our effective but not ontological basis, we might have thought to have a superimposed state as our initial state, but that of course never happens, all states we ever realise in the lab are the ontological ones, that later will collapse into states where classical observables take definite values, even if we cannot always predict these.

In the author's mind this resolution to the collapse problem, the measurement problem and the Schr\"odinger cat problem is actually one of the strongest arguments in favour of the Cellular Automaton Interpretation.

\subsecl{Born's probability rule}{born.sec}
\subsubsection{The use of templates\labell{templates.sub}}
	For the approach advocated in this book,  the notion of \emph{templates} was coined, as  introduced in section~\ref{basics}.
We argue that conventional quantum mechanics is arrived at if we perform some quite complicated basis transformation on the ontological basis states. These new basis elements so obtained will all be quite complex quantum superpositions of the ontological states. It is these states that we call ``template states"; they are the recognisable states we normally use in quantum mechanics. It is not excluded that the transformation may involve non-locality to some extent. 
	
	Upon inverting this transformation, one finds that, in turn, the ontological states will be complicated superpositions of the template states. The superpositions are complicated because they will involve many modes that are hardly visible to us. For instance, the vacuum state, our most elementary template state, will be a superposition of very many ontic (short for ontological) states. Why this is so, is immediately evident, if we realise that the vacuum is the lowest eigenstate of the Hamiltonian, while the Hamiltonian is not a beable but a `changeable' (see subsection~\ref{operators}).  Of course, if this holds for the vacuum (subsection~\ref{vac}), it will surely also hold for all other template states normally used. We know that some ontic states will transform into \emph{entangled} combinations of templates, since entangled states can be created in the laboratory.

\begin{figure}[h] 
\begin{center}\includegraphics[width=110 mm]{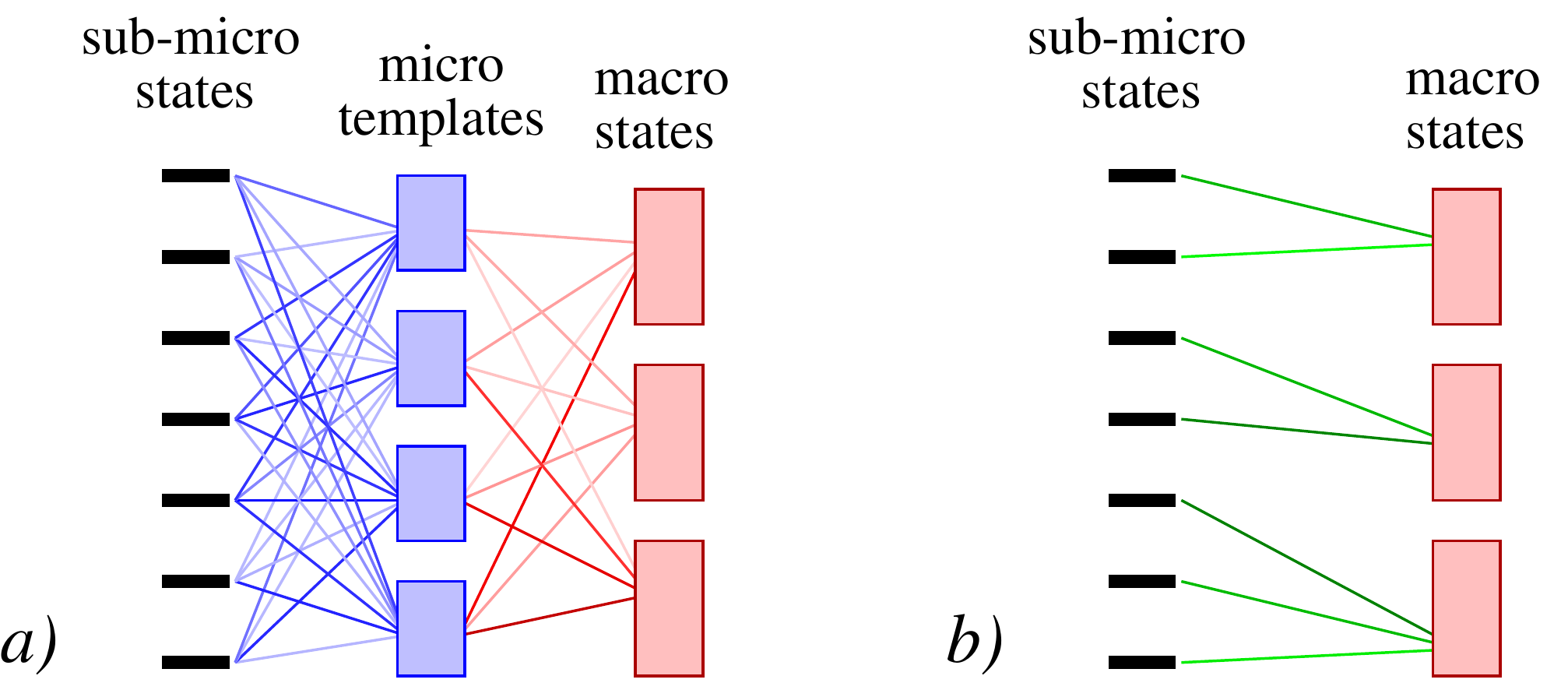} \end{center}      
 \begin{quote}  \begin{caption}[\small $a)$ The ontological sub-microscopic states, the templates and the classical states. $b)$ Classical states are (probabilistic) distributions of the sub-microscopic states.]
 {\small Classical and quantum states. $a)$ The sub-microscopic states are the ``hidden variables".  Atoms, molecules and fields are  templates, defined as quantum superpositions of the sub-microscopic states, and used at the conventional microscopic scale. The usual ``classical" objects, such as planets and people, are macroscopic, and again superpositions of the micro-templates. The lines here indicate quantum matrix elements.
 $b)$ The classical, macroscopic states are \emph{probabilistic} distributions of the sub-microscopic states. Here, the lines therefore indicate probabilities.  All states are astronomical in numbers, but the microscopic templates are more numerous than the classical states, while the sub-microscopic states are even more numerous.  \labell{states.fig} }
\end{caption}\end{quote}
\end{figure}

The macroscopic states, which are the classical states describing people and planets, but also the pointers of a measuring device, and of course live and dead cats, are again superpositions of the template states, in general, but they are usually not infinitely precisely defined, since we do not observe every atom inside these objects. Each macroscopic state is actually a composition of very many quantum states, but they are well-distinguishable from one another. 

In Figure~\ref{states.fig}, the fundamental ontological states are the sub-microscopic ones, then we see the microscopic states, which are the quantum states we usually consider, that is, the template states, and finally the macroscopic or classical states. The matrix elements relating these various states are indicated by lines of variable thickness.

What was argued in the previous section was that the classical, or macroscopic, states are diagonal in terms of the sub-microscopic states, so these are all ontic states. It is a curious fact of Nature that the states that are most appropriate for us to describe atoms, molecules and sub-atomic particles are the template states, requiring superpositions. So, when we observe a classical object, we are also looking at ontological things, which is why a template state we use to describe what we expect to see, ``collapses" into delta peeked probability distributions in terms of the classical states.

In discussions with colleagues the author noticed how surprised they were with the above statements about classical states being ontological. The reasoning above is however almost impossible to ignore, and indeed, our simple observation explains a lot about what we sometimes perceive as genuine `quantum mysteries'. So, it became an essential ingredient of our theory.

\subsubsection{Probabilities\labell{probs}}
	At first sight, it may seem that the notion of probability is lost in our treatment of quantum mechanics. Our theory is ontic, it describes certainties, not probabilities. 
	
	However, probabilities emerge naturally also in many classical systems. Think of how a \(19^\th\) century scientist would look at probabilities. Two beams of particles cross in an interaction area. How will the particles scatter? Of course, the particles will be too small to aim them so precisely that we would know in advance exactly how they meet one another, so we apply the laws of statistics. Without using quantum mechanics, the  \(19^{\th}\) century scientist would certainly know how to compute the angular distribution of the scattered particles, assuming some classical interaction potential. The origin of the statistical nature of the outcome of his calculations is simply traced to the uncertainty about the initial state.
	
	In conventional quantum mechanics, the initial state may seem to be \emph{precisely} known: we have two beams consisting of perfectly planar wave functions; the statistical distribution comes about because the wave functions of the \emph{final} state have a certain shape, and only there, the quantum physicist would begin to compute amplitudes and deduce the scattering probabilities from those. So this looks very different. We are now going to explain however, that the origin of the statistics in both cases is identical after all.
	
	In our theory, the transition from the classical notation to the quantum notation takes place when we decide to use a template state \(|\j(t)\ket\) to describe the state of the system.  At \(t=0\), the coefficients \(|\l_A|^2\), where (see the remarks following Eq.~\eqn{superposstate})
	\be \bra\ont(0)_A\,|\,\j(0)\ket = \l_A\ , \eel{alphai}
determine the probability that we are starting with ontological state \#$A$. We then use our Schr\"odinger equation to determine \(|\j(t)\ket\). When, at some time \(t_1\), the asymptotic out-state is reached, we compute \(\bra\ont(t_1)_A\,|\,\j(t_1)\ket\), where now the ontological state represents the outcome of a particular measurement, say, the particles hitting a detector at some given angle. According to quantum mechanics, using the Born rule, the absolute square of this amplitude is the probability of this outcome. But, according to our theory, the initial ontological states \(|\ont(0)\ket_A\) evolved into final ontological states \(|\ont(t_1)\ket_A\), so, \emph{we have to use the same coefficients \(\l_A\).} And now, these determine the probability of the given outcome. So indeed, we conclude that these probabilities coincide with the probabilities that we started with given ontological in-states.
	
The final ontological states are the ontological states that lead to a given outcome of the experiment. Note, that we used superpositions in calculating the transition amplitudes, but the final answers just correspond to the probability that we started with a given ontological in-state that, with certainty, evolved into a given final, classical out-state.
	
	Our template states form a very tiny subset of all ontological states, so that every time we repeat an experiment, the actual ontic state is a different one. The initial template state now does represent the probabilities of the initial ontic states, and because these are projected into the classical final states, the classical final states obey the Born rule if the initial states do. Therefore, we can prove that our theory obeys the Born rule if we know that the initial state does that regarding the ontic modes. If we now postulate that the template states used always reflect the relative probabilities of the ontic states of the theory, then the Born rule appears to be an inevitable consequence.\,\cite{GtHcollapse-2011}
	
	Most importantly, there is absolutely no reason to attempt to incorporate deviations from Born's probability interpretation of the Copenhagen interpretation into our theory. Born's rule will be exactly obeyed; there cannot be systematic, reproducible deviations. Thus, we argue, Born's rule follows from our requirement that the basis of template states that we use is related to the basis of ontological states by an orthonormal, or unitary, transformation.
	
Thus, we derived that:	\emph{as long as we use orthonormal transformations to go from one basis to another, Born's rule, including the use of absolute squares to represent probabilities, is the only correct expression for these probabilities.} 

\newsecl{Concise description of the CA Interpretation}{CAshort}

Of course we assume that the reader is familiar with the `Schr\"odinger representation' as well as the `Heisenberg representation' of conventional quantum mechanics. We often use these intermittently.

\subsecl{Time reversible cellular automata}{CAgeneral}

	The Cogwheel Model, which, as described in section~\ref{Cogwheelm}, can be mapped on a Zeeman atom with equally spaced energy levels. It is the prototype of an automaton. All our deterministic models can be characterised as `automata'. A \emph{cellular} automaton is an automaton where the data are imagined to form a discrete, \(d\)-dimensional lattice, in an \(n=d+1\) dimensional space-time. The elements of the lattice are called `cells', and each cell can contain a limited amount of information. The data \(Q(\vec x,t)\) in each cell \((\vec x,\,t)\) could be represented by an integer, or a set of integers, possibly but not necessarily limited by a maximal value \(\Bbb N\). An evolution law prescribes the values of the cells at time \(t+1\) if the values at time \(t\) (or \(t\) and \(t-1\)) are given. Typically, the evolution law for the data in a cell at the space-time position 
	\be (\vec x,\,t)\ ,\qquad \vec x=(x^1,\,x^2,\,\cdots,\,x^d)\ ,\qquad x^i,\, t\ \in\ \Bbb Z\ , \eel{Dlattice}
will only depend on the data in neighbouring cells at \((\vec x\,',\,t-1)\) and possibly those at \((\vec x,\,t-2)\). If, in this evolution law,  \(\|\vec x\,'-\vec x\|\) is limited by some bound, then the cellular automaton is said to obey \emph{locality}.  

Furthermore, the cellular automaton is said to be \emph{time-reversible} if the data in the past cells can be recovered from the data at later times, and if the rule for this is also a cellular automaton. Time reversibility can easily be guaranteed if the evolution law is assumed to be of the form
		\be Q(\vec x,\,t+1)=Q(\vec x,\,t-1)+F(\vec x,\,\{Q(t) \} )\ , \eel{CAevollaw}
where \(Q(\vec x,\,t)\) represent the data at a given point \((\vec x,\,t)\) of the space-time lattice, and \(F(\vec x,\,\{Q(t)\})\) is some given function of the data of all cells neighbouring the point \(\vec x\) at time \(t\). The \(+\) here stands for addition, addition \emph{modulo} some integer \(\Bbb N\), or some other simple, invertible operation in the space of variables \(Q\). Of course, we then have time reversibility:
		\be Q(\vec x,\,t-1)=Q(\vec x,\,t+1)-F(\vec x,\,\{ Q(t) \} )\ , \eel{CAreverse}
where \(-\) is the inverse of the operation \(+\).  \def\AA{{\mathsf{a}}} \def\BB{{\mathsf{b}}}

The simple cogwheel model allows for both a classical and a quantum mechanical description, without any modification of the physics. We can now do exactly the same thing for the cellular automaton. The classical states the automaton can be in, are regarded as an orthonormal set of basis elements of an ontological basis. The evolution operator \(U_\op(\d t)\) for one single time step whose duration is \(\d t\), is a unitary operator, so that all its eigen states \(|E_i\ket\) are unimodular:
	\be U_\op(\d t)|E_i\ket=e^{-i\w_i}|E_i\ket\ , \qquad 0\le\w<2\pi\ , \eel{Ueigenv}
and one can find an operator \(H_\op\) such that 
	\be U_\op(\d t)=e^{-iH_\op\d t}\ ,\qquad 0\le H_\op<2\pi/\d t\ . \eel{hamU}
However, one is free to add integer multiples of \(2\pi/\d t\) to any of the eigen values of this Hamiltonian without changing the expression for \(U_\op\), so that there is a lot of freedom in the definition of \(H_\op\). One may add arbitrary phase angles to the eigenstates, \(|E_i\ket\ra e^{i\vv_i}|E_i\ket\),  and these modifications may also depend on possible conserved quantities. Clearly, one can modify the Hamiltonian quite a bit without damaging its usefulness to generate the evolution operator \(U_\op\). In subsection~\ref{generalfinite} of section~\ref{Cogwheelm}, it is demonstrated how quite complex energy spectra can emerge this way in relatively simple generalisations of the cogwheel model.

	Modifications of this sort in the Hamiltonian may well be needed if we wish to reflect the locality property of the cellular automaton in the Hamiltonian:
	\be H_\op\qu\sum_{\vec x}\HH_\op\,(\vec x)\ ,\qquad [\,\HH_\op\,(\vec x\,')\,,\ \HH_\op\,(\vec x)\,]\ra 0\quad\hbox{if}\quad \|\vec x\,'-\vec x\|\gg 1\ , \eel{hamdensity}
but it is an important mathematical question whether a Hamiltonian obeying Eq.~\eqn{hamdensity} can be constructed even if the classical cellular automaton is local in the sense described above. This problem is discussed further in subsection~\ref{locy}, in section~\ref{posham}, and in part II, chapters~\ref{locality} and \ref{quloc}. There, we shall see that the situation can become quite complex. Very good approximations may exist for certain cellular automaton systems with possibly some modified locality properties and a Hamiltonian that approximately obeys a locality principle as in Eq.~\eqn{hamdensity}.

	Note, that a Hamiltonian that obeys \eqn{hamdensity} in combination with Poincar\'e invariance will correspond to a fully renormalized quantum field theory, with all complexities of that, and this assures that finding a complete mathematical solution to the problems sketched will not be easy.

\subsecl{The CAT and the CAI}{catcai}

	The Cellular Automaton \emph{Theory} (CAT) assumes that, once a universal Schr\"odinger equation has been identified that encapsulates all conceivable phenomena in the universe (a \emph{Grand Unified Theory}, or a \emph{Theory for Everything}), it will feature an ontological basis that maps the system into a classical automaton. It is possible, indeed quite likely, that the true automaton of the universe will be more complex than an `ordinary' cellular automaton as sketched here, but it may well share some of its main characteristics. The extent to which this is so, will have to be sorted out by further research.
	
	 How the \emph{symmetries} of nature will be reflected in these classical rules is also difficult to foresee; it is difficult to imagine how Lorentz invariance and diffeomorphism invariance can be realised in these classical rules. Probably, they will refer to more general quantum basis choices. This we will assume, for the time being, see in Part II, chapter~\ref{symm} on symmetries.

This theory then does seem to be what is usually called a `hidden variable theory'. Indeed, in a sense, our variables are hidden; if symmetry transformations exist that transform our basis into another one, which diagonalises different operators, then it will be almost impossible for us to tell which of these is the `true' ontological basis, and so we will have different candidates for the `hidden variables', which will be impossible to distinguish in practice.

The Cellular Automaton \emph{Interpretation} (CAI)\,\cite{GtHCA-1990} takes it for granted that this theory is correct, even if we will never be able to explicitly identify any ontological basis. We assume that the templates presently in use in quantum mechanics are to be regarded as superpositions of ontological states, and that the classical states that describe outcomes of observations and measurements are probabilistic distributions of ontological states. If two classical states are distinguishable, their probabilistic distributions have no ontological state in common (see Fig.~\ref{states.fig}~$\!b$). The universe is in a single ontological state, never in a superposition of such states, but, whenever we use our templates (that is, when we perform a conventional quantum mechanical calculation), we use superpositions just because they are mathematically convenient. Note that, because the superpositions are linear,  our templates obey the same Schr\"odinger equation as the ontological states do.

In principle, the transformation from a conventional quantum basis to an ontological basis may be quite complex. Only the eigenvalues of the Hamiltonian will be unaffected, and also in deterministic models, these can form quite general spectra, see Figure~\ref{generalfinite.fig}$c$. In group theory language, the Hamiltonians we obtain by transforming to different basis choices, will form one single conjugacy class, characterised by the set of eigenvalues.

Eventually, our quantum system should be directly related to some quantum field theory in the continuum limit. We describe quantum field theories in part~II, chapter~\ref{QFT}. At first sight, it may seem to be obvious that the Hamiltonian should take the form of Eq.~\eqn{hamdensity}, but we have to remember that the Hamiltonian definitions~\eqn{firstH}, \eqn{clockham}, \eqn{NHam} and the expressions illustrated in Fig.~\ref{generalfinite.fig}, are only well-defined \emph{modulo} \(2\pi/\d t\), so that, when different, non interacting systems are combined,  their Hamiltonians do not necessarily have to add up.

	These expressions do show that the Hamiltonian \(H_\op\) can be chosen in many ways, since one can add any operator that commutes with \(H_\op\). Therefore it is reasonable to expect that a Hamiltonian obeying Eq.~\eqn{hamdensity} can be defined. An approach is explained in much more detail in Part II, chapter~\ref{CAdetail}, but some ambiguity remains, and convergence of the procedure, even if we limit ourselves to low energy states, is far from obvious.

	What we do see is this: a Hamiltonian obeying Eq.~\eqn{hamdensity} is the sum of terms that each are finite and bounded below as well as above. Such a Hamiltonian must have a \emph{ground state}, that is, an eigenstate \(|\emptyset\ket\) with lowest eigenvalue, which could be normalised to zero. 
This eigenstate  should be identified with the `vacuum'. This vacuum is
stationary, even if the automaton itself may have no stationary solution. The next-to-lowest
eigenstate may be a one-particle state. In a Heisenberg picture, the fields \(Q(\vec x,t)\) will behave as operators \(Q_\op(\vec x,t)\) when we pass on to a basis of template states, and as such they may
create one-particle states out of the vacuum. Thus, we arrive at something that resembles a
genuine quantum field theory. The states are quantum states in complete accordance with a
Copenhagen interpretation. The fields \(Q_\op\,(\vec x,t)\)  should obey the
Wightman axioms. Quantum field theories will further be discussed in chapter~\ref{QFT}.

	However, if we start from just any cellular automaton, there are three ways in which the resulting theory will differ from conventional quantum field
theories. One is, of course, that space and time are discrete. Well, maybe there is an interesting
`continuum limit', in which the particle mass(es) is(are) considerably smaller than the inverse of
the time quantum, but, unless our models are chosen very carefully, this will not be the case.

	Secondly, the generic cellular automaton will not even remotely be Lorentz invariant. Not only will the one-particle states fail to exhibit Lorentz invariance, or even Galilei invariance; the states where particles may be moving with respect to the vacuum state will be completely different from the static one-particle states. Also, rotation symmetry will be reduced to some discrete lattice rotation group if anything at all. 
So, the familiar symmetries of relativistic quantum field theories will be totally absent.

	Thirdly, it is not clear that the cellular automaton can be associated to a single quantum model or perhaps many inequivalent ones. The addition or removal of other conserved operators to \(H_\op\,\) is akin to the addition of chemical potential terms. In the absence of Lorentz invariance, it will be difficult to distinguish the different types of `vacuum' states one thus obtains.

	For all these reasons, most cellular automaton models will be very different from the quantised field theories for elementary particles. The main issue discussed in this book, however, is not whether it is easy to mimic the Standard Model in a cellular automaton, but whether one can obtain quantum mechanics, and something resembling quantum field theory at least in principle. The origin of the continuous symmetries of the Standard Model will stay beyond what we can handle in this book, but we can discuss the question to what extent cellular automata can be used to approximate and understand the quantum nature of this world. See  our discussion of symmetry features in part~II, section~\ref{symm}.

	As will be explained later, it may well be that invariance under general coordinate transformations will be a crucial ingredient in the explanation of continuous symmetries, so it may well be that the ultimate explanation of quantum mechanics will also require the complete solution of the quantum gravity problem. We cannot pretend to have solved that.

	Many of the other models in this book will be explicitly integrable. The cellular automata that we started off with in the first section of this chapter illustrate that our general philosophy also applies to non-integrable systems. It is generally believed, however,  that time reversible cellular automata can be \emph{computationally universal}\,\cite{Fredkin-1982}. This means that any such automaton can be arranged in special subsets of states that obey the rule of any other computationally universal  cellular automaton. One would be tempted to argue then, that \emph{any} computationally universal cellular automaton can be used to mimic systems as complicated as the Standard Model of the subatomic particles. However, in that case, being physicists, we would ask for one single special model that is more efficient than any other, so that \emph{any} choice of initial state in this automaton describes a physically realisable configuration.

\subsecl{Motivation}{motives}
It is not too far-fetched to expect that, one day, quantum gravity will be completely solved, that is, a concise theory will be phrased that shows an air-tight description of the relevant physical degrees of freedom, and a simple model will be proposed that shows how these physical variables evolve. We might not even need a conventional time variable for that, but what we do need is an unambiguous prescription telling us how the physical degrees of freedom will look in a region of space-time that lies in the future of a region described earlier.

A complete theory explaining quantum mechanics can probably not be formulated without also addressing quantum gravity, but what we can do is to formulate our proposal, and to establish  the language that will have to be employed. Today, our description of molecules, atoms, fields and relativistic subatomic particles is interspersed with wave functions and operators. Our operators do not commute with operators describing other aspects of the same world, and we have learned not to be surprised by this, but just to choose a set of basis elements as we please, guess a reasonable looking Schr\"odinger equation, and calculate what we should find when we measure things.  We were told not to ask what reality is, and this turned out to be a useful advice: we calculate, and we observe that the calculations make sense. It is unlikely that any other observable aspects of fields and particles can ever be calculated, it will never be more than what we can derive from quantum mechanics. For example, given a radio-active particle, we cannot calculate exactly at what moment it will decay. The decay is controlled by a form of randomness that we cannot control, and this randomness seems to be far more perfect than what can be produced using programmed pseudo-random sequences. We are told to give up any hope to outsmart Nature in this respect.

The Cellular Automaton Interpretation (CAI)  suggests to us what it is that we actually do when we solve a Schr\"odinger equation. We thought that we are following an infinite set of different worlds, each with some given amplitude, and the final events that we deduce from our calculations depend on what happens in all these worlds. This, according to the CAI, is an illusion. There is no infinity of different worlds, there is just one, but we are using the ``wrong" basis to describe it. The word ``wrong" is here used not to criticise the founding fathers of quantum mechanics, who made marvellous discoveries, but to repeat what they of course also found out, which is that the basis they are using is not an ontological one. The terminology used to describe that basis does not disclose to us exactly how our world, our single world, `actually' evolves in time.

Many other proposed interpretations of quantum mechanics exist. These may either regard the endless numbers of different worlds all to be real, or they require some sort of modification, mutilation rather, to understand how a wave function can collapse to produce measured values of some observable without allowing for mysterious superpositions at the classical scale.

The CAI proposes to use the complete mathematical machinery that has been developed over the years to address quantum mechanical phenomena. It includes exactly the Copenhagen prescriptions to translate the calculations into precise predictions when experiments are done, so, at this point, definitely no modifications are required.

There is one important way in which we deviate from Copenhagen however. According to Copenhagen, certain questions are \emph{not} to be asked: 
\begin{quote} \emph{Can it be that our world is just a single world where things happen, according to evolution equations that might be fundamentally simpler than Schr\"odinger's equation, and are there ways to find out about this?  Can one remove the element of statistical probability distributions from the basic laws of quantum mechanics?}\end{quote}
According to Copenhagen, there exist no experiments that can answer such questions, and therefore it is silly even to ask them. But what I have attempted to show in this work is that not experimentally, but theoretically, we may find answers. We may be able to identify models that describe a single classical world, even if forbiddingly complex compared to what we are used to, and we may be able to identify its physical degrees of freedom with certain quantum variables that we already know about.

The cellular automaton described in the preceding sections, would be the prototype example; it is complicated yet quite possibly not complicated enough. It has symmetries, but in the real world there are much larger symmetry groups, such as the Lorentz or Poincar\'e group, showing relations between different kinds of events, and these are admittedly difficult to implement. The symmetry groups, think of space-time translation symmetry, may actually be at the roots of the mysterious features that were found in our quantum world, so that these may have natural explanations.

Why should we \emph{want} just a single world with classical equations describing its evolution? What's wrong with obeying Copenhagen's dictum about not asking questions, if they will not be experimentally accessible anyway?

According to the author, there will be overwhelming motivations: \emph{If} a classical model does exist, it will tremendously simplify our view of the world, it will help us once and forever to understand what really happens when a measurement is made and a wave function `collapses'. It will explain the Schr\"odinger cat paradox once and for good. 

Even more importantly, the quantum systems that allow for a classical interpretation form an extremely tiny subset of all quantum models. If it is indeed true that our world falls in that class, which one may consider to be likely after having read all this, then this restricts our set of allowable models so much that it may enable us to guess the right one, a guess that otherwise could never have been made. So indeed, what we are really after is a new approach towards guessing what the `Theory of the World' is. We strongly suspect that, without this superb guide, we will never even come close. Thus, our real motivation is not to be able to better predict the outcomes of experiments, which may not happen soon, but rather to predict which class of models we should scrutinise to find out about the truth.

Let us emphasise once more, that this means that the CAI / CAT will primarily be of importance if we wish to decipher nature's laws at the most fundamental time- and distance scales, that is, the Planck scale. Thus, an important new frontier where the empire of the quantum meets the classical world, is proclaimed to be near the Planckian dimensions. As we also brought forward repeatedly, the CAI requires a reformulation of our standard quantum language also when describing the other important border: that between the quantum empire and the `ordinary' classical world at distance scales larger than the sizes of atoms and molecules.

The CAI actually has more in common with the original Copenhagen doctrine than many other approaches, as will be explained. It will do away with the `many worlds', more radically than the De Broglie-Bohm interpretation. The CAI assumes the existence of one or more models of Nature that have not yet been discovered. We do discuss many toy models. These toy models are not good enough to come anywhere close to the Standard Model, but there is reason to hope that one day such a model will be found. In any case, the CAI will apply only to a tiny sub class of all quantum  mechanical models usually considered to explain the observed world, and for this reason we hope it might be helpful to pin down the right procedure to arrive at the correct theory.

Other models were exposed in this work, just to display the set of tools that one might choose to use.

\subsubsection{The wave function of the universe\labell{uniwave}}
	Standard quantum mechanics can confront us with a question that appears to be difficult to answer: \emph{Does the universe as a whole have a wave function?  Can we describe that wave function?} Several responses to this question can be envisaged:
	\bi{(1)} I do not know, and I do not care. Quantum mechanics is a theory about observations and  measurements. You can't measure the entire universe.
	\itm{(2)} I do care, but I do not know. Such a wave function might be so much entangled that it will be impossible to describe. Maybe the universe has a density matrix, not a wave function.
	\itm{(3)} The universe has no fixed wave function. Every time an observation or measurement is made, the wave function collapses, and this collapse phenomenon does not follow any Schr\"odinger equation.
	\itm{(4)} Yes, the universe has a wave function. It obeys the Schr\"odinger equation, and at all times the probability that some state \(|\j\ket\) is realised is given by the norm of the inner product squared. Whenever any `collapse' takes place, it always obeys the Schr\"odinger equation.
	\ei

\noindent The agnostic answers (1) and (2) are scientifically of course defensible. We should limit ourselves to observations, so don't ask silly questions. 
However, they do seem to admit that quantum mechanics may not have universal validity. It does not apply to the universe as a whole. Why not? Where exactly is the limit of the validity of quantum mechanics? The ideas expressed in this work were attacked because they allegedly do not agree with observations, but all observations ever made in atomic and subatomic science appear to confirm the validity of quantum mechanics, while the answers (1) and (2) suggest that quantum mechanics should break down at some point. In the CAI, we assume that the \emph{mathematical rules} for the application of quantum mechanics have absolute validity. This, we believe, is not a crazy assumption.

In the same vein, we also exclude answer (3). The collapse should not be regarded as a separate axiom of quantum mechanics, one that would invalidate the Schr\"odinger equation whenever an observation, a measurement, and hence a collapse takes place. Therefore, according to our theory, the only correct answer is  answer (4). An apparent problem with this would be that the collapse would require `conspiracy', a very special choice of the initial state because otherwise we might accidentally arrive at wave functions that are quantum superpositions of different collapsed states. This is where the ontological basis comes to the rescue. If the universe is in an ontological state, its wave function will collapse automatically, whenever needed. As a result, \emph{classical} configurations, such as the positions and velocities of the planets are always described by wave functions that are delta-peaked at these values of the data, whereas wave functions that are superpositions of planets in different locations, will \emph{never} be ontological.

The conclusion of this subsection is that, as long as we work with templates, our amplitudes are psi-epistemic, as they were in the Copenhagen view, but a psi-ontic wave function does exist: the wave function of the universe itself. It is epistemic and ontological at the same time.

Now, let us go back to Copenhagen, and formulate the rules. As the reader can see, in some respects we are even more conservative than the most obnoxious quantum dogmatics.

\subsecl{The rules}{rules}

As for the Copenhagen rules that we keep, we emphasise the ones most important to us: 
	\bi{(i)} \emph{To describe a physical phenomenon, the use of any basis is as legitimate as any other. We are free to perform any transformation we like, and rephrase the Schr\"odinger equation, or rather, the Hamiltonian employed in it, accordingly. In each basis we may find a useful description of variables, such as positions of particles, or the values of their momenta, or the energy states they are in, or the fields of which these particles are the energy quanta. All these descriptions are equally `real' .}\ei
But none of the usually employed descriptions are completely real. We often see that superpositions occur, and the phase angles of these superpositions can be measured, occasionally. In such cases, the basis used is not an ontological one. In practice, we have learned that this is just fine; we use all of these different basis choices in order to have \emph{templates}. We shall impose no restrictions on which template is `allowed', or which of them may represent the truth `better'  than others. Since these are mere templates, reality may well emerge to be a superposition of different templates.

Curiously, even among the diehards of quantum mechanics, this was often thought not to be self-evident. ``Photons are not particles, protons and electrons are", was what some investigators claimed. Photons must be regarded as energy quanta. They certainly thought it was silly to regard the \emph{phonon} as a particle; it is merely the  quantum of sound. Sometimes it is claimed that electric and magnetic fields are the "true" variables, while photons are merely abstract concepts. There were disputes on whether a particle is a true particle in the position representation or in the momentum representation, and so on. We do away with all this. All basis choices are equivalent. They are nothing more than a coordinate frame in Hilbert space. As long as the Hamiltonian employed appears to be such that  finite-time evolution operators turn diagonal operators (beables) into non-diagonal ones (superimposables), these basis choices are clearly non-ontological; none of them describes what is really happening. As for the \emph{energy  basis}, see subsection~\ref{energy}.

Note, that this means that it is not really the Hamiltonian that we will be interested in, but rather its conjugacy class. If a Hamiltonian \(H\) is transformed into a new Hamiltonian by the transformation
	\be \tl H=G\,H\,G^{-1}\ , \eel{Conjham}
where \(G\) is a unitary operator, then the new Hamiltonian \(\tl H\), in the new basis, is just as valid as the previous one. In practice, we shall seek the basis, or its operator \(G\), that gives the most concise possible expression for \(\tl H\).
	\bi{(ii)} \emph{Given a ket state \(|\j\ket\), the probability for the outcome of a measurement to be described by a given state \(|a\ket\)  in Hilbert space, is exactly given by the absolute square of the inner product \(\bra a|\j\ket\).}\ei 
	This is the well-known Born rule. We shall never modify its mathematical form; only the absolute squares of the inner products can be used. However, there is a limitation in principle: the bra state \(\bra a|\) must be an ontological state. In practice, this is always the case: the bras \(\bra a|\) usually are represented by the classical observations. The Born rule is often portrayed as a separate axiom of the Copenhagen Interpretation. In our view it is an inevitable consequence of the mathematical nature of quantum mechanics as a tool to perform calculations, see section~\ref{born.sec}.

The most important point where we depart from Copenhagen is that we make some fundamental assumptions:
	\bi{(a)} \emph{We postulate the existence of an ontological basis. It is an orthonormal basis of Hilbert space that is truly superior to the basis choices that we are familiar with. In terms of an ontological basis, the evolution operator for a sufficiently fine mesh of time variables,  does nothing more than permute the states. }\ei
	How exactly to define the mesh of time variables we do not know at present, and may well become a subject of debate, particularly in view of the known fact that space-time has a general coordinate invariance built in. We do not claim to know how to arrive at this construction -- it is too difficult. In any case, the system is expected to behave entirely as a classical construction. Our basic assumption is that a classical evolution law exists, dictating exactly how space-time and all its contents evolve.
	The evolution is deterministic in the sense that nothing is left to chance. Probabilities enter only if, due to our ignorance, we seek our refuge in some non-ontological basis. 
	
	\bi{(b)} \emph{When we perform a conventional quantum mechanical calculation, we employ a \emph{set of templates} for what we thought the wave function is like. These templates, such as the orthonormal set of solutions of the hydrogen atom, just happen to be the states for which we know how they evolve. However, they are in a basis that is a rather complicated unitary transformation of the ontological basis.}\ei
	 Humanity discovered that these templates obey Schr\"odinger equations, and we employ these to compute probabilities for the outcomes of experiments. These equations are correct to very high accuracies, but they falsely suggest that there is a `multiverse' of many different worlds that interfere with one another. Today, these templates are the best we have.
	
	\bi{(c)} \emph{Very probably, there are more than one different choices for the ontological basis, linked to one another by Nature's continuous symmetry transformations such as the elements of the Poincar\'e group, but possibly also by the local diffeomorphism group used in General Relativity. Only one of these ontological bases will be `truly' ontological. }\ei
	Which of them will be truly ontological will be difficult or impossible to determine. The fact that we shall not be able to distinguish the different possible ontological bases, will preclude the possibility of using this knowledge to perform predictions beyond the usual quantum mechanical ones. This was not our intention anyway. The motivation for this investigation has alway been that we are searching for new clues for constructing models more refined than the Standard Model.
	
	The symmetry transformations that link different (but often equivalent) ontological basis choices are likely to be truly quantum mechanical: operators that are diagonal in one of these ontological bases may be non diagonal in an other. However, in the very end we shall only use the `real' ontological basis. This will be evident in axiom (e).
	
	\bi{(d)} \emph{Classical states are ontological, which means that classical observables are always diagonal in the `truly' ontological basis.} \ei 
	This would be more difficult to `prove' from first principles, so we introduce it indeed as an axiom. 
	However, it seems to be very difficult to avoid: it is hard to imagine that two different classical states, whose future evolution will be entirely different in the end, could have non-vanishing inner products with the same ontological state. 
	\bi{(e)} \emph{From the very beginning onwards, the Universe was, and it always will be, in a single, evolving ontological state. This means that not only the observables are diagonal in the ontological basis, but the wave function always takes the simplest possible form: it is one of the elements of the basis itself, so this wave function only contains a single 1 and for the rest zeros. }\ei
Note that this singles out the `true' ontological basis from other choices, where the physical degrees of freedom can also be represented by `beables', that is, operators that at all times commute. So, henceforth, we only refer to this one `true' ontological basis as `the' ontological basis.
	
	Most importantly, the last two axioms completely solve the measurement problem \,\cite{Zeh-1970}, the collapse question and Schr\"odinger's cat paradox. The argument is now simply that Nature is always in a single ontological state, and therefore it has to evolve into a single classical state. 

\subsecl{Features of the Cellular Automaton Interpretation (CAI)}{features}
	
	A very special feature of the CAI is that ontological states do \emph{never} form superpositions. From day and time zero onwards, the universe must have been in a single ontological state that evolves. This is also the reason why it never evolves into a superposition of classical states. Now remember that the Schr\"odinger equation is obeyed by the ontological states the universe may be in. This then is the reason why this theory automatically generates `collapsed wave functions' that describe the results of a measurement, without ever parting from the Schr\"odinger equation. For the same reason, ontological states can never evolve into a superposition of a dead cat and a live cat. Regarded from this angle, it actually seems hard to see how any \emph{other} interpretation of quantum mechanics could have survived in the literature: quantum mechanics by itself would have predicted that if states \(|\,\psi\,\ket\) and \(|\,\chi\,\ket\) can be used as initial states, so can the state \(\a|\,\psi\,\ket+\b|\,\chi\,\ket\). Yet the superposition of a dead cat and a live cat cannot serve to describe the final state. If \(|\,\psi\,\ket\) evolves into a live cat and \(|\,\chi\,\ket\) into a dead one, then what does the state \(\a|\,\psi\,\ket+\b|\,\chi\,\ket\) evolve into? the usual answers to such questions cannot be correct.\fn{I here refer to the argument that \emph{decoherence}, some way or other,  does the job. See section~\ref{decoBorn}.}

The Cellular Automaton Interpretation adds some notions to quantum mechanics that do not have any distinguished meaning in the usual Copenhagen view. We introduced the ontological basis as being, in some sense, superior to any other choice of basis. One might naturally argue that this would be a step backwards in physics. Did Copenhagen, in section \ref{rules},  not emphasise that all choices of basis are equivalent? Why would one choice stand out?

Indeed, what was stated in rule \#i in section \ref{rules} was that all basis choices are equivalent, but what we really meant was that all basis choices \emph{normally employed} are equivalent. Once we adopt the Copenhagen doctrine, it does not matter anymore which basis we choose. Yet there is one issue in the Copenhagen formalism that has been heavily disputed in the literature and now is truly recognised as a weakness: the collapse of the wave function and the treatment of measurements. At these points the superposition axiom fails. As soon as we admit that one superior basis exists, this weakness disappears. \emph{All wave functions} may be used to describe a physical process, but then we have to tolerate the collapse axiom if we do \emph{not} work in the ontological basis.

The CAI allows us to use a basis that stands out. It stands out because, in this basis, we recognise wave functions that are special: the ontological wave functions. In the ontological basis, the ontological wave functions are the wave functions that correspond to the basis elements; each of these wave functions contains a one and for the rest zeros. The ontological wave functions exclusively evolve into ontological wave functions again. In this basis, there is no room for chance anymore, and superpositions can be completely avoided. 

The fact that, also in the usual formalism of quantum mechanics, states that start out with a classical description, such as beams of particles aimed at each other, end up as \emph{classical} probability distributions of particles coming out of the interaction region, in our view can be seen as evidence for an `ontology conservation law', a law that says that an ontological basis exists, such that true ontological states at one moment of time, always evolve into true ontological states at later times. This is a new conservation law. It is tempting to conclude that the  CAI is inevitable.

In the ontological basis, the evolution is deterministic. However, this term must be used with caution. ``Deterministic" cannot imply that the outcome of the evolution process can be foreseen. No human, nor even any other imaginable intelligent being, will be able to compute faster than Nature itself. The reason for this is obvious: our intelligent being would also have to employ Nature's laws, and we have no reason to expect that Nature can duplicate its own actions more efficiently than having them happen in the first place. This is how one may restore the concept of ``free will": whatever happens in our brains is unique and unforeseeable by anyone or anything.

\subsubsection{Beables, changeables and superimposables\labell{bechange}}
Having a special basis and special wave functions, also allows us to distinguish special observables or operators. In standard quantum mechanics, we learned that operators and observables are indistinguishable, so we use these concepts interchangeably. Now, we will have to learn to restore the distinctions. We repeat what was stated in subsection \ref{operators}, operators can be of three different forms:
\bi{(I)}  \emph{beables} \(\BBB_\op\): 
these denote a property of the ontological states, so that beables are diagonal in the ontological basis \(\{|A\ket,\,|B\ket,\cdots\}\) of Hilbert space:
 	\be\BBB_\op^{\,a}|A\ket = \BBB^{\,a}(A)|A\ket\ ,\qquad\hbox{(beable)}\ . \eel{beable1} \ei
Beables will always commute with one another, \emph{at all times}:
	\be[\BBB_\op^{\,a}(\vec x_1,t_1),\,\BBB_\op^{\,b}(\vec x_2,t_2)]=0\quad\forall\quad\vec x_1,\,\vec x_2,\,t_1,\,t_2. \eel{becomm}	
	Quantised fields, copiously present in all elementary particle theories, do obey Eq.~\eqn{becomm}, but only outside the light cone (where \(|t_1-t_2|<|\vec x_1-\vec x_2|\)), not inside that cone, where Eqs.~\eqn{equaltime}, see Part II,  do not hold, as can easily be derived from explicit calculations. Therefore, quantised fields are altogether different from beables.
\bi{(II)} \emph{changeables} \(\CCC_\op\): operators that replace  an ontological state \(|A\ket\) by another ontological state \(|B\ket\), such as a permutation operator:
	\be\CCC_\op|A\ket=|B\ket\ ,\qquad\hbox{(changeable)}\ . \eel{changeable1} \ei
Changeables do not commute, but they do have a special relationship with beables; they interchange them:
	\be\BBB_\op^{(1)}\CCC_\op=\CCC_\op\BBB_\op^{(2)}\ . \eel{changecomm}
We may want to make an exception for infinitesimal changeables, such as the Hamiltonian \(H_\op\):
	\be[\BBB_\op,\,H_\op]=-i{\pa\over\pa t}\BBB_\op\ . \eel{infchangecomm}
\bi{(III)} \emph{superimposables} \(\SSS_\op\): these map an ontological state onto any other, more general superposition of ontological states:
	\be\SSS_\op|A\ket=\l_1|A\ket+\l_2|B\ket+\cdots\ ,\qquad\hbox{(superimposable)}\ .\eel{superimposable1}
 \ei
All operators normally used are superimposables, even the simplest position or momentum operators of classroom quantum mechanics. This is easily verified by checking the time-dependent commutation rules (in the Heisenberg representation). In general\fn{A rare exception, for example, is the harmonic oscillator when the time span is an integer multiple of the period \(T\).}:
	\be [\vec x(t_1),\,\vec x(t_2)]\ne 0\ ,\ \hbox{ if }\ t_1\ne t_2\ . \eel{xxnoncomm}

\subsubsection{Observers and the observed\labell{observers}} 
	Standard quantum mechanics taught us a number of important lessons. One was that we should not imagine that an \emph{observation} can be made without disturbing the observed object. This warning is still valid in the CAI. If measuring the position of a particle means checking the wave function whether perhaps \(\vec x\qu\vec x^{(1)}\), this may be interpreted as having the operator \(P_\op(\vec x^{(1)})\) act on the state:
	\be|\j\ket\ra P_\op(\vec x^{(1)})|\j\ket\ , \quad P_\op(\vec x^{(1)})=\d(\vec x_\op-\vec x^{(1)})\ . \eel{xproj}
This modifies the state, and consequently all operators acting on it after that may yield results different from what they did before the ``measurement".

However, when a genuine beable acts on an ontological state, the state is simply multiplied with the value found, but will evolve in the same way as before (assuming we chose the `true' ontological basis, see axiom \#\(c\) in section~\ref{rules}). Thus, measurements of beables are, in a sense, classical measurements.

Other measurements, which seem to be completely legal according to conventional quantum mechanics, will not be possible in the CAI. In the CAI, as in ordinary quantum mechanics, we can consider any operator and study its expectation value. But since the class of physically realisable wave functions is now smaller than in standard QM, certain states can no longer be realised, and we cannot tell what the outcome of such a measurement could possibly be. Think of any (non infinitesimal) changeable \(\CCC_\op\). All ontological states will give the `expectation value' zero to such a changeable, but we can consider its eigenvalues, which in general will not yield the value zero. The corresponding eigenstates are definitely not ontological (see subsection~\ref{EarthMars}).

Does this mean that standard quantum mechanics is in conflict with the CAI? We emphasise that this is not the case. It must be realised that, also in conventional quantum mechanics, it may be considered acceptable to say that the Universe has been, and will always be, in one and the same quantum state, evolving in time in accordance with the Schr\"odinger equation (in the Schr\"odinger picture), or staying the same (in the Heisenberg picture). If this state happens to be one of our ontological states then it behaves exactly as it should. Ordinary quantum mechanics makes use of template states, most of which are not ontological, but then the real world in the end must be assumed to be in a superposition of these template states so that the ontological state resurfaces anyway, and our apparent disagreements disappear.

\subsubsection{Inner products of template  states\labell{inprod}} 

In doing technical calculations, we perform transformations and superpositions that lead to large sets of quantum states which we now regard as {templates}, or candidate models for (sub) atomic processes that can always be superimposed at a later stage to describe the phenomena we observe. Inner products of templates can be studied the usual way. 

A template state \(|\j\ket\) can be used to serve as a model of some actually observed phenomenon.
It can be any quantum superposition of ontological states \(|A\ket\). 
The inner product expressions \(|\bra A|\j\ket|^2\) then represent the probabilities that ontological state \(|A\ket\) is actually realised.

According to the Copenhagen rule \#ii, section~\ref{rules}, the probability that the template  state  \(|\j_1\ket\) is found to be equal to the state \(|\j_2\ket\), is given by \(|\,\bra\j_1|\j_2\ket|^2\). However, already at the beginning, section~\ref{basics}, we stated that the inner product \(\bra\j_1|\j_2\ket\) should not be interpreted this way. Even if their inner product vanishes, template  states \(|\j_1\ket\) and \(|\j_2\ket\) may both have a non vanishing coefficient with the same ontological state \(|A\ket\). This does not mean that we depart from Copenhagen rule \#ii, but that the true wave function cannot be just a generic template  state; it is always an ontological state \(|A\ket\).  This means that the inner product rule is only true if \emph{either \(|\j_1\ket\) or \(|\j_2\ket\)} is an ontological state while the other may be a template. We are then considering the probability that the template state  \emph{coincides exactly} with one of the ontological states \(|A\ket\). We may use the Born interpretation of the inner product if the template state in question is used to describe only a small subspace of all physical degrees of freedom. How the unobserved degrees of freedom are entangled with this state is then immaterial. Thus, the template state then can be assumed to represent a probability distribution of the ontological states of the universe. 

We see that the inner product rule can be used in two ways; one is to describe the probability distribution of the initial states of a system under consideration, and one is to describe the probability that a given classical state is reached at the end of a quantum process. If the Born rule is used to describe the initial probabilities, the same rule can be used to calculate the probabilities for the final states.

\subsubsection{Density matrices\labell{density}}
	Density matrices are used when we neither know exactly the ontological states nor the templates. One takes a set of templates \(|\j_i\ket\), and attributes to them probabilities \(W_i\ge 0\), such that \hbox{\(\sum_iW_i=1\)}.
This is called a mixed state. In standard quantum mechanics, one then finds for the expectation values of an operator \(\OO\):
	\be \bra\,\OO\,\ket=\sum_i\,W_i\,\bra\j_i|\OO|\j_i\ket=\Tr(\r\,\OO)\ ;\qquad\r=\sum_i\,W_i\,|\j_i\ket\,\bra\j_i|\ . \eel{densitytrace}
The operator \(\r\) is called the density matrix.

Note that, if the ontologic basis is known, and the operator \(\OO\) is a beable, then the probabilities are indistinguishable from those generated by a template,
	\be |\j\ket=\sum_i\l_i|\j^\ont_i\ket\ ,\qquad|\l_i|^2=W_i\ , \eel{probtemplate}
since in both cases the density matrix in that basis is diagonal:
	\be\r=\sum_iW_i\,|\j_i^\ont\ket\,\bra\J_i^\ont|\ . \eel{ontdens}
If \(\OO\) is not a beable, the off-diagonal elements of the density matrix may seem to be significant, but we have to remember that non-beable operators in principle cannot be measured, and this means that the formal distinction between density matrices and templates disappears. 

	Does this imply that mixed states from now on must be considered to be indistinguishable from pure template states? Not quite. If several ontological basis choices are possible, and we do not know the ontic states \(|\j^\ont\ket\), the beable operators are not a priori known, and this lack of knowledge will force us to use density matrices with non-diagonal elements in the ontological basis.
	
	Indeed, again, we hit the very core of our present understanding of quantum mechanics. If we do not know what the ontic states are, we have to keep open the option of using various choices of mutually non-commuting operators as our template states, and then the density matrix reflects our inability to choose the right ones.

We employ non-commuting sets of operators to characterise a state, before choosing the operator that will actually be used to measure things. According to the CAI, only that operator will represent the beables, the others won't. This feature was at the center of our considerations concerning Bell's inequalities as well as the question posed here about the density matrix.

\subsecl{The Hamiltonian}{hamil}

As was explained in the Introduction to this chapter, section~\ref{CAgeneral}, there are many ways to choose a Hamiltonian operator that correctly produces a Schr\"odinger equation for the time dependence of a (cellular) automaton. Yet the Hamiltonian for the quantum world as we know it, and in particular for the Standard Model, is quite unique. How does one derive the `correct'  Hamiltonian?

Of course there are conserved quantities in the Standard Model, such as chemical potentials, global and local charges, and kinematical quantities such as angular momentum. Those may be added to the Hamiltonian with arbitrary coefficients, but they are usually quite distinct from what we tend to call `energy', so it should be possible to dispose of them. Then, there are many non-local conserved quantities, which explains the large number of possible shifts \(\d E_i\) in Fig.~\ref{generalfinite.fig}, subsection\ \ref{generalfinite}.
Most of such ambiguities will be removed by demanding the Hamiltonian to be local.

\subsubsection{Locality\labell{locy}} Our starting expressions for the Hamiltonian of a deterministic system are Eqs.~\eqn{firstH} and \eqn{NHam}. These, however, converge only very slowly for large values of \(n\). If we apply such expansions to the cellular automaton, Eqs.~\eqn{CAevollaw} and \eqn{CAreverse}, we see that the \(n^\th\) term will involve interactions over neighbours that are  \(n\) steps separated. If we write the total Hamiltonian \(H\) as
	\be H=\sum_{\vec x}\HH(\vec x)\ ,\qquad \HH(\vec x)=\sum_{n=1}^\infty\HH_n(\vec x)\ , \eel{hamdens}
we see contributions  \(\HH_n(\vec x)\) that involve interactions over \(n\) neighbours, with coefficients dropping typically as \(1/n\). Typically,
	\be [\HH_n(\vec x),\,\HH_m(\vec x\,')]=0\qquad\hbox{only if}\qquad|\vec x-\vec x\,'|>n+m\ ,\eel{nmloc}
while in relativistic quantum field theories, we have \([\HH(\vec x),\,\HH(\vec x\,')]=0\) as soon as \(\vec x\ne\vec x\,'\).
Considering that the number of interacting neighbouring cells that fit in a \(d\)-dimensional sphere with radius \(n\), may grow rapidly with \(n\), while the leading terms start out being of the order of the Planck energy, we see that this convergence is too slow: large contributions spread over large distances. This is not the Hamiltonian that has the locality structure typical for the Standard Model.

Now this should not have been surprising. The eigenvalues of Hamiltonians \eqn{firstH} and \eqn{NHam} are both bounded to the region \((0,\,2\pi/\d t)\), while any Hamiltonian described by equations such as \eqn{hamdens}, should be extensive: their eigenvalues may grow proportionally to the volume of space.

A better construction for a cellular automaton is worked out further in part II, chapter \ref{CAdetail}. There, we first introduce the Baker Campbell Hausdorff expansion. In that, also, the lowest terms correspond to a completely local Hamiltonian density, while all terms are extensive. Unfortunately, this also comes at a price that is too expensive for us: the Baker Campbell Hausdorff series does not seem to converge at all in this case. One could argue that, if used only for states where the total energies are much less than the Planck energy, the series should converge, but we have no proof for that. Our problem is that, in the expressions used, the intermediate states can easily represent higher energies.

Several attempts were made to arrive at a positive Hamiltonian that is also a space integral over local Hamiltonian densities. The problem is that the cellular automaton only defines a local evolution operator over finite time steps, not a local Hamiltonian that specifies infinitesimal time evolution. Generically valid formal procedures seem to fail, but if we stick closer to procedures that resemble perturbative quantum field theories, we seem to approach interesting descriptions that almost solve the problem. In chapter~\ref{quloc}  of part II, we use second quantisation. This procedure can be summarised as follows: consider first a cellular automaton that describes various types of particles, all non-interacting. This automaton will be integrable, and its Hamiltonian, \(H_0\), will certainly obey locality properties, and have a lower bound. Next, one must introduce interactions as tiny perturbations. This should not be difficult in cellular automata; just introduce small deviations from the free particle evolution law. These small perturbations, even if discrete and deterministic, can be handled perturbatively, assuming that the perturbation occurs infrequently, at sufficiently separated spots in space and time. This should lead to something that can reproduce perturbative quantum field theories such as the Standard Model.

Note, that in most quantum field theories, perturbation expansions have been used with much success (think for instance of the calculation of the anomalous magnetic moment \(g-2\) of the electron, which could be calculated and compared with experiment in superb precision), while it is still suspected that the expansions eventually do not converge. The non-convergence, however, sets in at very high orders, way beyond today's practical limits of accuracy in experiments.

We now believe that this will be the best way to construct a Hamiltonian with properties that can be compared to experimentally established descriptions of particles. But, it is only a strategy; it was not possible to work out the details because the deterministic free particle theories needed are not yet sufficiently well understood. 

Thus, there is still a lot of work to be done in this domain. The questions are technically rather involved, and therefore we postpone the details to part II of this book.

\subsubsection{The double role of the Hamiltonian\labell{doubleham}}
	Without a Hamiltonian, theoretical physics would look completely different. In classical mechanics, we have the central issue that a mechanical system obeys an energy conservation law. The energy is a non negative, additive quantity that is locally well-defined. It is these properties that guarantee stability of mechanical systems against complete collapse or completely explosive solutions.
	
	The classical Hamiltonian principle is a superb way to implement this mechanism. All that is needed is to postulate an expression for the non-negative, conserved quantity called energy, which turns into a Hamiltonian \(H(\vec x,\,\vec p\,)\) if we carefully define the dynamical quantities on which it depends, being canonical pairs of positions \(x_i\) and momenta \(p_i\). The ingenious idea was to take as equation of motion the Hamilton equations
	\be {\dd\over\dd t} x_i(t)={\pa\over\pa p_i}H(\vec x,\,\vec p\,)\,\ ,\qquad  {\dd\over\dd t} p_i(t)=-{\pa\over\pa x_i}H(\vec x,\,\vec p\,)\ . \eel{Hamequs}
This guarantees that \(\fract\dd {\dd t}H(\vec x,\,\vec p\,)=0\) . The fact that the equations~\eqn{Hamequs} allow for a large set of mathematical transformations makes the principle even more powerful.

In quantum mechanics, as the reader should know, one can use the same Hamiltonian function \(H\) to define a Schr\"odinger equation with the same property: the operator \(H\) is conserved in time. If \(H\) is bounded from below, this guarantees the existence of a ground state, usually the vacuum,

Thus, both quantum and classical mechanics are based on this powerful principle, where a single physical variable, the Hamiltonian, does two things: it generates the equations of motion, and it gives a locally conserved energy function that stabilises the solutions of the equations of motion. This is how the Hamiltonian principle describes equations of motion, or evolution equations,  whose solutions are guaranteed to be stable. 

Now how does this work in discrete, deterministic systems of the kind studied here? Our problem is that, in a discrete, classical system, the energy must also be discrete, but the generator of the evolution must be an operator with continuous eigenvalues. The continuous differential equations \eqn{Hamequs} must be replaced by something else. In principle, this can be done,  we could attempt to recover a continuous time variable, and derive how our system  evolves in terms of that time variable. What we really need is and \emph{operator} \(H\), that partly represents a positive, conserved energy, and partly a generator of infinitesimal time changes. We elaborate this issue further in part II, chapter~\ref{Hamiltonform}, where, among other things, we construct a classical, discretised Hamiltonian in order to apply a cellular automaton version of the Hamilton principle.

\subsubsection{The energy basis\labell{energy}} 
	
Beables are operators that commute at all times, and ontological states are eigenstates of these beables. There is a trivial example of such operators and such states in the real world: the Hamiltonian and its eigenstates. According to our definition they form a set of beables, but unfortunately, they are trivial: there is only one Hamiltonian, and the energy eigenstates do not change in time at all. This describes a static classical world. What is wrong with it?

Since we declared that superpositions of ontological states are not ontological, this solution also tells us that, if the energy eigenstates would be considered as ontological, no superpositions of these would be admitted, while all conventional physics only re-emerges when we do consider superpositions of the energy eigenstates. Only superpositions of different energy states can be time-dependent, so yes, this is a solution, but no, it is not the solution we want.
The energy basis solution emerges for instance if we take the model of section~\ref{generalfinite}, Figures~\ref{genmod.fig} and \ref{generalfinite.fig}, where we replace all loops by trivial loops, having only a single state, while putting all the physics in the arbitrary numbers \(\d E_i\). It is in accordance with the rules, but not useful.

Thus, the choice of the energy basis represents an extreme limit that is often not useful in practice. We see this also happen when a very important procedure is considered: it seems that we will have to split energy into two parts: on the one hand, there is a large, classical component, where energy, equivalent to mass, acts as the source of a gravitational field and as such must be ontological, that is, classical. This part must probably be discretized. On the other hand, we have the smaller components of the energy that act as eigen values of the evolution operator, over sufficiently large time steps (much larger than the Planck time). These must form a continuous spectrum. 

If we would consider the energy basis as our ontological basis, we regard \emph{all} of the energy as classical, but then the part describing evolution disappears; that is not good physics. See Part II, Figure ~\ref{PQdomains.fig}, in subsection~\ref{oned}. The closed contours in that figure must be non-trivial.

\subsection{Miscellaneous}
\subsubsection{The Earth--Mars interchange operator\labell{EarthMars}} The CAI surmises that quantum models exist that can be regarded as classical systems in disguise. If one looks carefully at these classical systems, it seems as if \emph{any} classical system can be rephrased in quantum language: we simply postulate an element of a basis of Hilbert space to correspond to every classical physical  state  that is allowed in the system. The evolution operator is the permutator that replaces a state by its successor in time, and we may or may not decide later to consider the continuous time limit.

Naturally, therefore, we ask the question whether one can reverse the CAI, and construct quantum theories for systems that are normally considered classical. The answer is yes. To illustrate this, let us consider the planetary system. It is the prototype of a classical theory. We consider large planets orbiting around a sun, and we ignore non-Newtonian corrections such as special and general relativity, or indeed the actual quantum effects, all of which being negligible corrections. We regard the planets as point particles, even if they may feature complicated weather patterns, or life; we just look at their classical, Newtonian equations of motion.

The ontological states are defined by listing the positions \(\vec x_i\) and velocities \(\vec v_i\) of the planets (which commute), and the observables describing them are the beables. Yet also this system allows for the introduction of changeables and superimposables. The quantum Hamiltonian here is not the classical thing, but
	\be H^{\mathrm{quant}}=\sum_i\bigg(\vec p_{x,\,i}^{\ \op}\cdot \vec v_i+\vec p_{v,\,i}^{\ \op}\cdot\vec F_i(\mathbf x)/m_i\bigg)\ , \eel{planetham}
where
	\be 	\vec p_{x,\,i}^{\ \op}  =-i{\pa\over\pa \vec x_i}\ ,\qquad\vec p_{v,\,i}^{\ \op}=-i{\pa\over\pa\vec v_i}\ ,
	\qquad\hbox{and }\ \mathbf x=\{\vec x_i\}\ . \eel{planetmoms}
Here, \(\vec F_i(\mathbf x)\) are the classical forces on the planets, which depend on all positions. \\ Eq.~\eqn{planetham} can be written more elegantly as
	\be H^{\mathrm{quant}}=\sum_i\bigg(\vec p_{ x,\,i}^{\ \op}\cdot{\pa H^{\mathrm{class}}\over\pa \vec p_i}-
	\vec p_{ p,\,i}^{\ \op}\cdot{\pa H^{\mathrm{class}}\over\pa\vec x_i}\bigg)\ , \eel{planethamformal}
where \(p_{ p,\,i}^{\ \op}=m_i^{-1}p_{ v,\,i}^{\ \op}\). 
Clearly, \(\vec p_{x\,i}^{\ \op}\), \(\vec p_{v\,i}^{\ \op}\) and \(\vec p_{p\,i}^{\ \op}\) are infinitesimal changeables, and so is, of course, the Hamiltonian \(H^{\mathrm{quant}}\). The planets now span a Hilbert space and behave as if they were quantum objects. We did not modify the physics of the system.

We can continue to define more changeables, and ask how they evolve in time. One of the author's favourites is the \emph{Earth--Mars interchange operator}. It puts the Earth where Mars is now, and puts Mars where planet Earth is. The velocities are also interchanged\fn{We disregarded the moons for a moment; we could drag the moons along as well, or keep them where they are. They don't play a role in this argument.}. If Earth and Mars had the same mass then the planets would continue to evolve as if nothing happened. Since the masses are different however, this operator will have rather complicated properties when time evolves. It is not conserved in time.

The eigenvalues of the Earth--Mars interchange operator \(X_{EM}\) are easy to calculate:
	\be X_{EM}=\pm 1\ ,\eel{EMeigen}
simply because the square of this operator is one.
In standard QM language, \(X_{EM}\) is an observable. It does not commute with the Hamiltonian because of the mass differences, but, at a particular moment, \(t=t_1\), we can consider one of its eigenstates and ask how it evolves. 

Now why does all this sound so strange? How do we observe \(X_{EM}\)? No-one can physically interchange planet Earth with Mars. But then, no-one can interchange two electrons either, and yet, in quantum mechanics, this is an important operator. The answer to these questions is, that regarding the planetary system, we happen to know what the beables are: they are the positions and the velocities of the planets, and that turns them into ontological observables. The basis in which these observables are diagonal operators is our preferred basis. The elements of this basis are the ontological states of the planets. If, in a quantum world, investigators one day discover what the ontological beables are, everything will be expressed in terms of these, and any other operators are no longer relevant.

It is important to realise that, in spite of the fact that, in Copenhagen language,  \(X_{EM}\)  is an observable, we cannot measure it, to see whether it is \(+1\) or \(-1\). This is because we know the wave function \(|\ont\ket\). It is 1 for the actual positions of Earth and Mars, 0 when we interchange the two. This is the superposition 
	\be |\ont\ket=\fract 1{\sqrt 2}\bigg(|\,X_{EM}=1\,\ket +|\,X_{EM}=-1\,\ket \bigg)\ ; \ee
which is a superposition of two template states. According to Copenhagen, a measurement would yield \(\pm 1\) with \(50\%/50\%\) chances.

\subsubsection{Rejecting local counterfactual definiteness and free will\labell{counterfactualwill}}
	The arguments usually employed to conclude that local hidden variables cannot exist, begin with assuming that such hidden variables should imply local counterfactual definiteness. One imagines a set-up such as the EPR-Bell experiment that we exposed in section~\ref{Bell}. Alice and Bob are assumed to have the `free will' to choose the orientation of their polarisation filters anytime and anyway they wish, and there is no need for them to consult anyone or anything to reach their (arbitrary) decision. The quantum state of the photon that they are about to investigate should not depend on these choices, nor should the state of a photon depend on the choices made at the other side, or on the outcome of those measurements.
	
	This means that the outcomes of measurements should already be determined by some algorithm long before the actual measurements are made, and also long before the choice was made what to measure. It is this algorithm that generates conflicts with the expectations computed in a simple quantum calculation. It is counterfactual, which means that there may be one `factual' measurement but there would have been many possible alternative measurements, measurements that are not actually carried out, but whose results should also have been determined. This is what is usually called counterfactual definiteness,  and it has essentially been disproven by simple logic.
	
	Now, as has been pointed out repeatedly, the violation of counterfactual definiteness is not at all a feature that is limited to quantum theory. In our example of the planetary system, subsection~\ref{EarthMars}, there is no a priori answer to the question which of the two eigenstates of the Earth-Mars exchange operator, the eigenvalue \(+1\) or \(-1\), we are in. This is a forbidden, counterfactual question. But in case of the planetary system, we know what the beables are (the positions and velocities of the planets), whereas in the Standard Model we do not know this. There, the illegitimacy of counterfactual statements is not immediately obvious. In essence, we have to posit that Alice and Bob do not have the free will to change the orientation of their filters; or if we say that they do,  their decisions must have their roots in the past and, as such, they affect the possible states a photon can be in. In short, Alice and Bob make decisions that are correlated with the polarisations of the photons, as explained in section~\ref{Bell}.

\subsubsection{Entanglement and superdeterminism\labell{entangle}} 
	An objection often heard against the general philosophy advocated in this work, is that it would never enable accommodation for entangled particles. The careful reader, however, must realise by now that, in principle, there should be no problem of this sort. Any quantum state can be considered as a template, and the evolution of these templates will be governed by the real Schr\"odinger equation. If the relation between the ontological basis and the more conventional basis choices is sufficiently complex, we will encounter superimposed states of all sorts, so one definitely also expects states where particles behave as being `quantum-entangled'.
	
	Thus, in principle, it is easy to write down orthonormal transformations that turn ontic states into entangled template states.
	
	There are some problems and mysteries, however. The EPR paradox and Bell's theorem are examples. As explained in section~\ref{Bell}, the apparent contradictions can only be repaired if we assume rather extensive correlations among the `hidden variables' everywhere in the Universe. The mapping of ontic states into entangled states appears to depend on the settings chosen by Alice and Bob in the distant future.
	
	It seems as if \emph{conspiracy} takes place: due to some miraculous correlations in the events at time\,\(t=\,0\), a pair of photons `knows in advance' what the polarisation angles of the filters will be that they will encounter later, and how they should pass through them. Where and how did this enter in our formalism, and how does a sufficiently natural system, without any conspiracy at the classical level, generate this strange behaviour?
	
	It is not only a feature among entangled particles that may appear to be so problematic. The conceptual difficulty is already manifest at a much more basic level. Consider a single photon, regardless whether it is entangled with other particles or not. Our description of this photon in terms of the beables dictates that these beables behave classically. What happens later, at the polarisation filter(s), is also dictated by classical laws. These classical laws in fact dictate how myriads of variables fluctuate at what we call the Planck scale, or more precisely, the smallest scale where distinguishable physical degrees of freedom can be recognised, which may or may not be close to what is usually called the Planck scale. Since entangled particles occur in real experiments, we claim that the basis-transformations will be sufficiently complex to transform states that are ontic at the Planck scale, into entangled states.
	
	But this is not the answer to the question posed. The usual way to phrase the question is to ask how `information' is passed on. Is this information classical or quantum? If it is true that the templates are fundamentally quantum templates, we are tempted to say, well, the information passed on is quantum information. Yet it does reduce to classical information at the Planck scale, and this was generally thought not to be possible.  
	
	That must be a mistake. As we saw in section~\ref{Bell}, the fundamental technical contradiction goes away if we assume strong correlations between the quantum fluctuations -- including the vacuum fluctuations -- at all distance scales (also including correlations between the fluctuations generated in quasars that are separated by billions of light years). We think the point is the following. When we use templates, we do not know in advance which basis one should pick to make them look like the ontological degrees of freedom as well as possible. For a photon going through a polarisation filter, the basis closest to the ontological one is the basis where coordinates are chosen to be aligned with the filter. But this photon may have been emitted by a quasar billions of years ago, how did the quasar know what the ontological basis is?
	
	The answer is that indeed the quasar knows what the ontological basis is, because our theory extends to these quasars as well. The information, `this is an ontological state, and any set of superimposed states is not', is a piece of information that, according to our theory, is absolutely conserved in time. So, if that turns out to be the basis now, it was the basis a billion years ago. The quasars seem to conspire in a plot to make fools of our experimenters, but in reality they just observe a conservation law: \emph{the information as to which quantum states form the ontological basis, is conserved in time}. Much like the law of angular momentum, or any other exactly conserved entity, this conservation law tells us what this variable is in the future if it is known in the past, and \emph{vice-versa}.
	
	We must conclude that, if there seems to be conspiracy in our quantum description of reality, then that is to be considered as a feature of our quantum techniques, not of the physical system we are looking at. There is no conspiracy in the classical description of the cellular automaton. Apparent conspiracy is a feature, not a bug.
	
	The answer given here, is often called \emph{superdeterminism}. It is the idea that indeed Alice and Bob can only choose ontological states to be measured, never the superimposed states, which we use as templates. In a sense, their actions were predetermined, but of course in a totally unobservable way. Superdeterminism only looks weird if one adheres to the description of entangled particles as being quantum systems, described by their quantum states. The ontological description does not use quantum states. In that description, the particles behave in a normal, causal manner. However, we do have to keep in mind that these particles, and everything else, including Bob and Alice's minds, are all strongly correlated. They are correlated now as strongly as when the photons were emitted by the source, as was laid down\fn{The mouse-dropping function does not show a 100\% correlation. This is a feature due to the somewhat artificial nature of the model used for the detection. If we use more realistic models for the detection, the mouse-dropping function might consist of sharper peaks.} in the mouse-dropping function, Fig.~\ref{mousedrop.fig} in subsection~\ref{mousedr}.
	
	In subsection \ref{will}, it is explained in explicitly physical terms, what `free will' should really stand for, and why superdeterminism can clash with it.
	
\subsubsection{The superposition principle in Quantum Mechanics\labell{superposition}}
	What exactly happened to the superposition principle in the CA Interpretation of quantum mechanics? Critics of our work brought forward that the CAI disallows superposition, while obviously the superposition principle is serving quite well as a solid back bone of quantum mechanics. Numerous experiments confirm that if we have two different states, also any superposition of these states can be realised. Although the reader should have understood by now how to answer this question, let us attempt to clarify the situation once again.
	
	At the most basic level of physical law, we assume only ontological states to occur, and any superposition of these, in principle, does not correspond to an ontological state. At best, a superposition can be used to describe probabilistic distributions of states (we call these ``template  states", to be used when we do not have the exact information at hand to determine with absolute certainty which ontological state we are looking at). In our description of the Standard Model, or any other known physical system such as atoms and molecules, we do not use ontological states but templates, which can be regarded as superpositions of ontological states. The hydrogen atom is a template, all elementary particles we know about are templates, and this means that the wave function of the universe, which is an ontological state, must be a superposition of our templates. \emph{Which} superposition? Well, we will encounter many different superpositions when doing repeated experiments. This explains why we were led to believe that all superpositions are always allowed.
	
	But not literally all superpositions can occur. Superpositions are man-made. Our templates are superpositions, but that is because they represent only the very tiny sector of Hilbert space that we understand today. The entire universe is in only one ontological state at the time, and it of course cannot go into superpositions of itself. This fact now becomes manifestly evident when we consider the ``classical limit". In the classical limit we again deal with certainties. \emph{Classical states are also ontological}. When we do a measurement, by comparing the calculated ``template  state" with the ontological classical states that we expect in the end, we recover the probabilities by taking the norm squared of the amplitudes. Classical states also never go into superpositions of classical states. Such superpositions never occur. 
	
	It appears that for many scientists this is difficult to accept. During a whole century we have been brainwashed with the notion that superpositions occur everywhere in quantum mechanics. At the same time we were told that if you try to superimpose classical states, you will get probabilistic distributions instead. It is here that our present theory is more accurate: if we knew the wave function of the universe exactly, we would find that it always evolves into one classical state only, without uncertainties and without superpositions. 
	
	Of course this does not mean that standard quantum mechanics would be wrong. Our knowledge of the template states, and how these evolve, is very accurate today. It is only because it is not yet known how to relate these template states to the ontological states, that we have to perform superpositions all the time when we do quantum mechanical calculations. They do lead to statistical distributions in our final predictions, rather than certainties. This could only change if we would find the ontological states, but since even the vacuum state is expected to be a template, and as such a complicated superposition of uncountably many ontic states, we should expect quantum mechanics to stay with us forever -- but as a mathematical tool, not as a mystic departure from classical logic.
	
	\subsubsection{The vacuum state\labell{vac}}
Is the vacuum state an ontological state? The vacuum state is generally defined to be the state with lowest energy. This also means that no particles can be found in this state, simply because particles represent energy, and, in the vacuum state,  we have not enough energy even to allow for the presence of a single particle.

The discretised Hamiltonian is only introduced in section \ref{Hamiltonform} of part II. It is a beable, but being discrete it can only be a rough approximation of the quantum Hamiltonian at best, and its lowest energy state is highly degenerate. As such, this is not good enough to serve as a definition of a vacuum.  More to the point, the Hamiltonian defined in section~\ref{discHamproblem} is quantised in units that seem to be as large as the Planck mass. It will be clear that the Hamiltonian to be used in any realistic Schr\"odinger equation has a much more dense, basically continuous, eigenvalue spectrum. The quantum Hamiltonian is definitely not a beable, as we explained earlier above, in section~\ref{energy}. Therefore, the vacuum is not an ontological state.

In fact, according to quantum field theories, the vacuum contains many \emph{virtual} particles or particle-antiparticle pairs that fluctuate in- and out of existence everywhere in space-time. This is typical for quantum superpositions of ontological states. Furthermore, the lightest particles in our theories are much lighter than the Planck mass. They are not ontological, and demanding them to be absent in our vacuum state inevitably turns our vacuum itself also into a non-ontological state. 

This is remarkable, because our vacuum state has one more peculiar property: its energy density itself is almost perfectly vanishing. Due to the cosmological constant, there is energy in our vacuum state, but it is as small as about 6 protons per \(m^3\), an extremely tiny number indeed, considering the fact that most length scales in particle physics are indeed much smaller than a metre. This very tiny but non-vanishing number is one of Nature's greater mysteries.

And yet, the vacuum appears to be non-ontological, so that it must be a place full of activity. How to reconcile all these apparently conflicting features is not completely understood.

The vacuum fluctuations may be seen as one of the primary causes of non-vanishing, non-local correlations, such as the mouse dropping function	of section~\ref{Bell}. Without the vacuum fluctuations, it would be difficult to understand how these correlations could be sustained. 

\subsubsection{Exponential decay\labell{expodecay}}

Vacuum fluctuations must be the primary reason why isolated systems such as atoms and molecules in empty space, show typical quantum features. A very special quantum mechanical property of many particles is the way they can decay into two or more other particles. Nearly always, this decay follows a perfect exponential decay law: the probability \(P(t)\) that the particle of a given type has not yet decayed after elapsed time \(t\) obeys the rule
	\be {\dd P(t)\over\dd t}=-\l P(t)\ , \quad\ra\quad P(t)=P(0)e^{-\l t}\ ,\eel{expolaw}
where \(\l\) is a coefficient that often does not at all depend on external circumstances, or on time. If we start off with \(N_0\) particles, then the expectation value \(\bra N(t) \ket\) of the particle number \(N(t)\) after time \(t\) follows the same law: 
	\be \bra N(t)\ket=N_0\,e^{-\l t}\ . \eel{Ndecay}

If there are various modes in which the particle can decay, we have \(\l=\l_1+\l_2+\dots\,,\)  and the ratios of the \(\l_i\) equals the ratios of the modes of the decays observed.

Now how can this be explained in a deterministic theory such as a cellular automaton? In general, this would not be possible if the vacuum would be a single ontological state. Consider particles of a given type. Each individual particle will decay after a different amount of time, precisely in accordance to Eq.~\eqn{expolaw}. Also, the directions in which the decay products will fly will be different for every individual particle, while, if there are three or more decay products involved, also the energy of the various decay products will follow a probability distribution. For many existing particles, these distributions can be accurately calculated following the quantum mechanical prescriptions.

In a deterministic theory, all these different decay modes would have to correspond to distinct initial states. This would be hopeless to accommodate for if every individual particle would have to behave like a `glider solution' of a cellular automaton, since all these different decay features would have to be represented by different gliders. One would quickly find that millions, or billions of different glider modes would have to exist.

The only explanation of this feature must be that the particles are surrounded by a vacuum that is in a different ontological state every time. A radio-active particle is continuously hit by fluctuating features in its surrounding vacuum. These features represent information that flies around, and as such, must be represented by almost perfect random number generators. The decay law \eqn{expolaw} thus emerges naturally.

Thus it is inevitable that the vacuum state has to be regarded as a \emph{single template state}, which however is composed of \emph{infinitely many ontological states}. The states consisting of a single particle roaming in a vacuum form a simple set of different template states, all orthogonal to the vacuum template state, as dictated in the Fock space description of particle states in quantum field theory.

\subsubsection{A single photon passing through a sequence of polarisers\labell{singlephoton}}

It is always illustrative to reduce a difficulty to its most basic form. The conceptual difficulty one perceives in Bell's gedanken experiment, already shows up if we consider a single photon, passing through a sequence of polarisation filters, \(F_1,\,\dots,\ F_N\). Imagine these filters to be rotated by angles \(\vv_1,\ \vv_2,\ \dots,\ \vv_N\), and every time a photon hits one of these filters, say \(F_n\),  there is a probability equal to \(\sin^2\j_n\), with \(\j_n=\vv_{n-1}-\vv_n\), that the photon is absorbed by this filter. Thus, the photon may be absorbed by any of the \(N\) polarisers. What would an ontological description of such a setup be like?

Note that, the setup described here resembles our description of a radio-active particle, see the previous subsection. There, we suggested that the particle is continuously interacting with the surrounding vacuum. Here, it is easiest to assume that the photon interacts with all filters. The fact that the photon arrives at  filter  \(F_n\) as a superposition of two states, one that will pass through and one that will be absorbed, means that, in the language of the ontological theory, we have an initial state that we do not quite know; there is a probability \(\cos^2\j_n\)  that we have a photon of the pass-through type, and a probability \(\sin^2\j_n\) that it is of the type that will be absorbed. If the photon passes through, its polarisation is again well-defined to be \(\vv_n\). This will determine what the distribution is of the possible states that may or may not pass through the next filter.

We conclude that the filter, being essentially classical, can be in a very large number of distinct ontological states. The simplest of all ontological theories would have it that a photon arrives with polarisation angle \(\j_n\) with respect to the filter. Depending on the ontological state of the filter, we have probability \(\cos^2\j\) that the photon is allowed through, but rotated to the direction \(\vv_n\), and probability \(\sin^2\j\) that it is absorbed (or reflected and rotated).

Now, return to the two entangled photons observed by Alice and Bob in Bell's set-up. In the simplest of all ontological theories, this is what happens:
Alice's and Bob's filters act exactly as above. The two photons carry both the same information in the form of a single angle, \(c\). Alice's filter has angle \(a\), Bob's has angle \(b\).
As we saw in subsection~\ref{mousedr}, there is a 3-point correlation between \(a\), \(b\) and \(c\), given by the mouse-dropping function, \eqn{bellprobl} and Fig.~\ref{mousedrop.fig}. 

Now note, that the mouse-dropping function is invariant under rotations of \(a\), \(b\) and/or \(c\) by \(90^\circ\).  The nature of the ontological state depends very precisely on the angles \(a\), \(b\), and \(c\), but each of these states differs from the others by multiples of \(90^\circ\) in these angles. Therefore, 
as soon as we have picked the desired orthonormal basis, the basis elements will be entirely uncorrelated. This makes the correlations unobservable whenever we work with the templates. Assuming the ontological conservation law at work here, we find that the ontological nature of the angles \(a\), \(b\) and \(c\) is correlated, but not the physical observables. It is to be expected that correlations of this sort will pervade all of physics.

\subsection{The quantum computer\labell{qucomp}}
	Quantum mechanics is often endowed with mysterious features. There are vigorous attempts to turn some of these features to our advantage. One famous example is the quantum computer. The idea is to use entangled states of information carriers, which could be photons, electrons, or something else, to represent vastly more information than ordinary bits and bytes, and are therefore called qubits.
		 
	Since the machines that investigators plan to construct would obey ordinary quantum mechanics, they should behave completely in accordance with our theories and models. However, this seems to lead to contradictions. 
	
In contrast with ordinary computers, the amount of information that can be carried by qubits in a quantum computer, in principle, increases exponentially with the number of cells, and consequently, it is expected that quantum computers will be able to perform calculations that are fundamentally impossible in ordinary computers. An ordinary, classical computer would never be able to beat a quantum computer even if it took the size of the universe, in principle.

Our problem is then, that our models underlying quantum mechanics are classical, and therefore they can be mimicked by classical computers, even if an experimentalist would build a `quantum computer' in such a world. Something is wrong.

Quantum computers still have not been constructed however. There appear to be numerous practical difficulties. One difficulty is the almost inevitable phenomenon of decoherence. For a quantum computer to function impeccably, one needs to have \emph{perfect} qubits. 

It is generally agreed that one cannot make perfect qubits, but what can be done is correct them for the errors that may sometimes occur. In a regular computer, errors can easily be corrected, by using a slight surplus of information to check for faulty memory sites. Can the errors of qubits also be corrected? There are claims that this can be done, but in spite of that, we still don't have a functioning quantum computer, let alone a quantum computer that can beat all classical computers. Our theory comes with a firm prediction:
\begin{quote} \emph{Yes, by making good use of quantum features, it will be possible in principle, to build a computer vastly superior to conventional computers, but no, these will not be able to function better than a classical computer would do, if its memory sites would be scaled down to one per Planckian volume element (or, in view of the holographic principle, one memory site per Planckian surface element), and if its processing speed would increase accordingly, typically one operation per Planckian time unit of \(10^{-43}\) seconds.} \end{quote}
Such scaled classical computers can of course not be built, so that this quantum computer will still be allowed to perform computational miracles, but factoring a number with millions of digits into its prime factors will not be possible -- unless fundamentally improved classical algorithms turn out to exist.
If engineers ever succeed in making such quantum computers, it seems to me that the CAT is falsified; no classical theory can explain quantum mechanics.

\newsecl{Quantum gravity}{grav} \def\Pl{\mathrm{Pl}}

The \emph{Planck scale} has been mentioned many times already. It is the scale of time, lengths, masses, and energies, where three grand physical theories all play equally significant roles, being special relativity (where the speed of light \(c\) is essential), quantum mechanics (with Planck's constant \(\hbar\)) and Newton's theory of gravity (with Newton's constant \(G\)). Having
	\be c&=&299\,792\,458 \ m/s\ ,\nm \\ 
		\hbar&=&1.05457\times 10^{-34}\ \mathrm{kg}\,m^2/s\ ,\nm \\
	 	G&=&6.674\times 10^{-11}\  m^3\mathrm{kg}^{-1}s^{-2}\ , \eel{consts} one finds\\[-25pt]
\be \hbox{the Planck length,}&L_\Pl=\sqrt{{G\,\hbar\over c^3}}\ &=\ 1.616\times10^{-35}\ m\ , \\
\hbox {the Planck time,}& T_\Pl=\sqrt{G\,\hbar\over c^5}\ &=\ 5.391\times 10^{-44}\ s\ ,   \\
\hbox{the Planck mass,}&M_\Pl=\sqrt{c\,\hbar\over G}\ &=\  21.76\ \m g\ , \\
\hbox{and the Planck energy,}&E_\Pl=\sqrt{c^5\,\hbar\over G}\ &=\ 1.956\times 10^9\ J\ . \eel{Plunits}

In this domain of physics, one expects Special and General Relativity and Quantum Mechanics all to be relevant, but a complete synthesis of these three has not yet been achieved -- in fact, our continued struggle towards finding such a synthesis was one of the main motivations for this work.

It is not unreasonable to suspect that the Planck length is the smallest significant length scale in physics, and the Planck time is the smallest time scale at which things can happen, but there is more. General Relativity is known to cause space and time to be curved, so,  if one might talk of some ``lattice" in space and time, curvature may be expected to cause defects in this lattice. Alternatively, one might suspect that lattice-like behaviour can also be realised by imposing a cutoff in local momentum and energy scales (a so-called bandwidth cut-off\,\cite{Kempf-2000}); however, with such a cut-off deterministic models are difficult to construct.

It is also important to note that General Relativity is based on the local automorphism group. This means that time translations are locally defined, so that one may expect that gravity could be essential to realise locality requirements for the Hamiltonian. Mass, energy and momentum are local sources of gravitational fields, but there is more.

Gravitation is a destabilising force. Causing masses to attract one another, it generates greater masses and thus even stronger attraction. This may lead to gravitational implosion. In contrast, electric as well as magnetic charges act repulsively (if they have equal signs), which makes electromagnetism a lot more stable than gravity as a force system.

When gravitational implosion takes place, black holes may form. \emph{Microscopic black holes} must play an essential role at the Planck scale, as they may act as virtual particles, taking part in the vacuum fluctuations. When one tries to incorporate black holes in an all-embracing theory, difficulties arise. According to standard calculations, black holes emanate elementary particles, and this effect (Hawking effect\,\cite{Hawking-1975}) allows one to compute the density of quantum states associated to black holes. This density is very large, but as black holes increase in size, the number of states does not grow as fast as one might expect: it grows exponentially with the size of the surface, rather than the encapsulated volume. The quantum states that one might expect in the \emph{bulk} of a black hole mysteriously disappear, 

We expect all this to produce a profound effect on the putative deterministic models that could possibly lie at the basis of quantum theory. Discreteness of space and time comes for free, because one can also argue that the number of quantum states inside a volume \(V\) can never exceed that of a black hole occupying \(V\), so that the surface at the border of \(V\) dictates how many independent ontological states are allowed inside \(V\), an effect called the `holographic principle'\,\cite{Suss-1995}\cite{GtH-Salam93}. Locality may come naturally because of the automorphism group as mentioned. Yet space-time curvature causes problems. Nature's book keeping system is still very ill-understood.
	
	\newsecl{Information loss}{infoloss}
	
Gravity is perhaps not the only refinement that may guide us towards better models. An other interesting modification -- though possibly related --  might be of help.	We shall now discuss \emph{information loss}.\,\cite{BJV-2000}\cite{GtHdisdet-1999}

\subsecl{Cogwheels with information loss}{coginfoloss}
Let us return to the Cogwheel Model, discussed in section~\ref{Cogwheelm}. The most general automaton may have the property that two or more different initial states evolve into the same final state. For example, we may have the following evolution law involving 5 states:
	\be (4)\ra(5)\ra(1)\ra(2)\ra(3)\ra(1)\ . \eel{infolosscog}
The diagram for this law is a generalisation of Fig.~\ref{threestate.fig}, now shown in Fig.~\ref{infolosscog.fig}. We see that in this example state \#3 and state \#5 both evolve into state \#1. 
	\begin{figure}[htb!]  \begin{quotation}\begin{center}
	$a)$\lowerheightfig{0pt}{20mm}{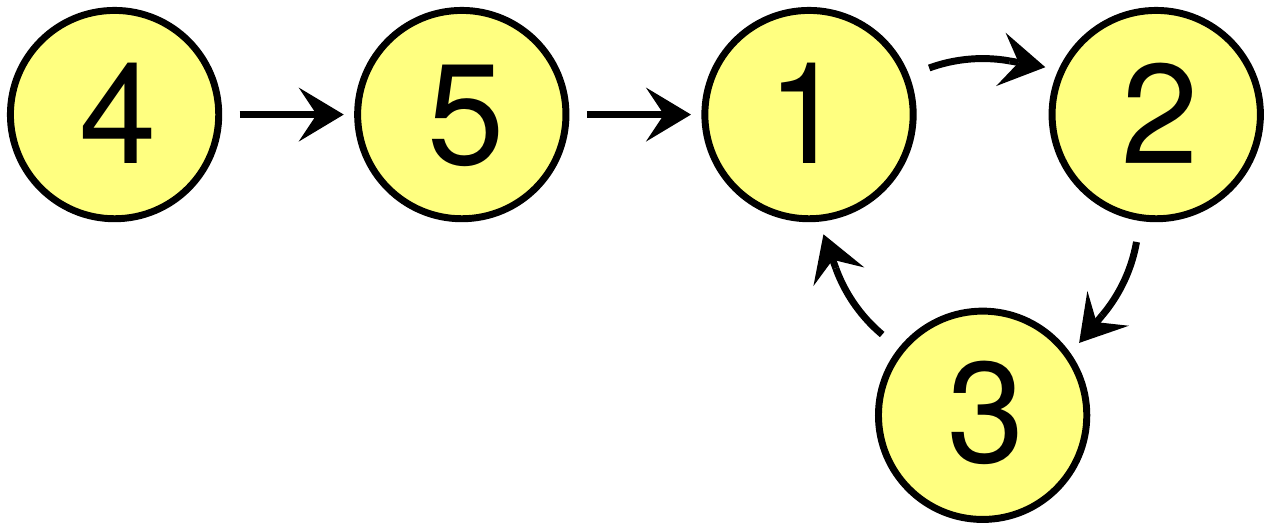}\qquad
	$b)$\lowerheightfig{0pt}{20mm}{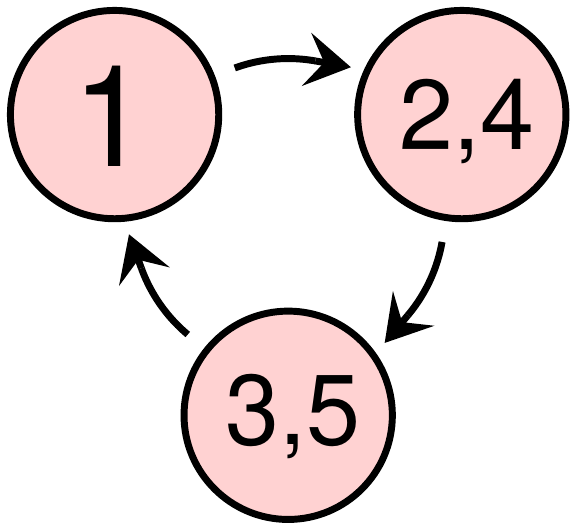}\qquad
	$c)$\lowerwidthfig{0pt}{20mm}{CAQu_clockham.pdf}
 \caption{\small $a)$ Simple 5-state automaton model with information loss.  $b)$ Its three equivalence classes.  $c)$ Its three energy levels.\labell{infolosscog.fig}}
\end{center} \end{quotation}
		\end{figure}

At first sight, one might imagine to choose the following operator as its evolution operator:
	\be U(\d t)\qu\pmatrix{0&0&1&0&1\cr 1&0&0&0&0\cr 0&1&0&0&0 \cr 0&0&0&0&0 \cr
		0&0&0&1&0}\ . \eel{infolossU}
However, since there are two states that transform into state \#1 whereas there are none that transform into state \#4, this matrix is not unitary, and it cannot be written as the exponent of \(-i\, \) times an Hamiltonian.

One could think of making tiny modifications in the evolution operator \eqn{infolossU}, since only infinitesimal changes suffice to find some sort of (non hermitean) Hamiltonian of which this then would be the exponent. This turns out not to be such a good idea. It is better to look at the physics of such models. Physically, of course, it is easy to see what will happen. States \#4 and 5 will only be realised once if ever. As soon as we are inside the cycle, we stay there. So it makes more sense simply to delete these two rather spurious states.

The problem with that is that, in practice, it might be quite difficult to decide which states are inside a closed cycle, and which states descend from a state with no past (``gardens of Eden"). Here, it is the sequence  \#4, \#5. but in many more realistic examples, the gardens of Eden will be very far in the past, and difficult to trace. Our proposal is, instead, to introduce the concept of \emph{information equivalence classes} (info-equivalence classes for short):
	\begin{quote} \emph{Two states \((a)\) and \((b)\) at time \(t_0\) are equivalent if there exists a time \(t_1>t_0\) such that, at time \(t_1\), state \((a)\) and \((b)\) have evolved into the same state \((c)\)}. \end{quote}
	This definition sends states \#5 and \#3 in our example into one equivalence class, and therefore also states \#4 and \#2 together form an equivalence class. Our example has just 3 equivalence classes, and these classes do evolve with a unitary evolution matrix, since, by construction, their evolution is time-reversible.
		Info-equivalence classes will show some resemblance with \emph{gauge equivalence classes}, and they may well actually be quite large. Also, the concept of \emph{locality} will be a bit more difficult to understand, since states that locally look quite different may nevertheless be in the same class. Of course, the original underlying classical model may still be completely local. Our pet example is Conway's \emph{game of life}\,\cite{conway-1970}: an arbitrary configuration of ones and zeros arranged on a two-dimensional grid evolve according to some especially chosen evolution law. The law is not time-reversible, and information is lost at a big scale. Therefore, the equivalence classes are all very big, but the total number of equivalence classes is also quite large, and the model is physically non-trivial.
		An example of a more general model with information loss is sketched in Figure~\ref{infolossgeneral.fig}. We see many equivalence classes that each may contain variable numbers of members.
			
				\begin{figure}[htb!] \begin{quotation}
\begin{center} \lowerwidthfig{0pt} {110mm}{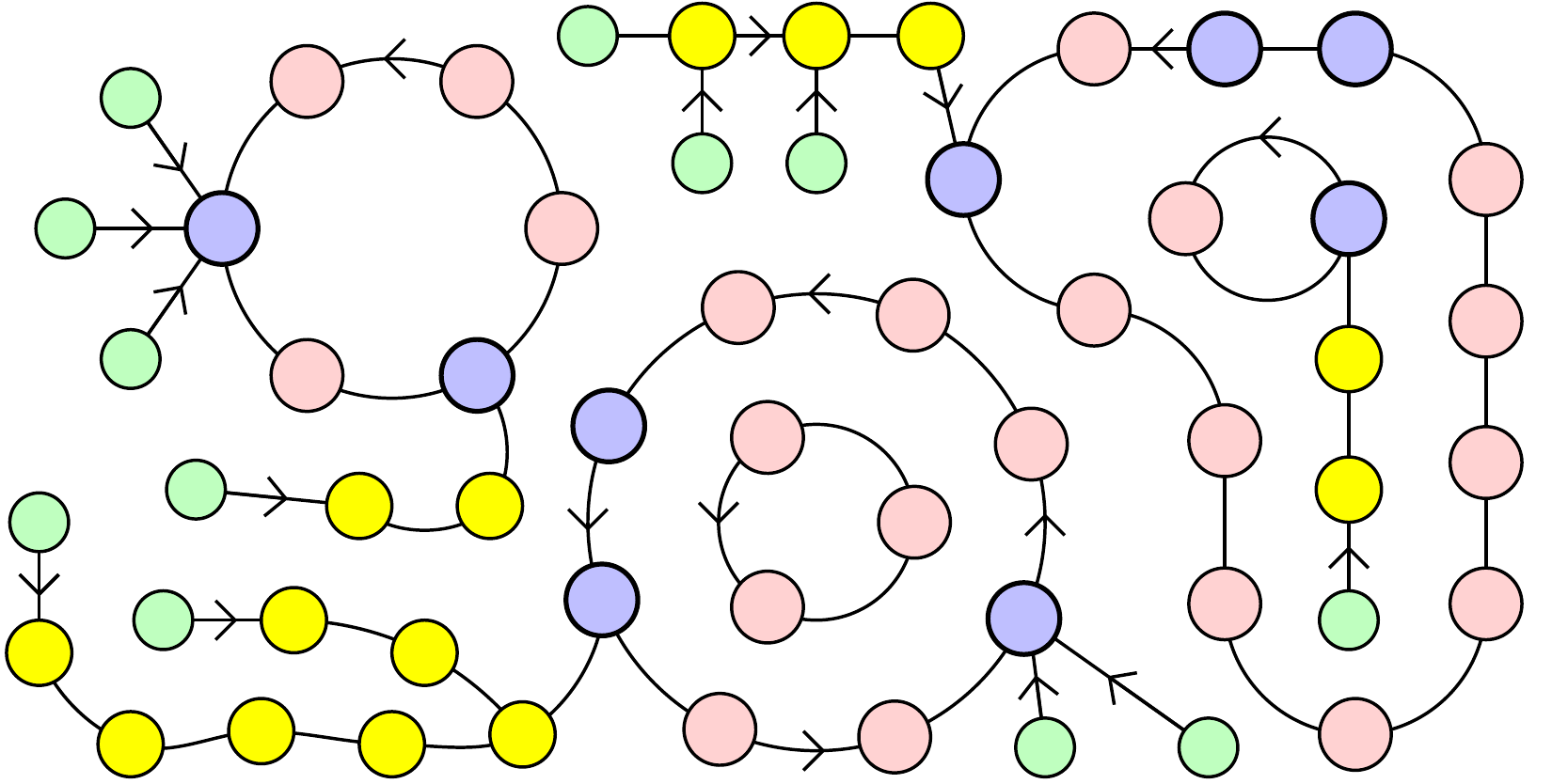}
\caption[\small Example of a more generic finite, deterministic, time non reversible model. ]{\small Example of a more generic finite, deterministic, time non reversible model. Largest 
(pink and blue) dots: these also represent the equivalence classes. Smallest (green) dots: ``gardens of Eden". Heavier dots (blue): equivalence classes that have ``merge sites" among their members. The info-equivalence classes and the energy spectrum are as in Figs.~\ref{genmod.fig} and 
\ref{generalfinite.fig}.\labell{infolossgeneral.fig}}\end{center}\end{quotation}
		\end{figure}	

Thus, we find how models with information loss may still be connected with quantum mechanics:
\begin{quote}\emph{Each info-equivalence class corresponds to an element of the ontological basis of a quantum theory.}\end{quote}

Can information loss be helpful? Intuitively, the idea might seem to be attractive. Consider the measurement process, where bits of information (``qubits") that originally were properties of single particles, are turned into macroscopic observables. These may be considered as being later in time, and all mergers that are likely to happen must have taken place. In other words, the classical states are obviously represented by the equivalence classes. However, when we were still dealing with individual qubits, the mergers have not yet taken place, and the equivalence classes may form very complex, in a sense ``entangled", sets of states. Locality is then difficult to incorporate in the quantum description, so,  in these models, it may be easier to expect  some rather peculiar features regarding locality -- perhaps just the thing we need.

What we also need is a better understanding of black holes. The idea that black holes, when emitting Hawking radiation, do still obey quantum unitarity, which means that the Hamiltonian is still hermitean, is gaining in acceptance by researchers of quantum gravity, even among string theorists. On the other hand, the classical black hole is surrounded by a horizon from which nothing seems to be able to escape. Now, we may be able to reconcile these apparently conflicting notions: the black hole is an example of a system with massive amounts of information loss at the classical level, while the quantum mechanics of its micro-states is nevertheless unitary. The micro states are not the individual classical states, but merely the equivalence classes of classical states. According to the holographic principle, these classes are distributed across the horizon in such a way that we have one bit of information for each area segment of roughly the Planck length squared. We now interpret this by saying that all information passing through a horizon disappears, with the exception of one bit per unit horizon area.

We return to Hawking radiation in part II, section~\ref{holo} .

\subsection{Time reversibility of theories with information loss\labell{timereverse}}

Now when we do quantum mechanics, there happens to be an elegant way to restore time reversibility. Let us start with the original evolution operator \(U(\d t)\),  such as the one shown in Eq.~\eqn{infolossU}. It has no inverse, but, instead of \(U^{-1}\), we could use \(U^\dag\) as the operator that brings us back in time. What it really does is the following: the operator \(U^\dag(\d t)\), when acting on an ontological state \(|\ont(t_0)\ket\) at time \(t_0\), gives us the additive quantum superposition of all states in the past of this state, at time \(t=t_0-\d t\). The norm is now not conserved: if there were \(N\) states in the past of a normalised state \(|\ont(t_0)\ket\), the state produced at time \(t_0-\d t\) now has norm \(\sqrt N\). If the state \(|\ont(t_0)\ket\) was a garden of Eden, then \(U^\dag|\ont(t_0)\ket=0\).

Now remember that, whenever we do quantum mechanics, we have the freedom to switch to another basis by using unitary transformations. It so happens that with any ontological evolution operator \(U_1\) that could be a generalisation of Eq.~\eqn{infolossU}, there exists a unitary matrix \(X\) with the property
	\be U_1^\dag\,X\iss X\,U_2\ , \eel{timereverseX}
where \(U_2\) again describes an ontological evolution with information loss.
This is not hard to prove. One sees right away that such a matrix \(X\) should exist by noting that \(U_1^\dag\) and \(U_2\) can be brought in the same normal form. Apart from the opposite time ordering, \(U_1\) and \(U_2\) have the same equivalence classes. 

Finding the unitary operator \(X\) is not quite so easy. We can show how to produce \(X\) in a very simple example. Suppose \(U_1\) is a very simple  \(N\times N\) dimensional matrix  \(D\) of the form
	\be D= \pmatrix{1&1&1&\cdots \cr 0&0&0&\cr 0&0&0&\cr \vdots &&} \ ,\quad  D^\dag= \pmatrix{1&0&0&\cdots \cr 1&0&0&\cr 1&0&0&\cr \vdots &&} \ ,\quad \eel{infolossblock}
so \(D\) has \(N\) 1's on the first row, and 0's elsewhere. This simply tells us that \(D\) sends all states \(|1\ket\,,\ \dots\ |N\ket\) to the same state \(|1\ket\).

The construction of a matrix \(Y\) obeying
	\be D^\dag Y=YD\ , \eel{Dtransform}
can be done explicitly. One finds
	\be Y_{k\ell}={1\over\sqrt N}\,e^{2\pi i k\ell/N}\ . \eel{Ymatrix}
\(Y\) is unitary and it satisfies Eq.~\eqn{Dtransform}, by inspection.

This result may come unexpected. Intuitively, one might think that information loss will make our models non-invariant under time reversal.  Yet our quantum mechanical tool does allow us to invert such a model in time. A ``quantum observer" in a model with information loss may well establish a perfect validity of symmetries such as \(CPT\) invariance. This is because, for a quantum observer, transformations with matrices \(X\) merely represent a transition to another orthonormal basis; the matrix \(Y\) is basically the discrete Fourier transform matrix. Note that merger states (see Fig.~\ref{infolossgeneral.fig}) transform into Gardens of Eden, and \emph{vice versa}.

\subsection{The arrow of time\labell{timearrow2}}

One of the surprising things that came out of this research is a new view on the arrow of time. It has been a long standing mystery that the local laws of physics appear to be perfectly time-reversible, while large-scale, classical physics is not at all time-reversible. Is this not a clash with the reduction principle? If large-scale physics can be deduced from small-scale physics, then how do we deduce the fact that the second law of thermodynamics dictates that entropy of a closed system can only increase and never decrease?

Most physicists are not really worried by this curious fact. In the past, this author always explained the `arrow of time' by observing that, although the small-scale laws of nature are time-reversible, the boundary conditions are not: the state of the universe was dictated at time \(t=0\), the Big Bang. The entropy of the initial state was very small, probably just zero. There cannot be any boundary condition at the Big Apocalypse, \(t=t_\infty\). So, there \emph{is} asymmetry in time and that is that. For some reason, some researchers are not content with such a simple answer.

We now have a more radical idea: the microscopic laws may not at all be time-reversible. The \emph{classical} theory underlying quantum mechanics does not have to be, see subsection~\ref{coginfoloss}. Then, in subsection~\ref{timereverse}, we showed that, even if the classical equations feature information loss at a great scale -- so that only tiny fractions of information are preserved -- the emerging quantum mechanical laws continue to be exactly time-reversible, so that, as long as we adhere to a description of things in terms of Hilbert space, we cannot understand the source of time asymmetry.

However, the \emph{classical}, ontological states are very asymmetric in time, because, as we stated, these are directly linked to the underlying classical degrees of freedom.

	
All this might make information loss acceptable in theories underlying quantum theory. Note, furthermore, that our distinction of the \emph{ontological states} should be kept, because classical states are ontological. Ontological states no \emph{not} transform into ontological states under time reversal, since the transformation operators \(X\) 	and \(Y\) involve quantum superpositions. In contrast, templates \emph{are} transformed into templates. This means automatically that \emph{classical states} (see section~\ref{classical}), are not invariant under time reversal. Indeed, they do not look invariant under time reversal, since classical states typically obey the rules of thermodynamics.

The quantum equations of our world are invariant (more precisely: covariant) under time reversal, but neither the \emph{sub-}microscopical world, where the most fundamental laws of nature reign, nor the classical world allow for time reversal.
	
Our introduction of information loss may have an other advantage:  two states may be seen to be in the same equivalence class even if we cannot follow the evolution very far back in time. In practice, one might suspect that the likelihood of two distinct states to actually be in one equivalence class, will diminish rapidly with time; these states will show more and more differences at different locations. This means that we expect physically relevant equivalence classes to be related by transformations that still look local at the particle scale that can presently be explored experimentally. This brings us to the observation that \emph{local gauge equivalence classes} might actually be identified with information equivalence classes. It is still (far) beyond our present mathematical skills to investigate this possibility.

Finally, note that info-equivalence classes may induce a subtle kind of \emph{apparent} non-locality in our effective quantum theory, a kind of non-locality that may help to accept the violation of Bell's theorem (section~\ref{Bell}).

The explicit models we studied so-far usually do not have information loss. This is because the mathematics will be a lot harder; we simply have not yet been able to use information loss in our more physically relevant examples.

\subsecl{Information loss and thermodynamics}{thermo}

There is yet another important novelty when we allow for information loss, in particular when it happens at a large scale (such as what we expect when black holes emerge, see above). Neither the operator \(U\) nor \(U^\dag\) are unitary now. In the example of Eq.~\eqn{infolossblock}, one finds
	\be D\,D^\dag=N|1\ket\bra 1|\ , \qquad D^\dag D=N|e\ket\bra e|\ , \eel{DDdag}
where \(|e\ket\) is the normalised state 
	\be |e\ket=\fract 1{\sqrt N}\,(\,|1\ket+\,\cdots\,X|N\ket\,) \eel{eketdef}
(which shows  that the matrix \(Y\) here must map the state \(|1\ket\) onto the state \(|e\ket\). Note, that \(D\) and \(D^\dag\) are in the same conjugacy class).
Thus, during the evolution, the state \(|1\ket\) may become more probable, while the probabilities of all other states dwindle to zero.
Some equivalence classes may gain lots of members this way, while others may stay quite small. In large systems, it is unlikely that the probability of a class vanishes altogether, so it might become possible to write the amplitudes as 
	\be e^{-iHt-\half\b E}\ , \eel{thermoprob}
where one might be tempted to interpret the quantity \(E\) as a (classical) energy, and \(\half\b\) as an imaginary component of time. This aspect of our theory is still highly speculative. Allowing time to obtain complex values can be an important instrument to help us understand the reasons why energy has a lower bound.

\newsecl{More problems}{problems}

It may take years, decades, perhaps centuries to arrive at a comprehensive theory of quantum gravity, combined with a theory of quantum matter that will be an elaborate extension of the Standard Model. Only then will we have the opportunity to isolate the beables that are the basic ingredients for an ontological theory; the templates of that theory will be projected onto the beable basis. Only then can we tell whether the CAI really works. Conceivably however, we may be able to use the CAI as a guide to arrive at such theories. That was the main motivation for writing this book.

\subsection{What will be the CA for the SM?\labell{CASM}}

There are numerous problems remaining. The first of these was encountered by the author right-away: how to convince a majority of researchers who have been working in this field for many decades, that the views expressed here may well be basically correct. In particular we state the most important conclusions:
	\bi{--} there's only a single, essentially classical, discrete, universe, not an infinity of distinct universes, as advocated in the Many World Interpretation, whether or not being guided by a pilot wave that is active in all those worlds.
	\itm{--} The Born probabilities hold exactly true, requiring no corrections of any form, since they were put in the state vectors (templates) to describe the probability distribution of the initial states. What has been put in, comes out unchanged: the Born probabilities. 
	\itm{--} The ``collapse of the wave function" takes place automatically, without requiring any corrections to Schr\"odinger's equation, such as non-linearities. This is because the universe is in a single ontological state from the Big Bang onwards, whereas the final result of an experiment will also always be a single ontological state. The final state can never be in a superposition, such as a live cat superimposed with a dead cat.
	\itm{--} The underlying theory may well be a local one, but the transformation of the classical equations into the much more efficient quantum equations, involves some degree of non-locality, which leaves no trace in the physical equations, apart from the well-known, apparent `quantum miracles'.
	\itm{--} It is very much worth-while to search for more models where the transformation can be worked out in detail; this could lead to a next generation of Standard Models, with `cellular automaton restrictions' that can be tested experimentally.\ei
The problem how to set up such searches in a systematic way is very challenging indeed. Presumably a procedure corresponding to second quantisation has to be employed, but as yet it is only clear how to do this for fermionic fields. The problem is then that we also must replace the Dirac equation for the first-quantised particles into deterministic ones. This could be done for free, massless particles, which is not a bad start, but it is also not good enough to proceed. Then, we have some rudimentary ideas about bosonic force-carrying fields, as well as a suggestive string-like excitation (worked out further in part II\,), but again, it is not known yet how to combine these into something that can compete with the Standard Model known today.
	
\subsection{The Hierarchy Problem\labell{hierarchy}}

There is a deeper reason why \emph{any} detailed theory involving the Planck scale, quantum mechanics and relativity, may be extremely difficult to formulate. This is the empirical fact that there are two or more radically different scales of very special significance: the Planck scale and the mass scale(s) of the most significant particles in the system. The amazing thing about our world as that these various scales are many orders of magnitude apart. The Planck scale is at \(\approx 10^{19}\) GeV, the nuclear scale is at \(\approx 1\) GeV, while there are also electrons, and finally neutrinos at some \(10^{-11}\) GeV.

The origin of these large numbers, which are essential for the universe to function the way it does, is still entirely obscure. We could add to this that there are very few experiments that reach accuracies better than 1 part in \(10^{11}\), let alone \(10^{19}\), so that it is questionable whether any of the fundamental principles pertaining to one scale, are still valid at an other -- they could well be, but everything could also be different. There is no lack of speculative ideas to explain the origins of these numbers. The simplest observation one can make is that fairly down-to-earth mathematics can easily generate numbers of such magnitudes, but to make them arise naturally in fundamental theories of nature is not easy at all.

Most theories of Planck scale physics, such as superstring theory and loop quantum gravity, make no mention of the origins of large numbers, whereas, we believe, good theories should\fn{An important exception is the theory of the \emph{anthropic principle}, the idea that some numbers are very large or very small, just because these would be the only values that can yield planets with civilised creatures (anthropoi) on them. This notion has been around for some time, but, understandably, it does not gather much adherence.}. In discrete cellular automata, one can certainly observe that, if lattices in space and time play any role, there may be special things happening when the lattice is exactly regular -- such lattices have zero curvature. The curvature of our universe is extremely tiny, being controlled by numbers even more extreme than the hierarchy scales mentioned: the cosmological constant is described by a dimensionless number of the order of \(10^{-122}\). This might mean that, indeed, we are dealing with a regular lattice, but it must accommodate for rare lattice defects.

In general, a universal theory must explain the occurrence of very rare events, such as the mass terms causing zitterbewegung in fermions such as electrons. We do believe that cellular automaton models are in a good position to allow for special events that are very rare, but it is far too early to try to understand these. 

In short, the most natural way to incorporate hierarchies of scales in our theory is not clear.

\newsecl{Alleys to be further investigated, and open questions}{miss}

\subsecl{Positivity of the Hamiltonian}{posham}

As we shall argue at length in Part II, it is quite likely that the gravitational force will be a crucial ingredient in resolving the remaining difficulties in the CAT theory. One of the various arguments for this is that gravity is partly based on the existence of local time translations, which are redefinitions of time that depend on the location in space. The generator for these transformations is the Hamiltonian density, which here must be a local operator. At the same time, it is important also for gravity theory to have a lower bound on energy density.  Apparently, gravity hinges exactly on those two important demands on the Hamiltonian operator that are causing us some troubles, and so, conceivably, the problem of quantising gravity and the problem of turning quantum mechanics into a deterministic theory for dynamics, will have to be solved together.

	On the other hand one might argue that the non-locality of the quantum-induced Hamiltonian is exactly what we need to explain away Bell's theorem. 
	
	The exact position of the gravitational force in our theory is not completely clear.
Therefore, one might hope that the inclusion of a gravitational force can be postponed, and that cellular automaton models exist also on flat space-time lattices.  In part II, we shall see that our \(PQ\) formalism, chapter~\ref{PQ}, allows us to split space-like coordinates into two parts: integers that specify the location of a point on a lattice, and fractional, or periodic, coordinates that could be used to position a point within one lattice cell, or else merely play the role of canonically conjugated variables associated to a discretised momentum variable. Here, hoever, accommodating for non-compact symmetries such as Lorentz invariance is extremely difficult. 
	
The most obnoxious, recurring question will be that the Hamiltonians reproduced in our models, more often than not, appear to lack a lower bound. This problem will be further studied in part II, chapters~\ref{locality}, \ref{quloc}, and in section~\ref{doubleHam}.
The properties that seem to raise conflicts when combined, are \bi{1.} \(H\) must be constant in time: \be{\dd\over\dd t} H=0\ , \ee
	\itm{2.} \(H\) must be bounded from below:
	\be\bra\,\j\,|\,H\,|\,\j\,\ket\ge 0\ , \eel{hambound}
	\itm{3.} \(H\) should be extensive, that is, it must be the sum (or integral) of local terms:
	\be H=\sum_{\vec x}\HH(\vec x)\ , \eel{Ham3}\\[-30pt]
	\itm{4.} {And} finally, it must generate the equations of motion:
		\be U(t)=e^{-iH\,t}\ , \ee 
	or equivalently, for all states \(|\,\j\,\ket\) in Hilbert space, 
	\be{\dd\over\dd t}|\,\j\,\ket=-iH\,|\,\j\,\ket\ . \eel{schro2} \ei
If we take just any classical model, we usually have no difficulty finding an operator \(H\) that generates the evolution law \eqn{schro2}, but then it often does not automatically obey both \eqn{hambound} and \eqn{Ham3}.	

	All these equations are absolutely crucial for understanding quantum mechanics.
In particular, the importance of the bound \eqn{hambound} is sometimes underestimated. If the bound would not have been there, none of the familiar solutions of the Schr\"odinger equation would be stable; any, infinitesimally tiny, perturbation in \(H\) would result in a solution with energy \(E\) that decays into a combination of two spatially separated solutions with energies \(E+\d E\) and \(-\d E\). 

\emph{All} solutions of the Schr\"odinger equation would have such instabilities, which would describe a world quite different from the quantum world we are used to. 

This is the reason why we always try to find an expression for \(H\) such that a lower bound exists (which we can subsequently normalise to be exactly zero). From a formal point of view, it should be easy to find such a Hamiltonian. Every classical model should allow for a description in the form of the simple models of section~\ref{cogwheelN}, a collection of cogwheels, and we see in Fig.~\ref{generalfinite.fig} that we can subsequently adjust the constants \(\d E_i\) so that the bound \eqn{hambound} exists, even if there are infinitely many cogwheels with unlimited numbers of teeth.

However, also the third condition is needed.  \(\HH(\vec x\,)\) is the Hamiltonian density. Locality amounts to the demand that, at distances exceeding some close distance limit, these Hamilton densities must commute (Eq.~\ref{nmloc}). 

One may suspect that the ultimate solution that obeys all our demands will come from quantising gravity, where we know that there must exist a local Hamiltonian density that generates local time diffeomorphisms. In other treatises on the interpretation of quantum mechanics, this important role that might be played by the gravitational force is rarely mentioned. 

Some authors do suspect that gravity is a new elementary source of `quantum decoherence', but such phrases are hardly convincing. In these arguments, gravity is treated perturbatively (Newton's law is handled as an additive force, while black holes and scattering between gravitational waves are ignored). As a perturbative, long-range force, gravity is in no fundamental way different from electromagnetic forces. Decoherence\,\cite{Zurek-1991} is a concept that we completely avoid here (see section~\ref{decoBorn}).

Since we have not solved the Hamiltonian positivity problem completely, we have no systematic procedure to control the kind of Hamiltonians that can be generated from cellular automata. Ideally, we should try to approximate the Hamiltonian density of the Standard Model. Having approximate solutions that only marginally violate our requirements will not be of much help, because of the \emph{hierarchy problem}, see subsection~\ref{hierarchy}: the Standard Model Hamiltonian (or what we usually put in its place, its Lagrangian), requires \emph{fine tuning}. This means that very tiny mismatches at the Planck scale will lead to very large errors at the Standard Model scale. The hierarchy problem has not been solved, so this indeed gives us an other obstacle.

	\subsecl{Second quantisation in a deterministic theory}{secondquant}
When Dirac arrived at the famous Dirac equation to describe the wave function of an electron, he realised that he had a problem: the equation allows for positive energy solutions, but these are mirrored by solutions where the energy, including the rest-mass energy, is negative. The relativistic expression for the energy of a particle with momentum \(\vec p\) (in units where the speed of light \(c=1\)), is
	\be E=\pm\sqrt{m^2+\vec p\,^2}\ . \eel{posnegenergy}
If a partial differential equation gives such an expression with a square root, it is practically impossible to impose conditions on the wave function such that the unwanted negative sign is excluded, unless such a condition would be allowed to be a non-local one, a price Dirac was not prepared to pay. He saw a more natural way out:
\begin{quote} \emph{There are very many electrons, and the \(N\)-electron solution obeys Pauli's principle: the wave function must switch sign under interchange of any pair of electrons. In practice this means that all electrons each must occupy \emph{different} energy levels. Levels occupied by one electron cannot be reached by other electrons. Thus, Dirac imagined that all negative energy levels are normally filled by electrons, so that others cannot get there. The vacuum state is by definition the lowest energy state, so the negative-energy levels are all occupied there. If you put an extra electron in a positive energy level, or of you remove an electron from a negative energy spot, then in both cases you get a higher energy state.} \end{quote}
If an electron is removed from a negative energy level, the empty spot there carries a net energy that is positive now. Its \emph{charge} would be opposite to that of an electron. Thus, Dirac predicted an \emph{antiparticle} associated to the electron\fn{Dirac first thought that this might be the \emph{proton}, but that was untenable; the mass had to be equal to the electron mass, and the positron and the electron should be able to annihilate one another when they come close together.}: a particle with mass \(m_e\) and charge \(+e\), where normal electrons have mass \(m_e\) and charge \(-e\). Thus, the positron was predicted.

In cellular automata we have the same problem. In section~\ref{locality}, it is explained why we cannot put the edge of the energy spectrum of an automaton where we would like to put it: at zero energy, the vacuum state, which would then naturally be the lowest energy state. We saw that locality demands a very smooth energy function.
If we symmetrise the spectrum, such that \(-\pi<E\,\d t<\pi\), we get the same problem that Dirac had to cope with, and indeed, we can use the same solution: second quantisation. How it works will be explained in section~\ref{fermions}. We take \(k\) fermionic particles, which each can occupy \(N\) states. If we diagonalise the \(U\) operator for each `particle', we find that half of the states have positive energy, half negative. If we choose \(k=\half N\), the lowest energy state has all negative energy levels filled, all positive energies empty; this is the lowest energy state, the vacuum. 

The excited states are obtained if we deviate slightly from the vacuum configuration. This means that we work with the energy levels close to the center of the spectrum, where we see that the Fourier expansions of section~\ref{locality} still converge rapidly. Thus, we obtain a description of a world where all net energies are positive, while rapid convergence of the Fourier expansion guarantees effective locality.

Is this then the perfect solution to our problem? Nearly, but not quite. First, we only find justifiable descriptions of second quantised fermions. The bosonic case will be more subtle, and is not yet quite understood\fn{But we made a good start: bosons are the energy quanta of harmonic oscillators, which we should first replace by harmonic rotators, see chapters~\ref{harmrot}---\ref{harmosc}. Our difficulty is to construct harmonically coupled chains of such rotators. Our procedures worked reasonably well in one space-, one time dimension, but we do not have a bosonic equivalent of the neutrino model (section~\ref{nu}), for example.} Secondly, we have to replace the Dirac equation by some deterministic evolution law, while a deterministic theory to be exposed in section~\ref{nu} describes sheets, not particles. We do not know how to describe local deterministic interactions between such sheets. What we have now, is a description of non-interacting particles. Before introducing interactions that are also deterministic, the sheet equations will have to be replaced by something different. 

Supposing this problem can be addressed, we work out the formalism for interactions in subsection~\ref{2ndquantCA}. The  interaction Hamiltonian is obtained from the deterministic law for the interactions using a BCH expansion, which is not guaranteed to converge. This may be not a problem if the interaction is weak. We bring forward arguments why, in that case, convergence may still be fast enough to obtain a useful theory. The theory is then not infinitely accurate, but this is not surprising. We could state that, indeed, that problem was with us all along in quantum field theory: the perturbative expansion of the theory is fine, and it gives answers that are much more precise that the numbers that can be obtained from any experiment, but they are not infinitely precise, just because perturbation expansion does not converge (it can be seen to be merely an \emph{asymptotic} expansion). Thus, our theory reproduces exactly what is known about quantum mechanics and quantum field theory, just telling us that if we want a more accurate description, we might have to look at the original automaton itself.

Needless to emphasise, that some of the ideas brought forward here are mostly speculation, they should still be corroborated by more explicit calculations and models.

\subsecl{Information loss and time inversion}{timeinv}
	
	A very important observation made in section~\ref{infoloss} is that, if we introduce information loss in the deterministic model, the total number of orthogonal basis elements of the ontological basis may be considerable reduced, while nevertheless the resulting quantum system will \emph{not} show any signs of time irreversibility. The \emph{classical} states however, referring to measurement results and the like, are linked to the original ontological states, and therefore \emph{do} possess a thermodynamical arrow of time.
	
	This may well explain why we have time non reversibility at large, classical scales while the microscopic laws as far they are known today, still seem to be perfectly time reversible.
	
	To handle the occurrence of information loss at the sub-microscopic level, we introduced the notion of info-equivalence classes: all states that, within a certain finite amount of time evolve into the same ontological state \(|\j(t)\ket\), are called info-equivalent, all being represented as the same quantum basis element \(|\j(0)\ket\). We already alluded to the similarity between the info-equivalence classes and the \emph{local gauge equivalence classes}. Could it be that we are therefore talking about the same thing?
	
	If so, this would mean that local gauge transformations of some state represented as an ontological state, may actually describe different ontological states at a given time \(t=t_0\), while two ontic states that differ from one another only by a local gauge transformation, may have the property that they both will evolve into the same final state, which naturally explains why observers will be unable to distinguish them. 
	
	Now we are aware of the fact that these statements are about something being fundamentally unobservable, so their relevance may certainly be questioned. 
	
	Nevertheless this suggestion is justifiable, as follows. One may observe that formulating quantum field theory \emph{without} employing the local gauge-equivalence principle, appears to be almost impossible\fn{The interactions would have to be kept quite weak, such as the electro-magnetic ones.}, so the existence of local gauge equivalence classes can be ascribed to the mathematical properties of these quantised fields. Only rarely can one replace a theory with local gauge equivalence by one where this feature is absent or of a totally different nature\fn{Examples are known, in the form of \emph{dual transformations}.}. Exactly the same can be said about ontological equivalence classes. They will be equally unobservable at large scales -- by definition. Yet rephrasing the deterministic theory while avoiding these equivalence classes altogether may be prohibitively difficult (even if it is not principally excluded). So our argument is simply this: these equivalence classes are so similar in nature, that they may have a common origin.
	
	This then leaves an exciting question: general relativity is also based on a local gauge principle: the equivalence of locally curved coordinate frames. Can we say the same thing about that gauge equivalence class? Could it also be due to information loss? This would mean that our underlying theory should be phrased in a \emph{fixed} local coordinate frame. General coordinate invariance would then be ascribed to the fact that the \emph{information that determines our local coordinates} is something that can get lost. Is such an idea viable? Should we investigate this?
	
	My answer is a resounding \emph{yes!} This could clarify some of the mysteries in today's general relativity and cosmology. Why is the cosmological constant so small? Why is the universe spatially flat (apart from local fluctuations)?  And, regarding our cellular automaton: How should we describe an automaton in curved space-time?
	
	The answers to these questions are then: yes, our universe is curved, but the curvature is limited to local effects. We \emph{do} have an important variable playing the role of a metric tensor \(g_{\m\n}(\vec x,t)\) in our automaton, but it lives in a well-define coordinate frame, which is flat. Take a gauge condition obeying \(g_{i0}=g^{i0}=0,\ i=1,2,3.\) Let then \(\l^i\) be the three eigenvalues of \(g^{ij}\), the space-like components of the inverse metric (so that \((\l^{i})^{-1}\equiv \l_i\) are the eigenvalues of \(g_{ij}\)). Let \(\l^0=|g^{00}|\). Then the local speed of light, in terms of the coordinates used,  is given by \(c^2={|\vec \l|/\l^0}\). We can impose an inequality on \(c\): assume that the values of the metric tensor are constrained to obey \(|c|\le 1\). If now we write \(g_{\m\n}=\w^2(\vec x,t)\^g_{\m\n}\), with a constraint on \(\^g_{\m\n}\) such as \(\det(\^g_{\m\n})=-1\) (See Appendix~\ref{confgrav}), then there is no limitation on the value of \(\w\). This means that the universe can inflate towards any size, but on the average it may stay flat.
	
We continue this subject in subsection~\ref{gravityrole}.

\subsecl{Holography and Hawking radiation}{holo}

There is an other reason to expect information loss to be an inescapable feature of an ultimate theory of physics, which concerns microscopic black holes. Classically, black holes absorb information, so, one may have to expect that our classical, or `pre-quantum' system also features information loss. Even more compelling is the original idea of \emph{holography}\,\cite{Suss-1995}\cite{GtH-Salam93}. It was Stephen Hawking's important derivation that black holes emit particles\,\cite{Hawking-1975}, due to quantum fluctuations near the horizon. However, his observation appeared to lead to a paradox: 
\begin{quote} The calculation suggests that the particles emerge in a thermal state, with perfect randomness, regardless how the black hole was formed. Not only in deterministic theories, but also in pure quantum theories, the arrangement of the particles coming out should depend to some extent on the particles that went in to form a black hole. \end{quote}
In fact, one expects that the state of all particles emerging from the black hole should be related to the state of all particles that formed the black hole by means of a unitary matrix, the \emph{scattering matrix \(S\).}\,\cite{GtH-BH-1985}\cite{GtH-BHfound-1988} 

Properties of this matrix \(S\) can be derived using physical arguments\,\cite{GtH-S-1996}. One uses the fact that particles coming out must have crossed all particles that went in, and this involves the conventional scattering matrix determined by their ``Standard Model" interactions.
Now since the centre-of-mass energies involved here are often much larger than anything that has been tested in laboratories, much stays uncertain about this theory, but in general one can make some important deductions:
\begin{quote} The only events that are relevant to the black hole's reaction upon the particles that went in, take place in the region very close to the event horizon. This horizon is two-dimensional.\end{quote}
This means that all \emph{information} that is processed in the vicinity of a black hole, must be effectively represented at the black hole horizon. Hawking's expression for the black hole entropy resulting from his analysis of the radiation
 clearly indicates that it leaves only one bit of information on every surface element of roughly the Planck length squared. In natural units:
 	\be S=\pi R^2=\log(W)=(\log2)\ {}^2\!\log W\ ;\qquad W=2^{\SS/4\log 2}\ , \eel{HawkingS}
 where \(\SS=4\pi R^2\) is the surface area of the horizon.
 
 Where did the information that went into the bulk of space-time inside a black hole go? We think it was lost. If we phrase the situation this way, we can have holography without losing locality of the physical evolution law. This evolution law apparently  is extremely effective in destroying information.\fn{Please do not confuse this statement with the question whether \emph{quantum information} is lost near a black hole horizon. According to the hypothesis phrased here, quantum information is what is left if we erase all redundant information by combining states in equivalence classes. The black hole micro states then correspond to these equivalence classes. By construction, equivalence classes do not get lost.}

Now we can add to this that we cannot conceive of a configuration of matter in space-time that is such that it contains \emph{more} information per unit of volume than a black with radius corresponding to the total energy of the matter inside. Therefore the black hole obeys the \emph{Bekenstein limit}\,\cite{Bekenstein-1981}: 

\begin{quote}the maximum amount of information that fits inside a (spherical) volume \(V\) is given by the entropy of the largest black hole that fits inside \(V\).\end{quote}
Information loss in a local field theory must then be regarded in the following way (``holography"):
	\begin{quote}\emph{In any finite, simply connected region of space, the information contained in the bulk gradually disappears, but what sits at the surface will continue to be accessible, so that the information at the surface can be used to characterise the info-equivalence classes.}\end{quote}

At first sight, using these info-equivalence classes to represent the basis elements of a quantum description may seem to be a big departure from our original theory, but we have to realise that, if information gets lost at the Planck scale, it will be much more difficult to lose any information at much larger scales; there are so many degrees of freedom that erasing information completely is very hard and improbable; rather, we are dealing with the question how exactly information is represented, and how exactly do we count bits of information in the info-equivalence classes.

Note that, in practice, when we study matter in the universe, the amount of energy considered is far less that what would correspond to a black hole occupying the entire volume of space. So in most practical cases, the Bekenstein limit is not significant, but we have to remember that, in those cases, we always consider matter that is still very close to the vacuum state.

Information loss is mostly a local feature; globally, information is preserved. This does mean that our identification of the basis elements of Hilbert space with info-equivalence classes appears to be not completely local. On the other hand, both the classical theory and the quantum theory naturally forbid information to be spread faster than the speed of light.

Let us end this subsection with our view on the origin of Hawking radiation. The physical laws near the horizon of a black hole should be derived from the laws controlling the vacuum state as seen by an observer falling in. This vacuum is in a single quantum state, but consists of myriads of ontological states, distinguishable by considering all conceivable fluctuations of the physical fields. Normally, all these states form a single equivalence class.

At the horizon, however, signals entering the black hole cannot return, so the mixture of the information that causes all these states to form a single equivalence class, is rearranged substantially by the presence of the black hole, so much so that, as seen by the distant observer, not a single equivalence class is experienced, but very many classes. Thus, the vacuum is replaced by the much larger Hilbert space spanned by all these classes. They together form the rich spectrum of physical particles seen to emerge from the black hole.

\newsecl{Conclusions}{conc1}

In an earlier version of this text,  a sub title was added to the title of this book: \emph{A view on the quantum nature of our universe}. This raised objections: ``your view seems to be more classical than anything we've seen before!" Actually, this can be disputed. We argue that classical underlying laws can be turned into quantum mechanical ones practically without leaving any trace. We insist that it is real quantum mechanics that comes out, including all ``quantum weirdness". The nature of our universe \emph{is} quantum mechanical, but it may have a classical explanation. The underlying classical laws may be seen to be \emph{completely} classical. We show how `quantum mechanical' probabilities can originate from completely classical probabilities.

It may seem odd that our theory, unlike most other approaches,  does not contain any strange kinds of stochastic differential equations, no ``quantum logic", not an infinity of other universes, no pilot wave, just completely ordinary equations of motion that we have hardly been able to specify, as they could be almost anything. Our most essential point is that we should not be deterred by `no go theorems' if these contain small print and emotional ingredients in their arguments. The small print that we detect is the assumed absence of strong local as well as non-local correlations in the initial state. Our models show that there should be such strong correlations. Correlations do not require superluminal signals, let alone signals going backwards in time. 

The emotional ingredient is the idea that the conservation of the ontological nature of a wave function would require some kind of `conspiracy', as it is deemed unbelievable that the laws of nature themselves can take care of that. Our point is that they obviously can. Once we realise this, we can consider studying very simple local theories.

In principle it is almost trivial to obtain ``quantum mechanics" out of classical theories. We demonstrated how it can be done with a system as classical as the Newtonian  planets moving around the sun. But then difficulties do arise, which of course explain why our proposal is not so simple after all. The positivity of the Hamiltonian is one of the prime stumbling blocks. We can enforce it, but then the plausibility of our models needs to be scrutinised. At the very end we have to concede that the issue will most likely involve the mysteries of quantum gravity. Our present understanding of quantum gravity suggests that discretised information is spread out in a curved space-time manifold; this is difficult to reconcile with nature's various continuous symmetry properties such as Lorentz invariance. So, yes, it is difficult to get these issues correct, but we suggest that these difficulties will only indirectly be linked to the problem of interpreting quantum mechanics.

This book is about these questions, but also about the tools needed to address them; they are the tools of conventional quantum mechanics, such as symmetry groups and Noether's theorem.

A distinction should be made between on the one hand explicit \emph{theories} concerning the fate of quantum mechanics at the tiniest meaningful distance scale in physics, and on the other hand proposals for the \emph{interpretation} of today's findings concerning quantum phenomena.	

Our theories concerning the smallest scale of natural phenomena are still very incomplete.  Superstring theory has come a long way, but seems to make our view more opaque than desired; in any case, in this book we investigated only rather simplistic models, of which it is at least clear what they say.

\subsecl{The CAI}{CAI}

What we found,  seems to be more than sufficient to extract a succinct interpretation of what quantum mechanics really is about. The technical details of the underlying theory do not make much difference here. All one needs to assume is that some ontological theory exists; it will be a theory that describes phenomena at a very tiny distance scale in terms of evolution laws that process bits and bytes of information. These evolution laws may be ``as local as possible", requiring only nearest neighbours to interact directly. The information is also strictly discrete, in the sense that every ``Planckian" volume of space may harbour only a few bits and bytes. We also suspect that the bits and bytes are processed as a function of local time, in the sense that only a finite amount of information processing can take place in a finite space-time 4-volume. On the other hand, one might suspect that some form of information loss takes place such that information may be regarded to occupy surface elements rather than volume elements, but this we could not elaborate very far.

In any case, in its most basic form, this local theory of information being processed, does not require any Hilbert space or superposition principles to be properly formulated. The bits and bytes we discuss are \emph{classical} bits and bytes; at the most basic level of physics (but \emph{only} there), \emph{qubits} do not play any role, in contrast with more standard approaches considered in today's literature.  Hilbert space only enters when we wish to apply powerful mathematical machinery to address the question how these evolution laws generate large scale behaviour, possibly collective behaviour, of the data. 

Our theory for the \emph{interpretation} of what we observe is now clear: humanity discovered that phenomena at the distance and energy scale of the Standard Model (which comprises distances vastly larger, and energies far smaller, than the Planck scale) can be captured by postulating the effectiveness of \emph{templates}. Templates are elements of Hilbert space that form a basis that can be chosen in numbers of ways (particles, fields, entangled objects, etc.), which allow us to compute the collective behaviour of solutions to the evolution equations that do require the use of Hilbert space and linear operations in that space. The original observables, the beables, can all be expressed as superpositions of our templates. \emph{Which} superpositions one should use, differs form place to place. This is weird but not inconceivable. Apparently there exists a powerful scheme of symmetry transformations allowing us to use the same templates under many different circumstances. The rule for transforming beables to templates and back is complex and not unambiguous; exactly how the rules are to be formulated, for all objects we know about in the universe, is not known or understood, but must be left for further research.

Most importantly, the original ontological beables do not allow for any superposition, just as we cannot superimpose planets, but the templates, with which we compare the beables, are elements of Hilbert space and require the well-known principles of superposition.

The second element in our CAI is that objects we normally call \emph{classical}, such as planets and people, but also the dials and all other detectable signals coming from measurement devices, can be directly derived from beables, in principle without the intervention of the templates. 

Of course, if we want to know how our measurement devices work, we use our templates, and this is the origin of the usual ``measurement problem". What is often portrayed as mysteries in quantum theory: the measurement problem, the `collapse of the wave function', and Schr\"odinger's cat, is completely clarified in the CAI. All wave functions that will ever occur in our world, may seem to be superpositions of our templates, but they are completely peaked, `collapsed', as soon as we use the beable basis. Since classical devices are also peaked in the beable basis, their wave functions are collapsed. No violation of Schr\"odinger's equation is required for that, on the contrary, the templates, and indirectly, also the beables, exactly obey the Schr\"odinger equation.

In short, it is not nature's degrees of freedom themselves that allow for superposition, it is the templates we normally use that are man-made superpositions of nature's ontological states. The fact that we hit upon the apparently inevitable paradoxes concerning superposition of the natural states, is due to our intuitive thinking that our templates represent reality in some way. If, instead, we start from the ontological states that we may one day succeed to characterise,  the so-called `quantum mysteries' will disappear.

\subsecl{Counterfactual definiteness}{counter}

Suppose we have an operator \(\OO^\op_1\) whose value cannot be measured since the value of another operator, \(\OO_2^\op\), has been measured, while \([\OO_1^\op\,,\,\OO_2^\op\,]\ne 0\). 		
\emph{Counterfactual} \hbox{\emph{reality}} is the assumption that, nevertheless, the operator \(\OO^\op_1\) takes \emph{some} value, even if we don't know it. It is often assumed that hidden variable theories imply counterfactual definiteness. We should emphasise categorically that no such assumption is made in the Cellular Automaton Interpretation. In this theory, the operator \(\OO_2^\op\), whose value has been measured, apparently turned out to be composed of ontological observables (beables). Operator \(\OO_1^\op\) is, by definition, not ontological and therefore has no well-defined value, for the same reason why, in the planetary system, the Earth-Mars interchange operator, whose eigenvalues are \(\pm 1\), has neither of these values; it is unspecified, in spite of the fact that planets evolve classically, and in spite of the fact that the Copenhagen doctrine would dictate that these eigenvalues \emph{are} observable!

The tricky thing about the CAI when applied to atoms and molecules, is that one often does not know a priori which of our operators are beables and which are changeables or superimposables (as defined in subsections~\ref{operators} and \ref{bechange}).
One only knows this a posteriori, and one might wonder why this is so. We are using \emph{templates} to describe atoms and molecules, and these templates give us such a thoroughly mixed up view of the complete set of observables in a theory that we are left in the dark, until someone decides to measure something. 

It looks as if the simple act of a measurement sends a signal backwards in time and/or with superluminal speed to other parts of the universe, to inform observers there which of their observables can be measured and which not. Of course, that is not what happens. What happens is that now we \emph{know} what can be measured accurately and which measurements will give uncertain results. The Bell and CHSH inequalities are violated as they should be in quantum field theory, while nevertheless quantum field theory forbids the possibility to send signals faster than light or back to the past.

\subsecl{Superdeterminism and conspiracy}{conspire}

Superdeterminism may be defined to imply that not only all physical phenomena are declared to be direct consequences of physical laws that do not leave anything anywhere to chance (which we refer to as `determinism'), but it also emphasises that the observers themselves behave in accordance with the same laws. They also cannot perform any whimsical act without any cause in the near past as well as in the distant past. By itself, this statement is so obvious that little discussion would be required to justify it, but what makes it more special is that it makes a difference. The fact that an observer cannot reset his or her measuring device without changing physical  states in the past is usually thought not to be relevant for our description of physical laws. The CAI claims that it is. Further explanations will be given in subsection~\ref{will}, where we attempt to demystify `free will'.

It is often argued that, if we want any superdeterministic phenomenon to lead to violations of the Bell-CHSH inequalities, this would require \emph{conspiracy} between the decaying atom observed and the processes occurring in the minds of Alice and Bob, which would be a conspiracy of a kind that should not be tolerated in any decent theory of natural law. The whole idea that a natural mechanism could exist that drives Alice's and Bob's behaviour is often found difficult to accept.

In the CAI, however, natural law forbids the emergence of states where beables are superimposed. Neither Alice nor Bob will ever be able to produce such states by rotating their polarisation filters. Indeed, the state their minds are in, are ontological in terms of the beables, and they will not be able to change that.

Superdeterminism has to be viewed in relation with correlations over space-like distances. We claim that not only there are correlations, but the correlations are also extremely strong. The state we call `vacuum state' is full of correlations. Quantum field theory (to be discussed in part~II, section~\ref{QFT}), must be a direct consequence of the underlying ontological theory. It explains these correlations. All 2-particle expectation values, called propagators, are non-vanishing outside the light cone. Also the many-particle expectation values are non-vanishing there; indeed, by analytic continuation, these amplitudes are seen to turn into the complete scattering matrix, which encapsulates all laws of physics that are implied by the theory. In chapter~\ref{Bell}, it is shown how a 3-point correlation can, in principle, generate the violation of the CHSH inequality, as  required by quantum mechanics.

In view of these correlation functions, and the law that says that beables will never superimpose, we now suspect that this law forbids Alice and Bob to both change their minds in such a way that these correlation functions would no longer hold.

\subsubsection{The role of entanglement\labell{entanglement}}

We shall not claim that these should be the last words on Bell's interesting theorem. Indeed, we could rephrase our observations in a somewhat different way. The reason why the standard, Copenhagen theory of quantum mechanics violates Bell is simply that the two photons (or electrons, or whatever the particles are that are being considered), are \emph{quantum entangled}. In the standard theory, it is assumed that Alice may change her mind about her setting without affecting the photon that is on its way to Bob. Is this still possible if Alice's photon and Bob's photon are entangled?

According to the CAI, we have been using templates to describe the entangled photons Alice and Bob are looking at, and this leads to the violation of the CHSH inequalities. In reality, these templates were reflecting the relative correlations of the ontological variables underlying these photons.  To describe entangled states as beables, their correlations are essential. We assume that this is the case when particles decay into entangled pairs because the decay has to be attributed to vacuum fluctuations (see section \ref{vac}), while also the vacuum cannot  be a single, translation invariant, ontological state.

Resetting Alice's experiment without making changes near Bob would lead to a state that, in quantum mechanical terms, is not orthogonal to the original one, and therefore not ontological. The fact that the new state is not orthogonal to the previous one is quite in line with the standard quantum mechanical descriptions; after all, Alice's photon was replaced by a superposition.

The question remains how this could be. If the cellular automaton stays as it is near Bob, why is the `counterfactual' state not orthogonal to it? The CAI says so, since the classical configuration of her apparatus has changed, and we stated that any change in the classical setting leads to a change in the ontological state, which is a transition to an orthogonal vector in Hilbert space.

We cannot exclude the possibility that the apparent non-locality in the ontic -- template mapping is related to the difficulty of identifying the right Hamiltonian for the standard quantum theory, in terms of the ontic states. We should find a Hamiltonian that is the integral of a local Hamiltonian density. Strictly speaking, there may be a complication here; the Hamiltonian we use in is often an approximation, a very good one, but it ignores some subtle non-local effects. This will be further explained in part~II.

\subsubsection{Choosing a basis\labell{basischoice}}

Some physicists of previous generations thought that distinguishing different basis sets is very important. Are particles `real particles' in momentum space or in configuration space? Which are the `true' variables of electro-magnetism, the photons or the electric and magnetic fields? The modern view is to emphasise that,  any basis serves our purposes as well as any other, since, usually, none of the conventionally chosen basis spaces is truly ontological.

In this respect, Hilbert space should be treated as any ordinary vector space, such as the space in which each of the coordinates of the planets in our planetary system are defined. It makes no difference which coordinate frame we use. Should the \(z\)-axis be orthogonal to the local surface of the earth? Parallel to the earth's axis of rotation? The ecliptic? Of course, the choice of coordinates is immaterial.
Clearly, this is exemplified in Dirac's notation. This beautiful notation is extremely general and  as such it is quite suitable for the discussions presented in this book.

But then, after we declared all sets of basis elements to be as good as any other, in as far as they describe some familiar physical process or event, we venture into speculating that a more special basis choice can be made. A basis could exist in which all \emph{super-microscopic} physical observables are \emph{beables}; these are the observable features at the Planck scale, they all must be diagonal in this basis. Also, in this basis, the wave function only consists of zeros and a single one. It is ontological. This special basis, called `ontological basis',  is hidden from us today but it should be possible to identify such a basis\fn{There may be more than one, non equivalent choices, as explained later (section~\ref{invisible}).}, in principle. This book is about the search for such a basis. It is not the basis of the particles, the fields of the particles, of atoms or molecules, but something considerably more difficult to put our hands on. 

The reader may be surprised when we emphasise that also all \emph{classical} states, describing stars and planets, automobiles, people, and eventually pointers on detectors, will be diagonal in the same ontological basis. Why is that? It is of crucial importance, and it was explained in section~\ref{classical}.

\subsubsection{Correlations and hidden information\labell{correlinfo}}

An essential element in our analysis may be the observations expressed in Subsection \ref{hiddeninfo}. It was noted that the details of the ontological basis must carry crucial information about the future, yet in a concealed manner: the information is non-local. It is a simple fact that, in all of our models, the ontological nature of a state the universe is in, will be conserved in time: once we are in a well-defined ontological state, as opposed to a template, this feature will be preserved in time (the ontology conservation law). It prevents Alice and Bob from choosing any setting that would `measure' a template that is not ontological. Thus, this feature prevents counterfactual definiteness. Even the droppings of a mouse cannot disobey this principle.

In adopting the CAI, we except the idea that all events in this universe are highly correlated, but in a way that we cannot exploit in practice. It would be fair to say that these features still carry a sense of mystery that needs to be investigated more, but the only way to do this is to search for more advanced models.

\subsubsection{Free Will\labell{will}}

The notion of ``free will" can be confusing. At some occasions, the discussion seems to border to a religious one. It should be quite clear that the theories of nature discussed in this book, have nothing to do with religion, and so we must formulate in a more concrete manner what is meant by ``free will". 

The idea that we call `free will' is actually extremely simple, in principle. What one might expect in a theory is that: 
	\begin{quote}  \emph{the theory predicts how its variables evolve, in an unambiguous way, from any chosen initial state. }\end{quote}
In a Bell-type experiment, suppose we start from a configuration with given settings \(a\) and \(b\) of Alice's and Bob's filters.
We see entangled particles moving from the source to the two detectors, What `free will' then means is that our theory not only yields a unique prediction for this setting, but it should also give a unique prediction of what happens when we look at a different initial state, such as the one we get if we make a slight modification in Alice's setting \(a\), without modifying anything in the approaching particles or Bob's setting \(b\).

We then don't care to check which modifications would be needed in the past events to realise this particular modification. The theory should produce a prediction. However, Bell derived his inequalities for the outcomes of different initial states that he chose, and these inequalities are violated by quantum mechanics.

We derived in subsection~\ref{mousedr} that, in order to reproduce the quantum mechanical result, the probabilities of the settings \(a,\ b\) and \(c\) must be correlated, and the correlation function associated to one simple model was calculated. Here we see how, in principle, the notion of free will as given above can be obstructed:
	\begin{quote}\emph{If a modification is made in the given values of the kinetic variables, they might have a much larger or  smaller probability than the original configuration.}\end{quote}
	The correlation function we found describes 3-point correlations. All two-point correlations vanish.
	
What happens if we follow a configuration back to the past, if it violates the Bell- or CHSH inequalities? In that case, the quantum state of the entangled photons will no longer be as it was originally prepared: a state where the total spin of the two particles is zero (an S-state). In the case of photons, a D-state with total spin s=2 is formed. Thus, by choosing a different setting, either Alice or Bob modified the states of the photons they are detecting. Following this \(s=2\) state back to the past, we do not see a simple decaying atom, but a much less probable state of photons bouncing off the atom, refusing to perform the  assumed decay process backwards in time. Thermodynamically, this is a much less probable initial state, let us call it the counterfactual initial state.

This counterfactual initial state will be an entirely legal one in terms of the microscopic laws of physics, but probably not at all in terms of the macroscopic laws, in particular, thermodynamics. What this argument shows is that, Bell's theorem requires more hidden assumptions than usually thought: \emph{The quantum theory only contradicts the classical one if we assume that the `counterfactual modification' does not violate the laws of thermodynamics.}

In our models, we must assume, it does. Inevitably, a more `probable' modification of the settings does turn the photon state into a different one. At first sight, this seems odd: the modification was made in one of the settings, not in the approaching photons. However, we must admit that the photons described in quantum mechanical language, are in template states; the ontological states, forming an orthonormal set, must involve many more ontological degrees of freedom than just these two photons, just in order to stay orthonormal.

\subsecl{The importance of second quantisation}{introsecond}

We fully realise that any attempts to explain the surprising outcome of Bell's Gedanken experiment will be received with suspicion unless more specific models can be constructed where solid calculations support our findings. We have specific models, but so-far, these have been less than ideal to explain our points. What would really be needed is a model or a set of models that obviously obey quantum mechanical equations, while they are classical and deterministic in their set-up. Ideally, we should find models that reproduce relativistic, quantised field theories with interactions, such as the Standard Model.

The use of second quantisation, will be explained further in subsections~\ref{secondneutrino} and \ref{2ndquantCA} of part II. We start with the free Hamiltonian and insert the interactions at a later stage, a procedure that is normally assumed to be possible using perturbation expansion. The trick used here is, that the free particle theory will be manifestly local, and also the interactions, represented by rare transitions, will be local. The interactions are introduced by postulating new transitions that create or annihilate particles. All terms in this expansion are local, so we have a local Hamiltonian. To handle the complete theory, one then has to do the full perturbation expansion. Obeying all the rules of quantum perturbation theory, we should obtain a description of the entire, interacting model in quantum terms. Indeed, we should reproduce a genuine quantum field theory this way.

Is this really true? Strictly speaking, the perturbation expansion does not converge, as is explained also in section~\ref{2ndquantCA}. However, then we can argue that this is a normal situation in quantum field theory. Perturbation expansions formally always diverge, but they are the best we have -- indeed they allow us to do extremely accurate calculations in practice. Therefore, reproducing these perturbative expansions, regardless how well they converge, is all we need to do in our quantum theories.

A fundamentally convergent expression for the Hamiltonian does exist, but it is entirely different. The differences are in non-local terms, which we normally do not observe. Look at the exact expression, Eq.~\eqn{firstH}, first treated in section~\ref{basics} of chapter~\ref{determm}:
	\be H_\op\,\d t=\pi-i\sum_{n=1}^\infty{1\over n}\left(U_\op(n\,\d t)-U_\op(-n\,\d t)\right)\ .\eel{firstH1}
	
	This converges, except for the vacuum state itself. Low energy states, states very close to the vacuum, are the states where convergence is excessively slow. Consequently, as was explained before, terms that are extremely non-local, sneak in. 
	
	This does \emph{not} mean that the cellular automaton would be non-local; it is as local as it can be, but it means that if we wish to describe it with infinite precision in quantum mechanical terms, the Hamiltonian may generate non-localities. One can view these non-localities as resulting from the fact that our Hamiltonian replaces time difference equations, linking instances separated by integral multiples of \(\d t\), by differential equations in time; the states in between the well-defined time spots are necessarily non-local functions of the physically relevant states.
	
	It is these non-localities that may well be responsible for being sensitive to Bob's and Alice's environment. Or in different words: if we modify Alice's settings by rotating her polariser, without modifying the state near Bob, we may well generate a state with excessively high energy. 
	
One might argue that there should be no reason to try to fill the gaps between integer times steps, but then there would not exist an additive energy function, which we need to stabilise the solutions of or equations. Possibly, we have to use a combination of a classical, integer-valued Hamilton function (chapter~\ref{Hamiltonform} of part II ), and the periodically defined Hamiltonian linking only integer-valued time steps, but exactly how to do this correctly is still under investigation. We have not yet found the best possible approach towards constructing the desired Hamilton operator for our template states. The second-quantised theory that will be further discussed is presently our best attempt, but we have not yet been able to reproduce the quite complex symmetry structure of the Standard Model, in particular Poincar\'e invariance.
	
	As long as we have no explicit model that, to the extent needed,  reproduces Standard Model-like interactions, we cannot verify the correctness of our approach. Lacking that, what has to be done next is the calculations in models that are as realistic as possible.

\bigskip\bigskip
 \part[\textbf{Calculation Techniques}]{\Large \textbf {Calculation Techniques}}
 
 \bigskip\bigskip

\newsecl{Introduction to part II}{intro2}   \def\k{\kappa}
Many of the technical calculations and arguments mentioned in part I of this book, were postponed to the second part, so as to make the first part easier to read while keeping it coherent, and to give some nice firework in the second part. The price we pay for this is that there will be a number of repetitions, for which we apologise.
\subsecl{Outline of part II } {outline2}
	One of our main themes is that quantum mechanics may be viewed as a mathematical tool rather than a new theory of physical phenomena. Indeed, in condensed matter theory, several models exist where the physical setup and the questions asked are fundamentally classical, yet the calculations are performed by regarding the system as a quantum mechanical one. The \emph{Ising Model} is a beautiful example of this.\cite{kauffman-1949}
	
	There is no better way to illustrate our approach than by actually showing how such calculations are done. The Cogwheel Model was already introduced in section \ref{Cogwheelm}. Now, in chapters \ref{harmrot}---\ref{harmosc}, we show some more of our mathematical tools, how to construct quantum Hamiltonians and how to approach continuum limits. Here, the cogwheel model is linked to the \emph{harmonic rotator}, but also other, notoriously `classical' structures, such as the planetary system, are transformed into models that appear to be quantum mechanical.
	
	The continuum limit of a single, periodic cogwheel is an important example. It approaches the ordinary quantum harmonic oscillator with the same period \(T\). The continuum cogwheel is actually a smoothly rotating wheel. Is the classical rotating wheel equivalent to a quantum harmonic oscillator? In a sense, yes, but there are some subtleties that one has to be aware of. This is why we decided to do this limit in two steps: first transform the cogwheel into a \emph{harmonic rotator}, allowing the teeth to form a representation of the group \(SU(2)\), and only then consider the continuum limit. This enables us to recognise the operators \(x\) and \(p\) of a genuine harmonic oscillator already in finite cogwheels.
	
	Like other technical calculations elsewhere in this book, they were done on order to check the internal consistency of the systems under study. It was fun to do these calculations, but they are not intended to discourage the reader. Just skip them, if you are more interested in the general picture. 
	
	The issue of the locality of the Hamiltonian is further treated in chapter~\ref{locality}. It will come up frequently in almost any deterministic model, and again the mathematics is interesting. We observe that a lot depends on the construction of the \emph{vacuum state}. It is the state of lowest energy, and the solution of the equation ``energy \(=\) lowest", generates non-localities indeed. In reality, as is well known in quantum field theories, signals will not go faster than the speed of light.
What will be shown in this chapter is that there is a way to avoid non-localities when objects move around surrounded by a vacuum, provided one uses a first-quantised theory where only the center part of the energy spectrum is used. Consequently, energy can be positive or negative there. Subsequently, one introduces anti-particles, such that the negative energy states actually represent holes of antiparticles. It is nothing but Dirac's trick to ensure that the physical vacuum has lowest possible energy.
	
Dirac first phrased his theory for fermionic particles. Indeed, fermions are easier to understand in this respect that bosons are. Therefore, we first introduce fermions as an essential element in our models, see chapter \ref{fermions}.

It so happens that Dirac's equation for the electron is well suited to demonstrate our prescription of searching for ``beables" in a quantum theory. 
Section \ref{nu} also begins at an easy pace but ends up in lengthy derivations. Here also, the reader is invited to enjoy the intricate features of the `neutrino' model, but they can just as well be skipped.

We take the simplified case of the Dirac equation for a two-component neutrino.  It is fundamentally simpler than the Dirac equation for the electron. Furthermore, we assume the absence of interactions. The math starts out simple, but the result is striking: neutrinos are configurations of flat membranes, or `sheets', rather than particles. the sheets move around classically. This is not a theory but a mathematical fact, as long as we keep mass terms and interactions out of the picture; these are left for later.

Having observed this, we asked the question how to go from the sheet variables back to the neutrino's quantum operators such as position \(\vec x\), momentum \(\vec p\), and spin \(\vec\s\). Here, the math does become complicated, and it is interesting as an exercise (subsections~\ref{neutrinoalgebra} and \ref{orthonormbeable}). The neutrinos are ideal for the application of second quantisation (subsection \ref{secondneutrino}), although, in this language, we cannot yet introduce interactions for them.

Our models, discussed in chapters~\ref{morecog}--\ref{2dnoint} and \ref{Hamiltonform}, have in common that they are local, realistic, and based on conventional procedures in physics. They also have in common that they are limited in scope, they do not capture all features known to exist in the real world, such as all particle species, all symmetry groups, and in particular special and general relativity. The models should be utterly transparent, they indicate directions that one should look at, and, as was our primary goal, they suggest a great approach towards \emph{interpreting} the quantum mechanical laws that are all so familiar to us.

\(PQ\) theory, chapter~\ref{PQ} is a first attempt to understand links between theories based on \emph{real} numbers and theories based on \emph{integer}, or \emph{discrete}, numbers. The idea is to set up a clean formalism connecting the two, so that it can be used in many instances.  Chapter~\ref{PQ} also shows some nice mathematical features, with good use of the elliptic theta functions. The calculations look more complicated than they should be, just because we searched for an elegant mechanism relating the real line to pairs of integers on the one hand and the torus on the other, keeping the symmetry between coordinates and momenta.

In chapter \ref{2dnoint}, we found some other interesting extensions of what was done in chapter~\ref{PQ}. A very straightforward argument drew our interest to String Theory and Superstring Theory. We are not strongly advocating the idea that the only way to do interesting physics at the Planck scale is to believe what string theoreticians tell us. It is not clear from our work that such theories are \emph{the} way to go, but we do notice that our program shows remarkable links with string theory. \emph{In the absence of interactions}, the local equations of string- and superstring theory appear to allow the construction of beables, exactly along the route that we advocate. The most striking feature exposed here, is that strings, written in the usual form of continuous quantum field theories in one space, and one time dimension, map onto classical string theories that are not defined in a continuous target space, but on a space-time lattice, where the lattice spacing \(a\) is given as \(a=2\pi\sqrt{\a'}\).

Symmetries, discussed in chapter~\ref{symm}, are difficult to understand in the CA Interpretation of quantum mechanics. However, in the CAI, symmetry considerations are as important as anywhere else in physics. Most of our symmetries are discrete, but in some cases, notably in string theory, continuous symmetries such as the Poincar\'e group, can be recovered.

In chapter~\ref{Hamiltonform},  we address the positivity problem of the Hamiltonian from a different perspective. There, the usual Hamiltonian formalism is extended to include discrete variables, again in pairs \(P_i,\,Q_i\), evolving in discrete time. When we first tried to study this, it seemed like a nightmare, but it so happens that the `discrete Hamilton formalism' comes out to be almost as elegant as the usual differential form. And indeed here, the Hamiltonian can easily be chosen to be bounded from below.

Eventually, we wish to reproduce effective laws of Nature that should take the form of today's quantum field theories. This is still quite difficult. It was the reason for setting up our procedures in a formal way, so that we will keep the flexibility to adapt our systems to what Nature seems to be telling us through the numerous ingenious experiments that have been performed. We explain some of the most important features of quantum field theory in chapter~\ref{QFT}. Most notably: in quantum field theories, no signal can carry useful information faster than the speed of light, and probabilities always add up to one. Quantum field theory is entirely local, in its own inimitable quantum way. These features we would like to reproduce in a deterministic quantum theory. 

To set up the \emph{Cellular Automaton Interpretation} in more detail, we first elaborate some technical issues in cellular automata in general (chapter~\ref{CAdetail}). These are not the technicalities encountered when computer programs are written for such systems; software experts will not understand much of our analysis. This is because we are aiming at understanding how such systems may generate quantum mechanics at the very large time and distance limit, and how we may be able to connect to elementary particle physics. What we find is a beautiful expression for a quantum Hamiltonian, in terms of an expansion called the BCH expansion. Everything would have been perfect if this were a convergent expansion. 

However, it is easy to see that the expansion is not convergent. We try a number of alternative approaches with some modest successes, but
not all issues will be resolved, and the suspicion is aired concerning the source of our difficulties: quantum gravitational effects may be of crucial importance, while it is exactly these effects that are still not understood as well as is needed here. We do propose to use the BCH expansion for many classes of cellular automata to demonstrate how they could be used to \emph{interpret} quantum mechanics. I know that the details are not yet quite right, but this probably has to be attributed to the simple fact that we left out lots of things, notably special and general relativity

\subsecl{Notation}{notationII}

It is difficult keep our notation completely unambiguous. In chapter~\ref{PQ}, we are dealing with many different types of variables and operators. When a dynamical variable is an integer there,  we shall use capitals \(A,\,B,\,\cdots, \,P,\,Q,\,\cdots\). Variables that are periodic with period \(2\pi\), or at least constrained to lie in an interval such as \((-\pi,\,\pi]\),  are angles, mostly denoted by Greek lower case letters \(\a,\,\b,\,\cdots,\,\k,\,\tht,\,\cdots\), whereas real variables will most often be denoted by lower case Latin letters \(a,\,b,\,\cdots,\,x,\,y,\,\cdots\). Yet sometimes we run out of symbols and deviate from this scheme, if it seems to be harmless to do so. For instance, indices will still be \(i,\,j,\,\cdots\) for space-like vector components,  \(\a,\,\b,\,\cdots\) for spinors and \(\m,\,\n,\,\cdots\) for Lorentz indices. The Greek letters \(\j\) and \(\vv\) will be used for wave functions as well.

Yet it is difficult to keep our notation completely consistent; in some chapters before chapter \ref{PQ}, we use the quantum numbers \(\ell\) and \(m\) of the \(SU(2)\) representations to denote the integers that earlier were denoted as \(k\) or \(k-m\), and later in chapter \ref{PQ} replaced by capitals.

As in part~I, we use a super- or subscript ``{\small{op}}" to distinguish an operator from an ordinary numerical variable. The caret (\(\^{}\)) will be reserved for vectors with length one, the arrow for more general vectors, not necessarily restricted to three dimensional space. Only in chapter~\ref{QFT}, where norms of vectors do not arise, we use the caret for the Fourier transform of a function.

Dirac's constant \(\hbar\) and the velocity of light \(c\) will nearly always be defined to be one in the units chosen. In previous work, we used a spacial symbol to denote \(e^{2\pi}\) as an alternative basis for exponential functions. This would indeed sometimes be useful for calculations, when we use fractions that lie between 0 and 1, rather than angles, and it would require that we normalise Planck's original constant \(h\) rather then \(\hbar\) to one, but in the present monograph we return to the more usual notation.

Concepts frequently discussed are the following:
	\bi{-} \emph{discrete} variables are variables such as the integer numbers, whose possible values can be counted. Opposed to continuous variables, which are typically represented by real or complex numbers.
	\itm{-} \emph{fractional} variables are variables that take values in a finite interval or on a circle. The interval may be \([0,1)\,,\ [0,2\pi)\,,\ (-\half,\,\half]\,,\) or \((-\pi,\,\pi]\). Here, the square bracket indicates a bound whose value itself may be included, a round bracket excludes that value. A real number can always be decomposed into an integer (or discrete) number and a fractional one.
	\itm{-} a theory is \emph{ontological}, or `ontic',  if it only describes `really existing' objects; it is simply a classical theory such as the planetary system, in the absence of quantum mechanics. The theory does not require the introduction of Hilbert space, although, as will be explained, Hilbert space might be very useful. But then, the theory is formulated in terms of observables that are commuting at all times.
	\itm{-} a feature is \emph{counterfactual} when it is assumed to exist even if, for fundamental reasons, it cannot actually be observed; if one would try to observe it, some other feature might no longer be observable and hence become counterfactual. This situation typically occurs if one considers the measurement of two or more operators that do not commute. More often, in our models, we shall encounter features that are not allowed to be counterfactual.
	\itm{-} We talk of \emph{templates} when we describe particles and fields as solutions of Schr\"o\-dinger's equation in an ontological model, as was explained in subsection~\ref{templates.sub}. Templates may be superpositions of ontic states and/or other templates, but the ontic states all form an orthonormal set; superpositions of ontic states are never ontic themselves.
	\ei

\subsecl{More on Dirac's notation for quantum mechanics}{Dirac2}

A denumerable set of states \(|e_i\ket\) is called an orthonormal basis of \(\HH\) if every state \(|\j\ket\in\HH\) can be approximated by a linear combination of a finite number of states \(|e_i\ket\) up to any required precision:
	\be |\j\ket=\sum_{i=1}^{N(\e)}\l_i|e_i\ket+|\e\ket\ ,\quad \|\e\|^2=\bra\e|\e\ket<\e^2\ , \quad\hbox{for any}\ \ \e>0\ . \eel{complete}
(a property called `completeness'), while
	\be \bra e_i|e_j\ket=\d_{ij} \eel{orthonorm}
(called `orthonormality'). From Eqs.~\eqn{complete} and \eqn{orthonorm}, one derives
	\be\l_i=\bra e_i|\j\ket\ ,\quad\sum_i|e_i\ket\bra e_i|=\Bbb I\ ,\eel{complete2}
where \(\Bbb I\) is the identity operator: \({\Bbb I}|\j\ket=|\j\ket\) for all \( |\j\ket\). 

In many cases, the discrete sum in Eqs.~\eqn{complete} and \eqn{complete2} will be replaced by an integral, and the Kronecker delta \(\d_{ij}\) in Eq.~\eqn{orthonorm} by a Dirac delta function, \(\d(x^{1}-x^{2})\). We shall still call the states \(|e_{(x)}\ket\) a basis, although it is not denumerable.

A typical example is the set of wave functions \(\j(\vec x)\) describing a particle in position space. They are regarded as vectors in Hilbert space where the set of delta peak wave functions \(|\vec x\ket\) is chosen to be the basis:
	\be \j(\vec x)\equiv\bra \vec x|\j\ket\ ,\quad \bra\vec x|\vec x\,'\ket=\d^3(\vec x-\vec x\,')\ . \eel{wavefnctn}
The Fourier transformation is now a simple rotation in Hilbert space, or a transition to the \emph{momentum basis}:
	\be\bra\vec x|\j\ket=\int\dd^3\vec p\,\,\bra\vec x|\vec p\,\ket\bra\vec p\,|\j\ket\ ;\quad \bra \vec x|\vec p\,\ket=\fract 1{(2\pi )^{3/2}}\,e^{i\vec p\cdot\vec x}\ . \eel{Fourier1}
	
Many special functions, such as the Hermite, Laguerre, Legendre, and Bessel functions, may be seen as generating different sets of basis elements of Hilbert space.

Often, we use product Hilbert spaces: \(\HH_1\otimes\HH_2=\HH_3\), which means that states  \(|\f\ket\) in \(\HH_3\) can be seen as normal products of states \(|\j^{(1)}\ket\) in \(\HH_1\) and \(|\j^{(2)}\ket\) in \(\HH_2\):
	\be|\f\ket=|\j^{(1)}\ket|\j^{(2)}\ket\ , \eel{prodstates}
and a basis for \(\HH_3\) can be obtained by combining a basis in \(\HH_1\) with one in \(\HH_2\):
	\be |e^{(3)}_{ij}\ket=|e^{(1)}_i\ket|e^{(2)}_j\ket\ . \eel{prodbasis}

Often, some or all of these factor Hilbert spaces are finite-dimensional vector spaces, which of course also allow all the above manipulations\fn{The term \emph{Hilbert space} is often restricted to apply to infinite dimensional vector spaces only; here we will also include the finite dimensional cases.}. We have, for example, the 2-dimensional vector space spanned by spin \(\half\) particles. A basis is formed by the two states \(|\uparrow\,\ket\) and \(|\downarrow\,\ket\). In this basis, the Pauli matrices \(\s^\op_{1,2,3}\) are defined as in part I, Eqs.~\ref{pauli}.
The states
	\be|\ra\,\ket=\fract1{\sqrt 2}\pmatrix{1\cr 1}\ ,\quad|\leftarrow\,\ket=\fract 1{\sqrt 2}\pmatrix{1\cr -1}\ , \eel{leftrightspin}
form the basis where the operator \(\s_1\) is diagonal: \(\s^\op_1\ra\pmatrix{1&0\cr 0&-1}\).

Dirac derived the words `bra' and `ket' from the fact that the expectation value for an operator \(\OO^\op\) can be written as the operator between brackets, or
	\be\bra\OO^\op\ket=\bra\j|\,\OO^\op|\j\ket\ . \eel{expectval}
More generally, we shall often need the matrix elements of an operator in a basis \(\{|e_i\ket\}\):
	\be \OO_{ij}=\bra e_i|\OO^\op|e_j\ket\ . \eel{matrixelem}
The transformation from one basis \(\{|e_i	\ket\}\) to another, \(\{|e'_i\ket\}\) is a unitary operator \(U_{ij}\):
	\be |e'_i\ket=\sum_jU_{ij}|e_j\ket\ ,\qquad U_{ij}=\bra e_j|e'_i\ket\ ;\nm \\
	\sum_kU_{ik}U_{jk}=\sum_k\bra e'_i|e_k\ket\bra e_k|e'_j\ket=\d_{ij}\ . \eel{unitarytrf}
This will be used frequently. For instance, the Fourier transform is unitary:
	\be\int\dd^3\vec p\,\bra \vec x|\vec p\,\ket\bra\vec p\,|\vec x'\ket \iss \fract1{(2\pi)^3}\int\dd^3\vec p\,e^{i\vec p\cdot\vec x-i\vec p\cdot\vec x'}\iss
	\d^3(\vec x-\vec x')\ . \eel{Fourierunitary}
	
The Schr\" odinger equation will be written as:
	\be & \fract{\dd}{\dd t}|\j(t)\ket=-iH^\op|\j(t)\ket\ , \qquad\fract{\dd}{\dd t}\bra\j(t)|=\bra\j(t)|iH^\op\ ;& \nm\\ 
	& |\j(t)\ket=e^{-iH^\op\, t}|\j(0)\ket\ ,& \eel{diracschro}
where \(H^\op\) is the Hamiltonian, defined by its matrix elements \(H_{ij}=\bra e_i|H^\op|e_j\ket\).	
	
Dirac's notation may be used to describe non-relativistic wave functions in three space dimensions, in position space, in momentum space or in some other basis, such as a partial wave expansion, it can be used for particles with spin, it can be used in many-particle systems, and also for quantised fields in solid state theory or in elementary particle theory. The transition from a Fock space notation, where the basis is spanned by states containing a fixed number \(N\) of particles (in position or in momentum space, possibly having spin as well), to a notation where the basis is spanned by the functions representing the fields of these particles, is simply a rotation in Hilbert space, from one basis into another.

		\def\op{{\mathrm{op}}}  \def\ont{{\mathrm{ont}}}  \def\tot{{\mathrm{tot}}}
	\newsecl{More on cogwheels}{morecog}
	\subsecl {the group $SU(2)$, and the harmonic rotator}{harmrot}
	Let us return to the original cogwheel with \(N\) teeth, as introduced in chapter~\ref{determm}, section \ref{Cogwheelm}. It may be very illuminating to  define the constant \(\ell=(N-1)/2\), and introduce the operators \(L_1,\ L_2\) and \(L_3\) as follows (\(k=0,1,\,\cdots, 2\,\ell\) is the energy quantum number; the time step is \(\d t=1\)):
		\be   &	L_3\iss \fract{N}{2\pi}H_\op-\ell\iss k-\ell \ ,& \nm\\
			&	L_1\iss \half(L_++L_-)\ ,\qquad L_2\iss -\half i\,(L_+-L_-)\ ,& \nm\\
			&L_+|k\ket_H\iss \sqrt{(k+1)(2\ell-k)}\,|k+1\ket_H\ ,& \nm\\
			& L_-|k\ket_H\iss \sqrt{k(2\ell+1-k)}\,|k-1\ket_H\ .&		\eel{Losc}
Using the quantum number \(m=k-\ell=L_3\), we get the more familiar expressions for the angular momentum operators \(L_a,\ a=1,2,3,\) obeying the commutation relations
	\be [L_a,\,L_b]=i\e_{abc} L_c\ . \eel{Lcomm}
The original ontological states \(|n\ket_\ont\) can be obtained from the angular momentum states by means of the transformation rules \eqn{Hstates} and \eqn{ontstates}. It is only these that evolve as ontological states. Other operators can be very useful, however. Take, for instance,
	\be x=\fract 1 {\sqrt{\ell}} L_1\ ,\qquad p=-\fract 1 {\sqrt{\ell }} L_2\ ,\qquad [x,p]=i(1-\fract{2\ell+1}{2\pi\ell}H_\op)\ ,\eel{qpmodcom}
then, for states where the energy \(\bra H_\op\ket\ll 1\), we have the familiar commutation rules for positions \(x\) and momenta \(p\), while the relation \(L_1^2+L_2^2+L_3^2=L^2=\ell(\ell+1)\) implies that, when \(\bra H_\op\ket\ll 1/\sqrt{\ell}\), 
	\be H\ra \fract{2\pi}{N}\,\half(p^2+x^2-1)\ , \eel{HharmoscL}
which is the Hamiltonian for the harmonic oscillator (the zero point energy has been subtracted, as the lowest energy eigen state was set at the value zero). Also, at low values for the energy quantum number \(k\), we see that \(L_\pm\) approach the creation and annihilation operators of the harmonic oscillator (see Eqs.~\eqn{Losc}):
	\be L_-\ra{\sqrt{2\ell+1}}\,a\ ,\qquad L_+\ra{\sqrt{2\ell+1}}\,a^\dag\ . \eel{Lpmraaadag}

Thus, we see that the lowest energy states of the cogwheel approach the lowest energy states of the harmonic oscillator. This will be a very useful observation if we wish to construct models for quantum field theories, starting from deterministic cogwheels. The model described by eqs.~\eqn{Losc}--\eqn{qpmodcom} will be referred to as the \emph{harmonic rotator}. The Zeeman atom of section~\ref{Cogwheelm} is a simple example with \(\ell=1\).

Note, that the spectrum of the Hamiltonian of the harmonic rotator is \emph{exactly} that of the harmonic oscillator, except that there is an upper limit, \(H_\op<2\pi\). By construction, the \emph{period} \(T=(2\,\ell+1)\d t\) of the harmonic rotator, as well as that of the harmonic oscillator, is exactly that of the periodic cogwheel.

The Hamiltonian that we associate to the harmonic rotator is also that for a spinning object that exhibits \emph{precession} due to a torque force on its axis. Thus, physically, we see that an oscillator drawing circles in its \((x,p)\) phase space is here replaced by a precessing top. At the lowest energy levels, they obey the same equations.

We conclude from this section that a cogwheel with \(N\) states can be regarded as a representation of the group SU(2) with total angular momentum \(\ell\), and \(N=2\,\ell+1\). The importance of this approach is that the representation is a unitary one, and that there is a natural ground state, the ground state of the harmonic oscillator. In contrast to the harmonic oscillator, the harmonic rotator also has an upper bound to its Hamiltonian. The usual annihilation and creation operators, \(a\) and \(a^\dag\), are replaced by \(L_-\) and \(L_+\), whose commutator is not longer constant but proportional to \(L_3\), and therefore changing sign for states \(|k\ket\) with \(\ell<k\le 2\:\!\ell\). This sign change assures that the spectrum is bounded from below as well as above, as a consequence of the modified algebra, \eqn{Lcomm}.

\subsecl{Infinite, discrete cogwheels}{infinitediscr}


	Discrete models with infinitely many states may have the new feature that some orbits may not be periodic. They then contain at least one non-periodic `rack'. There exists a universal definition of a quantum Hamiltonian for this general case, though it is not unique. Defining the time reversible evolution operator over the smallest discrete time step to be an operator \(U_\op(1)\), we now construct the simplest Hamiltonian \(H_\op\) such that \(U_\op(1)=e^{-iH_\op}\). For this, we use the evolution over \(n\) steps, where \(n\) is positive or negative:
		\be U_\op(n)=U_\op(1)^n=e^{-inH_\op}\ . \eel{nevolve} 
Let us assume that the eigenvalues \(\w\) of this Hamiltonian lie between 0 and \(2\pi\). We can then consider the Hamiltonian in the basis where
both \(U(1)\) and \(H\) are diagonal. Write 
		\be e^{-in\w}=\cos(n\w)-i\sin(n\w)\ , \eel{Udiag}
and then use Fourier transformations to derive  that, if \(-\pi<x<\pi\),
	\be x=2\sum_{n=1}^\infty{(-1)^{n-1}\sin(n x)\over n}\ . \eel{omegasumsin}
Next, write \(H_\op=\w=x+\pi\), to find that Eq.~\eqn{omegasumsin} gives
	\be\w=\pi-2\sum_{n=1}^\infty{\sin(n\w)\over n}. \eel{etasumsin}
Consequently, as in Eq.~\eqn{firstH},
	\be \w&=&\pi-\sum_{n-1}^\infty{i\over n}\,(\,U(n\,\d t)-U(-n\,\d t)\,)\qquad \hbox{and}\nm\\
	  H_\op&=&\pi-\sum_{n-1}^\infty{i\over n}\,(\,U_\op(n\,\d t)-U_\op(-n\,\d t)\,)\  . \eel{HfrUn}
Very often, we will not be content with this Hamiltonian, as it has no eigenvalues beyond the range \((0,\,2\pi)\). As soon as there are conserved quantities, one can add functions of these at will to the Hamiltonian, to be compared with what is often done with chemical potentials. Cellular automata in general will exhibit many such conservation laws. See Fig.~\ref{generalfinite.fig}, where every closed orbit represents something that is conserved: the label of the orbit.

In section~\ref{harmosc}, we consider the other continuum limit, which is the limit \(\d t\ra 0\) for the cogwheel model. First, we look at continuous theories more generally.

\subsecl{Automata that are continuous in time}{contont}
	
	In the physical world, we have no direct indication that time is truly discrete. It is therefore tempting to consider the limit \(\d t\ra 0\). At first sight, one might think that this limit should be the same as having a continuous degree of freedom obeying differential equations in time, but this is not quite so, as will be explained later in this chapter. First, in this section, we consider the strictly continuous deterministic systems. Then, we compare those with the continuum limit of discrete systems.

	Consider an ontological  theory described by having a continuous, multi-dimensional space of degrees of freedom \(\vec{q}\,(t)\),  depending on one continuous time variable \(t\), and its time evolution following  	
classical differential equations:
	\be {\dd \over\dd t}q_i(t)=f_i(\vec q\,)\ , \eel{classcont}
where \(f_i(\vec q\,)\) may be almost any function of the variables \(q_j\). 

An example is the description of massive objects obeying classical mechanics in \(N\) dimensions.	Let \(a=1,\cdots,N,\) and \( i=1,\cdots,2N\):
	\be \{i\}=\{a\}\oplus\{a+N\}\ ,\qquad q_a(t)=x_a(t)\ ,\qquad q_{a+N}(t)=p_a(t)\ , \nm \\[3pt] 
		f_a(\vec q\,)={\pa H_{\mathrm{class}}(\vec x,\vec  p\,)\over\pa p_a}\ , \qquad
		f_{a+N}(\vec q\,)=-{\pa H_{\mathrm{class}}(\vec x,\vec p\,)\over \pa x_a}\ , \eel{classmech}
where \(H_{\mathrm{class}}\) is the \emph{classical} Hamiltonian.
		
An other example is the quantum wave function of a particle in one dimension: 
	\be \{i\}=\{x\}\ ,\qquad q_i(t)=\j(x,t)\ ;\qquad f_i(\vec q\,)=-iH_S\,\j(x,t)\ , \eel{schroe1}
where now \(H_S\) is the Schr\" odinger Hamiltonian.
Note, however, that, in this case, the function \(\j(x,t)\) would be treated as an ontological object, so that the Schr\" odinger equation and the Hamiltonian eventually obtained will be quite different from the Schr\" odinger equation we start off with; actually it will look more like the corresponding \emph{second quantised} system (see later).

We are now interested in turning Eq.~\eqn{classcont} into a quantum system by changing the notation, not the physics. The ontological basis is then the set of states \(|\vec q\,\ket\), obeying the orthogonality property
	\be\bra\vec q\,|{\vec q}\,'\ket=\d^N(\vec q-{\vec q}\,')\  , \eel{contortho}
where \(\d\) is now the Dirac delta distribution, and \(N\) is the dimensionality of the vectors \(\vec q\).

If we wrote
	\be {\dd\over\dd t}\j(\vec q\,)\qu -f_i(\vec q\,){\pa\over{\pa q_i}}\j(\vec q\,)\qu -iH_\op\,\j(\vec q\,)\ , \eel{wrongdtpsiq}
where the index \(i\) is summed over, we would read off that
	\be H_\op\,\qu-if_i(\vec q\,){\pa\over\pa q_i}\iss f_i(\vec q\,)p_i\ ,\qquad p_i=-i{\pa\over\pa q_i}\ . \eel{wrongham}
This, however, is not quite the right Hamiltonian because it may violate hermiticity: \(H_\op\,\ne H^\dag_\op\,\).
The correct Hamiltonian is obtained if we impose that probabilities are preserved, so that, in case the Jacobian of \(\vec f(\vec q\,)\) does not vanish, the integral \(\int\dd^N\vec q\ \j^\dag(\vec q\,)\,\j(\vec q\,)\) is still conserved:
	\be \displaystyle {\dd\over\dd t}\j(\vec q\,&=& -f_i(\vec q\,){\pa\over{\pa q_i}}\j(\vec q\,)-\half\Big({\pa f_i(\vec q\,)\over\pa q_i}\Big)\j(\vec q\,)\iss -iH_\op\,\j(\vec q\,)\ , 
		\label{dtpsiq}\\[5pt]
	\displaystyle  H_\op &=& -if_i(\vec q\,){\pa\over\pa q_i}-\half i\Big({\pa f_i(\vec q\,)\over\pa q_i}\Big) \nm \\
	&=&\  \half ( f_i(\vec q\,) p_i + p_if_i(\vec q\,))
	 	  \ \equiv\ \half\{f_i(\vec q\,),\,p_i\}\  .  \eel{contham}
The \(\halff\) in Eq.~\eqn{dtpsiq} ensures that the product \(\j^\dag\,\j\) evolves with the right Jacobian. Note that this Hamiltonian is hermitean, and the evolution equation \eqn{classcont} follows immediately from the commutation rules
	\be[q_i,\,p_j]=i\d_{ij}\ ;\qquad{\dd\over\dd t}\OO_\op(t)=-i[\OO_\op\,,\,H_\op\,]\ . \eel{opevolv}
	
Now, however, we encounter a very important difficulty: this Hamiltonian has no lower bound. It therefore cannot be used to \emph{stabilise} the wave functions. Without lower bound, one cannot do thermodynamics. This feature would turn our model into something very unlike quantum mechanics as we know it. 

If we take \(\vec q\) space either one-dimensional, or in some cases two-dimensional, we can make our system periodic. Then let \(T\) be the smallest positive number such that 
 	\be \vec q\,(T)=\vec q\,(0)\ . \eel{qperiodic}
We have consequently
	\be e^{-iHT}|\vec q\,(0)\ket=|\vec q(0)\,\ket\ , \eel{eq35}
and therefore, on these states,
	\be H_\op\,|\vec q\,\ket=\sum_{n=-\infty}^\infty{2\pi n\over T}|n\ket\low{H\,H}\bra n|\vec q\,\ket\  .\eel{hdiscr}
Thus, the spectrum of eigenvalues of the energy eigenstates \(|n\ket_H\) runs over all integers from \(-\infty\) to \(\infty\).

 In the discrete case, the 
Hamiltonian has a finite number of eigenstates,  with eigenvalues \(2\pi k/(N\d t)+ \d E\) where \(k=0,\cdots,\,N-1\), which means that they lie in an interval \([\d E\,,\ 2\pi/T\,+\,\d E]\), where \(T\) is the period, and \(\d E\) can be freely chosen. So here, we always have a lower bound, and the state with that energy can be called `ground state' or `vacuum'.	
	
Depending on how the continuum limit is taken, we may or may not preserve this lower bound. The lower bound on the energy seems to be artificial, because all energy eigenstates look exactly alike.	 It is here that the \(SU(2)\) formulation for harmonic rotators, handled in subsection~\ref{harmrot}, may be hore useful
	
	An other remedy against this problem could be that we demand analyticity when time is chosen to be complex,	and boundedness of the wave functions in the lower half of the complex time frame. This would exclude the negative energy states, and still allow us to represent all probability distributions with wave functions. Equivalently, one could consider complex values for the variable(s) \(\vec q\) and demand the absence of singularities in the complex plane below the real axis. Such  analyticity constraints however seem  to be rather arbitrary; they are difficult to maintain as soon as interactions are introduced, so they would certainly have to be handled with caution. \label{contcaution}

One very promising approach to solve the ground state problem is Dirac's great idea of second quantisation: take an indefinite number of objects \(\vec q\), that is, a Hilbert space spanned by all states \(|\vec q^{\,(1)},\,\vec q^{\,(2)},\cdots \vec q^{\,(n)}\ket\), for all particle numbers \(n\), and regard the negative energy configurations as `holes' of antiparticles. This we propose to do in our `neutrino model', section~\ref{nu}, and in later chapters.

Alternatively, we might consider the continuum limit of a discrete theory more carefully. This we try first in the next chapter. Let us emphasise again: in general, excising the negative energy states just like that is not always a good idea, because any perturbation of the system might cause transitions to these negative energy states, and leaving these transitions out may violate unitarity.

The importance of the ground state of the Hamiltonian was discussed in  chapter~\ref{miss} of part I. The Hamiltonian \eqn{contham} is an important expression for fundamental discussions on quantum mechanics.

	As in the discrete case, also in the case of deterministic models with a continuous evolution law, one finds discrete and continuous eigenvalues, depending on whether or not a system is periodic. In the limit \(\d t\ra 0\) of the discrete periodic ontological model, the eigenvalues are integer multiples of \(2\pi/T\), and this is also the spectrum of the harmonic oscillator with period \(T\), as explained in chapter~\ref{harmosc}. The harmonic oscillator may  be regarded as a deterministic system in disguise. 
	
	The more general continuous model is then the system obtained first by having a (finite or infinite) number of harmonic oscillators, which means that our system consists of many periodic substructures, and secondly by admitting a (finite or infinite) number of conserved quantities on which the periods of the oscillators depend. An example is the field of non-interacting particles; quantum field theory then corresponds to having an infinite number of oscillating modes of this field. The particles may be fermionic or bosonic; the fermionic case is also a set of oscillators if the fermions are put in a box with periodic boundary conditions. \emph{Interacting quantum particles} will be encountered later (chapter~\ref{Hamiltonform} and onwards). 


\def\th{{\mathrm{th}}} 
\newsecl{The continuum limit of  cogwheels, harmonic rotators and oscillators}{harmosc}

	In the \(N\ra\infty\) limit, a cogwheel will have an infinite number of states. The  Hamiltonian will therefore also have infinitely many eigenstates. We have seen that there are two ways to take a continuum limit. As \(N\ra\infty\), we can keep the quantised time step  fixed, say it is 1. Then, in the Hamiltonian~\eqn{Hamiltonperiodic}, we have to allow the quantum number \(k\) to increase proportionally to \(N\), keeping \(\k=k/N\) fixed. Since the time step is one, the Hamiltonian eigenvalues, \(2\pi\k\), now lie on a circle, or, we can say that the energy  takes values in the continuous line segment \([0,\,2\pi)\) (including the point 0 but excluding the point \(2\pi\)). Again, one may add an arbitrary constant \(\d E\) to this continuum of eigenvalues. What we have then, is a model of an object moving on a lattice in one direction. At the beat of a clock, a state moves one step at a time to the right. This is the rack, introduced in section~\ref{infinitediscr}. The second-quantised version is handled in section~\ref{2dbosons}.
	
	The other option for a continuum limit is to keep the period \(T\) of the cogwheel constant, while the time quantum \(\d t\) tends to zero. This is also a cogwheel, now with infinitely many, microscopic teeth, but still circular. Since now \(N\,\d t=T\) is fixed, the ontological\fn{The words `ontological' and `deterministic' will be frequently used to indicate the same thing: a model without any non deterministic features, describing things that are really there.} states of the system can be described as an angle:
	\be 2\pi n/N\ra\vv\ ,\qquad {\dd\over\dd t}\vv(t)={2\pi\over T}\ . \eel{circlerotate}
The energy eigenvalues become
		\be E_k=2\pi\,k/T+\d E\ , \qquad k=0,1,\cdots,\,\infty\ .\eel{harmoscspectrum}
If \(\d E\) is chosen to be \(\pi/T\), we have the spectrum \(E_k=(2\pi/T)(k+\half)\). This is the spectrum of a harmonic oscillator. In fact,  \emph{any} periodic system with period \(T\), and a continuous time variable can be characterised by defining an angle \(\vv\) obeying the evolution equation~\eqn{circlerotate}, and we can attempt to apply a mapping onto a harmonic oscillator with the same period. 

Mappings of one model onto another one will be frequently considered in this book. It will be of importance to understand what we mean by this. If one does not put any constraint on the nature of the mapping, one would be able to map any model onto any other model; physically this might then be rather meaningless. Generally, what we are looking for are mappings that can be prescribed in a time-independent manner. This means that, once we know how to solve the evolution law for one model, then after applying the mapping, we have also solved the evolution equations of the other model. The physical data of one model will reveal to us how the physical data of the other one evolve.

This requirement might not completely suffice in case a model is exactly integrable. In that case, every integrable model can be mapped onto any other one, just by considering the exact solutions. In practice, however, time independence of the prescription can be easily verified, and if we now also require that the mapping of the physical data of one model onto those of the other is one-to-one, we can be confident that we have a relation of the kind we are looking for. If now a deterministic model is mapped onto a quantum model, we may demand that the classical states of the deterministic model map onto an orthonormal basis of the quantum model. Superpositions, which look natural in the quantum system, might look somewhat contrived and meaningless in the deterministic system, but they are certainly acceptable as describing probabilistic distributions in the latter. This book is about these mappings.

We have already seen how a periodic deterministic system can produce a discrete spectrum of energy eigenstates. The continuous system described in this subsection generates energy eigenstates that are equally spaced, and range from a lowest state to \(E\ra\infty\). Mapping this onto the harmonic oscillator, seems to be straightforward; all we have to do is map these energy eigenstates onto those of the oscillator, and since these are also equally spaced, both systems will evolve in the same way. Of course this is nothing to be surprised about: both systems are integrable and periodic with the same period  \(T\).

For the rest of this subsection, we will put \(T=2\pi\). The Hamiltonian of this harmonic oscillator can then be chosen to be\fn{In these expressions, \(x,\,p,\,a\), and \(a^\dag\) are all operators, but we omitted the subscript ``op" to keep the expressions readable.}
	\be H_\op=\half(p^2+x^2-1)\ ;\qquad H_\op\,|n\ket_H = n\,|n\ket_H\ ,\quad n=0,1,\cdots,\infty\ . \eel{harmham}
The subscript \({\scriptstyle{H}}\) reminds us that we are looking at the eigenstates of the Hamiltonian \(H_\op\). For later convenience, we subtracted the zero point energy\fn{Interestingly, this zero point energy would have the effect of flipping the sign of the amplitudes after exactly one period. Of course, this phase factor is not directly observable, but it may play some role in future considerations. In what we do now, it is better to avoid these complications.}.

In previous versions of this book's manuscript, we described a mapping that goes directly from a deterministic (but continuous) periodic system onto a harmonic oscillator. Some difficulties were encountered with unitarity of the mapping. At first sight, these difficulties seemed not to be very serious, although they made the exposition less than transparent. It turned out, however, that having a lower bound but not an upper bound on the energy spectrum does lead to pathologies that we wish to avoid.

It is much better to do the mapping in two steps: have as an intermediate model the harmonic rotator, as was introduced in chapter~\ref{harmrot}. The harmonic rotator differs from the harmonic oscillator by having not only a ground state, but also a ceiling. This makes it symmetric under sign switches of the Hamiltonian. The lower energy domain of the rotator maps perfectly onto the harmonic oscillator, while the transition from the rotator to the continuous periodic system is a straightforward limiting procedure.

Therefore, let us first identify the operators \(x\) and \(p\) of the harmonic rotator, with operators in the space of the ontological states  \(|\vv\ket_\ont\)  of our periodic system. This is straightforward, see Eqs.~\eqn{qpmodcom}.
In the energy basis of the rotator, we have the lowering operator \(L_-\) and the raising operator \(L_+\) which give to the operators \(x\) and \(p\)  the following matrix elements between eigenstates \(|m\ket\,,\ -\ell\le m\le \ell\,,\) of \(H_\op\):
	\be \bra m-1|x|m\ket\ =&\half \sqrt{(m+\ell)(\ell+1-m)\over\ell}\ =&\bra m|x|m-1\ket\ , \nn
	\bra m-1|p|m\ket\ =&\half i\sqrt{(M+\ell)(\ell+1-m)\over \ell}\ =& -\bra m|p|m-1\ket\ , \eel{xpmelements}
while all other matrix elements vanish.

 \subsection{The operator $\vv_\op$ in the harmonic rotator\labell{phiop.sec}}
As long as the harmonic rotator has finite \(\ell\), the operator \(\vv_\op\) is to be replaced by a discrete one:
	\be \vv_\op={2\pi\over 2\:\!\ell+1}\s\ ,\qquad\s=-\ell\,,\ -\ell+1\,,\ \cdots\ ,\ +\ell\ .\ \eel{discrphiop}
By discrete Fourier transformations, one derives that the discrete function \(\vv(\s)\) obeys the finite Fourier expansion,
	\be \s= i\sum_{k=1}^{2\ell}{(-1)^k\over 2\sin{\left({\pi k\over 2\ell+1}\right)}}\,e^{\textstyle{2\pi i k\s\over 2\ell+1}}\ . \eel{sigmafourier}
By modifying some phase factors, we replace the relations \eqn{Hstates} and \eqn{ontstates} by
	\be |m\ket_H&\equiv&{1\over\sqrt{2\ell+1}}\sum_{\s=-\ell}^\ell e^{\textstyle{2\pi i m\s\over 2\ell+1}}|\s\ket_\ont\ ,\labell{Hontell}\\ 
	|\s\ket_\ont&=&{1\over\sqrt{2\ell+1}}\sum_{m=-\ell}^\ell e^{\textstyle{-2\pi i m\s\over 2\ell+1}}|m\ket_H\  \eel{ontHell}
(which symmetrizes the Hamiltonian eigenvalues \(m\), now ranging from \(-\ell\) to \(\ell\)).

One now sees that the operator \(e^{\textstyle{2\pi i\over 2\ell+1}\s}\) increases the value \(m\) by one unit, with the exception of the state \(|m=\ell\ket\), which goes to \(|m=-\ell\ket\). Therefore, 
    \be e^{\textstyle{2\pi i\s\over 2\ell+1}}&=&(\ell+1-H)^{-1/2}\,L_+\,(\ell+1+H)^{-1/2}+|-\ell\,\ket\bra\,\ell\,|\ ;\nn
    		e^{\textstyle{-2\pi i\s\over 2\ell+1}}&=&(\ell+1+H)^{-1/2}\,L_-\,(\ell+1-H)^{-1/2}+|\,\ell\,\ket\bra-\ell\,|\ , \eel{expsigmaLpm}
and Eq.~\eqn{sigmafourier} can now be written as
	\be\bra m+k|\s|m\ket&=&{i(-1)^k\over 2\sin\left({\pi k\over 2\ell+1}\right)}\ ,\qquad k\ne 0\ , \nn
		\bra m|\s|m\ket &=&0\ . \eel{sigmamelements}
Eqs~\eqn{expsigmaLpm} can easily be seen to be unitary expressions, for all \(\ell\) in the harmonic rotator.  Care was taken to represent the square roots in the definitions of \(L_\pm\) correctly: since now \(|H|\le\ell\), one never encounters division by 0. The quantity \(m+k\) characterising the state in Eq.~\eqn{sigmamelements} must be read \emph{Modulo} \(2\ell+1\), while observing the fact that for half-odd-integer values of \(\ell\) the relations \eqn{Hontell} and \eqn{ontHell} are both anti periodic with period \(2\ell+1\).

The operator \(\s\) in Eqs~\eqn{expsigmaLpm} and \eqn{sigmamelements} can now be seen to evolve deterministically:
	\be\s(t)=\s(0)+(2\ell+1)\,t/ T\ . \eel{determsigma}

The eigenstates \(|\vv\ket\) of the operator \(\vv_\op\) are closely related to Glauber's coherent states in the harmonic oscillator\,\cite{Glauber1963}, but our operator \(\vv_\op\) is hermitean and its eigenstates are orthonormal; Glauber's states are eigenstates of the creation or annihilation operators, which were introduced by him following a  different philosophy. Orthonormality is a prerequisite for the ontological states that are used in this work.

\subsecl{The harmonic rotator in the $x$ frame}{rotx}
In harmonic oscillators, it is quite illuminating to see how the equations look in the coordinate frame. The energy eigen states are the Hermite functions. It is an interesting exercise in mathematical physics to investigate how the ontological operator (beable) \(\vv_\op\) can be constructed as a matrix in \(x\)-space. As was explained at the beginning of this chapter however, we refrain from exhibiting this calculation as it might lead to confusion.

The operator \(\vv_\op\) is represented by the integer \(\s\) in Eq.~\eqn{discrphiop}.  This operator is transformed to the energy basis by Eqs.~\eqn{Hontell} and \eqn{ontHell}, taking the form \eqn{expsigmaLpm}.
In order to transform these into \(x\) space, we first need the eigen states of \(L_x\) in the energy basis. This is a unitary transformation requiring the matrix elements \(\bra m_3|m_1\ket\), where \(m_3\) are the eigen values of \(L_3\) and \(m_1\) those of \(L_x\). Using the ladder operators \(L_\pm\), one finds the useful recursion relation
	\be 2m_1\bra m_1|m_3\ket &=& \sqrt{(\ell+m_3+1)(\ell-m_3)}\,\bra m_1|m_3+1\ket\ + \nm\\
		&&\sqrt{(\ell+m_3)(\ell+1-m_3)}\,\bra m_1|m_3-1\ket\ , \eel{recrelsm}
First remove the square roots by defining new states \(||m_3\ket\) and \(||m_1\ket\):
	\be ||m_3\ket&\equiv&\sqrt{(\ell+m_3)!\,(\ell-m_3)!\,}\,|m_3\ket\ ,\nm\\
	 ||m_1\ket&\equiv&\sqrt{(\ell+m_1)!\,(\ell-m_1)!\,}\,|m_1\ket\ . \eel{nonnormed}
For them, we have
	\be L_\pm\, ||m_3\ket=(\ell\mp m_3)\,||m_3\ket\ , \eel{Lpmnew}
so that the inner products of these new states obey
	 \be 2m_1\bra m_1||| m_3\ket&=&(\ell-m_3)\,\bra m_1|||m_3+1\ket+(\ell+m_3)\,\bra m_1|||m_3\ket\ ;\nm\\ 
	  \bra m_1|||m_3\ket&=&\bra m_3|||m_1\ket\ . \eel{recrelsm2}
	 \begin{figure}[htb!]\bcenter $a.$\includegraphics[width=70mm]{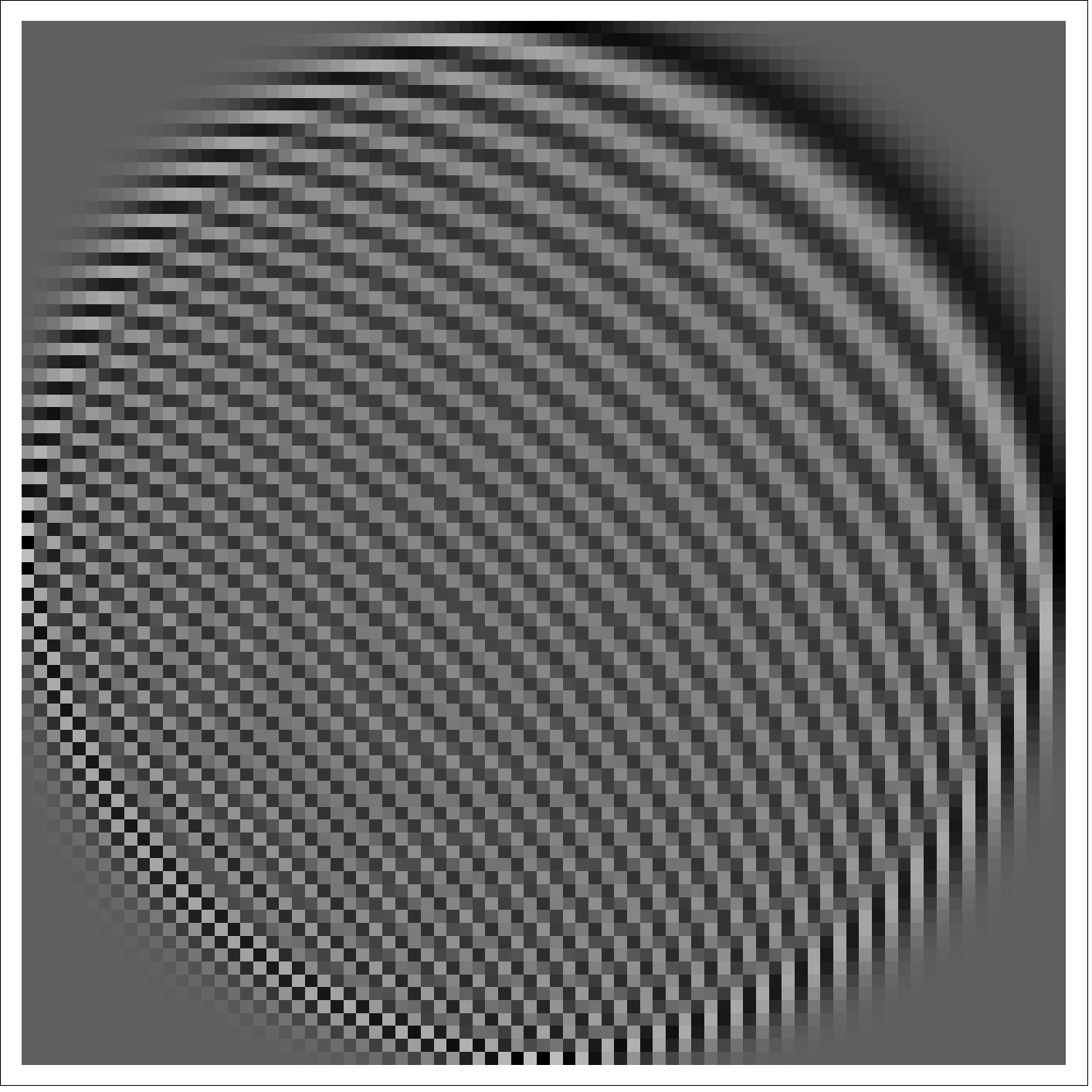}\ $b.$\includegraphics[width=70mm]{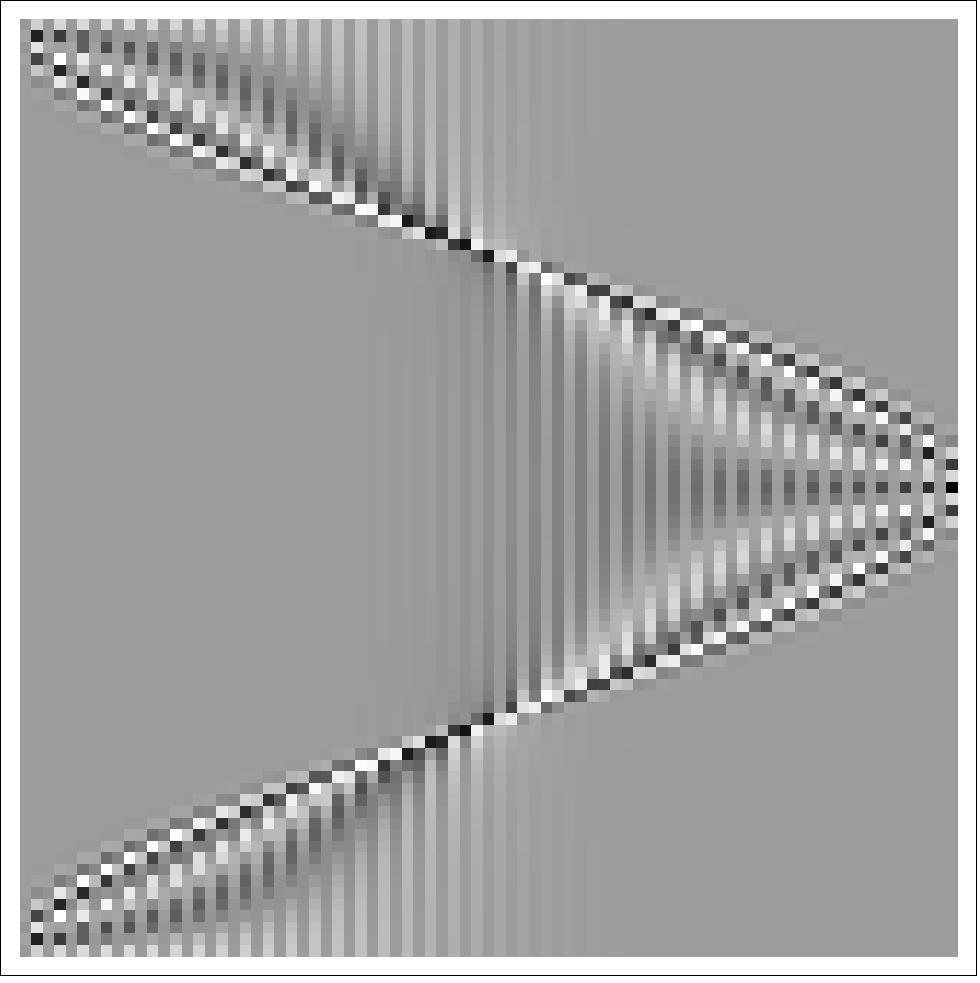}
	\begin{quotation}    \caption{\small a) Plot of the inner products  
	$\bra m_3|m_1\ket$; b) Plot of the transformation matrix $\bra m_1|\s\ket_\ont$ (real part). Horiz.: $m_1$, vert.: $\s$.
	\labell{rotontmatrix.fig}}\end{quotation}	 \ecenter \end{figure}
	
	These equations can easily be inserted in a numerical procedure to determine the matrix elements of the transformation to the `coordinate frame' \(L_x\). With Eqs.~\eqn{Hontell} and \eqn{ontHell}, we now find the elements \(\bra m_1|\s\ket\) of the matrix relating the beable eigen states \(|\s\ket_\ont\) to the \(x\) eigen states of \(L_x\). A graphic expression of the result (for \(\ell=40\)), is displayed in Fig.~\ref{rotontmatrix.fig}. We see in Fig.~\ref{rotontmatrix.fig}b that the ontological variable is loosely following the template degree of freedom \(x=m_1/\sqrt\ell\), just as it will follow the momentum \(p=-m_2/\sqrt\ell\), with a \(90^\circ\) phase shift.

	\newsecl{Locality}{locality}
Replacing harmonic oscillators by harmonic rotators may be the first step towards obtaining a Hamiltonian that describes deterministic processes on the one hand, and still obeys a lower bound on the other hand. Yet we did pay a price. We modified the commutation rules between the coordinate operators called \(x\) and the momentum operators \(p\). If we would apply this to field theories, we would find it difficult to decompose the fields into harmonic modes. These modes would no longer commute with one another, and that would constitute a serious blow to the concept of locality.
	
	We have also seen how a Hamiltonian can be constructed starting off from just any deterministic system, including systems that are entirely local in the usual sense. If time is continuous, that Hamiltonian tends to take the form of Eq,~\eqn{contham}, which has neither a lower nor an upper bound, but it does seem to be local. In contrast, the Hamiltonians of the discrete-time models, such as Eqs.~\eqn{firstH}, \eqn{NHam}, \eqn{genH}, and \eqn{HfrUn}, have in common that they are bounded, but they are expressed in terms of the evolution operators at fairly large times \(t=n\d t\). For cellular automaton models, discussed in section~\ref{CAgeneral}   and chapter~\ref{CAdetail}, the evolution operator over \(n\) time steps, involves interactions among neighbours that are \(n\) space-steps apart. If, instead, we wish to restrict ourselves to local expressions, this means that a cut-off will have to be introduced when defining \(H\), but this is only allowed if the sums in question converge sufficiently rapidly. It seems to be the combination of the positivity requirement and the locality requirement that is often difficult to obey.
	
	Can this conflict be avoided?  Should we search for different models, or should we search for different approximation methods in all models we are using? The author's present understanding is, that we will have to put constraints on the models to which we can apply our theories. Different models will be discussed later (section~\ref{secondquant}). Let us here concentrate on the nature of the conflict. 
	
	In part I, section~\ref{basics}, we introduced the concept of the templates. Let us see what happens when we impose a further constraint on the templates: \emph{Consider only those template states that are slowly varying in time.} We assume that the time dependence in the templates is much slower than the fundamental time interval \(\d t\) in the ontological evolution law. This means that we consider only those elements of Hilbert space where the eigenvalue \(E\) of \(H\) lies in an interval \(|E|\le\half\L\), or, when we add our free constant to the energy levels, we impose 
		\be 0\le E\le\L\ . \eel{energyconstraint}
States composed as superpositions of these energy eigenvalues will show probabilities \(|\bra\ont|\j(t)\ket|^2\) whose time dependence only contains terms \(e^{i\w t}\) with \(|\w|\le\L\). Templates obeying Ineq.~\eqn{energyconstraint} will be referred to as \emph{slow templates}.  

It is advised, however, to be reserved in the use of slow templates; in classical states, energies can easily reach values above the Planck energy (the kinetic energy of a small passenger airplane at cruise speed), and these would require faster templates.
 			
	Figure \ref{Hamiltonapprox.fig}$a$ shows the approximation obtained for the Hamiltonian, in the case we use the expansion \eqn{firstH}, with a smooth cut-off. We introduced a suppression factor \(e^{-k/R}\) for the \(k^\th\) term (in the Figure, \(R=30\)). What happens when we use this approximation for the Hamiltonian?

\begin{figure}[htb!]
 \begin{quotation} \noindent \begin{center}
$a)$\lowerwidthfig{40pt} {60mm}{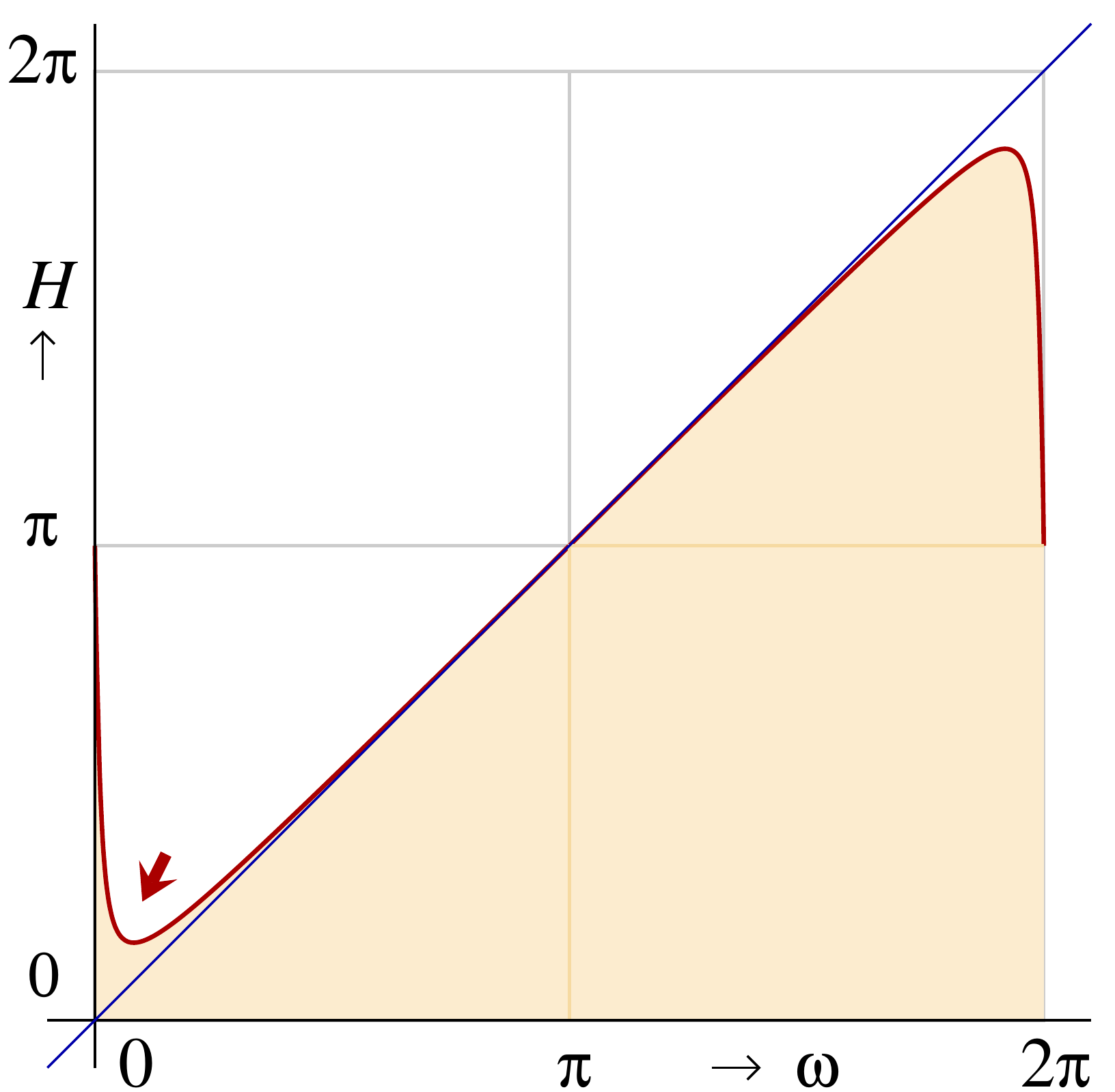}\quad $b)$ \lowerwidthfig{40pt} {60mm}{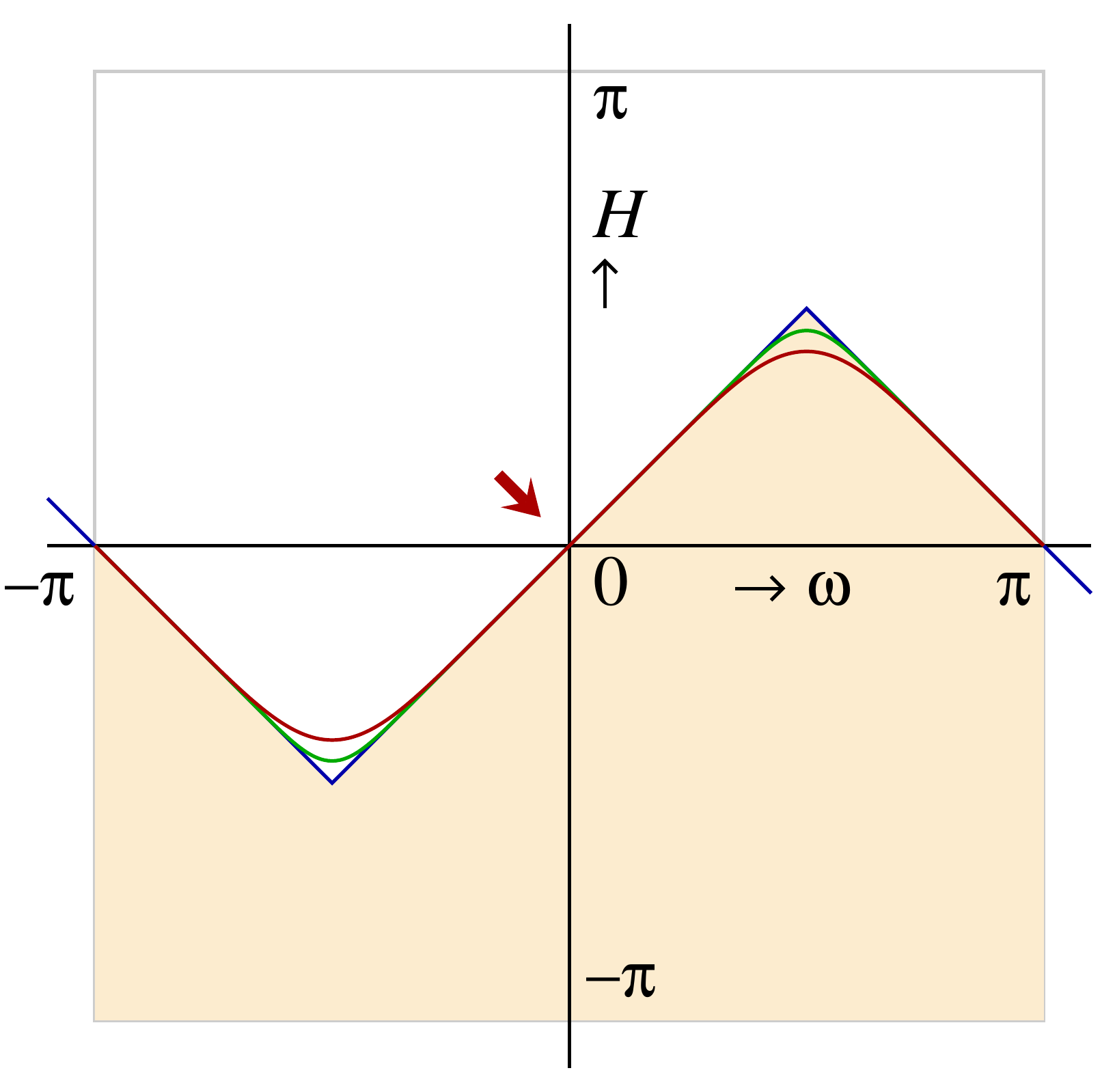}\\
$c)$ \lowerwidthfig{40pt} {60mm}{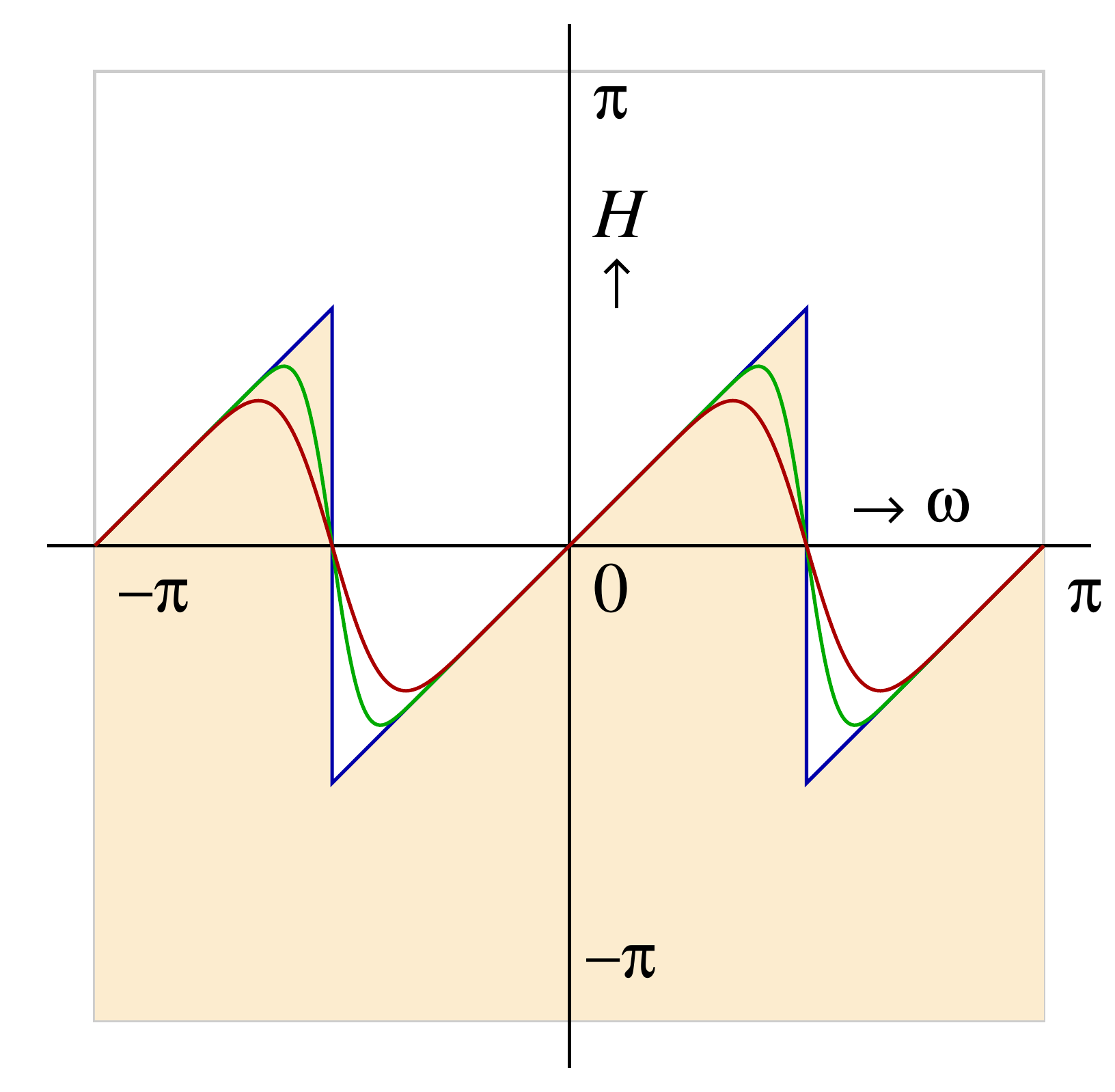}\quad $d)$ \lowerwidthfig{40pt} {60mm}{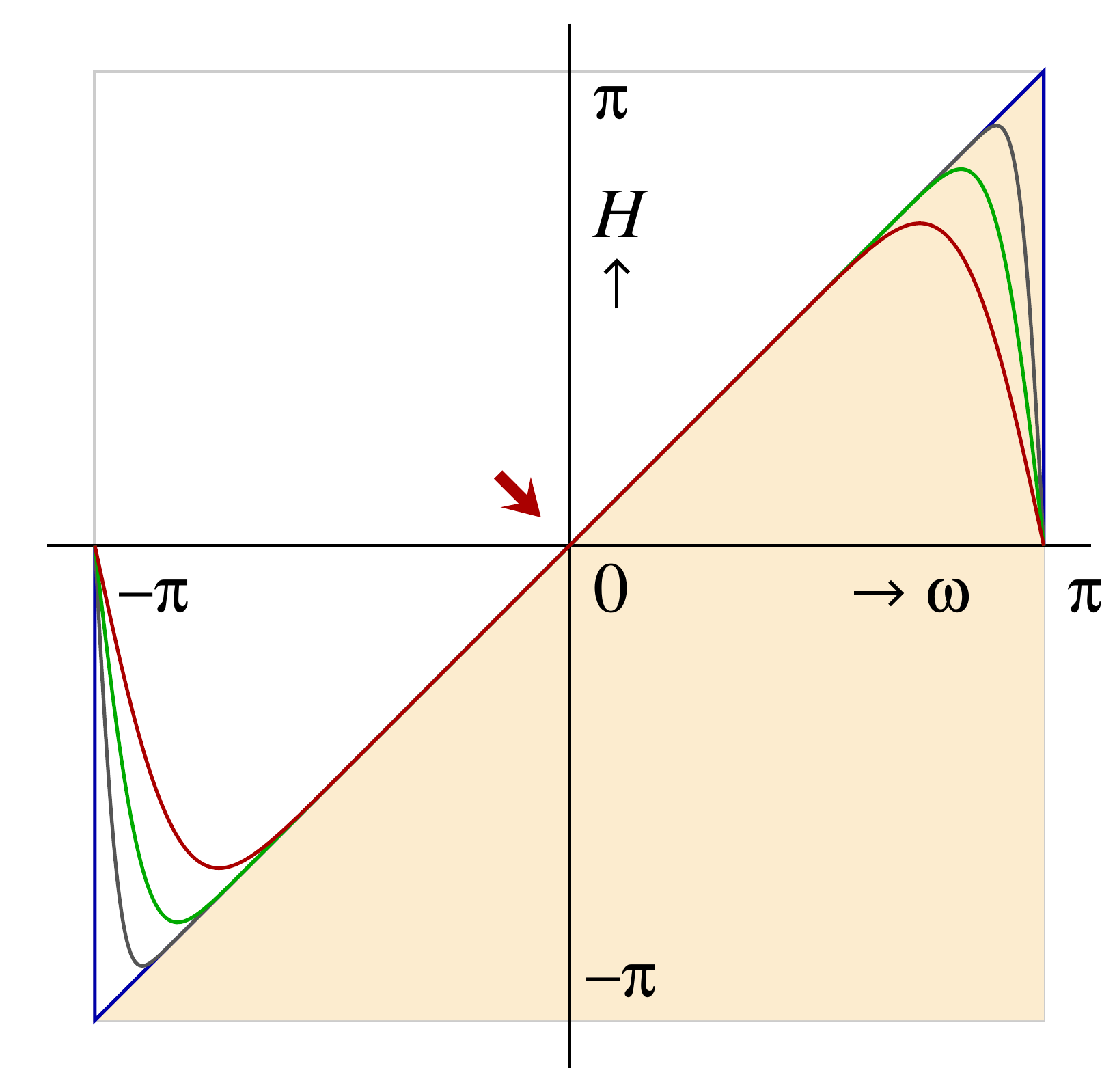} \end{center}
\caption[\small The spectrum of the Hamiltonian in various expansions.]{\small The spectrum of the Hamiltonian in various expansions. $a)$ The Fourier expansion \eqn{firstH} with suppression factor, where we chose $R=30$. The most important region, near the vacuum, shown by the arrow, is maximally distorted by the suppression factor. $b)$Using the expansion \eqn{omegapowers} for \(\arcsin(z)\) to get the most accurate expansion for \(H\) near the center of the spectrum. The curves for \(R=9\) and \(R=31\) are shown. $c$ Result after multiplying with \(\cos\w/|\cos\w|\). The curves shown go up to the powers 13 and 41. $d$. Stretching the previous curve by a factor 2 then removes the unwanted states (see text). Powers shown are  10, 30 and 120
(The difference between Figs. $a$ and $d$ is that in $d$, the straight line at the center is approached much more precisely). \labell{Hamiltonapprox.fig}}
\end{quotation}
		\end{figure}	

\def\appr{{\mathrm{approx}}}
	First, it is not quite local anymore. In a cellular automaton, where we have only nearest neighbour interactions, the Hamiltonian will feature `ghost interactions' between neighbours at \(k\) units of distance apart, assuming that the \(k^\th\) term contains the evolution operator \(U(\pm k\d t)\). With the suppression factor, we expect the Hamiltonian to have non-local features over distance scales of the order of \(R\d t\,c\), where \(c\) is the maximal velocity of information transfer in the automaton, since the exponential suppression factor strongly suppresses effects ranging further out.

On the other hand, if we use  the suppression factor, the lowest energy states in the spectrum will be altered, see the arrow in Fig.~\ref{Hamiltonapprox.fig}$a$. Unfortunately, this is exactly the physically most important region, near the vacuum state (the lowest energy state). It is not difficult to estimate the extent of the deformation close to the origin.   The sum with cut-off can be evaluated exactly. For large values of the cut-off \(R\), and \(0<\w<\pi\), the approximation \(\w_\appr\) for the true eigenvalues \(\w\) of the Hamiltonian will be:
	\be\w_\appr\iss\pi-2\sum_{n=1}^\infty{\sin(n\w)e^{-n/R}\over n}	
	&=&2 \arctan\bigg({e^{\ffract{1\,}R}-\cos \w\over\sin \w} \bigg)\nm\\[3pt]
	\iss 2\arctan\bigg({1-\cos\w+\ffract{1\,}R\over\sin\w}\bigg)&=&
	2\arctan\bigg({\sin\w\over 1+\cos\w}+{\ffract{1\,}R\over\sin\w}\bigg)\ ,\quad\hbox{\ }		\eel{truncsum}
where we replaced \(e^{1/R}\) by \(1+\ffract{1\,}R\) since \(R\) is large and arbitrary. 

Writing \({\sin\a\over 1+\cos\a}=\tan\half\a\), we see that the approximation becomes exact in the limit \(R\ra\infty\). 
We are interested in the states close to the vacuum, having  a small but positive energy \(H=\a\).
Then, at finite \(R\), the cut-off at \(R\) replaces the eigenvalues  \(H\) of the Hamiltonian \(H_\op\)  by 
	\be H\ra H+{2\over R\,H}\ , \eel{Happrox}
which has its minimum at \(H_0\approx \sqrt{2/R}\), where the value of the minimum is \(H\approx 2\sqrt{2/R}\). 	
This is only acceptable if
	\be R\gg M_{\mathrm{Pl}}/\bra H_\op\ket ^2\ . \eel{convradius}	
Here, \(M_{\mathrm{Pl}}\) is the ``Planck mass", or whatever the inverse is of the elementary time scale in the model. This cut-off radius \(R\) must therefore be chosen to be  very large, so that, indeed, the exact quantum description of our local model generates non-locality in the Hamiltonian. 
		
 Thus, if we want a Hamiltonian that represents the behaviour near the vacuum correctly, so that time scales of the order \(T\) are described correctly, the Hamiltonian generated by the model will be non-local over much larger distances, of the order of \(T^2M_{\mathrm{Pl}}\). Apparently, a deterministic automaton does generate quantum behaviour, but the quantum Hamiltonian features spurious non-local interactions.

It is not difficult to observe that the conflict between locality and positivity that we came across is caused by the fact that the spectrum of the energy had to be chosen such that a \(\tht\) jump occurs at \(\w=0\), exactly where we have the vacuum state (see Fig.~\ref{Hamiltonapprox.fig}$a$). The Fourier coefficients of a function with a \(\tht\) function jump will always converge only very slowly, and the sharper we want this discontinuity to be reproduced by our Hamiltonian, the more Fourier coefficients are needed. Indeed, the induced non-locality will be much greater than the size of the system we might want to study. Now, we stress that this non-locality is only apparent; the physics of the automaton itself is quite local, since only directly neighbouring cells influence one another. Yet the quantum mechanics obtained does not resemble what we see in the physical world, where the Hamiltonian can be seen as an integral over a Hamilton density \(\HH(\vec x)\), where \([\HH(\vec x),\,\HH(\vec x\,']=0\) as soon as \(|\vec x-\vec x\,'|>\e>0\). In standard quantum field theories, this \(\e\) tends to zero. If it stretches over millions of CA cells this would still be acceptable if that distance is not yet detectable in modern experiments, but what we just saw was unacceptable. Clearly, a better strategy has to be found.

Our best guess at present for resolving this difficulty is \emph{second quantisation}, which was introduced in section~\ref{secondquant}, and we return to it in sections~\ref{2ndquantisation} and \ref{2ndquantCA}. Here, we just mention that second quantisation will allow us to have the most important physics take place in the central region of this spectrum, rather than at the edges, see the arrow in Fig.~\ref{Hamiltonapprox.fig}$b$.
In this region, our effective Hamiltonians can be made to be very accurate, while still local. Suppose we expand the Hamiltonian in terms of Fourier coefficients that behave as
	\be(\sin\w)^n=(i/2)^n\bigg(U(\d t)-U(-\d t)\bigg)^n\ , \eel{Fourierpowers}
with a limit on the power \(n\). For small \(\w\), the most accurate approximation may seem to be
	\be \w=\sum_{n=0}^{(R-1)/2} a_n(\sin\w)^{2n+1}\ , \eel{omegapowers}
where \(a_n=1,\ \fract 16,\ \fract 3{40},\ \fract 5{112},\ \dots\) are the coefficients of the expansion of \(\arcsin(z)\) in powers of 
\(z\). If we continue that to the power \(R\), we get a very rapidly converging expression for the energy near the center of the spectrum, where \(\w\) is small. If we use that part of the spectrum, the error in the Hamiltonian will be of order \( (\w \d t)^{R+2} \), so that only a few neighbours suffice to give a sufficiently accurate Hamiltonian. 

However, Fig.~\ref{Hamiltonapprox.fig}$b$ is  still not quite what we want.  For all states that have \(\w\) near \(\pm\pi\), the energies are also low, while these are not the states that should be included, in particular when perturbations are added for generating interactions. To remove those states, we desire Fourier expansions that generate the curves 
of Fig~\ref{Hamiltonapprox.fig}$d$. Here, the states with \(\w\approx\pi\) still contribute, which is inevitable because, indeed, we cannot avoid a \(\tht\) jump there, but since the lines are now much steeper at that spot, the states at \(\w=\pm\pi\) may safely be neglected; 
the best expression we can generate will have a density of the spurious states that drops as  \(1/\sqrt R\) times the density of the allowed states. How does one generate these Fourier expansions?

To show how that is done, we return the Fig~\ref{Hamiltonapprox.fig}$b$, and notice that differentiating it with respect to \(\w\) gives us the Fourier expansion of a \(\tht\) function. Multiplying that with the original should give us the functions of Fig.~\ref{Hamiltonapprox.fig}$c$. The easiest way to see what happens is to observe that we multiply the limit curve (the zigzag line in \ref{Hamiltonapprox.fig}$b$) with \(\cos\w/|\cos\w|\), where the denominator is expanded in powers of \(\sin\w\). Then, we are given the functions
	\be H(\w)=\cos\w\sum_{k=0}^{(R-1)/2}b_k(\sin\w)^{2k+1}\ , \eel{sharperzigzag}
where \(\sum_k b_k\,z^{2k+1}\), with \(b_k=1,\ \fract23,\ \fract 8{15},\ \fract{16}{35},\ \fract{128}{315},\ \dots\),  is the power expansion of
	\be(\arcsin z)/\sqrt{1-z^2}\ , \eel{arcsinprime}
in powers of \(z\). 	\def\ol{\overline}

Finally, because Fig.~\ref{Hamiltonapprox.fig}$c$ is periodic with period \(\pi\), we can stretch it by multiplying \(\w\) by 2. 
This gives Fig.~\ref{Hamiltonapprox.fig}$d$, where the limit curve is approximated by
	\be H(\w)= \sin\w\sum_{k=0}^{R-1}b_k\bigg((1-\cos\w)/2\bigg)^k\ , \eel{sharpFourier}
with the same coefficients \(b_k\). Thus, it is now this equation that we use to determine the operator \(H\) from the one-time-step evolution operator \(U=U(\d t)\). Using \(\ol U\) to denote the inverse, \(\ol U=U(-\d t)=U^{-1}\), we substitute in Eq.~\eqn{sharpFourier}:
	\be \sin\w=\fract i 2(U-\ol U)\ , \qquad (1-\cos\w)/2=\quart(2-U-\ol U)\ . \ee

The trick we can then apply is to consider the negative energy states as representing antiparticles, after we apply second quantisation. This very important step, which we shall primarily apply to fermions, is introduced in the next section, while \emph{interactions} are postponed to section \ref{2ndquantCA}.
	
	\newsecl{Fermions}{fermions}
	\subsecl{The Jordan-Wigner transformation}{jordanwigner}
	
	In order to find more precise links between the real quantum world on the one hand and deterministic automaton models on the other,  much more mathematical machinery is needed. For starters, fermions can be handled in an elegant fashion. 

	Take a deterministic model with \(M\) states in total. The example described in Fig.~\ref{genmod.fig} (page~\pageref{genmod.fig}), is a model with \(M=31\) states, and the evolution law for one time step is an element \(P\)  of the permutation group for \(M=31\) elements: \(P\in P_{M}\). Let its states be indicated as \(|1\ket,\,\cdots,\,|M\ket\). We write the single time step evolution law as:
		\be |i\ket_t\ra |i\ket_{t+\d t}=|P(i)\ket_t=\sum_{j=1}^MP_{ij}|j\ket_t\ , \quad i=1,\cdots,M\ ,\eel{genevolve}
where the latter matrix \(P\) has matrix elements \(\bra j|P|i\ket\) that consists of 0s and 1s, with one 1 only in each row and in each column.	 As explained in section~\ref{generalfinite}, we assume that a Hamiltonian matrix \(H^\op_{ij}\) is found such that (when normalising the time step \(\d t\) to one)
	\be \bra j|P|i\ket = (e^{-iH^\op\,})_{ji}\ , \eel{PexpH}
where possibly a zero point energy \(\d E\) may be added that represents a conserved quantity: \(\d E\) only depends on the cycle to which the index \(i\) belongs, but not on the item inside the cycle.

		\begin{figure}[htb!] \begin{quotation}
\begin{center} \lowerwidthfig{0pt} {100mm}{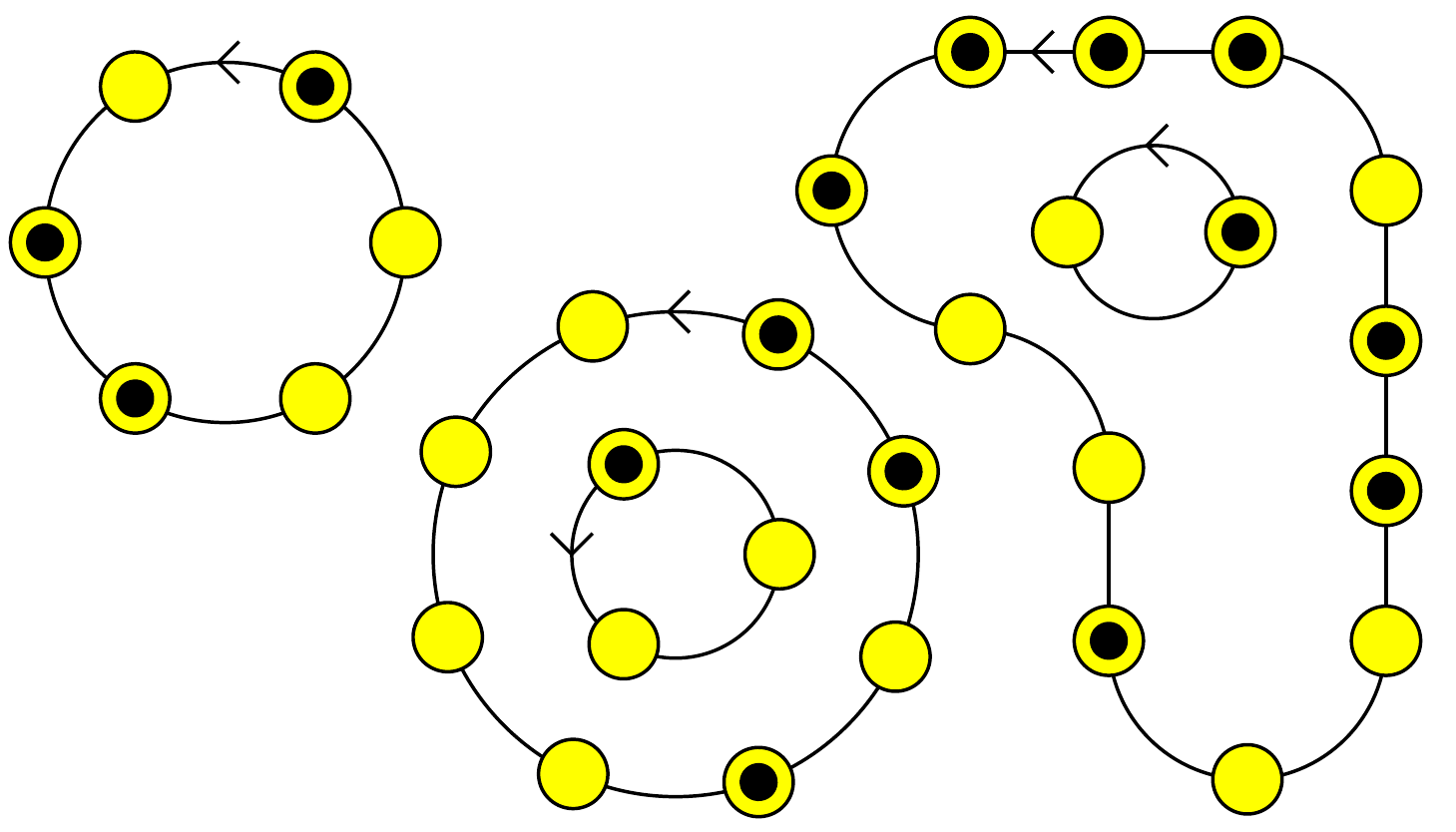}\end{center}
\caption[\small The ``second quantised" version of the multiple-cogwheel model of Fig.~\ref{genmod.fig}.  ]{\small The ``second quantised" version of the multiple-cogwheel model of Fig.~\ref{genmod.fig}. Black dots represent fermions.\label{fermionmod.fig}}\end{quotation}
		\end{figure}	

	We now associate to this model a different one, whose variables are Boolean ones, taking the values 0 or 1 (or equivalently, $+1$ or  $-1$) at every one of these \(M\) sites. This means that, in our example, we now have \(2^{M}=\)\hbox{$2,\! 147,\! 483,\! 648$} states, one of which is shown in Fig.~\ref{fermionmod.fig}. The evolution law is defined such that these Boolean numbers travel just as the sites in the original cogwheel model were dictated to move. Physically this means that, if in the original model, exactly one particle was moving as dictated, we now have \(N\) particles moving, where \(N\) can vary between 0 and \(M\). In particle physics, this is known as ``second quantisation". Since no two particles are allowed to sit at the same site, we have fermions, obeying Pauli's exclusion principle.
		
	To describe these deterministic fermions in a quantum mechanical notation, we first introduce operator fields \(\f_i^\op\), acting as annihilation operators, and their hermitean conjugates, \(\f_i^{\op\,\dag}\), which act as creation operators. Denoting our states now as \(|n_1,\,n_2,\,\cdots,\,n_{M}\ket\), where all \(n\)'s are 0 or 1, we postulate
		\be \f^\op_i\,|n_1,\,\cdots, n_i,\,\cdots,\,n_{M}\ket &=&n_i\, |n_1,\,\cdots,\,n_i-1,\,\cdots,\,n_{M}\ket \ , \nm \\
		 	\f_i^{\op\,\dag\,} |n_1,\,\cdots, n_i,\,\cdots,\,n_{M}\ket &=&(1-n_i)\, |n_1,\,\cdots,\,n_i+1,\,\cdots,\,n_{M}\ket \ , \eel{preJW}
At one given site \(i\), these fields obey (omitting the superscript `{\small op}' for brevity): 
		\be( \f_i)^2=0,\quad \f_i^\dag\f_i+\f_i\,\f_i^\dag=\Bbb{I}\ , \eel{preJWcomm}
where \(\Bbb I\) is the identity operator;	 at different sites, the fields commute: \(\f_i\,\f_j=\f_j\,\f_i\); \(\f_i^\dag\f_j=\f_j\f^\dag_i\), if \(i\ne j\).			

To turn these into completely anti-commuting (fermionic) fields, we apply			
the so-called Jordan-Wigner transformation\,\cite{JordanWigner-1928}:
		\be \j_i=(-1)^{n_1+\cdots+n_{i-1}}\f_i\ , \eel{JWtransf}
where \(n_i=\f^\dag_i\f_i=\j^\dag_i\j_i\) are the occupation numbers at the sites \(i\), 
\emph{i.e.}, we insert a minus sign if an odd number of sites \(j\) with \(j<i\) are occupied. As a consequence of this well-known procedure, one now has
		\be \j_i\,\j_j+\j_j\,\j_i=0\ ,\quad \j_i^\dag\j_j+\j_j\j_i^\dag=\d_{ij}\ ,\quad \forall(i,j)\ . \eel{psianticomm}
The virtue of this transformation is that the anti-commutation relations \eqn{psianticomm} stay unchanged after any linear, unitary transformation of the \(\j_i\) as vectors in our \(M\)-dimensional vector space, provided that \(\j_i^\dag\) transform as contra-vectors. Usually, the minus signs in Eq.~\eqn{JWtransf} do no harm, but some care is asked for.

Now consider the permutation matrix \(P\) and write the Hamiltonian  in Eq.~\eqn{PexpH} as a lower case \(h_{ij}\); it is an \(M\times M\) component matrix. Writing \(U_{ij}(t)= (e^{-iht})_{ij}\), we have, at integer time steps, \(P_\op^{\,t}=U^\op(t)\). We now claim that the permutation that moves the fermions around, is generated by the Hamiltonian \(H^\op_F\) defined as
	\be H^\op_F\,=\sum_{ij}\j^\dag_j\,h_{ji}\,\j_i\ . \eel{secondH}
This we prove as follows:\\
Let
		\be& \j_i(t)=e^{iH^\op_F\,t}\,\j_i\,e^{-iH^\op_F\, t}\ ,&\nm\\ [5pt]
	&	{\dd\over\dd t}e^{-iH^\op_F\,t}=-iH^\op_F\,\,e^{-iH^\op_F\,t}=-i\,e^{-iH^\op_F\,t}H^\op_F\,\ ; 	& \eel{psit}
Then
		\be{\dd\over\dd t}\j_k(t)\iss\qquad i\,e^{iH^\op_F\,t}\sum_{ij}[\j_j^\dag\,h_{ji}\,\j_i,\,\j_k]e^{-iH^\op_F\,t}\ &=&\\
	\hskip-20pt	i\,e^{iH^\op_F\,t}\sum_{ij}h_{ji}\Big(-\{\j^\dag_j,\,\j_k\}\j_i+\j^\dag_j\{\j_i,\j_k\}\Big)e^{-iH^\op_F\,t}&=&\labell{dpsidt2} \\
		-\,i\,e^{iH^\op_F\,t}\sum_i h_{ki}\j_i\,e^{-iH^\op_F\,t}\iss-\,i\sum_ih_{ki}\j_i(t)\ ,&& \eel{dpsidt}
where the anti-commutator is defined as \(\{A,\,B\}\equiv AB+BA\) (note that the second term in Eq.~\eqn{dpsidt2} vanishes).

	This is the same equation that describes the evolution of the states \(|k\ket\) of the original cogwheel model. So we see that, at integer time steps \(t\),  the fields \(\j_i(t)\) are permuted according to the permutation operator \(P^t\).
	Note now, that the empty state \(|0\ket\) (which is \emph{not} the vacuum state) does not evolve at all (and neither does the completely filled state). The \(N\) particle state (\(0\le N\le M\)), obtained by applying \(N\) copies of the field operators \(\j^\dag_i\), therefore evolves with the same permutator. The Jordan Wigner minus sign, \eqn{JWtransf}, gives the transformed state a minus sign if after \(t\) permutations the order of the \(N\) particles has become an odd permutation of their original relative positions. Although we have to be aware of the existence of this minus sign, it plays no significant role in most cases. Physically, this sign is not observable.

The importance of the procedure displayed here is that we can read off how anti-commuting fermionic field operators  \(\j_i\), or \(\j_i(x)\), can emerge from deterministic systems. The minus signs in their (anti-)commutators is due to the Jordan-Wigner transformation \eqn{JWtransf}, without which we would not have any commutator expressions at all, so that the derivation \eqn{dpsidt} would have failed.

The final step in this second quantisation procedure is that we now use our freedom to perform orthogonal transformations among the fields \(\j\) and \(\j^\dag\), such that we expand them in terms of the eigenstates \(\j(E_i)\) of the one-particle Hamiltonian \(h_{ij}\). Then the state \(|\emptyset\ket\) obeying
	\be \j(E_i)|\emptyset\ket=0 \ \hbox{ if }\ E_i>0\ ;\qquad \j^\dag(E_i)|\emptyset\ket=0 \ \hbox{ if }\ E_i<0\ , \eel{fermivac}
has the lowest energy of all. Now, \emph{that} is the vacuum state, as Dirac proposed. The negative energy states are interpreted as holes for antiparticles. The operators \(\j(E)\) annihilate particles if \(E>0\) or create antiparticles if \(E<0\).  For \(\j^\dag(E)\) it is the other way around. Particles and antiparticles now all carry positive energy. Is this then the resolution of the problem noted in chapter~\ref{locality}? This depends on how we handle interactions, see chapter~\ref{secondquant} in part I, and we discuss this important question further in section~\ref{2ndquantCA} and in chapter \ref{conc2}.

The conclusion of this section is that, if the Hamiltonian matrix \(h_{ij}\) describes a single or composite cogwheel model, leading to classical permutations of the states \(|i\ket\), \(i=1,\cdots,M\), at integer times, then the model with Hamiltonian \eqn{secondH} is related to a system where occupied states evolve according to the same permutations, the difference being that now the total number of states is \(2^M\) instead of \(M\). And the energy is \emph{always} bounded from below.

One might object that in most physical systems the Hamiltonian matrix \(h_{ij}\) would not lead to classical permutations at integer time steps, but our model is just a first step. A next step could be that \(h_{ij}\) is made to depend on the values of some local operator fields \(\vv(x)\). This is what we have in the physical world, and this may result if the permutation rules for the evolution of these fermionic particles are assumed to depend on other variables in the system.

In fact, there does exist a fairly realistic, simplified fermionic model where \(h_{ij}\) does appear to generate pure permutations. This will be exhibited in the next section.

A procedure for \emph{bosons} should go in analogous ways, if one deals with bosonic fields in quantum field theory. However, a relation with deterministic theories is not as straightforward as in the fermionic case, because arbitrarily large numbers of bosonic particles may occupy a single site. To mitigate this situation, the notion of harmonic rotators was introduced, which also for bosons only allows finite numbers of states. We can apply more conventional bosonic second quantisation in some special two-dimensional theories, see subsection~\ref{qft2d}.

How second quantisation is applied in standard quantum field theories is described in section~\ref{2ndquantisation}.

\subsecl{`Neutrinos' in three space dimensions}{nu}
\def\ont{\mathrm{ont}}
	
In some cases, it is worth-while to start at the other end. Given a typical quantum system, can one devise a deterministic classical automaton that would generate all its quantum states? We now show a new case of interest. 

One way to determine whether a quantum system may be mathematically  equivalent to a deterministic model is to search for a \emph{complete set of beables}. As defined in subsection~\ref{operators}, beables are operators that may describe classical observables, and as such they must commute with one another, always, at all times. Thus, for conventional quantum particles such as the electron in Bohr's hydrogen model, neither the operators \(x\) nor \(p\) are beables because \hbox{\([\,x(t)\),\, \(x(t')\,]\ne0\)} and \hbox{\([\,p(t),\,p(t')\,]\ne0\)} as soon as \(t\ne t'\). Typical models where we do have such beables are ones where the Hamiltonian is linear in the momenta, such as in section~\ref{contont}, Eq.~\eqn{contham},  rather than quadratic in \(p\). But are they the only ones? 

Maybe the beables only form a space-time grid, whereas the data on points in between the points on the grid do not commute. This would actually serve our purpose well, since it could be that the physical data characterising our universe really do form such a grid, while we have not yet been able to observe that, just because the grid is too fine for today's tools, and interpolations to include points in between the grid points could merely have been consequences of our ignorance.

Beables form a complete set if, in the basis where they are all diagonal, the collection of eigenvalues completely identify the elements of this basis.

No such systems of beables do occur in nature, as far as we know today; that is, if we take all known forces into account, all operators that we can construct today cease to commute at some point. We can, and should, try to search better, but, alternatively, we can produce simplified models describing only parts of what we see, which do allow transformations to a basis of beables. In chapter~\ref{harmrot}, we already discussed the harmonic rotator as an important example, which allowed for some interesting mathematics in chapter~\ref{harmosc}. Eventually, its large \(N\) limit should reproduce the conventional harmonic oscillator. Here, we discuss another such model: \emph{massless `neutrinos'}, in 3 space-like and one time-like dimension. 

A single quantised, non interacting Dirac fermion obeys the Hamiltonian\fn{Summation convention: repeated indices are usually summed over.}
	\be H^\op\,=\a_ip_i+\b m\ , \eel{massivefermion}
where \(\a_i,\,\b\) are Dirac \(4\times 4\) matrices obeying   \def\ddt{{\dd\over\dd t}}
	\be \a_i\a_j+\a_j\a_i=2\d_{ij}\ ;\quad \b^2=1\ ;\quad \a_i\,\b+\b\,\a_i=0\ .\eel{Diracmatrices}
		
Only in the case \(m=0\) can we construct a complete set of beables, in a straightforward manner\fn{Massive `neutrinos' could be looked upon as massless ones in a space with one or more extra dimensions, and that does also have a beable basis. Projecting this set back to 4 space-time dimensions however leads to a rather contrived construction.}. In that case, we can omit the matrix \(\b\), and replace \(\a_i\) by the three Pauli matrices, the \(2\times 2\) matrices \(\s_i\). The particle can then be looked upon as a massless (Majorana or chiral) ``neutrino", having only two components in its spinor wave function. The neutrino is entirely `sterile', as we ignore any of its interactions. This is why we call this the `neutrino' model, with `neutrino' between quotation marks. 

There are actually two choices here: the relative signs of the Pauli matrices could be chosen such that the particles have positive (left handed) helicity and the antiparticles are right handed, or they could be the other way around. We take the choice that particles have the right handed helicities, if our coordinate frame \((x,y,z)\) is oriented  as the fingers 1,2,3 of the right hand. The Pauli matrices \(\s_i\) obey
	\be\s_1\,\s_2=i\s_3\ ,\qquad\s_2\,\s_3=i\s_1\ ,\qquad\s_3\,\s_1=i\s_2\ ;\qquad \s_1^2=\s_2^2=\s_3^2=1\ .\eel{Paulimult}

The beables are:
	\be 
	& \{ \OO^\op_i\}=\{\,\hat q,\,s,\,r\}\ ,\quad\hbox{where}&\nm\\  & \^q_i\equiv \pm p_i/|p|\ ,\quad s\equiv \hat q\cdot\vec\s\ ,\quad r\equiv	\half(\hat q\cdot\vec x+\vec x\cdot\hat q)\ .&
	 \eel{neutrinobeables}
To be precise, \(\hat q\) is a unit vector defining the direction of the momentum, \emph{modulo} its sign. What this means is that we write the momentum \(\vec p\) as
	\be \vec p=p_r\,\hat q\ , \eel{pvec}
where  \(p_r\) can be a positive or negative real number. This is important, because we need its canonical commutation relation with the variable \(r\), being  \([\,r,\,p_r\,]=i\), without further restrictions on \(r\) or \(p_r\). If \(p_r\) would be limited to the positive numbers \(|p|\), this would imply analyticity constraints for wave functions \(\j( r )\). \def\intr{{\mathrm{int}}}

The caret \(\^{}\) on the operator \(\^q\) is there to remind us that it is a vector with length one, \(|\^q|=1\). To define its sign, one could use a condition such as  \(\hat q_z>0\). Alternatively, we may decide to keep the symmetry   \(P_\intr\) (for `internal parity'),

	\be \hat q\leftrightarrow -\hat q\ ,\quad p_r\leftrightarrow -p_r\ , \quad r\leftrightarrow -r\ , \quad s\leftrightarrow -s\ , \eel{pmsymmetry}
after which we would keep only the wave functions that are even under this reflection. The variable \(s\) can only take the values 
\(s=\pm 1\ , \) as one can check by taking the square of \(\hat q\cdot\vec\s\). In the sequel, the symbol \(\^p\) will be reserved for
\(\^p=+\vec p/|p|\), so that \(\^q=\pm\^p\).

The last operator in Eq.~\eqn{neutrinobeables}, the operator \(r\), was symmetrised so as to guarantee that it is hermitean. It can be simplified by using the following observations. In the \(\vec p\) basis, we have
	\be \vec x=i{\pa\over\pa\vec p}\ ;\quad  {\pa\over\pa\vec p}\, p_r=\hat q\ ;&& [\,x_i,p_r\,]=i\hat q_i\  ;\quad [\,x_i,\hat q_j\,]={i\over p_r}(\d_{ij}-\hat q_i\hat q_j)\ ; \qquad \labell{dilatep}\\[3pt]
	x_i\hat q_i-\hat q_ix_i={2i\over p_r}\ \ \ra&& \half(\hat q\cdot\vec x+\vec x\cdot\hat q)=\hat q\cdot\vec x+{i\over p_r}\ . 
		\eel{radialparam}
This can best be checked first by checking the case \(p_r=|p|>0,\ \hat q=\hat p\), and noting that all equations are preserved under the reflection symmetry \eqn{pmsymmetry}.

It is easy to check that the  operators \eqn{neutrinobeables} indeed form a completely commuting set. The only non-trivial commutator to be looked at carefully is \([r,\^ q]=[\,\hat q\cdot \vec x\,,\ \hat q\,]\). Consider again the \(\vec p\) basis, where \(\vec x=i\pa/\pa \vec p\): the operator \(\vec p\cdot\pa/\pa\vec p\) is the dilatation operator. But, since \(\hat q\) is scale invariant, it commutes with the dilatation operator:
	\be [\,\vec p\cdot{\pa\over\pa\vec p}\,,\ \hat q\,] =0 \ . \eel{dilatationcommute}
Therefore,
	\be[\,\hat q\cdot \vec x,\,\hat q\,]=i\,[\,p_r^{-1}\vec p\cdot{\pa\over\pa\vec p}\,,\ \hat q\,]=0\ , \eel{phatphatxcomm}
since also \([\,p_r,\,\^q\,]=0\), but of course we could also have used equation  \eqn{dilatep}, \# 4.

The unit vector \(\hat q\) lives on a sphere, characterised by two angles \(\tht\) and \(\vv\). If we decide to define \(\hat q\) such that \( q_z>0\) then the domains in which these angles must lie are:
	\be 0\le\tht\le\pi/2\ ,\qquad 0\le\vv<2\pi\ . \eel{thetaphidomains}
The other variables take the values
	\be s=\pm 1\ , \qquad -\infty<r<\infty\ .\eel{srdomains} 

An important question concerns the completeness of these beables and their relation to the more usual operates \(\vec x,\ \vec p\) and \(\vec\s\), which of course do not commute so that these themselves are no beables. This we discuss in the next subsection, which can be skipped at first reading. For now, we mention the more fundamental observation that these beables can describe ontological observables at all times, since the Hamiltonian \eqn{massivefermion}, which here reduces to
	\be H=\vec\s\cdot\vec p\ , \eel{masslessfermion}
generates the equations of motion \def\ds{\displaystyle}
	\be&\ds \ddt\vec x =-i[\vec x,\,H]= \vec\s\ ,\qquad\ddt\vec p=0\ ,\qquad	\ddt\s_i= 2\e_{ijk}\,p_j\,\s_k\ ;&  \crl{eomneutrinogen}
 &\ds	\ddt\hat p=0\ ;\qquad\ddt (\hat p\cdot\vec\s)=2\e_{ijk}\,  (p_i/|p|)\, p_j \s_k=0\ ,&\nm\\
 &\ds \ddt(\hat p\cdot\vec x)=\hat p\cdot\vec\s\ ,\qqquad\qquad&
	 \eel{eomneutrinobeables1}
where \(\^p=\vec p/|p|=\pm\^q\), and thus we have:
	\be\ddt\tht=0\ ,\qquad\ddt\vv=0\ ,\qquad\ddt s=0\ ,\qquad \ddt r=s=\pm\, 1\ . \eel{eomneutrinobeables2}	
	
The physical interpretation is simple: the variable \(r\) is the position of a `particle' projected along a predetermined direction \(\hat q\), given by the two angles \(\tht\) and \(\vv\), and the sign of \(s\) determines whether it moves with the speed of light towards larger or towards smaller \(r\) values. 

Note, that a rotation over \(180^\circ\) along an axis orthogonal to \(\hat q\) may turn \(s\) into \(-s\), which is characteristic for half-odd spin representations of the rotation group, so that we can still consider the neutrino as a spin \(\half\) particle.\fn{But rotations in the plane, or equivalently, around the axis \(\^q\), give rise to complications, which can be overcome, see later in this section.}

What we have here is a representation of the wave function for a single `neutrino'  in an unusual basis. As will be clear from the calculations presented in the subsection below, in this basis the `neutrino' is entirely non localised in the two transverse directions, but its direction of motion is entirely fixed by the unit vector \(\hat q\) and the Boolean variable \(s\). In terms of this basis, the `neutrino' is a deterministic object. Rather than saying that we have a particle here, we have a flat sheet, a plane. The unit vector \(\hat q\) describes the orientation of the plane, and the variable \(s\) tells us in which of the two possible directions the plane moves, always with the speed of light. Neutrinos are deterministic planes, or flat sheets. The basis in which the operators \(\^q,\ r\), and \(s\) are diagonal will serve as an ontological basis.

Finally, we could use the Boolean variable \(s\) to define the sign of \(\hat q\), so that it becomes a more familiar unit vector, but this can better be done after we studied the operators that flip the sign of the variable \(s\), because of a slight complication, which is discussed when we work out the algebra, in subsections~\ref{neutrinoalgebra} and \ref{orthonormbeable}.

\def\vq{\hat q\,'}

	\begin{figure}[htb!] \begin{quotation}
\begin{center} \lowerheightfig{20pt} {100mm}{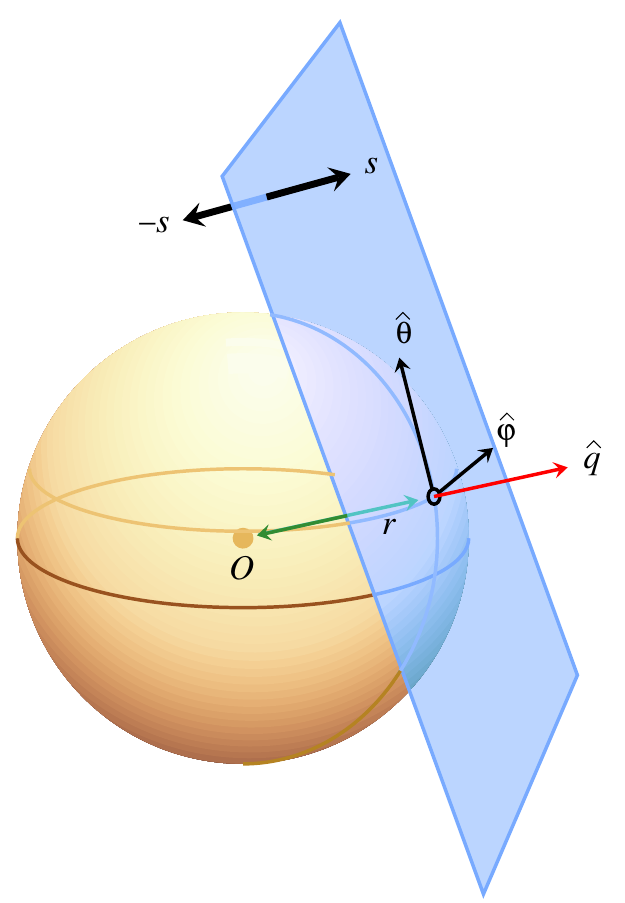}\end{center}
\caption[\small The beables for the ``neutrino".]{\small The beables for the ``neutrino", indicated as the scalar \(r\) (distance of the sheet from the origin), the Boolean \(s\), and the unit vectors \(\^q,\ \^\tht\), and \(\^\vv\).  \(O\) is the origin of 3-space.\label{neutrinosheet.fig}}\end{quotation}
		\end{figure}	

Clearly,  operators that flip the sign of \(s\) exist. For that, we take any vector \(\vq\) that is orthogonal to \(\hat q\). Then, the operator  \(\vq\cdot\vec \s\) obeys  
\((\vq\cdot\vec\s)\ s=-s\ (\vq\cdot\vec\s)\), as one can easily check. So, this operator flips the sign. The problem is that, at each point on the sphere of \(\hat q\) values, one can take any unit length superposition of \emph{two} such vectors \(\vq\) orthogonal to \(\hat q\). Which one should we take? Whatever our choice, it depends on the angles \(\tht\) and \(\vv\). This implies that we necessarily introduce some rather unpleasant angular dependence. This is inevitable; it is caused by the fact that the original neutrino had spin \(\half\), and we cannot mimic this behaviour in terms of the \(\hat q\) dependence because all wave functions have integral spin. One has to keep this in mind whenever the Pauli matrices are processed in our descriptions.

Thus, in order to complete our operator algebra in the basis determined by the eigenvalues \(\hat q,\,s,\) and \(r\), we introduce two new operators whose squares ore one. Define two vectors orthogonal to \(\hat q\), one in the \(\tht\)-direction and one in the \(\vv\)-direction:
	\be \hat q=\pmatrix{q_1\cr q_2\cr q_3}\ ,\qquad \hat\tht={1\over\sqrt{q_1^2+ q_2^2}}
	\pmatrix{ q_3\,q_1\cr q_3\, q_2\cr q_3^2-1}\ ,
	\qquad \hat\vv={1\over\sqrt{ q_1^2+ q_2^2}}\pmatrix{-q_2\cr  q_1\cr 0}\  . \eel{thetaphihat}
All three are normalised to one, as indicated by the caret. Their components obey 
	\be q_i=\e_{ijk}\,\tht_j\vv_k\ ,\qquad  \tht_i=\e_{ijk}\,\vv_j q_k\ ,\qquad\vv_i=\e_{ijk}\,q_j\tht_k\ . \eel{orthoframe}
Then we define two sign-flip operators: write \(s=s_3\), then
	\be s_1=\hat\tht\cdot\vec\s\ ,\qquad s_2=\hat\vv\cdot\vec\s\ ,\qquad s_3=s=\hat q\cdot\vec\s\ . \eel{sflipops}
They obey:
	\be s_i^2={\Bbb I}\ ,\qquad s_1\,s_2=is_3\ ,\qquad s_2\,s_3=is_1\ ,\qquad s_3\,s_1=is_2\ . \eel{flipprods}
	
Considering now the beable operators \(\^q,\ r,\) and \(s_3\), the translation operator \(p_r\) for the variable \(r\), the spin flip operators (``changeables") \(s_1\)   and \(s_2\), and the rotation operators for the unit vector \(\^q\), 
how do we transform back to the conventional neutrino operators \(\vec x\), \(\vec p\) and \(\vec\s\)?

Obtaining the momentum operators is straightforward:
	\be p_i=p_r\,\^q_i\ , \eel{mombeablemom}
and also the Pauli matrices \(\s_i\) can be expressed in terms of the \(s_i\), simply by inverting Eqs.~\eqn{sflipops}. Using Eqs~\eqn{thetaphihat} and the fact that \(q_1^2+q_2^2+q_3^2=1\), one easily verifies that
	\be \s_i=\tht_i s_1+\vv_i s_2+q_i s_3\ . \eel{paulis}

However, to obtain the operators \(x_i\) is quite a bit more tricky; they must commute with the \(\s_i\).
For this, we first need the rotation operators \(\vec L^\ont\). This is \emph{not} the standard orbital \emph{or} total angular momentum. 	Our transformation from standard variables to beable variables will not be quite rotationally invariant, just because we will be using either the operator \(s_1\) or the operator \(s_2\) to go from a left-moving neutrino to a right moving one.
Note, that in the standard picture, chiral neutrinos have spin
\(\half\). So flipping from one mode to the opposite one involves one unit \(\hbar\) of angular momentum in the
plane. The ontological basis does not refer to neutrino spin, and this is why our algebra gives some spurious
angular momentum violation. As long as neutrinos do not interact, this effect stays practically unnoticeable, but
care is needed when either interactions or mass are introduced.

The only rotation operators we can start off with in the beable frame, are the operators that rotate the planes with respect to 
the origin of our coordinates. These we call  \(\vec L^\ont\):
	\be L_i^\ont=-i\e_{ijk} q_j{\pa\over\pa  q_k}\ . \eel{angularmombeable}
By definition, they commute with the \(s_i\), but care must be taken at the equator, where we have a boundary condition, which can be best understood by imposing the symmetry condition~\eqn{pmsymmetry}.	
	
Note that the operators \(L^\ont_i\) defined in Eq.~\eqn{angularmombeable} do not coincide with any of the conventional angular momentum operators because those do not commute with the \(s_i\), as the latter depend on \(\^\tht\) and \(\^\vv\). One finds the following relation between the angular momentum \(\vec L\) of the neutrinos and \({\vec L}^{\,\ont}\):
	\be L_i^\ont\equiv L_i+\half\Big(\tht_i s_1+\vv_i s_2-{q_3\,\tht_i\over\sqrt{1-q_3^2}} s_3\Big)\ ;
	\eel{Lbeable}
the derivation of this equation is postponed to subsection~\ref{neutrinoalgebra}.
		
Since \(\vec J=\vec L+\half\vec\s\), one can also write, using Eqs.~\eqn{paulis} and \eqn{thetaphihat},
	\be L_i^\ont=J_i-{1\over 1-q_3^2}\pmatrix{q_1\cr q_2\cr 0}s_3\ . \eel{totalangmom}	
		
We then derive, in subsection~\ref{neutrinoalgebra}, Eq.~\eqn{xoponto}, the following expression for the operators \(x_i\) in the neutrino wave function, in terms of the beables \(\^q,\,r\) and \(s_3\), and the changeables\fn{See Eq.~\eqn{pinversedef} and the remarks made there concerning the definition of the operator \(1/p_r\) in the world of the beables, as well as in the end of section~\ref{orthonormbeable}.}\ \(L_k^\ont\,,\ p_r\,,\ s_1\) and \(s_2\):
	\be x_i= q_i\,(r-{i\over p_r})+\e_{ijk}\,q_j\,L^\ont_k/p_r +{1\over 2p_r}\Big(-\!\vv_i\,s_1+\tht_i\,s_2+
			{q_3\over\sqrt{1-q_3^2}}\,\vv_i\,s_3\Big)	\eel{xopont}	
(note that \(\tht_i\) and \(\vv_i\) are beables since they are functions of \(\^q\)).	
		
The complete transformation from the beable basis to one of the conventional bases for the neutrino can be derived from
	\be \bra\vec p,\a\,|\,\^q,\,p_r,\,s\ket =p_r\,\d^3(\vec p-\^q\,p_r)\chi_\a^s(\^q)\ , \eel{mommomtrf}
where \(\a\) is the spin index of the wave functions in the basis where \(\s_3\) is diagonal, and \(\chi^s_\a\) is a standard spinor solution for the equation \((\^q\cdot\vec\s_{\a\b})\,\chi^s_\b(\^q)=s\,\chi^s_\a(\^q)\).

In subsection~\ref{orthonormbeable}, we show how this equation can be used to derive the elements of the
 unitary transformation matrix mapping the beable basis to the standard coordinate frame of the neutrino wave function basis\fn{In this expression, there is no need to symmetrise \(\^q\cdot\vec x\), because both \(\^q\) and \(\vec x\) consist of C-numbers only.} (See Eq.~\ref{xbeablematrix}):
	\be \bra\,\vec x,\,\a\,|\,\^q,\,r,\,s\,\ket = {i\over 2\pi}\d\,'(r-\^q\cdot\vec x\,)\,\chi_\a^s(\^q)\ ,	\eel{xbeabletrf}
 where \(\d\,'(z)\equiv{\dd\over\dd z}\d(z)\). This derivative originates from the factor \(p_r\) in 
Eq.~\eqn{mommomtrf}, which is necessary for a proper normalisation of the states.

\subsubsection{Algebra of the beable `neutrino' operators\labell{neutrinoalgebra}}
		
This subsection is fairly technical and can be skipped at first reading. It derives the results mentioned in the previous section, by handling the algebra needed for the transformations from the \((\hat q,\,s,\,r)\) basis to the \((\vec x,\,\s_3)\) or \((\vec p,\,\s_3)\) basis and back.
This algebra is fairly complex, again, because, in the beable representation, no direct reference is made to neutrino spin. Chiral neutrinos are normally equipped with spin \(+\half\) or \(-\half\) with spin axis in the direction of motion. The flat planes that are moving along here, are invariant under rotations about an orthogonal axis, and the associated spin-angular momentum does not leave a trace in the non-interacting, beable picture.

This  forces us to introduce some axis inside each plane that defines the phases of the quantum states, and these (unobservable) phases explicitly break rotation invariance.
	
We consider the states specified by the variables \( s\) and \(r\), and the polar coordinates \(\tht\) and \(\vv\) of the beable \(\^q\), in the domains given by Eqs.~\eqn{thetaphidomains}, \eqn{srdomains}. Thus, we have the states \(|\tht,\,\vv,\,s,\,r\ket\). How can these be expressed in the more familiar states \(|\vec x,\,\s_z\ket\) and/or \(|\vec p,\,\s_z\ket\), where \(\s_z=\pm 1\) describes the neutrino spin in the \(z\)-direction, and \textit{vice versa}?

Our ontological states are specified in the ontological basis spanned by the operators \(\hat q,\ s\,(=s_3),\) and \(r\). We add the operators (changeables) \(s_1\) and \(s_2\) by specifying their algebra \eqn{flipprods}, and the operator
	\be p_r=-i\pa/\pa r\ ;\qquad [\,r,\,p_r]=i\ . \eel{rprcomm}
The original momentum operators are then easily retrieved. As in Eq,~\eqn{pvec}, define 
	\be\vec p=p_r\,\hat q\ . \eel{pvec1}
	
The next operators that we can reproduce from the beable operators \(\hat q,\ r\), and \(s_{1,2,3}\) are the Pauli operators \(\s_{1,2,3}\):
	\be \s_i=\tht_i s_1+\vv_i s_2+q_i s_3\ . \eel{paulibeable}
Note, that these now depend non-trivially on the angular parameters \(\tht\) and \(\vv\), since the vectors \(\^\tht\) and \(\^\vv\), defined in Eq.~\eqn{thetaphihat}, depend non-trivially on \(\^q\), which is the radial vector specified by the angles \(\tht\) and \(\vv\). One easily checks that the simple multiplication rules from Eqs.~\eqn{flipprods} and the right-handed orthonormality \eqn{orthoframe} assure that these Pauli matrices obey the correct multiplication rules also. Given the trivial commutation rules for the beables, \([q_i,\,\tht_j]=
[q_i,\vv_j]=0\), and \([p_r,\,q_i]=0\), one finds that \([p_i,\,\s_j]=0\), so here, we have no new complications.

Things are far more complicated and delicate for the \(\vec x\) operators.
To reconstruct an operator \(\vec x=i\pa/\pa\vec p\), obeying \([x_i,\,p_j]=i\d_{ij}\) and \([x_i,\,\s_j]=0\), we first introduce the orbital angular momentum operator
	\be L_i=\e_{ijk}x_ip_k=-i\e_{ijk} q_j{\pa\over\pa  q_k} \eel{angularmom}
(where \(\s_i\) are kept fixed), obeying the usual commutation rules
	\be[L_i,\,L_j]=i\e_{ijk}L_k\ , \quad[L_i,\, q_j]=i\e_{ijk} q_k\ ,\quad [L_i,\,p_j]=i\e_{ijk}p_k\ ,\ \hbox{etc.},\eel{angcomm}
while \([L_i,\,\s_j]=0\). Note, that these operators are not the same as the angular momenta in the ontological frame, the \(L_i^\ont\) of Eq.~\eqn{angularmombeable}, since those are demanded to commute with \(s_j\), while the orbital angular momenta \(L_i\) commute with \(\s_j\). In terms of the orbital angular momenta \eqn{angularmom}, we can now recover the original space operators 
\(x_i,\ i=1,2,3\), of the neutrinos: \def\sgn{\mathrm{sgn}}
	\be x_i=q_i\,(r-{i\over p_r})+	\e_{ijk}\,q_j\,L_k/p_r \ . \eel{xopbeable}
The operator \(1/p_r\), the inverse of the operator \(p_r=-i\pa/\pa r\), should be \(-i\) times the integration operator. This leaves the question of the integration constant. It is straightforward to define that in momentum space, 
but eventually, \(r\) is our beable operator. For wave functions in \(r\) space, \(\j(r,\cdots)=\bra r,\cdots|\j\ket\), where the ellipses stand for other beables (of course commuting with \(r\)), the most careful definition is:
	\be {1\over p_r}\j( r )\equiv\int_{-\infty}^\infty  \half i\, \sgn(r-r')\j(r')\dd r'\ , \eel{pinversedef}
which can easily be seen to return \(\j( r )\) when \(p_r\) acts on it. ``sgn(\textit{x})" stands for the sign of \(x\).
We do note that the integral must converge at \(r\ra\pm\infty\). This is a restriction on the class of allowed wave functions: in momentum space, \(\j\) must vanish at \(p_r\ra 0\). Restrictions of this sort will be encountered more frequently in this book. 

The anti hermitean term \(-i/p_r\) in Eq.~\eqn{xopbeable} arises automatically in a careful calculation, and it just compensates the non hermiticity of the last term, where \(q_j\) and \(L_k\) should be symmetrised to get a hermitean expression. \(L_k\) commutes with \(p_r\). The \(x_i\) defined here ends up being hermitean.

This perhaps did not look too hard, but we are not ready yet. The operators \(L_i\) commute with \(\s_j\), but not with the beable variables \(s_i\). Therefore, an observer of the beable states, in the beable basis, will find it difficult to identify our operators \(L_i\). It will be easy for such an observer to identify operators \(L_i^\ont\),
which generate rotations of the \(q_i\) variables while commuting with \(s_i\). He might also want to rotate the Pauli-like variables \(s_i\), employing a rotation operator such as \(\half s_i\), but that will not do, first, because they no longer obviously relate to spin, but foremost, because the \(s_i\) in the conventional basis have a much less trivial dependence on the angles \(\tht\) and \(\vv\), see Eqs.~\eqn{thetaphihat} and \eqn{sflipops}. 

Actually, the reconstruction of the \(\vec x\) operators from the beables will show a non-trivial dependence on the variables \(s_i\) and the angles \(\tht\) and \(\vv\). This is because \(\vec x\) and the \(s_i\) do not commute. From the definitions \eqn{thetaphihat} and the expressions \eqn{thetaphihat} for the the vectors \(\^\tht\) and \(\^\vv\), one derives, from judicious calculations:
	  \be 	{} [x_i,\,\tht_j]&=& {i\over p_r}\Big({q_3\over\sqrt{1-q_3^2}}\,\vv_i\,\vv_j- \tht_i\,q_j\Big)\ ,\crl{xthetacomm}
		{} [x_i,\,\vv_j]&=&{-i\vv_i\over p_r}\Big(q_j+{\tht_j\,q_3\over \sqrt{1-q_3^2}}\Big)\ \,, \crl{xphicomm}
		 {}[x_i,\,q_j]&=&{i\over p_r}(\d_{ij}-q_i\,q_j)\ \ .\eel{xqcomm}
The expression \(q_3/\sqrt{1-q_3^2}=\cot(\tht)\) emerging here is singular at the poles, clearly due to the vortices there in the definitions of the angular directions \(\tht\) and \(\vv\). 

From these expressions, we now deduce the commutators of \(x_i\) and \(s_{1,2,3}\):
	\be  {}[x_i,s_1]&=&{i\over p_r}\Big({ \vv_i\,q_3\over\sqrt{1-q_3^2}}\,s_2-\tht_i\,s_3\Big)\ ,\crl{xs1comm}	
	 {}[x_i,s_2]&=& {-i\over p_r} \vv_i\Big(s_3+{  q_3\over\sqrt{1-  q_3^2}}s_1\Big) \ ,\crl{xs2comm}
		 {}[x_i,s_3]&=&{i\over p_r}( \s_i-  q_i\,s_3)\iss{i\over p_r}( \tht_i\,s_1+ \vv_i\, s_2)\  . \eel{xs3comm}
In the last expression Eq.~\eqn{paulibeable} for \(\vec\s\) was used. Now, observe that these equations can be written more compactly:
	\be   [x_i,s_j]=\half\Big[{1\over p_r}\Big(- \vv_is_1+ \tht_i s_2+{ q_3\  \vv_i\over\sqrt{1-  q_3^2}}s_3\Big),\ s_j 	\Big] \ . \eel{xscommall}

To proceed correctly, we now need also to know how the angular momentum operators \(L_i\) commute with \(s_{1,2,3}\). Write 
\(L_i=\e_{ijk}x_j \^q_k\,p_r\), where only the functions \(x_i\) do not commute with the \(s_j\). It is then easy to use Eqs.~\eqn{xs1comm}---\eqn{xscommall} to find the desired commutators:
	\be [L_i,s_j]=\half\Big[-\tht_is_1-\vv_is_2+{q_3 \,\tht_i\over\sqrt{1-q_3^2}}s_3,\ s_j\Big]\ , \eel{Lscomm}
where we used the simple orthonormality relations \eqn{orthoframe} for the unit vectors \(\^\tht,\ \^\vv\), and \(\^q\). Now, this means that we can find new operators \(L_i^\ont\) that commute with all the \(s_j\):
	\be L_i^\ont\equiv L_i+\half\Big(\tht_i s_1+\vv_i s_2-{q_3\,\tht_i\over\sqrt{1-q_3^2}} s_3\Big)\ ,\qquad
		[L_i^\ont,s_j]=0\ , \eel{Lont}
as was anticipated in Eq.~\ref{Lbeable}. It is then of interest to check the commutator of two of the new ``angular momentum" operators. One thing we know: 
according to the Jacobi identity, the commutator of two operators \(L_i^\ont\) must also commute with all \(s_j\).  
Now, expression~\eqn{Lont} seems to be the only one that is of the form expected, and commutes with all 
\(s\) operators. It can therefore be anticipated that the commutator of two \(L^\ont\) 
operators should again yield an \(L^\ont\) operator, because other expressions could not possibly commute with all \(s\). 
The explicit calculation of the commutator is a bit awkward. For instance, one must not forget that \(L_i\) also commutes 
non-trivially with \(\cot(\tht)\):
	\be [L_i,\,{q_3\over\sqrt{1-q_3^2}}]={i\vv_i\over 1-q_3^2} \  . \eel{Lq3comm}
But then, indeed, one  finds	
	\be	\Big[L^\ont_i,L^\ont_j\Big]=i\e_{ijk}L^\ont_k\ . \eel{LontLontcomm}
The commutation rules with \(q_i\) and with \(r\) and \(p_r\) were not affected by the additional terms:
	\be [L^\ont_i,\,q_j]=i\e_{ijk}\,q_k\ ,\qquad [L_i^\ont,\,r]=[L_i^\ont,\,p_r]=0\ . \eel{Lontbeablecomm}
This confirms that we now indeed have the generator for rotations of the beables \(q_i\), while it does not affect the other beables \(s_i,\ r\) and \(p_r\).

Thus, to find the correct expression for the operators \(\vec x\) in terms of the beable variables, we replace \(L_i\) in Eq.~\eqn{xopbeable} by
\(L_i^\ont\), leading to
	\be x_i= q_i\,(r-{i\over p_r})+\e_{ijk}\,q_j\,L^\ont_k/p_r +{1\over 2p_r}\Big(-\!\vv_i\,s_1+\tht_i\,s_2+{q_3\over\sqrt{1-q_3^2}}\,\vv_i\,s_3\Big)\ .	\eel{xoponto}	
This remarkable expression shows that, in terms of the beable variables, the \(\vec x\) coordinates undergo finite, angle-dependent displacements proportional to our sign flip operators \(s_1,\ s_2\), and \(s_3\).  These displacements are in the plane. However, the operator \(1/p_r\) does something else. From Eq.~\eqn{pinversedef} we infer that, in the \(r\) variable,
	\be \bra r_1|\,{1\over p_r}\,|r_2\ket = \half i\,\sgn(r_1-r_2)\ . \eel{pinversematrix}
			
Returning now to a remark made earlier in this chapter, one might decide to use the sign operator \(s_3\) (or some combination of the three \(s\) variables) to distinguish opposite signs of the \(\^q\) operators. The angles \(\tht\) and \(\vv\) then occupy the domains that are more usual for an \(S_2\) sphere: \(0<\tht<\pi\),  and \(0<\vv\le 2\pi\). In that case, the operators \(s_{1,2,3}\) refer to the signs of \(\^q_3,\ r\) and \(p_r\). Not much would be gained by such a notation.

The Hamiltonian in the conventional basis is
	\be H=\vec\s\cdot\vec p\ . \eel{hamconv}
It is linear in the momenta \(p_i\), but it also depends on the non commuting Pauli matrices \(\s_i\). This is why the conventional basis cannot be used directly to see that this is a deterministic model. Now, in our ontological basis, this becomes
	\be H=s\,p_r\ . \eel{hamont}
Thus, it multiplies one momentum variable with the commuting operator \(s\). The Hamilton equation reads
	\be {\dd r\over\dd t} =s\ , \eel{hamonteq}
while all other beables stay constant. This is how our `neutrino' model became deterministic. In the basis of states \(|\,\^q,\,r,\,s\,\ket\) our model clearly describes planar sheets at distance \(r\) from the origin, oriented in the direction of the unit vector \(\^q\), moving with the velocity of light in a transverse direction, given by the sign of \(s\). 

Once we defined, in the basis of the two eigenvalues of \(s\), the two other operators \(s_1\) and \(s_2\), with (see Eqs.~\eqn{flipprods})
	\be s_1=\pmatrix{0&1\cr 1&0}\ ,\quad s_2=\pmatrix{0&-i\cr i&0}\ ,\quad s_3=s=\pmatrix{1&0\cr 0&-1}\ , \eel{smatrices}
in the basis of states \(|\,r\,\ket\) the operator \(p_r=-i\pa/\pa r\), and, in the basis \(|\,\^q\,\ket\) the operators \(L_i^\ont\) by
	\be\hskip-15pt  [L^\ont_i,\,L^\ont_j]=i\e_{ijk}L^\ont_k\ ,\quad[L_i^\ont,\,q_j]=i\e_{ijk}\,q_k\ ,\quad [L^\ont_i,\,r]=0\ , \quad[L^\ont_i,\,s_j]=0\ , \eel{Lontbeable}
we can write, in the `ontological' basis, the conventional `neutrino' operators  \(\vec\s\) (Eq.\ \eqn{paulibeable}), \(\vec x\) (Eq.~\eqn{xoponto}), and 
\(\vec p\) (Eq.~\eqn{pvec1}). By construction, these will obey the correct commutation relations. 
	
\subsubsection{Orthonormality and transformations of the `neutrino' beable states\labell{orthonormbeable}}		
						
The quantities that we now wish to determine are the inner products
	\be \bra\vec x,\s_z|\tht,\vv,s,r\ket\ ,\qquad\bra\vec p,\s_z|\tht,\vv,s,r\ket\ . \eel{inprodneutrino}
The states \(|\,\tht,\vv,s,r\ket\) will henceforth be written as \(|\,\hat q,s,r\ket\). The use of momentum variables \(\hat q\equiv\pm\vec p/|p|\ , q_z>0,\) together with a real parameter \(r\) inside a Dirac bracket will always denote a beable state in this subsection.

Special attention is required for the proper normalisation of the various sets of eigenstates.	We assume the following normalisations:
	\be	\bra\vec x,\a\,|\,\vec x\,',\b\ket&=&\d^3(\vec x-\vec x\,')\,\d_{\a\b}\ ,\labell{dormpos} \\[4pt]
				\bra\vec p,\a\,|\,\vec p\,',\b\ket&=&\d^3(\vec p-\vec p\,')\,\d_{\a\b}\ ,\labell{normmom}\\[4pt]
		\bra\vec x,\a\,|\,\vec p,\b\ket&=&(2\pi)^{-3/2}\,e^{i\,\vec p\cdot\vec x}\,\d_{\a\b}\ ;\labell{momposinpr}\\[4pt]
	\bra\hat q,\,r,\,s\,|\,\hat q\,',\,r',\,s'\,\ket&=&\d^2(\hat q,\,\hat q\,')\,\d(r-r')\,\d_{s\,s'}\ ,\\[4pt]
	 \d^2(\hat q,\, \hat q')
				&\equiv&{\d(\tht-\tht')\,\d(\vv-\vv') \over\sin\tht} \ , \eel{normbeables}
and \(\a\) and \(\b\) are eigenvalues of the Pauli matrix \(\s_3\); furthermore,
	\be	\int\d^3\vec x\sum_\a |\,\vec x,\a\ket\bra\vec x,\a\,|\  =&\Bbb I&=\ \int\dd^2\hat q\int_{-\infty}^\infty\dd r
				\sum_{s=\pm}|\,\hat q,r,s\ket\bra\hat q,r,s\,|\ ; \nm\\
		 \int\dd^2\hat q &\equiv& \int_0^{\pi/2}\sin\tht\,\dd\tht\int_0^{2\pi}\dd\vv\ .	\ee

\def\dsp{\displaystyle}
The various matrix elements are now straightforward to compute. First we define the spinors \(\chi_\a^\pm(\^q)\) by solving
	\be (\^q\cdot\vec\s_{\a\b})\chi^s_\b=s\,\chi^s_\a\ ;\qquad\pmatrix{q_3-s&q_1-iq_2\cr q_1+iq_2&-q_3-s}
			\pmatrix{\chi^s_1\cr\chi^s_2}=\,0\ , \eel{neutrinospinordef}
which gives, after normalising the spinors,
	\be \matrix{\dsp\chi^+_1(\^q)=\sqrt{\half(1+q_3)}\nm\\[2pt]\dsp \chi^+_2(\^q)={q_1+i\,q_2\over \sqrt{2(1+q_3)}}}\ ; \qquad
		\matrix{ \chi^-_1(\^q)= -\sqrt{\half(1-q_3)}     \nm\\[2pt]\dsp \chi^-_2(\^q)={\,q_1+i\,q_2\over \sqrt{2(1-q_3)}} }
				\ ,	\eel{spinorsolution}
where not only the equation \(s_3\,\chi^\pm_\a=\pm\,\chi^\pm_\a\) was imposed, but also
	\be s_{1\,\b}^{\ \,\a}\,\chi^\pm_\a=\chi^\mp_\b\ ,\qquad s_{2\,\b}^{\ \, \a}\,\chi^\pm_\a=\pm\, i\,\chi^\mp_\b\ , \eel{s2s3eqs}	
which implies a constraint on the relative phases of \(\chi^+_\a\) and \(\chi^-_\a\).	
The sign in the second of these equations is understood if we realise that the index \(s\) here, and later in 
Eq.~\eqn{mommommatrix},  is an upper index.	

Next, we need to know how the various Dirac deltas are normalised:
	\be \dd^3\vec p=p_r^2\,\dd^2\hat q\,\dd p_r\ ;\qquad\d^3(\^q\,p_r-\^q\,'\,p_r\!')={1\over p_r^2}\,\d^2(\hat q,\,\^q\,')\,\d(p_r-p_r\!')\ , \eel{momnorm}			
We demand completeness to mean 			
	\be \hskip-20pt  \int\dd^2\^q\int_{-\infty}^\infty\dd p_r\,\sum_{s=\pm}\bra \,\vec p,\,\a\,|\,\^q,\,p_r,\,s\,\ket
			\bra\,\^q,\,p_r,\,s\,|\,\vec p\,',\a'\ket&=&\d_{\a\,\a'}\d^3(\vec p-\vec p\,')\ ;\crl{compl1}
		\int\dd^3\vec p\,\sum_{\a=1}^2\bra\,\^q,\,p_r,\,s\,|\,\vec p,\,\a\,\ket\bra\,\vec p,\,\a\,|\,\^q\,',\,p_r\!',\,s'\,\ket&=&
			\d^2(\^q,\,\^q\,')\d(p_r-p_r\!')\,\d_{s s'}\ ,\qquad \eel{compl2}			
which can easily be seen to imply\fn{Note that the \emph{phases} in these matrix elements could be defined at will, so we could have chosen \(|p|\) in stead of \(p_r\). Our present choice is for future convenience.}
	\be \bra\vec p,\a\,|\,\^q,\,p_r,\,s\ket =p_r\,\d^3(\vec p-\^q\,p_r)\chi_\a^s(\^q)\ , \eel{mommommatrix}
since the norm \(p_r^2\) has to be divided over the two matrix terms in Eqs~\eqn{compl1} and \eqn{compl2}.
\vskip10pt
This brings us to derive, using \(\bra\, r\,|\, p_r\,\ket=(2\pi)^{-1/2}\,\ex{\,i\,p_r\,r}\),
	\be\bra\,\vec p,\,\a\,|\,\^q,\,r,\,s\,\ket ={1\over \sqrt{2\pi}}\,{1\over p_r}\,\d^2\Big({\pm\vec p\over |p|}\,,\ \^q\Big)\ 
	\ex{-i(\^q\cdot \vec p\,)\,r}\  \chi_\a^s(\^q)\ , \eel{mombeablematrix}
where the sign is the sign of \(p_3\).

The Dirac delta in here can also be denoted as
	\be \d^2\Big({\pm\vec p\over |p|}\,,\ \^q\Big)=(\^q\cdot\vec p\,)^2\,\d^2(\vec p\wedge\^q)\ , \eel{wedgedelta}
where the first term is a normalisation to ensure the expression to become scale invariant, and the second just forces \(\vec p\) and \(\^q\) to be parallel or antiparallel. In the case \(\^q=(0,0,1)\), this simply describes \(p_3^2\,\d(p_1)\,\d(p_2)\).

Finally then, we can derive the matrix elements \(\bra\,\vec x,\,\a\,|\,\^q,\,r,\,s\,\ket\). Just temporarily, we put \(\^q\) in the 3-direction: \(\^q=(0,0,1)\),
	\be \bra\,\vec x,\,\a\,|\,\^q,\,r,\,s\,\ket&=&{1\over\sqrt{2\pi}}(2\pi)^{-3/2}\int\dd^3\vec p\,{(\^q\cdot\vec p\,)^2\over p_r}\,
			\d^2(\vec p\wedge\^q)	\ex{-i(\^q\cdot \vec p\,)\,r+i\,\vec p\cdot\vec x}\  \chi_\a^s(\^q)\ = 	\nm\\
		&=&{1\over(2\pi)^2}\int\dd^3\vec p\,p_3\,\d(p_1)\,\d(p_2)\,\ex{i\,p_3(x_3-r)}\,\chi_\a^s(\^q)\ =\nm\\	
		&=&{1\over 2\pi}{i\dd\over\dd r}\d(r-\^q\cdot \vec x\,)\,\chi_\a^s(\^q)\ =\ 
			{i\over 2\pi}\d\,'(r-\^q\cdot\vec x\,)\,\chi_\a^s(\^q)\ .	\eel{xbeablematrix}
			
With these equations, our transformation laws are now complete. We have all matrix elements to show how to go from one basis to another. Note, that the states with vanishing \(p_r\), the momentum of the sheets, generate singularities.
Thus, we see that the states \(|\,\j\,\ket\) with \(\bra p_r=0\,|\,\j\,\ket	\ne 0\), or equivalently, \(\bra\,\vec p=0\,|\,\j\,\ket\ne 0\), must be excluded. We call such states `edge states', since they have wave functions that are constant in space (in \(r\) and also in \(\vec x\)), which means that they stretch to the `edge' of the universe. There is an issue here concerning the boundary conditions at infinity, which we will need to avoid. We see that the operator \(1/p_r\), Eq.~\eqn{pinversedef}, is ill defined for these states.

\subsubsection{Second quantisation of the `neutrinos'\labell{secondneutrino}}

	Being a relativistic Dirac fermion, the object described in this chapter so-far suffers from the problem that its Hamiltonian, \eqn{masslessfermion} and \eqn{hamont}, is not bounded from below. There are positive and negative energy states. The cure will be the same as the one used by Dirac, and we will use it again later: second quantisation. We follow the procedure described in section~\ref{jordanwigner}: for every given value of the unit vector \(\hat q\), we consider an unlimited number of `neutrinos', which can be in positive or negative energy states. To be more specific, one might, temporarily, put the variables \(r\) on a discrete lattice:
	\be r=r_n=n\,\d r\ , \eel{rlattice}
but often we ignore this, or in other words, we let \(\d r\) tend to zero. 

We now describe these particles, having spin \(\half\), by anti-commuting fermionic operators. We have operator fields
 \(\j_\a(\vec x\,)\) and \(\j_\a^\dag(\vec x\,)\) obeying anticommutation rules,
	\be \{\j_\a(\vec x\,),\, \j_\b^\dag(\vec x\,'\,)\}=\d^3(\vec x-\vec x\,')\d_{\a\b}\ . \eel{anticommfields}
Using the transformation rules of subsection~\ref{orthonormbeable}, we can transform these fields into fields \(\j(\^q,r,s)\) and \(\j^\dag(\^q,r,s)\) obeying
	\be \{\j(\^q,r,s),\,\j^\dag(\^q\,',r',s'\,)\}\iss\d^2(\^q,\^q\,'\,)\,\d(r-r')\d_{s\,s'}\ \ra\ 
		\d^2(\^q,\^q\,')\,\d_{n\,n'}\,\d_{s\,s'}\ . \eel{beableanticomm}	
	
At any given value of \(\^q\) (which could also be chosen discrete if so desired), we have a straight line of \(r\) values, limited to the lattice points \eqn{rlattice}. On a stretch of \(N\) sites of this lattice, we can imagine any number of fermions, ranging from \(0\) to \(N\). Each of these fermions obeys the same evolution law \eqn{hamonteq}, and therefore also the entire system is deterministic. 

There is no need to worry about the introduction of anti-commuting fermionic operators \eqn{anticommfields}, \eqn{beableanticomm}. The minus signs are handled through the Jordan-Wigner transformation, implying that the creation or annihilation of a fermion that has an odd number of fermions at one side of it, will be accompanied by an artificial minus sign. This minus sign has no physical origin but is exclusively introduced in order to facilitate the mathematics with anti-commuting fields. Because, at any given value of \(\^q\), the fermions propagate on a single line, and they all move with the same speed in one direction, the Jordan-Wigner transformation is without complications. Of course, we still have not introduced interactions among the fermions, which indeed would not be easy as yet.

This `second quantised' version of the neutrino model has one big advantage: we can describe it using a Hamiltonian that is bounded from below. The argument is identical to Dirac's own ingenious procedure. The Hamiltonian of the second quantised system is (compare the first quantised Hamiltonian \eqn{masslessfermion}):
	\be H=\int\dd^3\vec x\sum_\a\j^{*\,\a}(\vec x)h_\a^{\ \b}\j_\b(\vec x)\ ,\qquad h_\a^{\ \b}=-i\vec\s_\a^{\ \b}\cdot{\pa\over\pa\vec x}\ . \eel{secondquham}
Performing the transformation to the beable basis described in subsection~\ref{orthonormbeable}, we find
	\be H=\int\dd^2\^q\int\dd r\sum_s\,\j^*(\^q,r,s)\,(-i\,s)\,{\pa\over\pa r}\j(\^q, r,s)\ . \eel{fermham1}
 \def\stand{{\,\mathrm{stand}\,}}

Let us denote the field in the standard notation as \(\j^\stand_\a(\vec x)\) or \(\j^\stand_\a(\vec p)\), and the field in the `beable' basis as
\(\j^\ont_s(\hat q,\,r)\). Its Fourier transform is not a beable field, but to distinguish it from the standard notation we will sometimes indicate it nevertheless as \(\j^\ont_s(\hat q,\,p_r)\).

In  momentum space, we have (see Eq.~\ref{mommomtrf}): 
	\be \j^\stand_\a(\vec p\,)&=&{1\over p_r}\sum_s\chi^s_\a(\hat q)\,\j^\ont_s(\hat q,\,p_r)\ ; \\
		\qquad\j^\ont_s(\hat q,\,p_r)&=&p_r\sum_\a\chi^s_\a(\hat q)^*\,\j^\stand_\a(\vec p)\ ,\qquad \vec p\equiv\hat q\,p_r\ , \eel{standonttrf}
where `stand' stands for the standard representation, and `ont'  for the ontological one, although we did the Fourier transform replacing
the variable \(r\) by its momentum variable \(p_r\).	The normalisation is such that
	\be\sum_\a\int\dd^3\vec p\,|\j^\stand_\a(\vec p\,)|^2=\sum_s\int_{\^q_3>0}\dd^2\hat q\int_{-\pi/\d r}^{\pi/\d r}\dd p_r|\j^\ont_s(\hat q,p_r)|^2\ , \eel{standontnorm}
see Eqs.~\eqn{momnorm} -- \eqn{mommommatrix}.	

In our case, \(\j\) has only two spin modes, it is a Weyl field, but in all other respects it can be handled just as a massless Dirac field. Following Dirac, in momentum space, each momentum \(\vec p\) has two energy eigenmodes (eigenvectors of the operator \(h^{\ \b}_\a\) in the Hamiltonian \eqn{secondquham}), which we write, properly normalised,  as
	\be  	u^{\stand\pm}_\a(\vec p\,)={1\over\sqrt{2|p|(|p|\pm p_3)}}\pmatrix{\pm|p|+p_3\cr p_1+ip_2}\ ;\qquad E=\pm|p|\ . \eel{fermieigenmodes}
Here, the spinor lists the values for the index \(\a=1,2\). In the basis of the beables:
	\be & u^{\ont\,\pm}_s(\^q,p_r)=\pmatrix{1\cr 0}\ \hbox{ if }\ \pm p_r>0\ ,&\pmatrix{0\cr 1}\ \hbox{ if }\ \pm p_r<0\ ;\quad\labell{beableeigenmodes}\\[5pt]
& E=\pm|p_r|\ . &	\ee
Here, the spinor lists the values for the index \(s=+\) and \(-\). 
		
In both cases, we write
	\be  \j(\vec p)&=&u^+\,a_1(\vec p\,)+u^-\,a_2^\dag(-\vec p\,)\ ;\qquad\{a_1,\,a_2\}=\{a_1,\,a_2^\dag\}=0\ ,\crl{psicreateannihilate}
\hskip-30pt		\{a_1(\vec p\,),\,a_1^\dag(\vec p\,')\}&=&\{a_2(\vec p\,),\,a_2^\dag(\vec p\,')\}\iss\d^3(\vec p-\vec p\,')\ \hbox{ or }\ \d(p_r-p_r\!')
			\, \d^2(\hat q,\,\hat q\,') 	\  ;   \\[2pt]
		H^\op&=&|p|\,(a_1^\dag a_1+a_2^\dag a_2-1)\ , \eel{fermionmomham}
where \(a_1\) is the annihilation operator for a particle with momentum \(\vec p\), and \(a_2^\dag\) is the creation operator for an antiparticle with momentum \(-\vec p\). We drop the vacuum energy \(-1\). In case we have a lattice in \(r\) space, the momentum is limited to the values \(|\vec p\,|=|p_r|<\pi/\d r\).

\subsecl{The `neutrino' vacuum correlations}{neutrinocorr}
	
The \emph{vacuum state} \(|\emptyset\ket\) is the state of lowest energy. This means that, at each momentum value \(\vec p\) or equivalently, at each \((\hat q,\,p_r)\), we have
	\be a_i|\emptyset\ket=0\ , \eel{fermionvacuumstate}
where \(a_i\) is the annihilation operator for all states with \(H=\s\cdot\vec p=s\,p_r>0\), and the creation operator if \(H<0\).
The beable states are the states where, at each value of the set \((\^q,\, r,\,s)\) the number of `particles' is specified to be either 1 or 0.
This means, of course, that the vacuum state \eqn{fermionvacuumstate} is not a beable state; it is a superposition of all beable states.

One may for instance compute the correlation functions of right- and left moving `particles' (sheets, actually) in a given direction. In the beable (ontological)  basis, one finds that left-movers are not correlated to right movers, but two left-movers are correlated as follows:
	\be P( r_1,\,r_2) -P(r_1)P(r_2)&=&\d r^{\,2}\bra\emptyset|\j^*_s(r_1)\j_s(r_1)\ \j^*_s(r_2)\j_s(r_2)|\emptyset\ket_\mathrm {conn}\ =\nm\\[5pt]
		\Big|{\d r\over 2\pi}\int_0^{\pi/\d r}\dd p\,e^{ip(r_2-r_1)}\Big|^2&=&\left\{ \matrix{
		\displaystyle{{ \d r^2\over \pi^2\,|r_1-r_{2\vphantom{|_|}}|^2}}&\hbox{if }& \fract{r_1-r_2\vphantom{|_g}}{\d r}= \hbox{odd }, \cr
		 \quart\d_{r_1,\,r_2}                &\hbox{if }& \fract{r_1-r_2\vphantom{|_g}}	{\d r}= \hbox{even}\ ,}\right.  \eel{scorrel}
where \(\d r\,^2\), the unit of distance between two adjacent sheets squared,  was added for normalisation, and `conn' stands for the connected diagram contribution only, that is, the particle and antiparticle created at \(r_2\) are both annihilated at \(r_1\). The same applies to two right movers. In the case of a lattice, where \(\d r\) is not yet tuned to zero, this calculation is still exact if \(r_1-r_2\) is an integer multiple of \(\d r\). Note that, for the vacuum, \(P(r)=P(r,r)=\half\).

An important point about the second quantised Hamiltonian \eqn{secondquham}, \eqn{fermham1}: on the lattice, we wish to keep the Hamiltonian \eqn{fermionmomham} in momentum space. In position space, Eqs.~\eqn{secondquham} or \eqn{fermham1} cannot be valid since one cannot differentiate in the space variable \(r\). But we can have the induced evolution operator over finite integer time intervals \(T=n_t\,\d r\). This evolution operator then displaces left movers one step to the left and right movers one step to the right. The Hamiltonian ~\eqn{fermionmomham} does exactly that, while it can be used also for infinitesimal times; it is however not quite local when re-expressed in terms of the fields on the lattice coordinates, since now momentum is limited to stay within the Brillouin zone \(|p_r|<\pi/\d r\).
This feature, which here does not lead to serious complications, is further explained, for the bosonic case, in section~\ref{qft2d}.

Correlations of data at two points that are separated in space but not in time, or not sufficiently far in the time-like direction to allow light signals to connect these two points, are called space-like correlations. 
The space-like correlations found in Eq.~\eqn{scorrel} are important. They probably play an important  role in the mysterious behaviour of the beable models when Bell's inequalities are considered, see part~I, chapter~\ref{Bell} and beyond.

Note that we are dealing here with space-like correlations of the ontological degrees of freedom. The correlations are a consequence of the fact that we are looking at a special state that is preserved in time, a state we call the vacuum. All physical  states normally encountered are template states, deviating only very slightly from this vacuum state, so we will always have such correlations.

In the chapters about Bell inequalities and the Cellular Automaton Interpretation (section~\ref{catcai} and chapter~\ref{intpr} of part~I\,), it is argued that the ontological theories proposed in this book must feature strong, space-like correlations throughout the universe. This would be the only way to explain how the Bell, or CHSH inequalities can be so strongly violated in these models. Now since our `neutrinos' are non interacting, one cannot really do EPR-like experiments with them, so already for that reason, there is no direct contradiction. However, we also see that strong space-like correlations are present in this model.

Indeed, one's first impression might be that the ontological `neutrino sheet' model of the previous section is entirely non local. The sheets stretch out infinitely far in two directions, and if a sheet moves here, we immediately have some information about what it does elsewhere. But on closer inspection one should concede that the \emph{equations of motion} are entirely local. These equations tell us that if we have a sheet going through a space-time point \(x\), carrying a sign function \(s\), and oriented in the direction \(\^q\), then, at the point \(x\), the sheet will move with the speed of light in the direction dictated by \(\^q\) and \(\s\). No information is needed from points elsewhere in the universe. This is locality. 

The thing that is non local is the ubiquitous correlations in this model. If we have a sheet at \((\vec x,\, t)\), oriented in a direction \(\^q\), we instantly know that the same sheet will occur at other points \((\vec y,\,t)\), if \(\^q\cdot(\vec y-\vec x\,)=0\), and it has the same values for \(\^q\) and \(\s\). It will be explained in chapter~\ref{QFT},  section~\ref{commsignals}, that space-like correlations are normal in physics, both in classical systems such as crystals or star clusters and in quantum mechanical ones such as quantised fields. In the neutrino sheets, the correlations are even stronger, so that assuming their absence is a big mistake when one tries to draw conclusions from Bell's theorem.

	\newsecl{$PQ$ theory}{PQ}
Most quantum theories describe operators whose eigenvalues form continua of real numbers. Examples are one or more particles in some potential well, harmonic oscillators, but also bosonic quantum fields and strings. If we want to relate these to deterministic systems we could consider ontological observables that take values in the real numbers as well. There is however an	other option.

One important application of the transformations described in this book could be our attempts to produce fundamental theories about nature at the Planck scale. Here, we have the holographic principle at work, and the Bekenstein bound\,\cite{Bekenstein-1981}. What these tell us is that Hilbert space assigned to any small domain of space should be finite-dimensional. In contrast, real numbers are described by unlimited sequences of digits and therefore require infinite dimensional Hilbert spaces. That's too large. One may suspect that uncertainty relations, or non-commutativity, add some blur to these numbers. In this chapter, we outline a mathematical procedure for a systematic approach.

For PQ theory, as our approach will be called, we employed a new notation in earlier work\,\cite{GtH-2012}\cite{GtH-duality-2012}, where not \(\hbar\) but \(h\) is normalised to one. Wave functions then take the form  \(e^{\,2\pi i px}=\ep^{\,ipx}\), where \(\ep=e^{2\pi}=535.5\). This notation was very useful to avoid factors \(1/\sqrt{2\pi}\) for the normalisation of wave functions on a circle. Yet we decided not to use such notation in this book, so as to avoid clashes with other discussions in the standard notation in various other chapters. Therefore, we return to the normalisation \(\hbar=1\). Factors \(\sqrt{2\pi}\) (for normalised states) will now occur more frequently, and hopefully they won't deter the reader.

In this chapter, and part of the following ones, dynamical variables can be real numbers, indicated with lower case letters: \(p,\ q,\ r,\ x,\ \cdots\), they can be integers indicated by capitals: \(N,\ P,\ Q,\ X,\ \cdots\), or they are angles (numbers defined on a circle), indicated by Greek letters \(\a,\ \eta,\ \k,\ \r,\ \cdots\), usually obeying \(-\pi<\a\le\pi\), or sometimes defined merely \textit{modulo} \(2\pi\).

A real number \(r\), for example the number \(r=137.035999074\cdots\), is composed of an integer, here \(R=137\), and an angle, \(\r /2\pi=0.035999074\cdots\). In examples such as a quantum particle on a line, Hilbert space is spanned by a basis defined on the line: \(\{\,|\,r\,\ket\,\}\). In PQ theory, we regard such a Hilbert space as the product of Hilbert space spanned by the integers \(|\,R\,\ket\) and Hilbert space spanned by the angles, \(|\,\r\,\ket\). So, we have
	\be \fract 1{\sqrt{2\pi}}\,|\,r\,\ket \iss |\,R,\r\,\ket \iss |\,R\,\ket\,|\,\r\,\ket\ . \eel{splithilbert}
Note that \emph{continuity} of a wave function \(|\j\ket\) implies twisted boundary conditions:
	\be\bra\, R+1,\,\r\,|\j\ket\iss\bra\,R,\,\r+2\pi\,|\j\ket\ . \eel{twistedboundary}
The fractional part, or angle, is defined unambiguously, but the definition of the integral part depends on how the angle is projected on the segment \([0,\,2\pi]\), or: how exactly do we round a real number to its integer value? We'll take care of that when the question arises.

 So far our introduction; it suggests that we can split up a coordinate \(q\) into an integer part \(Q\) and a fractional part \(\xi/2\pi\) and its momentum into an integer part \(2\pi P\) and a fractional part \(\k\). Now we claim that this can be done in such a way that both \([P,Q]=0\) and \([\xi,\k]=0\).
 
Let us set up our algebra as systematically as possible.

\subsecl{The algebra of finite displacements}{PQalgebra}

Let there be given an operator \(q_\op\) with non-degenerate eigenstates \(|q\ket\) having eigenvalues \(q\) spanning the entire real line. The associated momentum operator is \(p_\op\) with eigenstates \(|p\ket\) having eigenvalues \(p\), also spanning the real line. The usual quantum mechanical notation, now with \(\hbar=1\), is
	\be &&\qquad q_\op|q\ket=q|q\ket\ ;\qquad p_\op|p\ket=p|p\ket\ ;\qquad[q_\op,\,p_\op]=i\ ; \nm\\
		&& \bra q|q\,'\ket=\d(q-q\,')\ ;\quad \bra p|p\,'\ket=\d(p-p\,')\ ;\quad\bra q|p\ket=\fract 1{\sqrt{2\pi}}e^{ipq}
		\eel{pqnotation}
(often, we will omit the subscript ``op" denoting that we refer to an operator, if this should be clear from the context)

Consider now the displacement operators \(e^{-ip\lowtje{\op}a}\) in position space, and \(e^{iq\lowtje{\op}b}\) in momentum space, where \(a\) and \(b\) are real numbers:
	\be e^{-ip_\op\,a}|q\ket = |q+a\ket\ ;\quad e^{iq_\op\,b}|p\ket=|p+b\ket\ . \eel{displops}
We have 
	\be && [q_\op\,,p_\op\,]=i\ ,\quad e^{ i q_\op\,b}\,p_\op\,=(p_\op- b)\,e^{ i q_\op\,b} \ ;\nm\\[3pt]
	 && e^{ iq_\op\,b} \,e^{-ip_\op\,a}	 =e^{-ip_\op\,a}\ e^{ i ab}\ e^{iq_\op\,b}\nm\\[3pt]
	 && = e^{-ip_\op\,a}\, e^{iq_\op\,b} \ , 
	 \ \hbox{ if }\ ab=2\pi\times\hbox{ integer}\ .  \eel{qpcomm}
Let us consider the displacement operator in position space for \(a=1\). It is unitary, and therefore can be written
uniquely as \(e^{-i\k}\), where \(\k\) is a hermitean operator with eigenvalues \(\k\) obeying \(-\pi<\k\le\pi\). As we see in Eq.~\eqn{displops}, \(\k\) also represents the momentum \textit{modulo} \(2\pi\). Similarly, \(e^{i\xi}\), with \(-\pi<\xi\le\pi\), is defined to be an operator that causes a shift in the momentum by a step \(b=2\pi\). This means that \(\xi/2\pi\) is the position operator \(q\) \textit{modulo} one. We write
	\be p=2\pi K+\k\ ,\quad q=X+\xi/2\pi\ , \eel{pqsplit}
where both \(K\) and \(X\) are integers and \(\k\) and \(\xi\) are angles. We suggest that the reader ignore the factors \(2\pi\) at first reading; these will only be needed when doing precise calculations.

\subsubsection{From the one-dimensional infinite line to the two-dimensional torus\labell{torus}}

As should be clear from Eqs.~\eqn{qpcomm}, we can regard the angle \(\k\) as the generator of a shift  in the integer \(X\), 
	and the angle  \(\xi\) as generating a shift in \(K\):
	\be e^{-i\k}\,|\,X\,\ket=|\,X+1\,\ket\ ,\qquad e^{i\xi}\,|\,K\,\ket=|\,K+1\,\ket\ . \eel{shifts1}

Since \(\k\) and \(\xi\) are uniquely defined as generating these elementary shifts, we deduce from Eqs.~\eqn{qpcomm} and \eqn{shifts1} that 
	\be [\xi,\,\k]=0\ . \eel{xikappacomm}
Thus, consider the torus spanned by the eigenvalues of the operators \(\k\) and \(\xi\). We now claim that the Hilbert space generated by the eigenstates \(|\k,\,\xi\ket\) is equivalent to the Hilbert space spanned by the eigenstates \(|q\ket\) of the operator \(q_\op\), or equivalently, the eigenstates \(|p\ket\) of the operator \(p_\op\) (with the exception of exactly one state, see later).

It is easiest now to consider the states defined on this torus, but we must proceed with care.
If we start with a wave function \(|\j\ket\) that is continuous in \(q\), we have twisted periodic boundary conditions, as indicated in Eq.~\eqn{twistedboundary}. Here, in the \(\xi\) coordinate,
			\be \bra\,X+1,\,\xi\,|\j\ket&=&\bra\,X,\,\xi+2\pi\,|\j\ket\ ,\quad\hbox{or} \nm\\[1pt]
					\bra\,\k,\,\xi+2\pi\,|\j\ket&=&\bra\,\k,\,\xi\,|e^{i\k}|\j\ket\ ,			\eel{twistedxi}
whereas, since this wave function assumes \(X\) to be integer, we have strict periodicity in \(\k\):
						\be \bra\,\k+2\pi,\,\xi\,|\j\ket\iss\bra\,\k,\,\xi\,|\j\ket\ . \eel{nontwistkappa}
If we would consider the same state in momentum space, the periodic boundary conditions would be the other way around, and this is why, in the expression used here, the transformations from position space to momentum space and back are non-trivial. For our subsequent calculations, it is much better to transform first to a strictly periodic torus. To this end, we introduce a phase function \(\f(\k,\,\xi)\) with the following properties:
	\be&\f(\k,\xi+2\pi)=\f(\k,\xi)+\k  ;\qquad\f(\k+2\pi,\xi)=\f(\k,\xi)\ ;& \crl{pseudoperiods}
		&\f(\k,\xi) = -\f(-\k,\xi) =  -\f(\k,-\xi)\ ;\quad \f(\k,\,\xi)+\f(\xi,\,\k)=\k\xi/2\pi\ . &\eel{phiproperties}
An explicit expression for such a function is derived in section~\ref{PQmatrix} and summarised in section~\ref{resumephase}. Here, we just note that this function suffers from a singularity at the point \((\k=\pm\pi,\ \xi=\pm\pi)\). This singularity is an inevitable consequence of the demands \eqn{pseudoperiods} and \eqn{phiproperties}. It is a topological defect that can be moved around on the torus, but not disposed of.

Transforming \(|\j\ket\) now with the unitary transformation 
	\be \bra\k,\xi|\j\ket=\bra\k,\xi|U(\k,\xi)|\tl\j\ket\ ;\qquad U(\k,\,\xi)=e^{i\f(\k,\,\xi)}=e^{i\k\xi/2\pi-i\f(\xi,\,\k)}\ , \eel{xikappaunitary}
turns the boundary conditions \eqn{twistedxi} and \eqn{nontwistkappa} both into strictly periodic boundaries for \(|\tl\j\ket\).

For the old wave function, we had \(X=i\pa/\pa\k\), so, \(q_\op=i	\pa/\pa\k+\xi/2\pi\). The operator \(p_\op\) would simply be \(-2\pi i\pa/\pa\xi\), assuming that the boundary condition \eqn{twistedxi} ensures that this reduces to the usual differential operator. Our new, transformed wave function now requires a modification of these operators to accommodate for the unusual phase factor \(\f(\k,\,\xi)\). Now our two operators become	
	\be q_\op&=& i{\pa\over\pa\k}+{\xi\over 2\pi}-
			\Big({\pa\over\pa\k} \f(\k,\,\xi)\Big)\iss i{\pa\over\pa\k}+\Big({\pa\over\pa\k}\f(\xi,\,\k)\Big) \ ;  \labell{qopkappaxi}  \\[3pt] 
		p_\op&=&-2\pi i{\pa\over\pa\xi}+2\pi\Big({\pa\over\pa\xi}\f(\k,\,\xi)\Big) \iss -2\pi i{\pa\over\pa\xi}+{\k}-
			2\pi\Big({\pa\over\pa\xi}\f(\xi,\,\k)\Big)\ . \eel{popkappaxi}
This is how the introduction of a phase factor \(\f(\k,\,\xi)\) can restore the symmetry between the operators \(q_\op\) and \(p_\op\). Note that, although \(\f\) is not periodic, the derivative \(\pa\f(\k,\,\xi)/\pa\xi\) is periodic, and therefore, both \(q_\op\) and \(p_\op\) are strictly periodic in \(\k\) and \(\xi\) (beware the reflections \(\xi\leftrightarrow \k\) in Eqs.~\eqn{qopkappaxi} and \eqn{popkappaxi}).

We check that they obey the correct commutation rule:
	\be[q_\op,\,p_\op]=i\ . \eel{commcheck}
It is very important that these operators are periodic. It implies that we have no theta jumps in their definitions. If we had not introduced the phase function \(\f(\k,\,\xi)\), we would have such theta jumps and in such descriptions the matrix elements in \(Q,\,P\) space would be much less convergent.
	
The operators \(i\pa/\pa\k\) and \(-i\pa/\pa\xi\) now do not exactly correspond to the operators X and K anymore, because of the last terms in Eqs.~\eqn{qopkappaxi} and \eqn{popkappaxi}. They are integers however, which obviously commute, and these we shall call \(Q\) and \(P\). To obtain the operators \(q_\op\) and \(p_\op\) in the basis of the states \(|Q,\,P\ket\), we simply expand the wave functions in \(\k,\,\xi\) space in terms of the Fourier modes,
	\be\bra\,\k,\,\xi\,|\,Q,\,P\,\ket={1\over2\pi}e^{iP\xi-iQ\k}\ . \eel{kappaxiFourier}
We now need the Fourier coefficients of the phase function \(\f(\k,\,\xi)\). They are given in section~\ref{PQmatrix}, where we also derive the explicit expressions for the operators \(q_\op\) and \(p_\op\) in the \(Q,\,P\) basis:	
	\be q_\op=Q_\op+a_\op\ ; \qquad \bra Q_1,P_1|\,Q_\op\,|Q_2,P_2\ket&=&Q_1\,\d_{Q_1\,Q_2}\,\d_{P_1\,P_2}\ ,\nm\\[3pt]
		 \bra Q_1,P_1|\,a_\op\,|Q_2,P_2\ket&=&{(-1)^{P+Q+1}\,iP\over 2\pi(P^2+Q^2)}\ , \eel{qopmatrix}
where \(Q\) stands short for \(Q_2-Q_1\), and \(P=P_2-P_1\).
	
For the \(p\) operator, it is derived analogously, 
	\be  &p_\op=2\pi P_\op+b_\op\ ,& \ \bra Q_1,P_1|P_\op|Q_2,P_2\ket=P_1\,\d_{Q_1Q_2}\d_{P_1P_2}\ ;\crl{popmatrix}
	&\bra Q_1,P_1|\,b_\op\,|Q_2,P_2\ket &=\ {(-1)^{P+Q}\,iQ\over P^2+Q^2}\  . \eel{bmatrix}
	
And now for some surprise. Let us inspect the commutator, \([\,q_\op,p_\op\,]\), in the basis of the integers \(Q\) and \(P\). We have
	\be&[Q_\op,\,P_\op]=0\ ;\quad [\,a_\op,\,b_\op\,]=0\ ;\qquad[q_\op,\,p_\op]=[Q_\op,\,b_\op\,]+2\pi [\,a_\op,\, P_\op]\ ; & \nm\\[3pt]
	&	\bra Q_1,P_1|\,[\,q_\op,p_\op\,]\,|Q_2,P_2\ket \iss -i(-1)^{P+Q)}(1-\d_{Q_1Q_2}\d_{P_1P_2})\ .&	\eel{qpQPcomm}
Here, the delta function is inserted because the commutator vanishes if \(Q_1=Q_2\) and \(P_1=P_2\).
So, the commutator is not equal to \(i\) times the identity, but it can be written as	
	\be [\,q_\op,\,p_\op\,]=i \,(\,{\Bbb I}-|\,\j_e\,\ket\bra\,\j_e\,|\,)\ ,\quad \hbox{where }\ \bra  Q,P|\,\j_e\,\ket=(-1)^{P+Q}\ . 
		\eel{PQedge}
Apparently, there is one state \(|\j_e\ket\) (with infinite norm), for which the standard commutator rule is violated. We encounter more such states in this book, to be referred to as \emph{edge states}, that have to be factored out of our Hilbert space. From a physical point of view it will usually be easy to ignore the edge states, but when we do mathematical calculations it is important to understand their nature. The edge state here coincides with the state \(\d(\k-\pi)\,\d(\xi-\pi)\), so its mathematical origin is easy to spot: it is located at the singularity of our auxiliary phase function \(\f(\k,\,\xi)\), the one we observed following Eqs.\ \eqn{pseudoperiods} and \eqn{phiproperties}; apparently, we must limit ourselves to wave functions that vanish at that spot in \((\k,\,\xi\)) space.
	
	\subsubsection{The states $|Q,P\ket$ in the $q$ basis\labell{QPwaves}}
	
As in other chapters, we now wish to identify the transformation matrix enabling us to transform from one basis to an other. Thus, we wish to find the matrix elements connecting states \(|Q,P\ket\) to states \(|q\ket\) or to states \(|p\ket\). If we can find the function \(\bra q |0,0\ket\), which gives the wave function in \(q\) space of the state with \(Q=P=0\) , finding the rest will be easy. Section~\ref{zerozerowave}, shows the derivation of this wave function. In \((\k,\,\xi)\) space, the state \(|q\ket\) is 
	\be\bra\,\k,\xi\,|q\ket=e^{i\,\f(\k,2\pi q)}\,\d(\xi-\xi_q)\ , \eel{qkappaximatrix}
if \(q\) is written as \(q=X+\xi_q/2\pi\), and \(X\) is an integer. Section\ \ref{zerozerowave} shows that then the wave function for \(P=Q=0\) is
	\be\bra q|0,0\ket =\fract 1{2\pi}\int_0^{2\pi}\,\dd\k\,e^{-i\f(\k,2\pi q)}\ . \eel{qmatrix00}
The general wave function is obtained by shifting \(P\) and \(Q\) by integer amounts:
	\be\bra q|Q,P\ket= \fract 1{2\pi}\,e^{2\pi iP q}\int_0^{2\pi}\dd\k\,e^{-i\f(\k,\,2\pi(q-Q))}\ . \eel{qQPmatrix}

\begin{figure}[htb!]
\begin{center} \lowerwidthfig{0pt} {120mm}{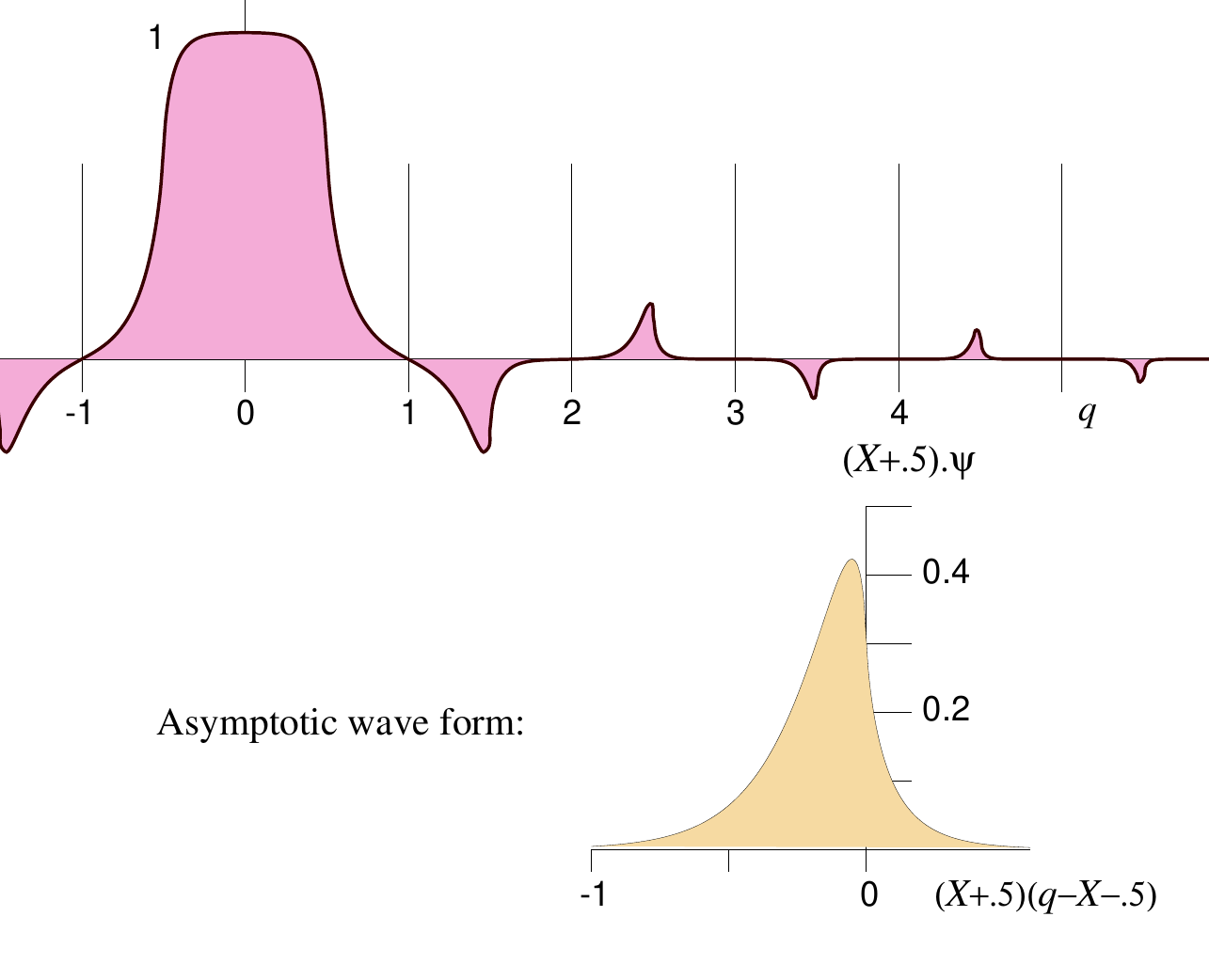}  \begin{quotation}
\caption[\small The wave function of the state $(P,Q)=(0,0)$, and the asymptotic form of some small peaks.]{\small The wave function of the state \(|Q,P\ket\) when \(P=Q=0\).  Below, the asymptotic form of the little peaks far from the centre, scaled up.\label{wave00.fig}}\end{quotation} \end{center}
		\end{figure}

The wave function \eqn{qmatrix00}, which is equal to its own Fourier transform, is special because it is close to a block wave, having just very small 
tails outside the domain \(|q|<\half\), see Fig.~\ref{wave00.fig}

The wave \(\bra q|Q,P\ket\) is similar to the so-called \emph{wavelets}\,\cite{wavelets-1997}, which are sometimes used to describe pulsed waves, but this one has two extra features. Not only is it equal to its own Fourier transform, it is also orthogonal to itself when shifted by an integer. This makes the set of waves \(|Q,P\ket\) in Eq.~\eqn{qQPmatrix} an orthonormal basis.

This \(PQ\) formalism is intended to be used to transform systems based on integer numbers only, to systems based on real numbers, and back. The integers
may be assumed to undergo switches described by a permutation operator \(\mathcal{P_\op}\). After identifying some useful expression for a Hamiltonian \(H_\op\), with \(\mathcal{P_\op}=e^{-iH_\op\,\d t}\), one can now transform this to a quantum system with that Hamiltonian in its new basis.
	
For a single \(PQ\) pair, constructing a deterministic model whose evolution operator resembles a realistic quantum Hamiltonian is difficult. A precise, canonical, discrete Hamiltonian formalism \emph{is} possible in the \(PQ\) scheme, but it requires some more work that we postpone to section~\ref{Hamiltonform}. Interesting Hamiltonians are obtained in the multidimensional case: \(P_i,\,Q_i\). Such a system is considered in chapter~\ref{2dnoint}.	
			
\subsection{Transformations in the $PQ$ theory\labell{PQmatrix}}

The following sections, with which we end chapter~\ref{PQ}, can be read as Appendices to chapter~\ref{PQ}. They contain technicalities most readers might not be interested in, but they are needed to understand  the details of some features that were encountered,
in particular when explicit calculations have to be done, connecting the basis sets of the real numbers \(q\), their Fourier transforms \(p\), the integers \((Q,\,P)\) and the torus \((\k,\,\xi)\). 

	Let us first construct a solution to the boundary conditions~\eqn{pseudoperiods} and \eqn{phiproperties} for a phase function \(\f(\k,\,\xi)\),
		\be&\f(\k,\xi+2\pi)=\f(\k,\xi)+\k  ;\qquad\f(\k+2\pi,\xi)=\f(\k,\xi)\ ;& \crl{pseudoperiodsA}
		&\f(\k,\xi) = -\f(-\k,\xi) =  -\f(\k,-\xi)\ ;\quad \f(\k,\,\xi)+\f(\xi,\,\k)=\k\xi/2\pi\ . &\eel{phipropertiesA}
At first sight, these conditions appear to be contradictory. If we follow a closed contour \((\k,\,\xi)=
(0,0)\ra(2\pi,0)\ra(2\pi,2\pi)\ra(0,2\pi)\ra(0,0)\), we pick up a term \(2\pi\). This implies that a function that is single valued everywhere does not exist, hence, we must have a singularity. We can write down an amplitude \(\j(\k,\xi)=r\,e^{i\f}\) of which this is the phase, but this function must have a zero or a pole. Let's assume it has a zero, and \(r\) is simply periodic. Then one can find the smoothest solution. If \(r\) and \(\f\) are real functions:
	\be r(\k,\,\xi)\,e^{i\f(\k,\,\xi)}=\sum_{N=-\infty}^\infty e^{-\pi (N-{\xi\vphantom{_g}\over 2\pi})^2+ iN\k}\ , \eel{rphidef}
one finds that this is obviously periodic in \(\k\), while the substitution
	\be \xi\ra\xi+2\pi\ ,\quad N\ra N+1\ ,\eel{pseudoper}
gives the first part of Eq.~\eqn{pseudoperiodsA}.

The sum in Eq.~\eqn{rphidef}, which fortunately converges rapidly, is a special case of the elliptic function  \(\vartheta_3\), and it can also be written as a product\,\cite{GR}:
 		\be\hskip-10pt  r(\k,\xi)\,e^{i\f(\k,\xi)}=e^{-{\xi^2\vphantom{_g}\over 4\pi}}\prod_{N=1}^\infty(1-e^{-2\pi N})\ \prod_{N=0}^\infty(1+e^{\,\xi+i\k-2\pi N-\pi}) (1+e^{-\xi-i\k-2\pi N-\pi})\ ,\quad \eel{theta}
from which we can easily read off the zeros: they are at \((\k,\,\xi)=(2\pi N_1+\pi,\,2\pi N_2+\pi)\).
We deliberately chose these to be at the corners \((\pm\pi,\,\pm\pi)\) of the torus, but it does not really matter where they are; they are simply unavoidable.

The matrix elements \(\bra Q_1,\,P_1|\,q_\op\,|Q_2,\,P_2\ket\) are obtained by first calculating them on the torus. We have\fn{With apologies for interchanging the \(\k\) and \(\xi\) variables at some places, which was unavoidable, please beware.} 
	\be q_\op=Q_\op+a(\xi,\k)\ ;\qquad Q_\op= i{\pa\over\pa\k}\ ,\quad a(\xi,\k)=\Big({\pa\over\pa\k}\f(\xi,\k)\Big)\ . \eel{qoptorus}
To calculate \(a(\xi,\k)\) we can best take the product formula \eqn{theta}:
	\be a(\xi,\k)&=&\sum_{N=0}^\infty a_N(\xi,\k)\ , \nm\\
		a_N(\xi,\k)&=&{\pa\over\pa\k}\Big(\arg(1+e^{\k+i\xi-2\pi(N+\half)})+\arg(1+e^{-\k-i\xi-2\pi(N+\half)})\Big)\ . \qquad
			\eel{astep1}
Evaluation gives (note the interchange of \(\k\) and \(\xi\)):
	\be a_N(\xi,\k)\ ={\half\sin\xi\over\cos\xi+\cosh(\k-2\pi N-\pi)}+{\half\sin\xi\over\cos\xi+\cosh(\k+2\pi N+\pi)}\ ,\eel{astep2}
which now allows us to rewrite it as a single sum for \(N\) running from \(-\infty\) to \(\infty\) instead of \(0\) to \(\infty\).

Let us first transform from the \(\xi\) basis to the \(P\) basis, leaving \(\k\) unchanged. This turns \(a(\xi,\k)\) into an operator \(a_\op\). With
	\be\bra P_1|\xi\ket\bra\xi|P_2\ket=\fract 1{2\pi} e^{iP\xi}\ ,\qquad P=P_2-P_1\ ,\eel{Pxibasis}
we get the matrix elements of the operator \(a_\op\) in the \((P,\,\k)\) frame: 
	\be \bra P_1|a_\op(\k)|P_2\ket=\sum_{N=-\infty}^\infty a_N(P,\,\k)\ ,\qquad P\equiv P_2-P_1\ , \ee
and writing \(z=e^{\,i\xi}\), we find
	\be a_N(P,\,\k)&=&\oint{\dd z\over 2\pi i z}z^P{-\half i(z-1/z)\over z+1/z+e^R+e^{-R}}\ ,\quad R=\k+2\pi N+\pi\ ,\nm\\
	      a_N(P,\,\k)&=&\half\sgn(P)(-1)^{P-1}ie^{-\big|\,P(\k+2\pi N+\pi)\,\big|}\ ,\eel{aNkappaP}
where \ sgn\(\!(P)\) is defined to be \(\pm 1\) if \(P\glt 0\) and \,0\, if \(P=0\). The absolute value taken in the exponent indeed means that we always have a negative exponent there; it originated when the contour integral over the unit circle forced us to choose a pole inside the unit circle.

Next, we find the \((Q,P)\) matrix elements by integrating this over \(\k\) with a factor \(\bra Q_1|\k\ket\bra\k|Q_2\ket=\fract 1{2\pi}e^{iQ\,\k}\), with \(Q=Q_2-Q_1\). The sum over \(N\) and the integral over \(\k\) from 0 to \(2\pi\) combine into an integral over all real values of \(\k\), to obtain the remarkably simple expression
	\be \bra Q_1,P_1|\,a_\op\,\,|Q_2,P_2\ket\iss{(-1)^{P+Q+1}\,iP\over 2\pi(P^2+Q^2)}\  .\eel{aQmatrix}
In Eq.~\eqn{qopkappaxi}, this gives for the \(q\) operator:
	 \be{}\hskip-22pt  q_\op&=&Q_\op+a_\op\ ;\nm\\[4pt]
	  \bra Q_1,P_1|\,q_\op\,|Q_2,P_2\ket&=&Q_1\,\d_{Q_1\,Q_2}\,\d_{P_1\,P_2}+\bra Q_1,P_1|\,a_\op\,|Q_2,P_2\ket\ . \eel{qmatrix}
	
For the \(p\) operator, the role of Eq.~\eqn{qoptorus} is played by
	\be p_\op=2\pi P+b(\k,\xi)\ ;\qquad P=-i{\pa\over\pa\xi}\ ,\quad b(\k,\xi)=2\pi\bigg({\pa\over\pa\xi}\f(\k,\xi)\bigg)\ , \eel{poptorus}
and one obtains analogously, writing \(P\equiv P_2-P_1\),
	\be  &p_\op=2\pi P_\op+b_\op\ ,& \ \bra Q_1,P_1|P_\op|Q_2,P_2\ket=P_1\,\d_{Q_1Q_2}\d_{P_1P_2}\ ;\crl{pmatrix}
	&\bra Q_1,P_1|\,b_\op\,|Q_2,P_2\ket &=\ {(-1)^{P+Q}\,iQ\over P^2+Q^2}\  . \eel{bPmatrix}
Note, in all these expressions, we have the symmetry under the combined interchange
	\be& p_{\,\op}\leftrightarrow 2\pi\,q_{\,\op}\ ,\quad P\leftrightarrow Q\ ,\quad X\leftrightarrow K\ ,&\nm\\[4pt] & \xi\leftrightarrow \k\ ,
	\quad i\leftrightarrow -i \ ,\quad 2\pi\,a_{\,\op} \leftrightarrow b_{\,\op}\ .&\ee

\subsection{Resume of the quasi-periodic phase function $\f(\xi,\k)$\labell{resumephase}	}
	Let \(Q\) and \(P\) be integers while \(\xi\) and \(\k\) obey \(-\pi<\xi<\pi\,,\ -\pi<\k<\pi\).
The operators obey 
	\be[Q_\op,P_\op]=[\xi_\op,P_\op]=[Q_\op,\k_\op]=[\xi_\op,\k_\op]=0\ .\eel{quasiphaseops}
Inner products:
	\be\bra\k|Q\ket=\fract 1{\sqrt{2\pi}}\,e^{-iQ\k}\ ,\qquad\bra\xi|P\ket=\fract 1{\sqrt{2\pi}}\,e^{iP\xi}\ . \eel{quasiinner}
The phase angle functions \(\f\) and \(\tl\f\) are defined obeying \eqn{pseudoperiodsA} and \eqn{phipropertiesA}, and
	\be\tl\f(\xi,\k)=\f(\k,\xi)\ ,\quad \f(\xi,\k)+\tl\f(\xi,\k)=\xi\k/2\pi\ . \eel{quasitilde}
When computing matrix elements, we should not take the operator \(\f\) itself, since it is pseudo periodic instead of periodic, so that the edge state \(\d(\k\pm\pi)\) gives singularities. Instead, the operator \(\pa\f(\xi,\k)/\pa\k\equiv  a (\xi,\k)\) is periodic, so we start from that. Note that, on the \((\xi,\k)\)-torus, our calculations often force us to interchange the order of the variables \(\xi\) and \(\k\). Thus, in a slightly modified notation,
	\be  a (\xi,\k)=\sum_{N=-\infty}^\infty a_N(\xi,\k)\ ;\qquad  a _N(\xi,\k)={\half\sin\xi\over \cos\xi+\cosh(\k+2\pi(N+\half))}\ . \ee
With \(P_2-P_1\equiv P\), we write in the \((\k,P)\) basis
	\be \bra \k_1,P_1|{ a }^\op_N|\k_2,P_2\ket&=&\d(\k_1-\k_2) a _N(\k_1,P)\nm\\[2pt]
	 a _N(\k,P)\ \equiv\  \fract 1{2\pi}\int_0^{2\pi}\dd\xi\, e^{iP\xi}a_N(\xi,\k) &=& \half\sgn(P)(-1)^{P-1} ie^{-|P(\k+2\pi N+\pi)|}\  ,\nm\\
	 a (\k,P)&=& \half i(-1)^{P-1}\,{\cosh(P\k)\over \sinh(P\pi)}\ . \eel{phasemix1}	
Conversely, we can derive in the \((Q,\xi)\) basis, with \(Q_2-Q_1\equiv Q\):
	\be \bra Q_1,\xi_1|{ a }^\op_N|Q_2,\xi_2\ket&=&\d(\xi_1-\xi_2) a _N(Q,\xi_1)\nm\\[2pt]
	 a _N(Q,\xi) &=& {1\over 4\pi}\sin\xi\int_{-\pi}^\pi\dd\k{e^{iQ\k}\over \cos\xi+\cosh(\k+2\pi N+\pi)} \ ;\nm\\[2pt]
	 a (Q,\xi)\iss\sum_N a _N(Q,\xi)&=&(-1)^Q{\sinh(Q\xi)\over 2\sinh(Q\pi)}  \eel{phasemix2}
(the latter expression is found by contour integration).\\[8pt]
Finally, in the \((Q,P)\) basis, we have, either by Fourier transforming \eqn{phasemix1} or \eqn{phasemix2}:
	\be \bra Q_1,P_1|{ a }^\op|Q_2,P_2\ket = {-i\over 2\pi}(-1)^{Q+P}{P\over Q^2+P^2}\ .	\eel{phasePQ}
	
The operators \(q_\op\)  and \(p_\op\) are now defined on the torus as
	\be q_\op(\k,\xi)=Q_\op+ a (\xi,\k)\ , \quad p_\op(\k,\xi)=2\pi(P_\op+ a (\k,\xi))\ , \eel{qopop}
with \(Q_\op=i\pa/\pa\k\ ,\quad P_\op=-i\pa/\pa\xi\). This reproduces Eqs.~\eqn{aQmatrix} and \eqn{bPmatrix}.

\subsection{The wave function of the state $|0,0\ket$\labell{zerozerowave}}
	We calculate the state \(|q\ket\) in the \((\k,\xi)\) torus. Its wave equation is (see Eq.~\eqn{qopkappaxi})
		\be q_\op|q\ket=i\,e^{i\f(\xi,\k)}{\pa\over\pa\k}\Big(e^{-i\f(\xi,\k)}\,|q\ket\Big)=q|q\ket\ . \eel{qeqkappaxi}
This equation is easy to solve:
	\be \bra\k,\xi|q\ket=C(\xi)\,e^{i\f(\xi,\k)-iq\k}\ . \eel{qkappaxi}
Since the solution must be periodic in \(\k\) and \(\xi\), while we have the periodicity conditions~\eqn{pseudoperiodsA} for \(\f\), we deduce that this
only has a solution if \(\xi/2\pi\) is the fractional part of \(q\), or, \(q=X+\xi/2\pi\), where \(X\) is integer. In that case, we can write
	\be \bra\k,\xi|q\ket=C(\xi)\,e^{-iX\k-i\f(\k,\xi)}\iss C(\xi)\,e^{-i\f(\k,2\pi q)}\ . \eel{qkappaxi2}

The complete matrix element is then, writing \(q=X+\xi_q/2\pi\), 
	\be\bra\k,\xi|q\ket=C\,e^{-i\f(\k,2\pi q)}\,\d(\xi-\xi_q) \eel{qkappaximatr}
(note that the phase \(\f(\k,\xi)\) is not periodic in its second entry \(\xi\); the entries are reversed compared to Eqs~\eqn{pseudoperiodsA}). 
The normalisation follows from requiring
	\be \int_0^{2\pi}\!\dd\k\int_0^{2\pi}\!\dd\xi\,\bra q_1|\k,\xi\ket\bra\k,\xi|q_2\ket &=& \d(q_1-q_2)\ ; \nm\\
	\int_{-\infty}^{\infty}\dd q\,\bra\k_1,\xi_1|q\ket\bra q|\k_2,\xi_2\ket &=& \d(\k_1-\k_2)\d(\xi_1-\xi_2)\ , \eel{normcondkxq}
from which
	\be C=1\ . \eel{normkxq}
Note that we chose the phase to be \(+1\). As often in this book, phases can be chosen freely.

Since
	\be\bra\k,\xi|Q,P\ket=\fract 1{2\pi}e^{iP\xi-iQ\k}\ , \eel{xikappaPQ}
we have
	\be\bra q|Q,P\ket=\fract 1{2\pi}\int_0^{2\pi}\dd\k\,e^{iP\xi-iQ\k+i\f(\k,2\pi q)}\ =\fract 1{2\pi}e^{2\pi iPq}\int_0^{2\pi}\dd\k\,e^{i\f(\k,2\pi(q-Q))}\ .\eel{QPqmatrix}
	
\newsecl{Models in two space-time dimensions without interactions}{2dnoint}	
\subsecl{Two dimensional model of massless bosons}{2dbosons}
	
An important motivation for our step from real numbers to integers is that we require deterministic theories to be infinitely precise. Any system based on a classical action, requires real numbers for its basic variables, but this also introduces limited precision. If, as one might be inclined to suspect, the ultimate physical degrees of freedom merely form bits and bytes, then these can only be discrete, and the prototypes of discrete systems are the integers. Perhaps later one might want to replace these by integers with a maximal size, such as integers \emph{modulo} a prime number \(p\), the elements of \(\Bbb Z/p\Bbb Z\).

The question is how to phrase a systematic approach. For instance, how do we mimic a quantum field theory? If such a theory is based on perturbative expansions, can we mimic such expansions in terms of integers? Needless to observe that standard perturbation expansions seem to be impossible for discrete systems, but various special kinds of expansions can still be imagined, such as \(1/N\) expansions, where \(N\) could be some characteristic of an underlying algebra.
	
We shall not be able to do this in this book, but we make a start at formulating systematic approaches.  In this chapter, we consider a quantised field, whose field variables, of course, are operators with continua of eigenvalues in the real numbers. If we want to open the door to perturbative field theories, we first need to understand free particles. One example was treated in section~\ref{nu}. These were fermions. Now, we try to introduce free bosons.

Such theories obey linear field equations, such as
	\be\pa_t^2\,\f(\vec x,t)=\sum_{i=1}^d\,\pa^2_{i}\,\f(\vec x,t)-m^2\,\f(\vec x,t)\ . \eel{ddimfreefield}
In the case of fermions, we succeeded, to some extent, to formulate the massless case in three space dimensions (section~\ref{nu}, subsection~ \ref{secondneutrino}), but applying \(PQ\) theory to bosonic fields in more than two dimensions has not  been successful. The problem is that equations such as Eq.~\eqn{ddimfreefield} are difficult to apply to integers, even if we may fill the gaps between the integers with generators of displacements.

In our search for systems where this can be done, we chose, as a compromise, massless fields in one space-like dimension only. The reason why this special case can be handled with \(PQ\) theory is, that the field equation, Eq.~\eqn{ddimfreefield}, can be reduced to first order equations by distinguishing left-movers and right-movers. Let us first briefly summarise the continuum quantum field theory for this case. 
	
\subsubsection{Second-quantised massless bosons in two dimensions\labell{qft2d}}
	
We consider a single, scalar, non interacting, massless field \(q(x,t)\). Both \(x\) and \(t\) are one-dimensional. The Lagrangian and  Hamiltonian are:
	\be\LL=\half(\pa_t q^2-\pa_x q^2)\ ;\qquad H_\op=\int\dd x(\half p^2+\half\pa_x q^2)\ , \eel{LH}
where we use the symbol \( p(x)\) to denote the canonical momentum field associated to the scalar field \(q(x)\), which, in the absence of interactions,  obeys \( p(x)=\pa_t\, q(x)\). The fields \(q(x)\) and \(p(x)\) are operator fields. 
The equal-time commutation rules are, as usual:
	\be[ q(x), q(y)]=[ p(x), p(y)]=0\ ;\qquad[ q(x), p(y)]=i\d(x-y)\ .\eel{standcomm}

Let us regard the time variable in \( q(x,t)\) and \( p(x,t)\) to be in Heisenberg notation. We then have the field equations:
	\be\pa_t^2q=\pa_x^2q\ , \eel{fieldeqs}
and the solution of the field equations can be written as follows:
	\be a^L(x,t)&=& p(x,t)+\pa_x q(x,t)\iss a^L(x+t)\ ;\crl{leftcont}
	 a^R(x,t)&=& p(x,t)-\pa_x q(x,t)\iss a^R(x-t)\ .\eel{rightcont}
The equations force the operators  \(a^L\) to move to the left and \(a^R\) to move to the right. In terms of these variables, the Hamiltonian (at a given time \(t\)) is
	\be H_\op=\int\dd x\,\quart\bigg((a^L(x))^2+(a^R(x))^2\bigg)\ . \eel{HaLaRcont}
The commutation rules for \(a^{L,R}\) are:
	\be [a^L,\,a^R]=0\ ,&& [a^L(x_1),\,a^L(x_2)]=2i\,\d\,'(x_1-x_2)\ , \nm\\
		&& [a^R(x_1),\,a^R(x_2)]=-2i\,\d\,'(x_1-x_2)\ , \eel{aLaRcomm}
where \(\d\,'(z)=\fract\pa{\pa z}\d(z)\).

Now let us Fourier transform in the space direction, by moving to momentum space variables \(k\), and subtract the vacuum energy. 
We have in momentum space (note that we restrict ourselves to \emph{positive} values of \(k\)):
	\be &&a^{L,R}(k)=\fract 1{\sqrt{2\pi}}\int\dd x\, e^{-ikx}\,a^{L,R}(x)\ ,  \qquad a^\dag(k)=a(-k)\ ;\\
	 &&H_\op=\int_0^\infty\dd k\,\half\bigg( (a^{L\dag}(-k)a^L(-k)+  a^{R\dag}(k)a^R(k)\bigg)\ , \crl{HaLaRk}
	k,k'>0:	&& [a^L(-k_1),a^L(-k_2)]=0\ ,\nm\\[3pt]
	&& [a^L(-k_1),a^{L\dag}(-k_2)]=2k_1\d(k_1-k_2)\ . \crl{aLaLdagcont}
		&& [a^R(k_1),a^R(k_2)]=0\ ,\nm\\[3pt] 
		&& [a^R(k_1),a^{R\dag}(k_2)]=2k_1\d(k_1-k_2)\ . \eel{aRaRdagcont}
In this notation, \(a^{L,R}(\pm k)\) are the annihilation and creation operators, apart from a factor \(\sqrt{2k}\), so the Hamiltonian~\eqn{HaLaRcont} can be written as
	\be H_\op=\int_0^\infty\dd k\,(kN^L(-k)+kN^R(k))\ . \eel{contoccnumbers}
where \(N^{L,R}(\mp k)\dd k\) are the occupation numbers counting the left and right moving particles. 
The energies of these particles are equal to the absolute values of their momentum. All of this is completely standard and can be found in all the text books about this subject.

Inserting a lattice cut-off for the UV divergences in quantum field theories is also standard practice. Restricting ourselves to integer values of the \(x\) coordinate, and using the lattice length in \(x\) space as our unit of length,
we replace the commutation rules ~\eqn{standcomm} by
	\be[ q(x), q(y)]=[   p^+  (x),   p^+  (y)]=0\ ;\qquad[ q(x),   p^+  (y)]=i\,\d_{x,y}\eel{latticecomm}
(the reason for the superscript + will be explained later, Eqs.~\eqn{pdef} --- \eqn{pevolve}).
The exact form of the Hamiltonian on the lattice depends on how we wish to deal with the lattice artefacts. The choices made below might seem somewhat artificial or special, but it can be verified that most alternative choices one can think of can be transformed to these by simple lattice field transformations, so not much generality is lost. It is important however that we wish to keep the expression~\eqn{contoccnumbers} for the Hamiltonian; also on the lattice, we wish to keep the same dispersion law as in the continuum, so that all excitations must move left or right exactly with the same speed of light (which of course will be normalised to \(c=1\)).
	
The lattice expression for the left- and right movers will be
	\be a^L(x+t)&=&   p^+  (x,t)+ q(x,t)- q(x-1,t)\ ;\crl{aL}
	     a^R(x-t)&=&   p^+  (x,t)+ q(x,t)- q(x+1,t)\ . \eel{aR} 
 They obey the commutation rules 
	    \be[a^L,a^R]=0\ ;&& [a^L(x),\,a^L(y)]=\pm\, i\ \hbox{ if }\ y=x\pm 1\ ;\quad\hbox{else }\ 0\ ; \crl{aLaLlatticecomm}
				&& [a^R(x),\,a^R(y)]=\mp\,i\ \hbox{ if }\ y=x\pm 1\ ;\quad\hbox{else }\ 0\ . \eel{aRaRlatticecomm}
They can be seen to be the lattice form of the commutators \eqn{aLaRcomm}. In momentum space, writing
	\be  a^{L,R}(x)\equiv\fract{1}{\sqrt{2\pi}}\int_{-\pi}^{\,\pi}\dd \k\ a^{L,R}(\k)\,e^{i\k x}\ ,\qquad a^{L,R\,\dag}(\k)=a^{L,R}(-\k)\ ; \eel{Fouriera}
the commutation rules \eqn{aLaLdagcont} and \eqn{aRaRdagcont} are now
	 \be  [a^L(-\k_1),\,a^{L\dag}(-\k_2)]&=&2\sin \k_1\,\d(\k_1-\k_2)\ ; \qquad\crl{aLkcomm} 
      			 [a^R( \k_1),\,a^{R\dag}(\k_2)]&=&2\sin \k_1\,\d(\k_1-\k_2)\ ,\qquad\eel{aRkcomm} 
so our operators \(a^{L,R}(\mp \k)\) and \(a^{L,R\,\dag}(\mp \k)\) are the usual annihilation and creation operators, multiplied by the factor
	\be \sqrt{2|\sin \k|}\ . \eel{creationannihilation}

If we want the Hamiltonian to take the form~\eqn{contoccnumbers}, 
then, in terms of the creation and annihilation operators \eqn{aLkcomm} and \eqn{aRkcomm}, the Hamiltonian must be
	\be H_\op=\int_0^\pi\dd \k{ \k\over 2\sin \k}\bigg(a^{L\dag}(-\k)\,a^L(- \k)+a^{\dag R}(\k)\,a^R(\k)\bigg)\ . \eel{Hamiltonian}
Since, in momentum space, Eqs.~\eqn{aL} and \eqn{aR} take the form
	\be a^L(\k)=   p^+(\k)+(1-e^{-i \k}) q( \k)\ ,\qquad  a^R( \k)=   p^+ (\k)+(1-e^{i \k}) q( \k)\ , \eel{aLRk}	
after some shuffling, we find the Hamiltonian (ignoring the vacuum term)
	\be H_\op=\half\int_0^\pi\dd \k\,\bigg({ \k\over\tan \half \k}\ |   p^+( \k) |^2\ +\ 4 k\tan\half \k\ | q(\k)+\half p^+(\k) |^2\bigg)\ , \eel{hampf} 
where \(|   p^+(\k)|^2\) stands for \(   p^+(\k)\,   p^+(-\k)\). 
	Since the field redefinition \( q(x)+\half   p^+(x)\ra q(x)\) does not affect the commutation rules, and
	\be \lim_{\k\ra 0}\,{\k\over2\tan(\half\k)}=1\ ,\qquad 4\sin^2(\half\k)| q( \k)|^2\ra|(\pa_x q)(\k)|^2\ , \eel{limtan}
 we see that the continuum limit~\eqn{LH}, \eqn{HaLaRk}  is obtained when the lattice length scale is sent to zero.
 
Because of the factor \eqn{creationannihilation}, the expression \eqn{Hamiltonian} for our Hamiltonian shows that the operators \(a\) and \(a^\dag\), annihilate and create energies of the amount \(|\k|\), as usual, and the Hamilton equations for \(a^{L,R}\) are
 	\be {\dd\over\dd t}a^L(-\k,t)\iss -i[\,a^L(-\k,t),\,H_\op\,]&=&{-i\,\k\over 2\sin \k}\,2\sin \k\  a^L(-\k)\iss -i\k\,a^L(-\k,t)\ ; \nm\\ 
		{\dd\over\dd t} a^R(\k,t)&=&-i\k\, a^R(\k,t)\ . \ee
Consequently, 
	\be a^L(-\k,t)\,e^{-i\k x}=a^L(-\k,0)e^{-i\k x-i\k t}\ ;\quad a^R(\k,t)e^{i\k x}=a^R(\k,0)e^{i\k x-i\k t}\ . \ee
We now notice that the operators \(a^L(x,t)=a^L(x+t)\) and \(a^R(x,t)=a^R(x-t)\) move exactly one position after one unit time step. Therefore,
\emph{even on the lattice},
	\be a^L(x,1)=a^L(x+1,0)\ ,\qquad a^R(x,1)=a^R(x-1,0)\ ,\quad\hbox{etc.}\eel{timesteps}
and now we can use this to eliminate \(   p^+  (x,t)\) and \( q(x,t)\) from these equations. Writing
	\be    p^+  (x,t)\equiv p(x,t+\half)\ , \eel{pdef}
one arrives at the equations
	\be  q(x,t+1)&=& q(x,t)\ +\ p(x,t+\half)\ ; \crl{qevolve}
	p(x,t+\half)&=&p(x,t-\half)\ \,+\,  q(x-1,t)-2 q(x,t)+ q(x+1,t)\ . \eel{pevolve}
We now see why we had to shift the field \( q(x,t)\) by half the field momentum in Eq.~\eqn{hampf}:  it puts the field at the same position \(t+\half\) as the momentum variable \(   p^+  (x,t)\).

Thus, we end up with a quantum field theory where not only space but also time is on a lattice. The momentum values \(p(x,t+\half)\) can be viewed as variables on the time-like links of the lattice.

At small values of \(\k\), the Hamiltonian~\eqn{Hamiltonian}, \eqn{hampf} closely approaches that of the continuum theory, and so it obeys locality conditions there. For this reason, the model would be interesting indeed, if this is what can be matched with a cellular automaton. However, there is a problem with it. At values of \(\k\) approaching \(\k\ra\pm\pi\), the kernels diverge. Suppose we would like to write the expression~\eqn{Hamiltonian} in position space as
	\be H_\op=\half\sum_{x,\,s}M_{|s|}\,\bigg(a^L(x)\,a^L(x+s)+a^R(x)\,a^R(x+s)\bigg)\ , \eel{kernel}
then \(M_s\) would be obtained by Fourier transforming the coefficient \(\k/2\sin(\k)\) on the interval \([-\pi,\,\pi]\) for \(k\). The factor \(\half\) comes from symmetrising the expression \eqn{kernel} for positive and negative \(s\). One obtains
	\be M_s=\fract{1}{2\pi}\int_{-\pi+\l}^{\pi-\l}{\k\dd \k\over2\sin \k}\,e^{- i s \k}\ =\half\left\{\matrix{\log{2\over\l}-\sum_{k=0}^{s/2-1}{1\over k+1/2}\ \hbox{ if }\ s= 
		 \hbox{ even}\nm \\[2pt] \log( 2\l)+\sum_{k=1}^{(s-1)/2}{1\over k}\ \hbox{ if }\ s=\hbox{ odd}}\right. \eel{Fourierham}
where \(\l\) is a tiny cut-off parameter. The divergent part of \(H_\op\) is
	\be\quart\big(\log{1\over\l}\big)\,\sum_{x,y}(-1)^{x-y}\bigg(a^L(x)a^L(y)+a^R(x)a^R(y)\bigg)&=&\nm\\
	\quart\big(\log{1\over\l}\big)\bigg(\bigg(\sum_x(-1)^xa^L(x)\bigg)^2+\bigg(\sum_x(-1)^xa^R(x)\bigg)^2\bigg)\ .&&\eel{divpartham}

Also the kernel \(4\k\tan\half \k\) in Eq.~\eqn{hampf} diverges as \(\k\ra\pm\pi\). Keeping the divergence would make the Hamiltonian non-local, as Eq.~\eqn{divpartham} shows. We can't just argue that the largest \(\k\) values require infinite energies to excite them because they do not; according to Eq.~\eqn{contoccnumbers}, the energies of excitations at momentum \(\k\) are merely proportional to \(\k\) itself.
We therefore propose to make a smooth cut-off, replacing the divergent kernels such as \(4\k\tan\half \k\) by expressions such as
	\be (4\k\,\tan \half \k)(1-e^{-\L^2(\pi-\k)^2})\ , \eel{momcutoff}
where \(\L\) can be taken to be arbitrarily large but not infinite.

We can also say that we keep only those excitations that are orthogonal to plane waves where \(a^L(x)\) or \(a^R(x)\) are of the form \(C(-1)^x\). Also these states we refer to as \emph{edge states}.

What we now have is a lattice theory where the Hamiltonian takes the form \eqn{contoccnumbers}, where \(N^L(-\k)\) and \(N^R(\k)\) (for positive \(\k\)) count excitations in the left- and the right movers, both of which move with the same speed of light at all modes. It is this system that we can now transform to a cellular automaton. Note, that even though the lattice model may look rather contrived, it has a smooth continuum limit, which would correspond to a very dense automaton, and in theories of physics, it is  usually only the continuum limit that one can compare with actual observations, such as particles in field theories. We emphasise that, up this point, our system can be seen as  a conventional quantum model.

\subsubsection{The cellular automaton with integers in 2 dimensions\labell{CA2d}}

The cellular automaton that will be matched with the quantum model of the previous subsection, is a model defined on a square lattice with one space dimension \(x\) and one time coordinate \(t\), where both \(x\) and \(t\) are restricted to be integers. The variables are two sets of integers, one set being integer numbers \(Q(x,t)\) defined on the lattice sites, and the other being defined on the links connecting the point \((x,t)\) with \((x,t+1)\). These will be called \(P(x,\,t+\half)\), but they may sometimes be indicated as 
		\be P^+(x,t)\equiv P^-(x,t+1)\equiv P(x,t+\half)\ . \eel{Ppmdef}
The automaton obeys the following time evolution laws:
		\be Q(x,t+1)&=&Q(x,t)\ +\ P(x,t+\half)\ ; \crl{Qevolve}
	P(x,t+\half)&=&P(x,t-\half)\ +\ Q(x-1,t)-2Q(x,t)+Q(x+1,t)\ , \eel{Pevolve}
just analogously to Eqs.~\eqn{qevolve} and \eqn{pevolve}.
It is also a discrete version of a \emph{classical} field theory where \(Q(x,t)\) are the field variables and \(P(x,t)={\pa\over\pa t}Q(x,t)\) are the \emph{classical} field momenta.

Alternatively, one can write Eq.~\eqn{Pevolve} as
	\be Q(x,\,t+1)=Q(x-1,\,t)+Q(x+1,\,t)-Q(x,\,t-1)\ , \eel{discrQeq}
which, incidentally, shows that the even lattice sites evolve independently from the odd ones. Later, this will become important.

As the reader must understand by now,  Hilbert space for this system is just introduced as a tool. The basis elements of this Hilbert space are the \emph{states} \(\bigg|\,\{Q(x,0),\,P^+(x,0)\}\bigg\ket\).
If we consider \emph{templates} as superpositions of such states, we will simply \emph{define} the squares of the amplitudes to represent the probabilities. The total probability is  the length-squared of the state, which will usually be taken to be one. At this stage, superpositions mean \emph{nothing} more than this, and it is obvious that any chosen superposition, whose total length is one, may represent a reasonable set of probabilities. The basis elements all evolve in terms of a permutation operator that permutes the basis elements in accordance with the evolution equations~\eqn{Qevolve} and \eqn{Pevolve}. As a matrix in Hilbert space, this permutation operator only contains ones and zeros, and it is trivial to ascertain that statistical distributions, written as ``quantum" superpositions, evolve with the same evolution matrix.

As operators in this Hilbert space, we shall introduce shift generators that are angles, defined exactly as in Eq.~\eqn{shifts1}, but now at each point \(x_1\) at time \(t=0\), we have an operator \(\k(x_1)\) that generates an integer shift in the variable \(Q(x_1)\) and an operator \(\xi^+(x_1)\) generating a shift in the integer \(P^+(x_1)\):
	\be\hskip-20pt     e^{-i\k(x_1)}\,|\,\{Q,P^+\}\,\ket&=&|\,\{Q\,'(x),P^+(x)\}\,\ket\ ;\qquad Q\,'(x)=Q(x)+\d_{x\,x_1}\ ;\label{etaQx}\\
	\hskip-20pt 	      e^{\,i\xi^+(x_1)}\,|\,\{Q,P^+\}\,\ket&=&|\,\{Q(x),{P^+}'(x)\}\,\ket\ ;\qquad {P^+}'(x)=P^+(x)+\d_{x\,x_1}\ ; \eel{etaPx}

The time variable \(t\) is an integer, so what our evolution equations \eqn{Qevolve} and \eqn{Pevolve} generate is an operator \(U_\op\,(t)\) obeying \(U_\op\,(t_1+t_2)=U_\op\,(t_1)\,U_\op\,(t_2)\), but only for integer time steps.
In section~\ref{infinitediscr}, Eq.~\eqn{HfrUn}, a Hamiltonian \(H_\op\) was found that obeys \(U_\op\,(t)=e^{-iH_\op\,t}\), by Fourier analysis.
The problem with that Hamiltonian is that
	\bi{1.} It is not unique: one may add \(2\pi\) times any integer to any of its eigenvalues; and
	\itm{2.} It is not extensive: if two parts of a system are space-like separated, we would like the Hamiltonian to be the sum of the two separate Hamiltonians, but then it will quickly take values more than \(\pi\), whereas, by construction, the Hamiltonian~\eqn{HfrUn} will obey \(|H|\le \pi\). \ei
Thus, by adding appropriate multiples of real numbers to its eigenvalues, we would like to transform our Hamiltonian into an extensive one. The question is how to do this.

Indeed, this is one of the central questions that forced us to do the investigations described in this book; the Hamiltonian of the quantum field theory considered here is an extensive one, and also naturally bounded from below. 

At first sight,  the similarity between the automaton described by the equations \eqn{Qevolve} and \eqn{Pevolve}, and the quantum field theory of section~\eqn{qft2d} may seem to be superficial at best. Quantum physicists will insist that the quantum theory is fundamentally different.

However, we claim that there is an \emph{exact mapping} between the basis elements of the quantised field theory of subsection~\ref{qft2d} and the states of the cellular automaton (again, with an exception for possible edge states). We shall show this by concentrating on the left-movers and the right-movers separately.

Our procedure will  force us first to compare the left-movers and the right-movers separately in both theories. The automaton equations \eqn{Qevolve} and \eqn{Pevolve} ensure that, if we start with integers at \(t=0\) and \(t=\half\), all entries at later times will be integers as well. So this is a discrete automaton. We now introduce the combinations \(A^L(x,t)\) and \(A^R(x,t)\) as follows (all these capital letter variables take integer values only):
	\be A^L(x,t)&=&P^+(x,t)+Q(x,t)-Q(x-1,t)\ ;\crl{ALdef}
	     A^R(x,t)&=&P^+(x,t)+Q(x,t)-Q(x+1,t)\ , \eel{ARdef}  
and we derive
	\be A^L(x,t+1)&=&P^+(x,t)+Q(x-1,\,t+1)-2Q(x,\,t+1)+Q(x-1,\,t+1)\qquad \nm\\
		&&\quad +\,Q(x,\,t+1)-Q(x-1,\,t+1)	\quad =	\nm\\
		&& P^+(x,t)+Q(x-1,\,t+1)-Q(x,\,t+1) \quad = \nm\\
		&&P^+(x,t)+Q(x-1,\,t)+P^+(x-1,\,t)-Q(x,t)-P^+(x,\,t) \ = \nm\\
		&&P^+(x-1,\,t)+Q(x-1,\,t)-Q(x,t)\ =\ A^L(x-1,t)\ . \quad \eel{CAleft}
So, we have 
	\be A^L(x-1,\,t+1)=A^L(x,t)=A^L(x+t)\ ;\qquad A^R(x,t)=A^R(x-t)\ , \eel{CAleftright}
which shows that \(A^L\) is a left-mover and \(A^R\) is a right mover. All this is completely analogous to Eqs.~\eqn{aL} and \eqn{aR}.

\subsubsection{The mapping between the boson theory and the automaton\labell{bosonCAmapping}}
	The states of the quantised field theory on the lattice were generated by the left- and right moving operators \(a^L(x+t)\) and \(a^R(x-t)\), where \(x\) and \(t\) are integers, but \(a^L\) and \(a^R\) have continua of eigenvalues, and they obey the commutation rules \eqn{aLaLlatticecomm} and \eqn{aRaRlatticecomm}:
	 \be[a^L,a^R]=0\ ;&& [a^L(x),\,a^L(y)]=\pm\, i\ \hbox{ if }\ y=x\pm 1\ ;\quad\hbox{else }\ 0\ ; \crl{aLaLlatcomm}
				&& [a^R(x),\,a^R(y)]=\mp\,i\ \hbox{ if }\ y=x\pm 1\ ;\quad\hbox{else }\ 0\ . \eel{aRaRlatcomm}
In contrast, the automaton has integer variables \(A^L(x+t)\) and \(A^R(x-t)\), Eqs,~\eqn{ALdef} and \eqn{ARdef}. They live on the same space-time lattice, but they are integers, and they commute.

Now, PQ theory suggests what we have to do. The shift generators \(\k(x_1)\) and \(\xi(x_1)\) (Eqs.~\ref{etaQx} and \ref{etaPx}) can be combined to define shift operators \(\eta^L(x_1)\) and \(\eta^R(x_1)\) for the integers \(A^L(x_1,t)\) and \(A^R(x_1,t)\). Define
	\be e^{\,i\eta^L(x_1)}|\{A^L,A^R\}\ket=|\{{A^L}',A^R\}\ket\ ,&& {A^L}'(x)=A^L(x)+\d_{x,x_1}\ , \label {etaLdef}\\
		e^{\,i\eta^R(x_1)}|\{A^L,A^R\}\ket=|\{A^L,{A^R}'\}\ket\ ,&& {A^R}'(x)=A^R(x)+\d_{x,x_1}\ . \eel{etaRdef}
These then have to obey the following equations:
	\be\xi(x)&=&\eta^L(x)+\eta^R(x)\ ; \nm\\
		-\k(x)&=&\eta^L(x)+\eta^R(x)-\eta^L(x+1)-\eta^R(x-1)\ . \eel{xikappax}
The first of these tells us that, according to Eqs.~\eqn{ALdef} and \eqn{ARdef}, raising \(P^+(x)\) by one unit, while keeping all others fixed, implies raising this combination of \(A^L\) and \(A^R\). The second tells us what the effect is of raising only \(Q(x)\) by one unit while keeping the others fixed. Of course, the additions and subtractions in Eqs.~\eqn{xikappax} are \textit{modulo} \(2\pi\).

Inverting Eqs.~\eqn{xikappax} leads to
	\be \eta^L(x+1)-\eta^L(x-1)   &=&\xi(x)+\k(x)-\xi(x-1)\ , \nm\\
		\eta^R(x-1)-\eta^R(x+1) &=&\xi(x)+\k(x)-\xi(x+1)\ . \eel{etaLRx}
These are difference equations whose solutions involve infinite sums with a boundary assumption. This has no further consequences; we take the theory to be defined by the operators \(\eta^{L,R}(x)\), not the \(\xi(x)\) and \(\k(x)\). As we have encountered many times before, there are some non-local modes of measure zero, \(\eta^L(x+2n)=\mathrm{constant}\), and \(\eta^R(x+2n)=\mathrm{constant}\).

What we have learned from the PQ theory, is that, in a sector of Hilbert space that is orthogonal to the edge state, an integer variable \(A\), and its shift operator \(\eta\), obey the commutation rules
	\be A\,e^{\,i\eta}=e^{\,i\eta}(A+1)\ ;\quad [\eta,\,A]=i\ . \eel{etaAcomm}
This gives us the possibility to generate operators that obey the commutation rules \eqn{aLaLlatcomm} and \eqn{aRaRlatcomm} of the quantum field theory:
	\be a^L(x)&\qu&\sqrt{2\pi}\,A^L(x)-\frac  1{\sqrt{2\pi}}\,\eta^L(x-1)\ ; \label{aAetaL}\\
		 a^R(x)&\qu&\sqrt{2\pi}\,A^R(x)-\frac  1{\sqrt{2\pi}}\,\eta^R(x+1)\ . \eel{aAetaR}
The factors \(\sqrt{2\pi}\) are essential here. They ensure that the spectrum is not larger or smaller than the real line, that is, without gaps or overlaps (degeneracies).

The procedure can be improved. In the expressions \eqn{aAetaL} and \eqn{aAetaR}, we have an edge state whenever \(\eta^{L,R}\) take on the values \(\pm\pi\). This is an unwanted situation: these edge states make all wave functions discontinuous on the points \(a^{L,R}(x)=\sqrt{2\pi}\,(N(x)+\half)\). Fortunately, we can cancel most of these edge states by repeating more precisely the procedure explained in our treatment of PQ theory: these states were due to vortices in two dimensional planes of the tori spanned by the \(\eta\) variables. Let us transform, by means of standard Fourier transforming the \(A\) lattices to the \(\eta\) circles, so that we get a multi-dimensional space of circles -- one circle at every point \(x\). 

As in the simple PQ theory (see Eqs.~\eqn{qopkappaxi} and \eqn{popkappaxi}), we can introduce a phase function \(\vv(\{\eta^L\})\) and a \(\vv(\{\eta^R\})\), so that  Eqs.~\eqn{aAetaL} and \eqn{aAetaR} can be replaced with
	\be a^L(x)&=&-\,i\sqrt{2\pi}\frac{\pa}{\pa\eta^L(x)}+ {\sqrt{2\pi}}\Big(\frac{\pa}{\pa\eta^L(x)}\vv(\{\eta^L\})\Big)
			-\frac 1{\sqrt{2\pi}}\,\eta^L(x-1)\ , \label{aLphiL}\\
 		a^R(x)&=&-\,i\sqrt{2\pi}\frac{\pa}{\pa\eta^R(x)}+ {\sqrt{2\pi}}\Big(\frac{\pa}{\pa\eta^R(x)}\vv(\{\eta^R\})\Big)
			-\frac 1{\sqrt{2\pi}}\,\eta^R(x+1)\ , \eel{aRphiR}
where \(\vv(\{\eta(x)\})\) is a phase function with the properties

	\be \vv(\{\eta^L(x)+2\pi\d_{x,x_1}\}&=&\vv(\{\eta^L(x)\})+\eta^L(x_1+1)  \ ;\label{phiLprop}\\
		\vv(\{\eta^R(x)+2\pi\d_{x,x_1}\}&=&\vv(\{\eta^R(x)\})+\eta^R(x_1-1)\ .\eel{phiRprop}
Now, as one can easily check, the operators \(a^{L,R}(x)\) are exactly periodic for all \(\eta\) variables, just as we had in section~\ref{PQ}.

A phase function with exactly these properties can be written down. We start with the elementary function \(\f(\k,\,\xi)\) derived in section~\ref{PQmatrix}, Eq.~\eqn{rphidef}, having the properties 
	\be \f(\k,\xi+2\pi)=\f(\k,\xi)+\k\  ;&&\f(\k+2\pi,\xi)=\f(\k,\xi)\ ;  \crl{phiperiods2d}
		 \f(\k,\xi) = -\f(-\k,\xi) =  -\f(\k,-\xi)\ ;&& \f(\k,\,\xi)+\f(\xi,\,\k)=\k\xi/2\pi\ .  \eel{phiproperties2d}
The function obeying Eqs.~\eqn{phiLprop} and \eqn{phiRprop} is now not difficult to construct:
   	\be \vv(\{\eta^L\})&=&\sum_x\f\bigg(\eta^L(x+1),\,\eta^L(x)\bigg)\ ;	\nm\\
		 \vv(\{\eta^R\})&=&\sum_x\f\bigg(\eta^R(x-1),\,\eta^R(x)\bigg)\ ,\eel{vhidef1}
and as was derived in section~\ref{PQmatrix}, a phase function with these properties can be given as the phase of an elliptic function, 
	\be  r(\k,\xi) e^{i\f(\k,\xi)}\equiv\sum_{N=-\infty}^\infty e^{-\pi(N-\fract{\xi}{2\pi})^2-iN\k}\ ,&&\eel{vhidef2} 
where \(r\) and \(\f\) are both real functions of \(\k\) and \(\xi\).

We still have edge states, but now these only sit at the corners where two consecutive \(\eta\) variables take the values \(\pm\pi\). This is where the phase function \(\f\), and therefore also \(\vv\), become singular. We suspect that we can simply ignore them.

We then reach an important conclusion. The states of the cellular automaton can be used as a basis for the description of the quantum field theory. These models are equivalent. This is an astounding result. For generations we have been told by our physics teachers, and we explained to our students, that quantum theories are fundamentally different from classical theories. No-one should dare to compare a simple computer model such as a cellular automaton based on the integers, with a fully quantised field theory. Yet here we find a quantum field system and an automaton that are based on states that neatly correspond to each other, they evolve identically. If we describe some probabilistic distribution of possible automaton states using Hilbert space as a mathematical device, we can use \emph{any} wave function, certainly also waves in which the particles are \emph{entangled}, and yet these states evolve exactly the same way.

Physically, using $19^\th$ century logic, this should have been easy to understand: when quantising a classical field theory, we get energy packets that are quantised and behave as particles, but exactly the same are generated in a cellular automaton based on the integers; these behave as particles as well. Why shouldn't there be a mapping?

Of course one can, and should, be skeptical. Our field theory was not only constructed without interactions and without masses, but also the wave function was devised in such a way that  it cannot spread, so it should not come as a surprise that no problems are encountered with interference effects, so yes, all we have is a primitive model, not very representative for the real world. Or is this just a beginning?

\underline{Note:} being exactly integrable, the model discussed in this section has infinitely many conservation laws. For instance, one may remark that the equation of motion \eqn{discrQeq} does not mix even sites with odd sites of the lattice; similar equations select out sub-lattices with \(x+t=4k\) and \(x\) and \(t\) even, from other sub-lattices.

\subsubsection{An alternative ontological basis: the compactified model\labell{compact}}

In the above chapters and sections of this chapter, we have seen various examples of deterministic models that can be mapped onto quantum models and back. The reader may have noticed that, in many cases, these mappings are not unique. Modifying the choices of the constant energy shifts \(\d E_i\) in the composite cogwheel model, section~\ref{generalfinite}, we saw that many apparently different quantum theories can be mapped onto the same set of cogwheels, although there, the \(\d E_i\) could have been regarded as various chemical potentials, having no effect on the evolution law. In our \(PQ\) theory, one is free to add fractional constants to \(Q\) and \(P\), thus modifying the mapping somewhat. Here, the effect of this would be that the ontological states obtained from one mapping do not quite coincide with those of the other, they are superpositions, and this is an example of the occurrence of sets of ontological states that are not equivalent, but all equally legal. 

The emergence of inequivalent choices of an ontological basis is particularly evident if the quantum system in question has symmetry groups that are larger than those of the ontological system. If the ontological system is based on a lattice, it can only have some of the discrete lattice groups as its symmetries, whereas the quantum system, based on real coordinates, can have continuous symmetry groups such as the rotation, translation and Lorentz group. Performing a symmetry transformation that is not a symmetry of the ontological model then leads to a new set of ontological states (or ``wave functions") that are superpositions of the other states. Only one of these sets will be the ``real" ontological states. For our theory, and in particular the cellular automaton interpretation, chapters~\hbox{\ref{CAshort} and \ref{CAdetail}}, this has no further consequences, except that it will be almost impossible to single out the ``true" ontological basis as opposed to the apparent ones, obtained after quantum symmetry transformations.

In this subsection, we point out that even more can happen. Two (or perhaps more) systems of ontological basis elements may exist that are entirely different from one another. This is the case for the model of the previous subsection, which handles the mathematics of non-interacting massless bosons in \(1+1\) dimensions. We argued that an ontological basis is spanned by all states where the field \(q(x,t)\) is replaced by integers \(Q(x,t)\). A lattice in \(x,t\) was introduced, but this was a temporary lattice; we could send the mesh size to zero in the end.

In subsection~\ref{CA2d}, we introduced the integers \(A^L(x)\) and \(A^R(x)\), which are the integer-valued left movers and right movers; they span an ontological basis. Equivalently, one could have taken the integers \(Q(x,t)\) and \(P^+(x,t)\) at a given time \(t\), but this is just a reformulation of the same ontological system.

But why not take the continuous degrees of freedom \(\eta^L(x)\) and \(\eta^R(x)\)  (or equivalently, \(\xi(x,t)\) and \(\k(x,t)\))? At each \(x\), these variables take values between \(-\pi\) and \(\pi\). Since they are also left- and right movers, their evolution law is exactly as deterministic as that of the integers \(A^L\) and \(A^R\):
	\be{\pa\over\pa t}\eta^L(x,t)= {\pa\over\pa x}\eta^L(x,t)\ ,\qquad {\pa\over\pa t}\eta^R(x,t)=- {\pa\over\pa x}\eta^R(x,t)\ ,\eel{etaLRleftright}
while all \(\eta\)'s commute.

Actually, for the \(\eta\) fields, it is much easier now to regard the continuum limit for the space--time lattice. The \emph{quantum operators} \(a^{L,R}\) are still given by Eqs,~\eqn{aLphiL} and \eqn{aRphiR}. There is a singularity when two consecutive \(\eta\) fields take the values \(\pm\pi\), but if they don't take such values  at \(t=t_0\), they never reach those points at other times.

There exists a somewhat superior way to rephrase the mapping by making use of the fact that the \(\eta\) fields are continuous, so that we can do away with, or hide, the lattice. This is shown in more detail in subsection~\ref{detstr}, where these ideas are applied in string theory.

What we conclude from this subsection is that \emph{both} our quantum model of bosons \emph{and} the model of left and right moving integers are mathematically equivalent to a classical theory of scalar fields where the values are only defined \textit{modulo} \(2\pi\). From the ontological point of view, this new model is entirely different from both previous models. 

Because the variables of the classical model only take values on the circle, we call the classical model a \emph{compactified} classical field theory. At other places in this book, the author warned that classical, continuous theories may not be the best ontological systems to assume for describing Nature, because they tend to be \emph{chaotic}: as time continues, more and more decimal places of the real numbers describing the initial state will become relevant, and this appears to involve unbounded sets of digital data. To our present continuous field theory in \(1+1\) dimensions, this objection does not apply, because there is no chaos; the theory is entirely integrable. Of course, in more complete models of the real world we do not expect integrability, so there this objection against continuum models does apply.

\subsubsection{The quantum ground state\labell{ground} }
Nevertheless, the mappings we found are delicate ones, and not always easy to implement. For instance, one would like to ask which solution of the cellular automaton, or the compactified field theory,  corresponds to the quantum ground state of the quantised field theory. First, we answer the question: if you have the ground state, how do we add a single particle to it?

Now this should be easy. We have the creation and annihilation operators for left movers and right movers, which are the Fourier transforms of the operators \(a^{L,R}(x)\) of Eqs.~\eqn{aLphiL} and \eqn{aRphiR}. When the Fourier parameter, the lattice momentum  \(\k\), is in the interval \(-\pi<\k<0\), then \(a^L(\k)\) is an annihilation operator and \(a^R(\k)\) is a creation operator. When \(0<\k<\pi\), this is the other way around.
Since nothing can be annihilated from the vacuum state, the annihilation operators vanish, so \(a^{L,R}(x)\) acting on the vacuum can only give a superposition of one--particle states.

To see how a single left-moving particle is added to a classical cellular automaton state, we consider the expression \eqn{aLphiL} for \(a^L(x)\), acting on the left-mover's coordinate \(x+t\ra x\) when \(t=0\). The operator \(\pa/\pa\eta^L(x)\) multiplies the amplitude for the state with \(iA^L(x)\); the other operators in Eq.~\eqn{aLphiL} are just functions of \(\eta^L\) at the point \(x\) and the point \(x-1\). Fourier transforming these functions gives us the operators \(e^{\pm Ni\eta^L}\) multiplied with the Fourier coefficients found, acting on our original state. According to Eq.~\eqn{etaLdef}, this means that, at the two locations \(x\) and \(x-1\), we add or subtract \(N\) units to the number \(A^L\) there, and then we multiply the new state with the appropriate Fourier coefficient. Since the functions in question are bounded, we expect the Fourier expansion to converge reasonably well, so we can regard the above as being a reasonable answer to the question how to add a particle. Of course, our explicit construction added a particle at the point \(x\). Fourier transforming it, gives us a particle with momentum \(-\k\) and energy \(\k\).

In the compactified field model, the action of the operators \eqn{aLphiL} and \eqn{aRphiR} is straightforward; we find the states with one or more particles added, provided that the wave function is differentiable. The \emph{ontological} wave functions are not differentiable -- they are delta peaks, so particles can only be added as templates, which are to be regarded as probabilistic distributions of ontological states.

Finding the vacuum state, \emph{i.e.} the quantum ground state itself, is harder. It is that particular superposition of ontological states from which no particles can be removed. Selecting out the annihilation parts of the operators \(a^{L,R}(x)\) means that we have to apply the projection operator \(\mathcal P^-\) on the function \(a^L(x)\) and \(\mathcal P^+\) on \(a^R(x)\), where the projection operators \(\mathcal P^\pm\) are given by
	\be\mathcal P^\pm a(x)=\sum_{x'}\mathcal P^\pm(x-x')a(x')\ , \ee  
where the functions \(P^\pm(y)\) are defined by
	 \be 	 \mathcal P^\pm(y)&=&{1\over 2\pi}\int_0^\pi\dd\k\,
		e^{\pm iy\k}={\pm i\over\pi y}\quad \hbox{if }\ y\hbox{ odd ,}\qquad \half\d_{y,0}\ \hbox{ if }\ y\hbox{ even .}\eel{Fourierprojops}
		
The state for which the operators \(\mathcal P^+a^L(x)\) and \(\mathcal P^-a^R(x)\) vanish at all \(x\) is the quantum ground state. It is a superposition of all cellular automaton states. 

Note that the theory has a Goldstone mode\,\cite{Gold-1961}, which means that an excitation in which all fields \(q(x,t)\), or all automaton variables \(Q(x,t)\), get the same constant \(Q_0\) added to them, does not affect the total energy. This is an artefact of this particular model\fn{Paradoxically, models in two space-time dimensions are known not to allow for Goldstone modes; this theorem\,\cite{Coleman-1973}, however, only applies when there are interactions. Ours is a free particle model.}.  Note also that the projection operators \(\mathcal P^\pm(x)\) are not well-defined on \(x\)- independent fields; for these fields, the vacuum is ill-defined.

\subsecl{Bosonic theories in higher dimensions?}{highd}

At first sight, it seems that the model described in previous sections may perhaps be generalised to higher dimensions. In this section, we begin with setting up a scheme that should serve as an approach towards handling bosons in a multiply dimensional space as a cellular automaton. Right-away, we emphasise that  a mapping in the same spirit as what was achieved in previous sections and chapters will not be achieved. Nevertheless, we will exhibit here the initial mathematical steps, which start out as deceptively beautiful. Then, we will exhibit, with as much clarity as we can, what, in this case, the obstacles are, and why we call this a failure in the end. As it seems today, what we have here is a loose end, but it could also be the beginning of a theory where, as yet, we were forced to stop half-way.

\subsubsection{Instability\labell{instability}}

We would have been happy with either a discretised automaton or a compactified classical field theory, and for the time being, we keep both options open.

Take the number of space-like dimensions to be  \(d\), and suppose that we replace Eqs.~\eqn{Qevolve} and \eqn{Pevolve} by
	\be\hskip-25pt  Q(\vec x,\,t+1)&=&Q(\vec x,t)+P(\vec x,\,t+\half)\ ;\crl{QDdim}
	\hskip-25pt	P(\vec x,\,t+\half)&=&P(\vec x,\,t-\half)+\sum_{i=1}^d\Big(Q(\vec x-\^e_i,\,t)-2Q(\vec x,\,t)+Q(\vec x+\^e_i,\,t)\Big)\ , \eel{PDdim}
where \(\^e_i\) are unit vectors in the \(i^{\,\th}\) direction in space.	

Next, consider a given time \(t\). We will need to localise operators in time, and can do this only by choosing the time at which an operator acts such that, at that particular time, the effect of the operator is as concise as is possible. This was why, for the \(P\) operators, in Eqs.~\eqn{QDdim} and \eqn{PDdim}, we chose to indicate time as \(t\pm\half\) (where \(t\) is integer).  Let 	\(\k^{\op\,}(\vec x,\,t+\half)\) be the generator of a shift of \(Q(\vec x,t) \) and 
the same shift in \hbox{\(Q(\vec x,\,t+1)\)} (so that \(P(\vec x,\,t+\half)\) does not shift), while \(\xi^{\op\,}(\vec x,t)\) generates identical, negative shifts of 
\(P(\vec x,\,t+\half)\)
and of \(P(\vec x,\,t-\half)\), without shifting \(Q(\vec x,t)\) at the same \(t\), and with the signs both as dictated in Eqs~\eqn{etaQx} and \eqn{etaPx}. Surprisingly, one finds that these operators obey the same equations~\eqn{QDdim} and \eqn{PDdim}: the operation
	\be  \k^{\op\,}(\vec x,\,t+\half) && \hbox{has the same effect as} \nn
	&&\k^{\op\,}(\vec x,\,t+\fract 32)-\sum_{i=1}^{2d}\Big( \xi^{\op\,}(\vec x+\^e_i,\,t+1)-\xi^{\op\,}(\vec x,\,t+1)\Big) \ ,\quad{ } \nm\\
	\hbox{and }\ \xi^{\op\,}(\vec x,t) && \hbox{has the same effect as }\nn
	&& \xi^{\op\,}(\vec x,\,t+1)-\k^{\op\,}(\vec x,\,t+\half)  \eel{kappaxieqs}
(where the sum is the same as in Eq.~\eqn{PDdim} but in a more compact notation) and therefore, 
	\be  \xi^{\op\,}(\vec x,\,t+1)&=&\xi^{\op\,}(\vec x,t)+\k^{\op\,}(\vec x,\,t+\half)\ ;\crl{xiddim}
		\k^{\op\,}(\vec x,\,t+\half)&=&\k^{\op\,}(\vec x,\,t-\half)\ +\cr
		&+&\sum_{i=1}^d\Big(\xi^{\op\,}(\vec x-\^e_i,\,t)
	-2\xi^{\op\,}(\vec x,\,t)+\xi^{\op\,}(\vec x+\^e_i,\,t)\Big)\ , \eel{kappaddim}
and, noting that \(Q\) and \(P\) are integers, while \(\k^{\op}\) and \(\xi^{\op}\) are confined to the interval \([0,\,2\pi)\),  we conclude that again the same equations are obeyed by the real number operators
	\be q^\op\,(\vec x,t)&\qu&Q(\vec x,t)+\fract 1{2\pi}\xi^{\op\,}(\vec x,t)\ , \ \hbox{ and}\nm\\
		p^\op\,(\vec x,\,t+\half)&\qu&2\pi P(\vec x,\,t+\half)+\k^{\op\,} (\vec x,\,t+\half)\ . \eel{ddimpq}
		
	These operators, however, do not obey the correct commutation rules. There even appears to be a factor 2 wrong, if we would insert the equations \([Q(\vec x),\k^{\op\,}(\vec x')]\qu i\d_{\vec x,\vec x'}\), \([\xi^{\op\,}(\vec x),P(\vec x')]\qu i\d_{\vec x,\vec x'}\). Of course, the reason for this failure is that we have the edge states, and we have not yet restored the correct boundary conditions in \(\xi,\k\) space by inserting the phase factors \(\vv\), as in Eqs.~\eqn{aLphiL}, \eqn{aRphiR}, or \(\f\) in \eqn{qopkappaxi}, \eqn{popkappaxi}. This is where our difficulties begin. These phase factors also aught to obey the correct field equations, and this seems to be impossible to realise.
	
	In fact, there is an other difficulty with the equations of motion, \eqn{QDdim}, \eqn{PDdim}: they are unstable. It is true that, in the continuum limit, these equations generate the usual field equations for smooth functions \(q(\vec x,t)\) and \(p(\vec x,t)\), but we now have lattice equations. Fourier transforming the equations in the space variables \(\vec x\) and time \(t\), one finds
	\be -2i(\sin\half \w) q(\vec k,\w)&=&p(\vec k, \w)\ , \nm\\
	   -2i(\sin\half \w) p(\vec k,\w)&=&\sum_{i=1}^d2(\cos k_i-1)\,q(\vec k,\w)\ . \eel{ddimFourier}
This gives the dispersion relation
	\be 4\sin^2\half\w=2(1-\cos\w)=\sum_{i=1}^d2(1-\cos k_i)\ .\eel{latticeddimdisp}
In one space-like dimension, this just means that \(\w=\pm k\), which would be fine, but if \(d>1\), and
 \(k_i\) take on values close to \(\pm\pi\), the r.h.s. of this equation exceeds the limit value 4, the cosine becomes an hyperbolic cosine, and thus we find modes that oscillate out of control, exponentially with time.
 
 To mitigate this problem, we would somehow have to constrain the momenta \(k_i\) towards small values only, but, both in a cellular automation where all variables are integers, and in the compactified field model, where we will need to respect the intervals \((-\pi,\,\pi)\), this is hard to accomplish. Note that we used Fourier transforms on functions such as \(Q\) and \(P\) in Eqs.~\eqn{QDdim} and \eqn{PDdim} that take integer values. In itself, that procedure is fine, but it shows the existence of exponentially exploding solutions. These solutions can also be attributed to the non-existence of an energy function that is conserved and bounded from below. Such an energy function does exist in one dimension:
 	\be\hskip-35pt  E&=&\half\sum_xP^+(x,t)^2\ +\nm\\
		&&\half\sum_x\Big(Q(x,t)+P^+(x,t)\Big)\Big(2Q(x,t)-Q(x-1,t)-Q(x+1,t)\Big)\ , \eel{energy1dx}
or in momentum space, after rewriting the second term as the difference of two squares,
	\be E=\half\int_{-\infty}^\infty\dd k\Big((\cos\half k)^2\,|P^+(k)|^2+4(\sin\half k)^2\,|Q(k)+\half P^+(k)|^2\Big)\ . \eel{energy1dk}
 Up to a factor \(\sin k\,/\,k\), this is the Hamiltonian \eqn{hampf} (Since the equations of motion at different \(k\) values are independent, conservation of one of these Hamiltonians implies conservation of the other).
 
 In higher dimensions, models of this sort cannot have a non-negative, conserved energy function, and so these will be unstable.

\subsubsection{Abstract formalism for the multidimensional harmonic oscillator\labell{oscilform}}

Our \(PQ\) procedure for coupled harmonic oscillators can be formalised more succinctly and elegantly. Let us write a time-reversible harmonic model with integer degrees of freedom as follows. In stead of Eqs.~\eqn{QDdim} and \eqn{PDdim} we write

	\be Q_i(t+1)&=&Q_i(t)+\sum_jT_{ij}P_j(t+\half)\ ; \crl{Qgeneral}
		P_i(t+\half)&=&P_i(t-\half)-\sum_jV_{ij}Q_j(t)\ . \eel{Pgeneral}
Here, \(t\) is an integer-valued time variable. It is very important that both matrices \(T\) and \(V\) are real and symmetric:
	\be T_{ij}=T_{ji}\ ;\qquad V_{ij}=V_{ji}\ . \eel{TVsymm}
\(T_{ij}\) would often, but not always, be taken to be the Kronecker delta \(\d_{ij}\), and \(V_{ij}\) would be the second derivative of a potential function, here being constant coefficients. Since we want \(Q_i\) and \(P_i\) both to remain integer-valued, the coefficients \(T_{ij}\) and \(V_{ij}\) will also be taken to be integer-valued. In principle, any integer-valued matrix would do; in practice, we will find severe restrictions.

Henceforth, we shall omit the summation symbol \(\sum_j\), as its presence can be taken to be implied by summation convention: every repeated index is summed over.
	
When we define the translation generators for \(Q_i\) and \(P_i\), we find that, in a Heisenberg picture, it is best to use an operator \(\k^{\op\,}_i(t+\half)\) to add one unit to \(Q_i(t)\) while all other integers \(Q_j(t)\)  with \(j\ne i\) and all \(P_j(t+\half)\) are kept fixed. Note that, according to the evolution equation \eqn{Qgeneral}, this also adds one unit to \(Q_i(t+1)\) while all other \(Q_j(t+1)\) are kept fixed as well, so that we have symmetry around the time value \(t+\half\). Similarly, we define an operator \(\xi^{\op\,}_i(t)\) that shifts the value of both \(P_i(t-\half)\) and \(P_i(t+\half)\), 
while all other \(Q\) and \(P\) operators at \(t-\half\) and at \(t+\half\) are kept unaffected; all this was also explained in the text between Eqs.~\eqn{PDdim} and \eqn{kappaxieqs}.
	
So, we define the action by operators \(\k^{\op\,}_i\) and	 \(\xi^{\op\,}_i\) by \pagebreak[2]
	\be   e^{-i\k^{\op\,}_i(t+\halfje)}|\{Q_j(t),\,P_j(t+\half)\}\ket &=& |\{Q_j'(t),\,P_j(t+\half)\}\ket\qquad\ \hbox{ or}\cr
	\hskip-25pt	   e^{-i\k^{\op\,}_i(t+\halfje)}|\{Q_j(t+1),\,P_j(t+\half)\}\ket^{ \vphantom{\big|}}&=& |\{Q_j'(t+1),\,P_j(t+\half)\}\ket\quad\hbox{with}\crl{kappait1}
				Q_j'(t)=Q_j(t)+\d_{ji}\ ,&& Q_j'(t+1)=q_j(t+1)+\d_{ji}\ ;\nm\ee
	\be   e^{i\xi^{\op\,}_i(t)}|\{Q_j(t),\,P_j(t+\half)\}\ket &=& |\{Q_j(t),\,P_j'(t+\half)\}\ket\qquad\hbox{or}\cr
		  e^{i\xi^{\op\,}_i(t)}|\{Q_j(t),\,P_j(t-\half)\}\ket &=& |\{Q_j(t),\,P_j'(t-\half)\}\ket\qquad\hbox{with}\crl{xiit1}
				&&P_j'(t\pm\half)=P_j(t\pm\half)+\d_{ji}\  . \nm\ee
	
The operators \(\xi^{\op\,}_i(t)\) and \(\k^{\op\,}_i(t+\half)\) then obey exactly the same equations as \(Q_i(t)\) and \(P_i(t+\half)\), as given in Eqs~\eqn{Qgeneral} and \eqn{Pgeneral}:
	\be \xi^{\op\,}_i(t+1)&=&\xi^{\op\,}_i(t)+ T_{ij}\,\k^{\op\,}_j(t+\half)\ ; \crl{xigeneral}
		\k^{\op\,}_i(t+\half)&=&\k^{\op\,}_i(t-\half)- V_{ij}\,\xi^{\op\,}_j(t)\ . \eel{kappageneral}
		
The stability question can be investigated by writing down the conserved energy function. After careful inspection, we find that this energy function can be defined at integer times:
	\be H_1(t)=\half T_{ij}\,P_i(t+\half)\,P_j(t-\half)+\half V_{ij}\,Q_i(t)\,Q_j(t)\ , \eel{ham1int}
and at half-odd integer times:
	\be H_2(t+\half)=\half T_{ij}\,P_i(t+\half)\,P_j(t+\half)+\half V_{ij}\,Q_i(t)\,Q_j(t+1)\ . \eel{ham2odd}
Note that \(H_1\) contains a pure square of the \(Q\) fields but a mixed product of the \(P\) fields while \(H_2\) has that the other way around.
It is not difficult to check that \(H_1(t)=H_2(t+\half)\):
	\be H_2-H_1=-\half T_{ij}\,P_i(t+\half)\,V_{jk}\,Q_k(t)+\half V_{ij}\,Q_i(t)\,T_{jk}\,P_k(t+\half)=0\ . \eel{H1equH2}
Similarly, we find that the Hamiltonian stays the same at all times. Thus, we have a conserved energy, and that could guarantee stability of the evolution equations.

However, we still need to check whether this energy function is indeed non-negative. This we do by rewriting it as the sum of two squares.
In \(H_1\), we write the momentum part (kinetic energy) as
	\be \half T_{ij}\Big(P_i(t+\half)+\half V_{ik}\,Q_k(t)\Big)\Big(P_j(t+\half)+\half V_{j\ell}\,Q_\ell(t)\Big)&&\nm\\
	-\ \fract18 T_{ij}V_{ik}\,V_{j\ell}\,Q_k(t)\,Q_\ell(t)&,&\eel{H1twosquares}
so that we get at integer times (in short-hand notation)
	\be H_1=\vec Q(\half V-\fract18 V\,T\,V)\vec Q+\half (\vec P^++\half\vec Q\,V)\,T\,(\vec P^++\half V\,\vec Q)\ , \eel{ham2squares}
and at half-odd integer times:
	\be H_2=\vec P(\half T-\fract18 T\,V\,T)\vec P+\half (\vec Q^-+\half\vec P\,T)\,V\,(\vec Q^-+\half T\,\vec P)\ , \eel{ham2squaresalt}
where \(P^+(t)\) stands for \(P(t+\half)\), and \(Q_-(t+\half)\) stands for \(Q(t)\)\ .

The expression \eqn{ham2odd} for \(H_2\) was the one used in Eq.~\eqn{energy1dx} above. It was turned into Eq.~\eqn{ham2squaresalt} in the next expression, Eq.~\eqn{energy1dk}.
	
	Stability now requires that the coefficients for these squares are all non-negative. This has to be checked for the first term in Eq.~\eqn{ham2squares} and in \eqn{ham2squaresalt}. If \(V\) and/or \(T\) have one or several vanishing eigenvectors, this does not seem to generate real problems, and we replace these by infinitesimal numbers \(\e>0\). Then, we find that, on the one hand one must demand
	\be \bra \,T\,\ket>0\ ,\qquad \bra\, V\,\ket>0\ ; \eel{VTpos}
while on the other had, by multiplying left and right by \(V^{-1}\) and \(T^{-1}\):
	\be \bra\, 4V^{-1}-T\,\ket\ge 0\ ,\qquad \bra\, 4T^{-1}-V\,\ket\ge 0\ . \eel{VTupperbound}
	
Unfortunately, there are not so many integer-valued matrices \(V\) and \(T\) with these properties. Limiting ourselves momentarily to the case \(T_{ij}=\d_{ij}\), we find that the matrix \(V_{ij}\) can have at most a series of 2's on its diagonal and sequences of \(\pm1\) 's on both sides of the diagonal. Or, the model displayed above in Eq.~\eqn{QDdim} and \eqn{PDdim}, on a lattice with periodicity \(N\), is the most general multi-oscillator model that can be kept stable by a nonnegative energy function.

If we want more general, less trivial models, we have to search for a more advanced discrete Hamiltonian formalism (see Sect.~\ref{Hamiltonform}).

If it were not for this stability problem, we could have continued to construct real-valued operators \(q^{\op\,}_i\) and \(p^{\op\,}_i\) by combining \(Q_i\) with \(\xi^{\op\,}_i\) and \(P_i\) with \(\k^{\op\,}_i\). The operators \(e^{iQ\xi^{\op\,}_i}\) and \(e^{-iP\k^{\op\,}_i}\) give us the states \(|\{Q_i,\,P_i\}\ket\) from the `zero-state' \(|\{0,0\}\ket\). This means that only one wave function remains to be calculated by some other means, after which all functions can be mapped by using the operators
\(e^{ia_ip_i}\) and \(e^{ib_jq_j}\). But since we cannot obtain stable models in more than 1 space-dimensions, this procedure is as yet of limited value. It so happens, however, that the one-dimensional model is yet going to play a very important role in this work: (super) string theory, see the next section.

\subsecl{(Super)strings}{sustr}

So-far, most of our models represented non-interacting massless particles in a limited number of space dimensions. Readers who are still convinced that quantum mechanical systems will never be explained in terms of classical underlying models, will not be shocked by what they have read until now. After all, one cannot do Gedanken experiments with particles that do not interact, and anyway, massless particles in one spacial dimension do not exhibit any dispersion, so here especially, interference experiments would be difficult to imagine. This next chapter however might make him/her frown a bit: we argue that the bulk region of the (super)string equations can be mapped onto a deterministic, ontological theory. The reason for this can be traced to the fact that string theory, in a flat background, is essentially just a one-space, one-time massless quantum field theory, without interactions, exactly as was described in previous sections.

As yet, however, our (super)strings will not interact, so the string solutions will act as non-interacting particles; for theories with interactions, go to chapters~\ref{miss}, \ref{Hamiltonform}, and \ref{CAdetail}.

Superstring theory started off as an apparently rather esoteric and formal approach to the question of unifying the gravitational force with other forces. The starting point was a dynamical system of relativistic string-like objects, subject to the rules of quantum mechanics. As the earliest versions of these models were beset by anomalies, they appeared to violate Lorentz invariance, and also featured excitation modes with negative mass-squared, referred to as ``tachyons". These tachyons would have seriously destabilised the vacuum and for that reason had to be disposed of. It turned out however, that by carefully choosing the total number of transverse string modes, or, the dimensions of the space-time occupied by these strings, and then by carefully choosing the value of the intercept \(a(0)\), which fixes the mass spectrum of the excitations, and finally by imposing symmetry constraints on the spectrum as well, one could make the tachyons disappear and repair Lorentz invariance\,\cite{GSW-1987}\cite{Polchinski-1998}.  It was then found that, while most excitation modes of the string would describe objects whose rest mass would be close to the Planck scale, a very specific set of excitation modes would be massless or nearly massless. It is these modes that are now identified as the set of fundamental particles of the Standard Model, together with possible extensions of the Standard Model at mass scales that are too large for being detected in today's laboratory experiments, yet small compared to the Planck mass.

A string is a structure that is described by a sheet wiped out in space-time, the string `world sheet'. The sheet requires two coordinates that describe it, usually called \(\s\) and \(\t\). The coordinates occupied in an \(n=d+1\) dimensional space-time, temporarily taken to be flat Minkowski space-time, are described by the symbols\fn{To stay in line with most literature on string theory, we chose here capital \(X^\m\) to denote the (real) space-time coordinates. Later, these will be specified either to be real, or to be integers.} \(X^\m(\s,\t),\ \m=0,1,\cdots d\).

Precise mathematical descriptions of a classical relativistic string and its quantum counterpart are given in several excellent text books\,\cite{GSW-1987}\cite{Polchinski-1998},  and they will not be repeated here, but we give a brief summary. We emphasise, and we shall repeat doing so, that our description of a superstring will not deviate from the standard picture, when discussing the fully quantised theory. We do restrict ourselves to standard \emph{perturbative} string theory, which means that we begin with a simply connected piece of world sheet, while topologically non-trivial configurations occur at higher orders of the string coupling constant \(g_s\). We restrict ourselves to the case \(g_s=0\).

Also, we do have to restrict ourselves to a flat Minkowski background. These may well be important restrictions, but we do have speculations concerning the back reaction of strings on their background; the graviton mode, after all, is as dictated in the standard theory, and these gravitons do represent infinitesimal curvatures in the background. Strings in black hole or (anti)-de Sitter backgrounds are as yet beyond what we can do.

\subsubsection{String basics\labell{strbasics}}

An infinitesimal segment \(\dd \ell\) of the string at fixed time, multiplied by an infinitesimal time segment \(\dd t\), defines an infinitesimal surface element \(\dd\SS=\dd\ell\wedge\dd t\). A Lorentz invariant description of \(\dd\SS\) is
	\be\dd\SS^{\m\n}={\pa X^\m\over\pa\s}\,{\pa X^\n\over\pa\t}-{\pa X^\n\over\pa\s}\,{\pa X^\m\over\pa\t}\ . \eel{stringsurfelem}
Its absolute value \(\dd\SS\) is then given by
	\be\pm\,\dd\SS^2=\half\SS^{\m\n}\dd\SS^{\m\n}=(\pa_\s X^\m)^2(\pa_\t X^\n)^2-(\pa_\s X^\m \pa_\t X^\m)^2\ , \eel{Lorinvsurfelem}
where the sign distinguishes space-like surfaces (+) from time-like ones \((-)\). The string world sheet is supposed to be time-like.

The string evolution equations are obtained by finding the extremes of the Nambu Goto action,
	\be S\iss-T\int\dd\SS\iss-T\int\dd\s\dd\t\sqrt{(\pa_\s X^\m \pa_\t X^\m)^2-(\pa_\s X^\m)^2(\pa_\t X^\n)^2} \ ,\eel{NambuGoto}
where \(T\) is the string tension constant; \(T=1/(2\pi\a')\).
	
The light cone gauge is defined to be the coordinate frame \((\s,\,\t)\) on the string world sheet where the curves \(\s\ =\) const. and the curves \(\t\ =\) const. both represent light rays confined to the world sheet. More precisely:
	\be  (\pa_\s X^\m)^2\iss(\pa_\t X^\m)^2\iss 0\ . \eel{lightconegauge}
In this gauge, the Nambu-Goto action takes the simple form
	\be S=\,T\,(\pa_\s X^\m)(\pa_\t X^\m)\ \eel{NGlightcone}
(the sign being chosen such that if \(\s\) and \(\t\) are both pointing in the positive time direction, and our metric is \((-,+,+,+)\)),
the action is negative.	
Imposing \emph{both} light cone conditions \eqn{lightconegauge} is important to ensure that also the infinitesimal variations of the action \eqn{NGlightcone} yield the same equations as the variations of \eqn{NambuGoto}. They give:
	\be\pa_\s\pa_\t\,X^\m=0\ ,
	\eel{stringequs}
but we must remember that these solutions must always be subject to the non-linear constraint equations \eqn{lightconegauge} as well.	
	
The solutions to these equations are left- and right movers:
	\be X^\m(\s,\,\t)=X^\m_L(\s)+X^\m_R(\t)\ ;\qquad	 (\pa_\s X^\m_L)^2=0\ ,\quad(\pa_\t X^\m_R)^2=0  \eel{stringLR}
(indeed, one might decide here to rename the coordinates \(\s=\s^+\) and \(\t=\s^-\)).
We now will leave the boundary conditions of the string free, while concentrating on the bulk properties. 

The re-parametrisation invariance on the world sheet has not yet been removed completely by the gauge conditions \eqn{lightconegauge}, since we can still transform
	\be \s\ra \tl\s(\s)\ ;\qquad\t\ra\tl\t(\t)\ . \eel{onshellgaugeinv}
The \(\s\) and \(\t\) coordinates are usually fixed by using one of the space-time variables; one can choose
	\be X^\pm=(X^0\pm X^d)/\sqrt 2\ , \eel{lcstringcoords}
to define
	\be \s=\s^+=X^+_L\ ,\qquad\t=\s^-=X^+_R\ . \eel{onshellgaugefix}
Substituting this in the gauge condition	\eqn{lightconegauge}, one finds:
	\be \pa_\s X^+_L\pa_\s X^-_L&=&\pa_\s X^-_L\iss \half\sum_{i=1}^{d-1}(\pa_\s X_L^i)^2 \ , \crl{leftminuscoord}
		\pa_\t X^+_R\pa_\t X^-_R&=&\pa_\t X^-_R\iss \half\sum_{i=1}^{d-1}(\pa_\t X_R^i)^2 \ .\eel{rightminuscoord}
So, the longitudinal variables \(X^\pm\), or, \(X^0\) and \(X^d\) are both fixed in terms of the \(d-1=n-2\) transverse variables \(X^i(\s,\,\t)\).	
	
The boundary conditions for an open string are then \def\tot{\mathrm{tot}}\def\openstr{\mathrm{open}}\def\closedstr{\mathrm{closed}}
	\be X_L^\m(\s+\ell)\iss X_L^\m(\s)+u^\m\ ,	\qquad X^\m_R(\s)=X^\m_L(\s)\ , \eel{openboundary}
while for a closed string,
	\be X_L^\m(\s+\ell)\iss X^\m_L(\s)+u^\m\, \qquad X^\m_R(\t+\ell)=X^\m_R(\t)+u^\m\ , \eel{closedboundary}
where \(\ell\) and \(u^\m\) are constants. \(u^\m\) is the 4-velocity. One often takes \(\ell\) to be fixed, like \(2\pi\), but it may be instructive to see how things depend on this free world-sheet coordinate parameter \(\ell\). 
One finds that, for an open string, the action over one period is\fn{The factor \(\halff\) originates from the fact that, over one period, only half the given domain is covered. Do note, however, that the string's orientation is reversed after one period.}
	\be S_\openstr=\half T\int_0^\ell\dd\s\int_0^\ell\dd\t\,\pa_\s X^\m\,\pa_\t X^\m=\half T u^2\ . \eel{actionperiodopen}
For a particle, the action is \(S=p^\m u^\m\), and from that, one derives that the open string's momentum is
	\be p^\m_\openstr=\half T\,u^\m\ . \eel{momopenstr}
For a closed string, the action over one period is
	\be S_\closedstr=T\int_0^\ell\dd\s\int_0^\ell\dd\t\,\pa_\s X^\m\,\pa_\t X^\m= T u^2\ , \eel{actionperiodclosed}
and we derive that the closed string's momentum is
	\be p^\m_\closedstr=T\,u^\m\ . \eel{momclosedstr}
Note that the length \(\ell\) of the period of the two world sheet parameters does not enter in the final expressions. This is because we have invariance under re-parametrisation of these world sheet coordinates.

Now, in a flat background, the \emph{quantisation} is obtained by first looking at the independent variables. These are the transverse components of the fields, being \(X^i(\s,\,\t)\), with \(i=1,2,\cdots, d-1\). This means that these components are promoted to being quantum operators. Everything  is exactly as in section \ref{2dbosons}. \(X^i_L\) are the left movers, \(X^\m_R\) are the right movers. One subsequently postulates that \(X^+(\s,\t)\) is given by the gauge fixing equation \eqn{onshellgaugefix}, or
	\be X^+_L(\s)=\s\ ,\qquad X^+_R(\t)=\t\ , \eel{onshellgaugefix2}
while finally the coordinate \(X^-(\s,\,\t)\) is given by the constraint equations \eqn{leftminuscoord} and \eqn{rightminuscoord}. 

The theory obtained this way is manifestly invariant under rotations among the transverse degrees of freedom, \(X^i(\s,\t)\) in coordinate space, forming the space-like group \(SO(d-1)\). To see that it is also invariant under other space-like rotations, involving the \(d^{\,\th}\) direction, and Lorentz boosts, is less straightforward. To see what happens, one has to work out the complete operator algebra of all fields \(X^\m\), the generators of a Lorentz transformation, and finally their commutation algebra. After a lengthy but straightforward calculation, one obtains the result that the theory is indeed Lorentz invariant provided that certain conditions are met:
	\bi{-} the sequences \(J=a(s)\) of string excited modes (``Regge trajectories") must show an intercept \(a(0)\) that must be limited to the value \(a(0)=1\) (for open strings), and 
	\itm{-} the number of transverse dimensions must be 24 (for a bosonic string) or 8 (for a superstring), so that \(d=25\) or \(9\), and the total number of space-time dimensions \(D\) is 26 or 10. \ei
So, one then ends up with a completely Lorentz invariant theory. It is this theory that we will study, and compare with a deterministic system. As stated at the beginning of this section, many more aspects of this quantised relativistic string theory can be found in the literature. 

The operators \(X^+(\s,\t)\) and \(X^-(\s,\t)\) are needed to prove Lorentz invariance, and, in principle, they play no role in the dynamical properties of the transverse variables \(X^i(\s,\t)\). It is the quantum states of the theory of the transverse modes that we plan to compare with classical states in a deterministic theory. At the end, however, we will need \(X^+\) and \(X^-\) as well. Of these, \(X^+\) can be regarded as the independent target time variable for the theory, without any further dynamical properties, but then \(X^-(\s,\t)\) may well give us trouble. It is not an independent variable, so it does not affect our states, but this variable does control where in space-time our string is. We return to this question in subsection~\ref{detstr}.

\subsubsection{Strings on a lattice\labell{latticestrings}}

To relate this theory to a deterministic system\,\cite{GtHsuperstring-2012}, one more step is needed: the world sheet must be put on a lattice\,\cite{Suss-1988}, as we saw in section~\ref{qft2d}. How big or how small should we choose the meshes to be? It will be wise to choose these meshes small enough. Later, we will see how small; most importantly, most of our results will turn out to be totally independent of the mesh size \(a_\mathrm{worldsheet}\). This is because the dispersion properties of the Hamiltonian \eqn{hampf} have been deliberately chosen in such a way that the lattice artefacts disappear there: the left- and right movers always go with the local speed of light. Moreover, since we have re-parametrisation invariance on the world sheet, in stead of sending \(a_\mathrm{worldsheet}\) to zero, we could decide to send \(\ell\) to infinity. This way, we can keep \(a_\mathrm{worldsheet}=1\) throughout the rest of the procedure. Remember that, the quantity \(\ell\) did not enter in our final expressions for the physical properties of the string, not even if they obey boundary conditions ensuring that we talk of open or closed strings. Thus, the physical limit will be the limit \(\ell\ra\infty\), for open and for closed strings.

We now proceed as in sections \ref{qft2d} and \ref{CA2d}. Assuming that the coordinates \(x\) and \(t\) used there, are related to \(\s\) and \(\t\) by\fn{With apologies for a somewhat inconsistent treatment of the sign of time variables for the right-movers; we preferred to have \(\t\) go in the \(+t\) direction while keeping the notation of section~\ref{2dbosons} for left- and right movers on the world sheet.}
	\be\hskip-15pt \s=\fract 1{\sqrt2}(x+t)\ ,\quad \t=\fract 1{\sqrt 2}(t-x)\ ;\qquad x=\fract 1{\sqrt 2}(\s-\t)\ ,\quad t=\fract 1{\sqrt 2}(\s+\t)\ , \eel{sigmatauxt}
we find that the Nambu Goto action \eqn{NGlightcone} amounts to \(d-1\) copies of the two-dimensional action obtained by integrating the Lagrangian \eqn{LH}:
	\be\LL=\sum_{i-1}^{d-1}\pa_\s X^i\,\pa_\t X^i\ , \eel{stringtrLagrangian}
\emph{provided the string constant \(T\) is normalised to one.} (Since all \(d+1\) modes of the string evolve independently as soon as the on shell constraints \eqn{onshellgaugefix} --- \eqn{rightminuscoord} are obeyed, and we are now only interested in the transverse modes, we may here safely omit the 2 longitudinal modes).

If our units are chosen such that \(T=1\), so that \(\a'=1/2\pi\), we can use the lattice rules \eqn{ALdef} and \eqn{ARdef}, for the transverse modes, or
	\be X^i_L(x,t)&=&p^i(x,t)+X^i(x,t)-X^i(x-1,t)\ ; \crl{XLdef}
	 	X^i_R(x,t)&=&p^i(x,t)+X^i(x,t)-X^i(x+1,t)\ . \eel{XRdef}
where \(P^i(x,t)=\pa_t X^i(x,t)\) (cf Eqs.~\eqn{aL} and \eqn{aR}, or   \eqn{ALdef} and \eqn{ARdef}).
Since these obey the commutation rules \eqn{aLaLlatticecomm} and \eqn{aRaRlatticecomm}, or	
	\be\hskip-15pt [X^i_L,X^j_R]=0\ ;&& [X^i_L(x),\,X^j_L(y)]=\pm\, i\,\d^{ij}\ \hbox{ if }\ y=x\pm 1\ ,\quad\hbox{else }\ 0\ , \crl{XLXLlatticecomm}
				&& [X^i_R(x),\,X^j_R(y)]=\mp\,i\,\d^{ij}\ \hbox{ if }\ y=x\pm 1\ ,\quad\hbox{else }\ 0\ , \eel{XRXRlatticecomm}
we can write these operators in terms of integer-valued operators \(A^i_{L,R}(x)\) and their associated shift generators \(\eta^i_{L,R}\), as in Eqs.~\eqn{aLphiL} and \eqn{aRphiR}. There, the \(\eta\) basis was used, so that the integer operators \(A^i_{L,R}\) are to be written as \(-i\pa/\pa\eta^i_{L,R}\).  To make an important point, let us, momentarily, replace the coefficients there by \(\a,\,\b,\) and \(\g\):
	\be X^i_L(x)&=&-\,i\a\,\frac{\pa}{\pa\eta^i_L(x)}+ {\b}\Big(\frac{\pa}{\pa\eta^i_L(x)}\vv(\{\eta^i_L\})\Big)
			-\g\,\,\eta^i_L(x-1)\ , \eel{XLphiL}
and similarly for \(X^i_R\); here \(\vv(\{\eta(x)\})\) is the phase function introduced in Eqs.~\eqn{phiLprop} and \eqn{phiRprop}.

What fixes the coefficients \(\a,\,\b\) and \(\g\) in these expressions? First, we must have the right commutation relations \eqn{XLXLlatticecomm} and \eqn{XRXRlatticecomm}. This fixes the product \(\a\,\g=1\). Next, we insist that the operators \(X^i_L\) are periodic in all variables \(\eta^i_L(x)\). This was why the phase function \(\vv(\{\eta\})\) was introduced. It itself is pseudo periodic, see Eq.~\eqn{phiLprop}. Exact periodicity of \(X^i_L\) requires \(\b=2\pi\g\). Finally, and this is very important, we demand that the spectrum of values of the operators \(X^-_{L,R}\) runs smoothly from \(-\infty\) to \(\infty\) without overlaps or gaps; this fixes the ratios of the coefficients \(\a\) and \(\g\): we have \(\a=2\pi\g\). Thus, we retrieve the coefficients:
	\be\a=\b=\sqrt{2\pi}\ ;\qquad \g=1/\sqrt{2\pi}\ . \eel{3coefficients}

The reason why we emphasise the fixed values of these coefficients is that we have to conclude that, in our units, the coordinate functions \(X^i_{L,R}(x,\,t)\) of the cellular automaton are \(\sqrt{2\pi}\) times some integers. In our units, \(T=1/(2\pi\a')=1\,;\ \a'=1/(2\pi)\). In arbitrary length units, one gets that the variables \(X^i_{L,R}\) are integer multiples of a \emph{space-time lattice mesh length} \(a_\mathrm{spacetime}\), with
	\be a_\mathrm{spacetime}=\sqrt{2\pi/T}=2\pi\sqrt{\a'}\ . \eel{meshsize}
In Figure~\ref{stringlattice.fig}, the spectrum of the allowed string target space coordinates	in the quantum theory is sketched. Only if Eq.~\eqn{meshsize} is exactly obeyed, the classical system exactly matches the quantum theory, otherwise false voids or overlappings appear\fn{If the mesh size would be chosen exactly half that of Eq.~\eqn{meshsize}, a universal overlap factor of \(2^{d-1}\) would emerge, a situation that can perhaps be accounted for in superstring theory.}.

		\begin{figure}[htb!]
\begin{center} \lowerwidthfig{0pt} {150mm}{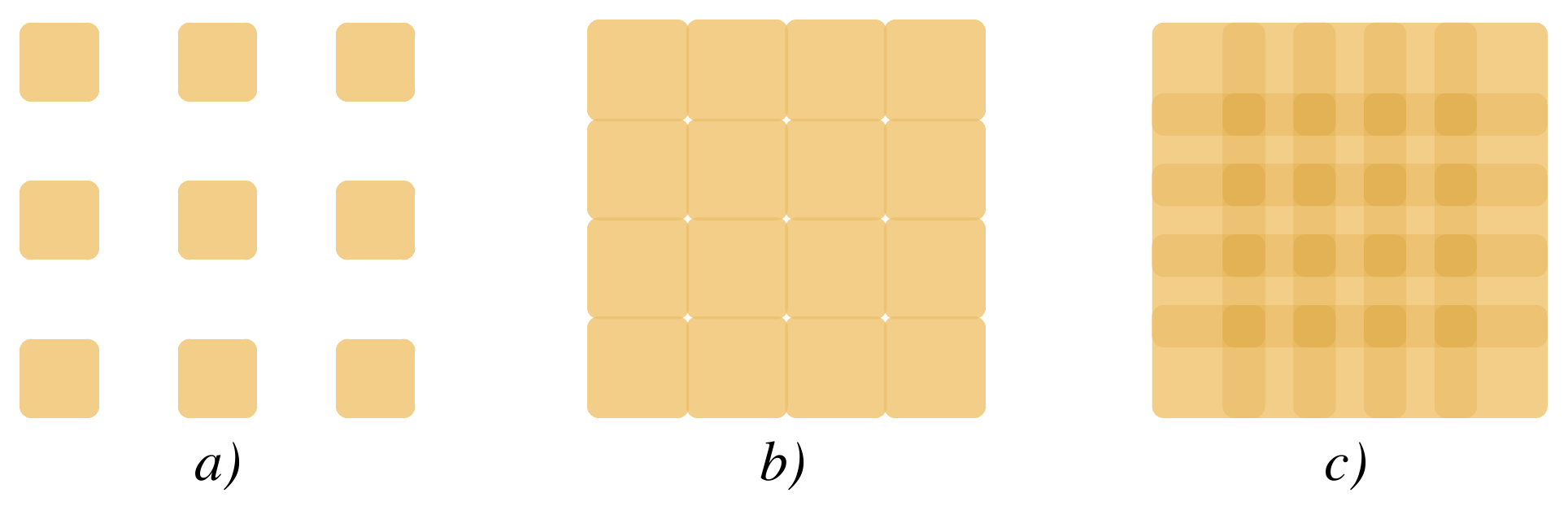}  \begin{quotation}
\caption  [\small The spectrum of allowed values of the quantum string coordinates.]
{\small The spectrum of allowed values of the quantum string coordinates $X^\m$. 
$a)$ The case $a_{\mathrm{spacetime}}>2\pi\sqrt{\a'}$, $b)$  $a_{\mathrm{spacetime}}=2\pi\sqrt{\a'}$, $c)$,  $a_{\mathrm{spacetime}}<2\pi\sqrt{\a'}$. The squares, representing the ranges of the $\eta$ parameters, were rounded a bit so as to show the location of passible edge states.
\labell{stringlattice.fig}}\end{quotation} \end{center}
		\end{figure}	
What we find is that our classical string lives on a square lattice with mesh size \(a_\mathrm{spacetime}\). According to the theory explained in the last few sections of this chapter, the fully quantised bosonic string is entirely equivalent to this classical string; there is a dual mapping between the two. The condition \eqn{meshsize} on the value of the lattice parameter \(a_\mathrm{spacetime}\) is essential for this mapping. If string theoreticians can be persuaded to limit themselves to string coordinates that live on this lattice, they will see that the complete set of quantum states of the bosonic string still spans the entire Hilbert space they are used to, while now all basis elements of this Hilbert space propagate classically, according to the discrete analogues of the classical string equations.

Intuitively, in the above, we embraced the lattice theory as the natural ontological system corresponding to a non-interacting string theory in Minkowski space. However, in principle, we could just as well have chosen the compactified theory. This theory would assert that the transverse degrees of freedom of the string do not live on \(\Bbb R^{\,d-1}\), but on \(\Bbb T^{\,d-1}\), a (continuous) torus in \(d-1\) dimensions, again with periodicity conditions over lengths \(2\pi\sqrt{\a'}\), and these degrees of freedom would move about classically.

In subsection~\ref{detstr} we elaborate further on the nature of the deterministic string versions.

\subsubsection{The lowest string excitations\labell{tachyon}}
	String theory is a quantum field theory on the \(1+1\) dimensional world sheet of the string. If this quantum field theory is in its ground state, the corresponding string mode describes the lightest possible particle in this theory. As soon as we put excited states in the world sheet theory, the string goes into excited states, which means that we are describing heavier particles. This way, one describes the mass- or energy spectrum of the string. 
	
In the original versions of the theory, the lightest particle turned out to have a negative mass squared; it would behave as a tachyon, which would be an unwanted feature of a theory. 

The more sophisticated, modern versions of the theory are rearranged in such a way that the tachyon mode can be declared to be unphysical, but it still acts as a description of the formal string vacuum. To get the string spectrum, one starts with this unphysical tachyon state and then creates descriptions of the other states by considering the action of creation operators. 

To relate these string modes to the ontological states at the deterministic, classical sides of our mapping equation, we again consider the ground state, as it was described in subsection~\ref{ground}, to describe the tachyon solution. Thus, the same procedure as in that subsection will have to be applied. Similarly one can get the physical particles by having the various creation operators act on the tachyon (ground) state. This way, we get our description of the photon (the first spin one state of the open string) and the graviton (the spin 2 excited state of the closed string).

\subsubsection{The Superstring  \labell{sustr1}}

To construct theories containing fermions, it was proposed to plant fermionic degrees of freedom on the string world sheet. Again, anomalies were encountered, unless the bosonic and fermionic degrees of freedom can be united to form super multiplets. Each bosonic coordinate degree of freedom \(X^\m(\s,\t)\) would have to be associated with a fermionic degree of freedom \(\j^\m(\s,\t)\). This should be done for the left-moving modes independently of the right-moving ones. A further twist can be given to the theory by \emph{only} adding fermionic modes to the left-movers, not to the right movers (or vice versa); this way, chirality can be introduced in string theory, not unlike the chirality that is clearly present in the Standard Model. Such a theory is called a \emph{heterotic string theory}.


Since the world sheet is strictly two-dimensional, we have no problems with spin and helicity within the world sheet, so, here, the quantisation of fermionic fields --- at least at the level of the world sheet --- is simpler than in the case of the `neutrinos' discussed in subsection~\ref{secondneutrino}.

Earlier, we used the coordinates \(\s\) and \(\t\) as light cone coordinates on the world sheet; now, temporarily, we want to use there a space-like coordinate and a time-like one, which we shall call \(x^1=x\) and \(x^0=t\).

On the world sheet, spinors are 2-dimensional rather than 4-dimensional, and we take them to be hermitean operators, called Majorana fields, which we write as
	\be\j^{\m*}(x,\,t)=\j^\m(x,\,t)\ ,\qquad\hbox{and}\qquad\j^{\m\dag}(x,\,t)={\j^\m}^{\,*T}(x,\,t) \eel{Majorana}
(assuming a real Minkowskian target space; the superscripts \({}^{*T}\) mean that if  \(\j=\big({\j_1\atop\j_2}\big)\)
then \(\j^{*T}=(\j_1^*,\,\j_2^*)\)).

There are only two Dirac matrices, call them \(\r^0\) and \(\r^1\), or, after a Wick rotation, \(\r^1\) and \(\r^2\). They obey
	\be \r^0=i\r^2\ ,\qquad \{\r^\a,\,\r^\b\}=\r^\a\r^\b+\r^\b\r^\a=2\eta^{\a\b}\ ,\quad(\a,\b=1,2)\ . \eel{2ddiracmatrices}
A useful representation is
	\be \r^1=\s_x=\pmatrix{0&1\cr 1&0}\ ,\quad \r^2=\s_y=\pmatrix{0&-i\cr i&0}\ ;\quad\r^0=\pmatrix{0&1\cr -1&0}\ . \eel{2ddiracrepr}
The spinor fields conjugated to the fields \(\j^\m(x,\,t)\) are \(\overline {\j}^\m(x,\,t)\), here defined by
	\be \overline{\j}^\m={\j^\m}^T\r^2\ , \eel{conjugpsi}
Skipping a few minor steps, concerning gauge fixing, which can be found in the text books about superstrings,
we find the fermionic part of the Lagrangian:
	\be\LL(x,\,t)=-\sum_{\m=0}^d\overline{\j}^\m(x,\,t)\,\r^\a\pa_\a\,\j^\m (x,\,t)\ . \eel{fermionL}

Since the \(\r^2\) is antisymmetric and fermion fields anti-commute, two different spinors \(\j\) and \(\chi\) obey
	\be\overline\chi\j=\chi^T\r^2\j=i(-\chi_1\j_2+\chi_2\j_1)=i(\j_2\chi_1-\j_1\chi_2)=\j^T\r^2\chi=\overline\j\chi\ . \eel{fermexchsymm}
Also we have 
	\be\overline\chi\r^\m\j=-\overline\j\r^\m\chi\ , 	\eel{eq205}
an antisymmetry that explains why the Lagrangian \eqn{fermionL} is not a pure derivative.
The Dirac equation on the string world sheet is found to be
	\be 	\sum_{\a=1}^2\r^\a\pa_\a\j^\m=0\ . \eel{worlddirac}
In the world sheet light cone frame, one writes	
	\be(\r_+\pa_-+\r_-\pa_+)\j=0\ ,\eel{lsdiraceq}
where, in the representation \eqn{2ddiracrepr}, 
	\be \r^\pm=\fract 1{\sqrt 2}(\r^0\pm\r^1)\ ,&&\hbox{and}\crl{lcframe}
		\r_+=-\r^-={\sqrt 2}\pmatrix{0&0\cr 1&0}\ ,&&\r_-=-\r^+=\sqrt 2\pmatrix{0&-1\cr 0&0}\ . \eel{lcdiracm}
		
The solution of Eq.~\eqn{worlddirac} is simply
	\be\j^\m(\s,\t)=\pmatrix{\j^\m_L(\s)\cr\j^{\m\vphantom{|^|}}_R(\t)}\ . \eel{leftrightfermions}
Thus, one finds that the fermionic left-movers and right-movers have no further spinor indices.

As is the case for the bosonic coordinate fields \(X^\m_{L,R}(x)\), also the fermionic field components \(\j^\m_{L,R}\) have two longitudinal modes, \(\m=\pm\), that are determined by constraint equations. These equations are dictated by supersymmetry. So for the fermions also, we only keep the \(d-1\) transverse components as independent dynamical fields (\(d\) is the number of space-like dimensions in target space).

The second-quantised theory for such fermionic fields has already briefly been discussed in our treatment of the second-quantised `neutrino' system, in section~\ref{secondneutrino}. Let us repeat here how it goes for these string world sheet fermions. Again, we assume a lattice on the world sheet, while the Dirac equation on the lattice now reduces to a finite-step equation, so chosen as to yield exactly the same solutions \eqn{leftrightfermions}:
	\be&\hskip-20pt \j^i(x,t)=-\fract 1{\sqrt 2}\r^0\Big(\r^+\j^i(x-1,t-1)+\r^-\j^i(x+1,t-1)\Big)\ ,&\nm\\  
	&i=1,\cdots,d-1\ .& \eel{discrdirac}
The deterministic counterpart is a Boolean variable set \(s^i(x,t)\), which we assume to be taking the values \(\pm 1\). One may write their evolution equation as 
	\be s^i(x,t)=s^i(x-1,t-1)\,s^i(x+1,t-1)\,s^i(x,t-2)\ .\eel{Booleansi}
of which the solution can be written as
	\be s^i(x,t)=s^i_L(x,t)\,s^i_R(r,t)\ ,\eel{sLR}
where \(s^i_L\) and \(s^i_R\) obey
	\be   s^i_L(x,t)=s^i_L(x+1,t-1)\ ;\qquad  s_R^i(x,t)=s_R^i(x-1,t-1) \ , \eel{sevolve}
which is the Boolean analogue of the Dirac equation \eqn{discrdirac}.

One can see right away that all basis elements of the Hilbert space for the Dirac equation can be mapped one-to-one onto the states of our Boolean variables. If we would start with these states, there is a straightforward way to construct the anti-commuting field operators \(\j^i_{L,R}(x,t)\) of our fermionic system, the {Jordan-Wigner transformation}\,\cite{JordanWigner-1928}, also alluded to in section~\ref{secondneutrino}. At every allowed value of the parameter set \((x,i,\a)\), where \(\a\) stands for \(L\) or \(R\), we have an operator \(a^i_\a(x)\) acting on the Boolean variable \(s^i_\a(x)\) as follows:
	\be a|+\ket =|-\ket\ ;\qquad a|-\ket=0\ . \eel{JWstep1}
These operators, and their hermitean conjugates \(a^\dag\) obey the mixed commutation - anti-commutation rules
	\be \{a^i_\a(x),\,a^i_\a(x)\}=0\ ,&& \{a^i_\a(x),\,a^{i\dag}_\a(x)\}=\mathbb I\  , \crl{cac1}
	  [a^i_\a(x_1),\,a^j_\b(x_2)]=0\ ,&&  [a^i_\a(x_1),a^{j\dag}_\b(x_2)]=0\quad\hbox{if}\quad
	  \Bigg\{\matrix{&\hskip-10pt x_1\ne x_2 \cr
	  			\hbox{and/or}& i\ne j \cr
				\hbox{and/or}&\a\ne\b\ .} 
	   \eel{cac2}

Turning the commutators in Eq.~\eqn{cac2} into anti-commutators is easy, if one can put the entire list of variables \(x,\,i\) and \(\a\) in some order.
Call them \(y\) and consider the ordering \(y_1<y_2\). Then, the operators \(\j(y)\) can be defined by:
	\be\j(y)\equiv\Big( \prod_{y_1<y} s(y_1)\Big) a(y)\ . \eel{psidef}
This turns the rules \eqn{cac1} and \eqn{cac2} into the anti-commutation rules for fermionic fields:
	\be\{\j(y_1),\,\j(y_2)\}=0\ ;\qquad\{\j(y_1),\,\j^\dag(y_2)\}=\d(y_1,\,y_2)\ . \eel{JWcommrules}
In terms of the original variables, the latter rule is written as 
	\be \{\j^i_\a(x_1),\,\j^{j\dag}_\b(x_2)\}=\d(x_1-x_2)\d^{ij}\d_{a\b}\ . \eel{fermac}
Let us denote the left-movers \(L\) by \(\a=1\) or \(\b=1\), and the right movers \(R\) by \(\a=2\) or \(\b=2\). 
Then, we choose our ordering procedure for the variables  \(y_1=(x_1,i,\a)\) and \(y_2=(x_2,j,\b)\) to be defined by
	\be \hbox{if }\ \a<\b & \hbox{then }\  y_1<y_2\ ;\cr
		\hbox{if }\ \a=\b  \hbox{ and } x_1<x_2 & \hbox{then }\ y_1<y_2\ ;\cr
		\hbox{if }\ \a=\b \hbox{ and } x_1=x_2\hbox{ and }i<j & \hbox{then }\ y_1<y_2\ ,\cr
		\hbox{else }\ y_1=y_2\hbox{ or }y_1>y_2\ . & \eel{yordering}
This ordering is time-independent, since all left movers are arranged before all right movers. Consequently, the solution \eqn{leftrightfermions} of the `quantum' Dirac equation holds without modifications in the Hilbert space here introduced.

It is very important to define these orderings of the fermionic fields meticulously, as in Eqs.\ \eqn{yordering}.
The sign function between brackets in Eq.~\eqn{psidef}, which depends on the ordering,  is typical for a Jordan-Wigner transformation. We find it here to be harmless, but this is not always the case. Such sign functions can be an obstruction against more complicated procedures one might wish to perform, such as interactions between several fermions, between right-movers and left-movers, or in attempts to go to higher dimensions (such as in \(k\)-branes, where \(k>2\)).

At this point, we may safely conclude that our dual mapping between quantised strings and classical lattice strings continues to hold in case of the superstring.

\subsubsection{Deterministic strings and the longitudinal modes\labell{detstr}} 

The transverse modes of the (non interacting) quantum bosonic and superstrings (in flat Minkowski space-time) could be mapped onto a deterministic theory of strings moving along a target space lattice. How do we add the longitudinal coordinates, and how do we check Lorentz invariance? The correct way to proceed is first to look at the quantum theory, where these questions are answered routinely in terms of quantum operators.

Now we did have to replace the continuum of the world sheet by a lattice, but we claim that this has no physical effect because we can choose this lattice as fine as we please whereas rescaling of the world sheet has no effect on the physics since this is just a coordinate transformation on the world sheet. We do have to take the limit \(\ell\ra\infty\) but this seems not to be difficult.

Let us first eliminate the effects of this lattice as much as possible. Rewrite Eqs.~\eqn{leftcont} and \eqn{rightcont} as:
	\be p(x,t)=\pa_x k(x,t)\ , && a^{L,R}(x,t)=\pa_x b^{L,R}(x,t)\ ; \labell{kbdef}\\
		b^L(x+t)= k(x,t)+q(x,t) \ ;&& b^R(x-t)= k(x,t)-q(x,t)\ ;		\eel{modified}
the new fields now obey the equal time commutation rules
	\be	[q(x,t),\,k(x',t)]&=&\half i \,\sgn(x-x'\,)\ ;\labell{qkcomm} \\	\ 
		 [b^L(x),\,b^L(y)]&=&-\,[b^R(x),\,b^R(y)]\iss -i\,  \sgn(x-y)\ , \eel{bcomm}
where \(\sgn(x)=1\) if \(x>0\,,\ \sgn(x)=-1\)	if \(x<0\) and \(\sgn(0)=0\). 

Staying with the continuum for the moment, we cannot distinguish two ``adjacent" sites, so there will be no improvement when we try to replace an edge state that is singular at \(\eta(x_)=\pm\pi\) by one that is singular when this value is reached at two adjacent sites; in the continuum, we expect our fields to be continuous. In any case, we now drop the attempt that gave us the expressions \eqn{aLphiL} and \eqn{aRphiR}, but just accept that there is a single edge state at every point. This means that, now, we replace these mapping equations by
	\be b^L(x)&=&\sqrt{2\pi}B{^L_\op\,}(x)+{1\over\sqrt{2\pi}}\z^L_\op(x)\ , \\ 
	 	b^R(x)&=&\sqrt{2\pi}B{^R_\op\,}(x)+{1\over\sqrt{2\pi}}\z^R_\op(x)\ ,\eel{bBzeta}
	where the functions \(B^{L,R}\) will actually play the role of the integer parts of the coordinates of the string, and
\(\z^{L,R}_{\,\op}(x)\) are defined by their action on the integer valued functions \(B^{L,R}(x)\), as follows:
	\be\hskip-20pt    e^{i\z^L(x_1)}|\{B^{L,R}(x)\}\ket =	|\{{B'}^{L,R}(x)\}\ket\ ,\quad \bigg\{
		\matrix{{B'}^L(x)&=&B^L(x)+\tht(x-x_1\,)\ ,\cr 
		{B'}^R(x)&=&B^R(x)\ ;\qquad} \label{zetaL}\\[5pt]  
   		\hskip-20pt    e^{i\z^R(x_1)}|\{B^{L,R}(x)\}\ket =	|\{{B''}^{L,R}(x)\}\ket\ ,\quad \bigg\{
		\matrix{{B''}^L(x)&=&B^L(x)\ ,\qquad \cr  {B''}^R(x)&=&B^R(x)+\tht(x_1-x)\ ,} \eel{zetaR} 
so that, disregarding the edge state,
	\be[B^L(x),\,\z^L(y)]=-i\tht(x-y)\ , \qquad [B^R(x),\,\z^R(y)]=-i\tht(y-x)\ . \eel{Bzetacomm}
This gives the commutation rules \eqn{bcomm}. If we consider again a lattice in \(x\) space, where the states are given in the \(\z\) basis, then the operator \(B^L_\op(x)\) obeying commutation rule \eqn{Bzetacomm} can be written as
	\be B^L_\op(x_1)=\sum_{y<x_1}-i{\pa\over\pa\z^L(y)}\ . \eel{BLop}

Now the equations of motion of the transverse string states are clear. These just separate into left-movers and right-movers, both for the discrete lattice sites \(X^i(\s,\t)\) and for the periodic \(\eta^i(\s,\t)\) functions, where \(i=1,\cdots,d-1\). Also, the longitudinal modes split up into left moving ones and right moving ones. These, however, are fixed by the gauge constraints. In standard string theory, we can use the light cone gauge to postulate that the coordinate variable \(X^+\) is given in an arbitrary way by the world sheet coordinates, and one typically chooses the constraint equations \eqn{onshellgaugefix}. 

This means that
	\be a^+_L(\s)=1\ ,\qquad a^+_R(\t)=1\ , \eel{gaugeconstramu}
but by simple coordinate transformations \(\s\ra\s_1(\s)\) and \(\t\ra\t_1(\t)\), one can choose any other positive function of the coordinate \(\s\) (left-mover) or \(\t\) (right mover). Now, Eqs.~\eqn{lightconegauge} here mean that
	\be(a^\m_L(\s))^2\iss (a^\m_R(\t))^2\iss 0\ , \ee
so that, as in Eqs.~\eqn{leftminuscoord} and \eqn{rightminuscoord}, we have the constraints
	\be a^+_L(\s)=\half\sum_{i=1}^{d-1}(a^i_L(\s))^2\ ,\qquad a^+_R(\t)=\half\sum_{i=1}^{d-1}(a^i_R(t))^2\ ,\ee
where \(a^i_{L,R}(x)=\pa_x b^i_{R,L}(x)\) (see the definitions \ref{kbdef}).

In view of Eq.~\eqn{BLop} for the operator \(B^i_L(x)\), it is now tempting to write for the longitudinal coordinate \(X^+_L\): \ \(\pa_\s B^i_L(\s)= -i{\pa\over\pa\z^i_L(\s)}\), so that
	\be\pa_\s X^+_L(\s)\qu\half\sum_{i=1}^{d-1}\bigg(-{2\pi\,\pa^2\over\pa\z(\s)^2}+{1\over 2\pi}(\pa_\s\z(\s))^2-2i\{\
	{\pa\over\pa\z(\s)},\,\pa_\s\z(\s)\}\bigg)\ , \ee
but the reader may have noticed that we now disregarded the edge states, which here may cause problems: they occur whenever the functions \(\eta^i\) cross the values \(\pm\pi\), where we must postulate periodicity. 

We see that we do encounter problems if we want to define the longitudinal coordinates in the compactified classical field theory. Similarly, this is also hard in the discrete automaton model, where we only keep the \(B^i_{L,R}\) as our independent ontological variables. How do we take their partial derivatives in \(\s\) and \(\t\)?

Here, we can bring forward that the gauge conditions \eqn{gaugeconstramu} may have to be replaced by Dirac delta functions, so as to reflect our choice of a world sheet lattice. 

These aspects of the string models we have been considering are not well understood. This  subsection was added to demonstrate briefly what happens if we study the gauge constraints of the theory to get some understanding of the longitudinal modes, in terms of the ontological states. At first sight it seemed that the compactified deterministic theory would offer better chances to allow us to rigorously derive what these modes look like; it seems as if we can replace the world sheet lattice by a continuum, but the difficulties are not entirely resolved.

If we adopt the cellular automaton based on the integers \(B^i_{L,R}\), use of a world sheet lattice is almost inevitable. On the world sheet, the continuum limit has to be taken with much care.

\subsubsection{Some brief remarks on (super) string interactions\labell{sustrint}}

As long as our (super) strings do not interact, the effects of the constraints are minor. They tell us what the coordinates \(X^-(\s,\t)\) are if we know all other coordinates on the world sheet. In the previous section, our point was that the evolution of these coordinates on the world sheet is deterministic. Our mappings from the deterministic string states onto the quantum string states is one-to-one, apart from the edge states that we choose to ignore. In the text books on string theory, superstring interactions are described by allowing topologically non-trivial world sheets. In practice, this means that strings may exchange arms when they meet at one point, or their end points may join or tear apart. All this is then controlled by a string coupling constant \(g_s\); an expansion in powers of \(g_s\) yields string world sheet diagrams with successively higher topologies.

Curiously, one may very well imagine a string interaction that is deterministic, exactly as the bulk theory obeys deterministic equations. Since, in previous sections, we did not refer to topological boundary conditions, we regard the deterministic description obtained there as a property of the string's `bulk'.

A natural-looking string interaction would be obtained if we postulate the following:
\begin{quotation} \noindent Whenever two strings meet at one point on the space-time lattice, they exchange arms, as depicted in Fig.~\ref{stringint.fig}. \end{quotation}
This ``law of motion" is deterministic, and unambiguous, provided that both strings are \emph{oriented} strings. The deterministic version of the interaction would not involve any freely adjustable string constant \(g_s\).

If we did not have the problem how exactly to define the longitudinal components of the space-time coordinates, this would complete our description of the deterministic string laws. Now, however, we do have the problem that the longitudinal coordinates are `quantum'; they are obtained from constraints that are non-linear in the other fields \(X^\m(\s,\t)\), each of which contain integer parts and fractional parts that do not commute. 
		\begin{figure}[h] \begin{quotation}
\begin{center}\hskip-5mm \lowerheightfig{0pt}{25mm}{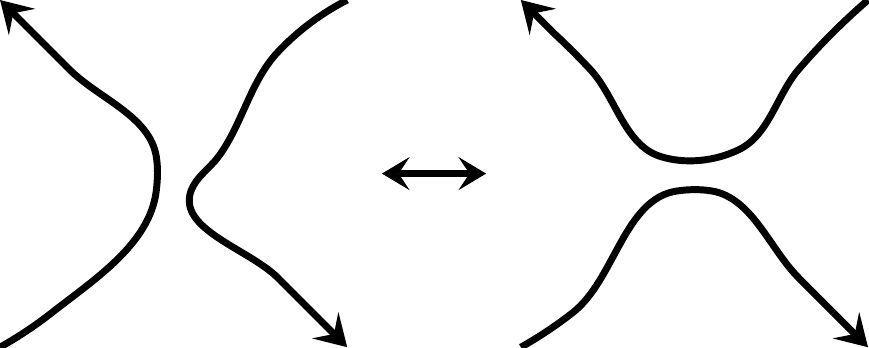}\qquad 
 \caption[\small Deterministic string interaction. ]{\small Deterministic string interaction. This interaction takes place whenever two pieces of string meet at one space-time point. \labell{stringint.fig}}
\end{center}\end{quotation}
		\end{figure}

This problem, unfortunately, is significant. It appears to imply that, in terms of the `deterministic' variables, we cannot exactly specify where on the world sheet the exchange depicted in Figure~\ref{stringint.fig} takes place. This difficulty has not been resolved, so as yet we cannot produce a `deterministic' model of interacting `quantum' strings. 

We conclude from our exercise in string theory that strings appear to admit a description in terms of ontological objects, but just not yet quite. The most severe difficulties lie in the longitudinal modes. They are needed to understand how the theory can be made Lorentz invariant. It so happens that local Lorentz invariance is a problem for every theory that attempts to describe the laws of nature at the Planck scale, so it should not come as a surprise that we have these problems here as well. We suspect that today's incomplete understanding of Lorentz invariance at the Planck scale needs to be repaired, but it may well be that this can only be done in full harmony with the Cellular Automaton Interpretation. What this section suggests us is that this cannot be done solely within the framework of string theory, although strings may perhaps be helpful to lead us to further ideas.

An example of a corner of string theory that has to be swept clean is the black hole issue. Here also, strings seem to capture the physical properties of black holes partly but not completely; as long as this is the case one should not expect us to be able to formulate a concise ontological theory. This is why most parts of this book concentrate on the general philosophy of the CAI rather than attempting to construct a complete model.

\newsecl{Symmetries}{symm}

In classical and in quantum systems, we have Noether's theorem\,\cite{noether-1918}: 
\begin{quote}\emph{Whenever there is a continuous symmetry in a system, there exists a conserved quantity associated with it.}\end{quote} 
Examples are conservation of momentum (translation symmetry), conservation of energy (symmetry with respect to time translations), and angular momentum (rotation symmetry). In classical systems, Noether's theorem is limited to continuous symmetries, in quantum systems, this theorem is even more universal: here also discrete symmetries have their associated conservation laws: parity \(P=\pm 1\) of a system or particle (mirror symmetry), non commuting discrete quantum numbers associated with more general discrete permutations, etc. Also, in a quantum system, one can reverse the theorem: 
\begin{quote}\emph{Every conserved quantity is associated to a symmetry,} \end{quote}
for instance isospin symmetry follows from the conservation of the isospin vector \(\vec I=(I_1,\,I_2,\,I_3)\), baryon number conservation leads to a symmetry with respect to \(U(1)\) rotations of baryonic wave functions, and so on.

\subsecl{Classical and quantum symmetries}{classquantsym}

We now claim that this more generalised Noether theorem can also be applied to classical systems, simply by attaching a basis element of Hilbert space to every state the classical system can be in. If, for instance, the evolution law \(U_{t,\,t+\d t}\) is independent of time \(t\), we have a conserved energy. This energy is obtained from the eigenvalue of  \(U_{t,\,t+\d t}\) for the smallest admissible value of \(\d t\). Now since an energy eigenstate will usually not be an ontological state of the system, this energy conservation law only emerges in our quantum procedure; it does not show up in standard classical considerations. For us, this is very important: if \(\d t\) is as small as the Planck time, the energy eigenstates, all the way to the Planck energy, are superpositions of ontological states. If, as we usually do, we limit ourselves to quantum systems with much lower energies, we are singling out a section of Hilbert space that is not represented by individual ontological states, and for this reason we should not expect recognisable classical features in the quantum systems that we are usually looking at: atoms, molecules, elementary particles.

Often, our deterministic models are based on a lattice rather than a space-time continuum. The classical space-time symmetries on a lattice are more restricted than those of a continuum. It is here that our mappings onto quantum systems may help. If we allow ontological states to have symmetry relations with superimposed states, much more general symmetry groups may be encountered. This is further illustrated in this chapter.

Since we often work with models having only finite amounts of data in the form of bits and bytes in given volume elements, we are naturally led to systems defined on a lattice. There are many ways in which points can be arranged in a lattice configuration, as is well known from the study of the arrangement of atoms in crystalline minerals. The symmetry properties of the minerals are characterised by the set of crystallographic point groups,  of which there are 32 in three dimensions.\,\cite{crystalgroups-1994}

The simplest of these is the cubic symmetry group generated by a cubic lattice:
	\be\vec x=(n_1,\,n_2,\,n_3)\ , \eel{cubiclattice}
where \(n_1,\ n_2\) and \(n_3\) are integers. What we call the cubic group here, is the set of all 48 orthogonal rotations including the reflections of these three integers into \(\pm\) each other (6 permutations and \(2^3\) signs). This group, called \(O(3,\mathbb Z)\), is obviously much smaller than the group \(O(3, \mathbb R)\) of all orthonormal rotations. The cubic group is a finite subgroup of the infinite orthogonal group.
	
Yet in string theory, section~\ref{latticestrings}, something peculiar seems to happen: even though the string theory is equivalent to a lattice model, it nevertheless appears not to lose its full orthogonal rotation symmetry. How can this be explained?

\subsecl{Continuous transformations on a lattice}{controtlattice}
	
Consider a classical model whose states are defined by data that can be arranged in a \(d\) dimensional cubic lattice. Rotation symmetry is then usually limited by the group \(O(d,\mathbb Z)\). If now we introduce our Hilbert space, such that every state of the classical system is a basis element of that, then we can introduce superpositions, and much more symmetry groups are possible. There are several ways now to introduce \emph{continuous} translations and rotations.

To this end, it is, again, very instructive to do the Fourier  transformation:
	\be \bra \vec x|\j\ket = (2\pi)^{-d/2}\int_{ |\k_i|<\pi   }\dd^d\vec \k\bra\vec \k|\j\ket e^{i\vec \k\cdot\vec x}\ , \eel{etaagain}
Here, \(|\j\ket\)  describes a single particle living on the lattice, but we could also take it as the operator field of a second-quantised system, as is usual in quantum field theories. Of course, as is usual in physics notation, \(\bra\vec x|\) are the bras in \(x\) space (where \(\vec x\) is Eq.~\eqn{cubiclattice}, the lattice), whereas \(\bra\vec \k|\) are the bras in momentum space, where \(\vec\k\) are continuous, and all its components \(\k_i\) obey \(|\k_i|<\pi\).

The inverse of the Fourier transform \eqn{etaagain} is:
	\be \bra\vec\k|\j\ket=(2\pi)^{-d/2}\sum_{\vec x\in\mathbb{Z}^d}\bra\vec x|\j\ket\,e^{-i\vec\k\cdot\vec x}\ . \eel{etainverse}

\subsubsection{Continuous translations\labell{conttransl.sec}}

Translations over \emph{any} distance \(\vec a\) can now be defined as the operation
	\be\bra\vec\k|\j\ket\ra\bra\vec\k|\j\ket \, e^{-i\vec\k\cdot\vec a}\ , \eel{contshift}
although only if \(\vec a\) has integer components, this represents an ontological shift
	\be\bra\vec x|\j\ket\ra\bra\,\vec x-\vec a\,|\j\ket\ , \eel{latticeshift}
since \(\vec x\) must sit in the lattice, otherwise this would represent a non ontological state.

If \(\vec a\) has fractional components, the translation in \(x\) space can still be defined. Take for instance a fractional value for \(a_x\), or, \(\vec a=(a_x,0,0)\). Then
	\be &\displaystyle\bra\k_x|\j\ket\ra\bra\k_x|\j\ket \,e^{-i\k_xa_x}\ ,\qquad \bra x|\j\ket\ra\sum_{x'}\bra x'|\j\ket \D_{a_x}(x-x')\ , &\nn
	&\displaystyle \D_{a_x}(x_1)\iss (2\pi)^{-1}\int_{-\pi}^\pi\dd\k_x\,e^{-ia_x\k_x+i\k_x\,x_1}\iss{\sin\pi(x_1-a_x)\over\pi\,(x_1-a_x)}\ ,&	\eel{fracshift}
where we also used Eq.~\eqn{etainverse} for the inverse of Eq.~\eqn{etaagain}.

One easily observes that  Eq.~\eqn{fracshift} reduces to Eq.~\eqn{latticeshift} if \(a_x\) tends to an integer. Translations over a completely arbitrary vector \(\vec a\) are 
obtained as the product of fractional translations over \((a_x,0,0)\), \((0,a_y,0)\) and \((0,0,a_z)\):
	\be\bra \vec x|\j\ket\ra\sum_{x',\,y',\,z'}\bra \vec x\,'|\j\ket\D_{\vec a}(\vec x-\vec x\,')\ ,\qquad \D_{\vec a}(\vec x_1)=\D_{a_x}(x_1)\D_{a_y}(y_1)\D_{a_z}(z_1)\ . \eel{generalshift}
Notice that the kernel function \(\D_{\vec a}(\vec x_1)\) maximises for the values of \(\vec x_1\) closest to \(\vec a\), so, even for translations over fractional values of the components of \(\vec a\), the translation operation involves only the components of \(|\j\ket\) closest to the target value \(\vec x-\vec a\).

The \emph{generator} for infinitesimal translations is the operator \(\vec\eta_{\,\op}\). Translations over a finite distance \(\vec a\) can then be described by the operator \(e^{i\vec a\cdot\vec\eta_{\,\op}}\). Writing \(\vec\eta_{\,\op}=(\eta_x,\,\eta_y,\,\eta_z)\), and taking \(\eta_x,\ \eta_y,\) and \(\eta_z\) each to act only in one dimension, we have
	\be &\bra\vec\k|\,\vec\eta_{\,\op}\,|\j\ket = -\vec\k\,\bra\vec\k|\j\ket\ ,& \crl{etakappa}
	&\displaystyle \bra x|\,e^{i\eta_x\,a_x}|\j\ket = \sum_{x'}\bra x'|\j\ket\,{\sin\pi(x-x'-a_x)\over\pi(x-x'-a_x)}\ ,&\eel{etax}
and when \(a_x\) is taken to be infinitesimal, while \(x\) and \(x'\) are integers, one finds
	\be \bra x|(\mathbb I+i\eta_x a_x)|\j\ket&=&\sum_{x'}\bra x'|\j\ket\Big(\d_{xx'}+(1-\d_{xx'}){(-1)^{x-x'}(-\pi a_x)\over\pi(x-x')}\Big)\ ; \crl{etainfinitesimal}
	\bra x|\,\eta_x\,|x\ket\iss 0\ ,&& \bra x|\,\eta_x\,|x'\ket\iss{i\,(-1)^{x-x'}\over x-x'}\ \hbox{ if }\ x\ne x'\ .\eel{etaxmatrix}
The eigenstates \(|\eta_x\ket\) of the operator \(\eta_x\) can be found:  \(\bra x|\eta_x\ket=e^{-i\eta_xx}\).

The expressions we found for this generator are the most natural ones but not the only possible choices;  we must always remember that one may add multiples of \(2\pi\) to its eigenvalues. This modifies the matrix elements \eqn{etaxmatrix} while the effects of translations over integer distances remain the same.

An important feature of our definition of fractional translations on a lattice is their commutation rules. These translations are entirely commutative (as we can deduce from the definition Eq.~\eqn{contshift}):
	\be [\vec\eta,\,\vec\eta\,']=0\ . \eel{tranlscomm}

\subsubsection{Continuous rotations 1: Covering the Brillouin zone with circular regions\labell{circles.subsec}}

What can be done with translations on a lattice, can also be done for rotations, in various ways. Let us first show how to obtain a  perfect general rotation operator on a lattice, in principle. 
Again, we start from the Fourier modes, \(e^{i\k x}\), Eq.~\eqn{etaagain}. How do we generate arbitrary rotations? 

Taking again the cubic lattice as our prototype, we immediately see the difficulty: the space of allowed values for \(\vec\k\) is a square (in 2 dimensions) or a cube (in 3 dimensions). This square and this cube are only invariant under the discrete rotation group \(O(d,\mathbb Z)\). Therefore, rotations over other angles can at best be approximate, it seems.  We illustrate the situation for a two-dimensional square lattice, but extrapolation to \(d>2\) space dimensions and/or other lattice configurations is straightforward.

The space of allowed momentum values is called the Brillouin zone, and it is the square in Figure~\ref{circles.fig}$a$. A first approximation for a rotation of the lattice by any angle \(\vv\) is obtained by drawing the largest possible circle in the Brillouin zone (or the largest possible sphere in the 3 or higher dimensional case) and rotate the region inside that. The data on the remainder of the Brillouin zone, outside the circle,  are ignored or replaced by zero.

This procedure perhaps looks good for the lower frequency modes, but it does not rotate everything, and it would clearly disobey the desired group properties of rotations and translations, so we must do something better with the remainder of the Brillouin zone. This is possible, see Fig~\ref{circles.fig}$b$. The rotation operator could then be defined as follows. 

	\begin{figure}[h] \begin{quotation}
\begin{center}\hskip-5mm \lowerwidthfig{0pt}{120mm}{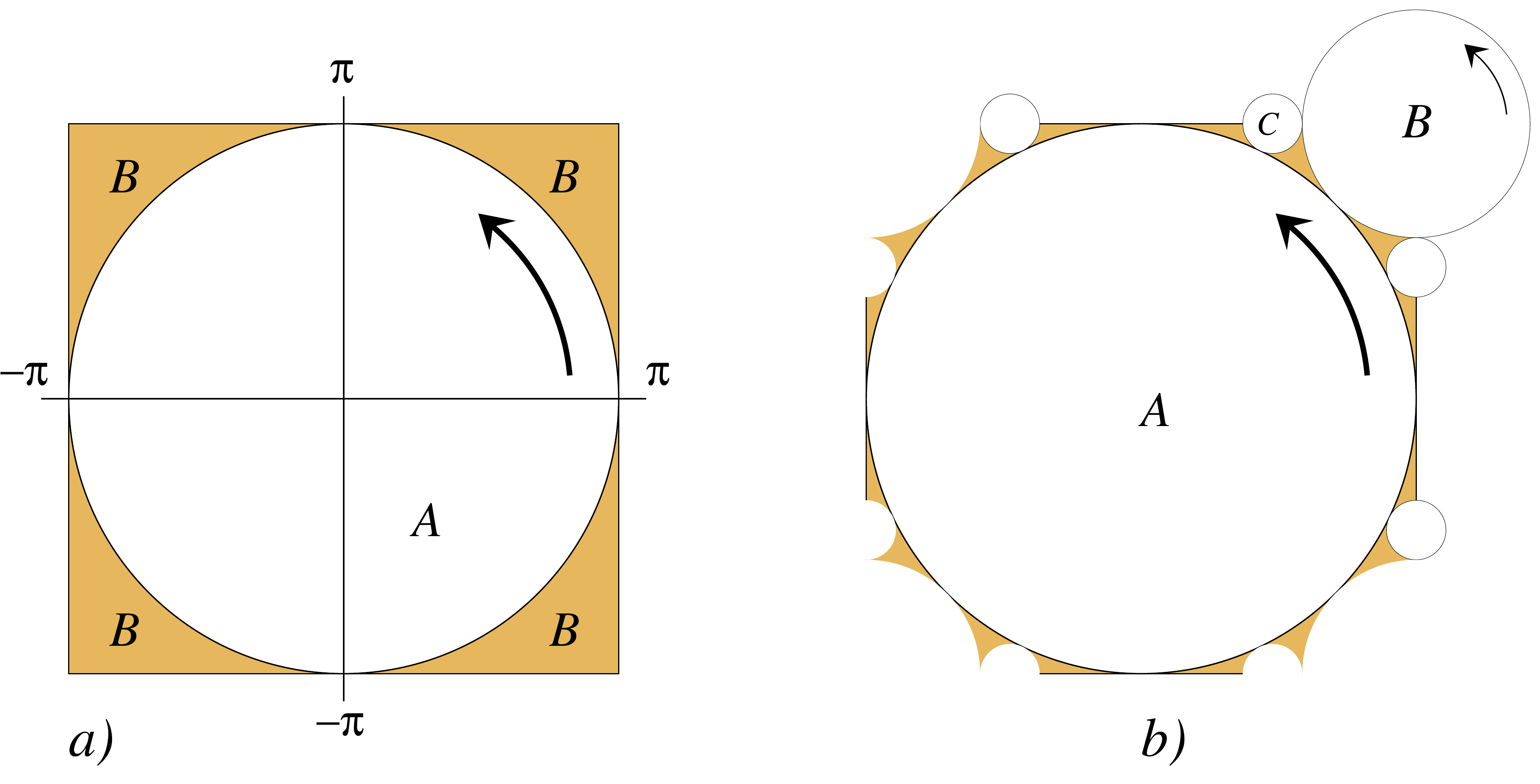}\qquad 
 \caption[\small Rotations in the Brillouin zone of a rectangular lattice.]{\small Rotations in the Brillouin zone of a rectangular lattice. $a)$ We can limit ourselves to the region inside the largest circle that fits in the Brillouin zone (A). The shaded region (B) is neglected and the amplitude there replaced by zero. This is good if strongly fluctuating modes in \(\vec x\) space may be ignored, such as in a photograph with a rectangular grid of pixels. $b)$ Unitarity is restored if we also fill the remainder of the Brillouin zone also with circles, \(B,\ C,\) etc., the larger the better (as explained in the text), but never overlapping.  In the picture, the shaded regions should also be filled with circles. The rotation operator must rotate every circle by the same angle \(\vv\) (arrows). \labell{circles.fig}}
\end{center}\end{quotation}
		\end{figure}

We fill the entire Brillouin zone with circular regions, such that they completely cover the entire space without overlappings. As will be explained shortly, we prefer to keep these circles as large as possible to get the best\fn{``best" here means that the effect of the rotation is maximally local, as will be seen in the sequel.} result. The action of the rotation operator will now be defined to correspond to a \emph{rotation over the same angle \(\vv\) inside all of these circles} (arrows in Figs.~\ref{circles.fig}$a$ and $b$). With ``circles" we here mean circular regions, or, if \(d>2\), regions bounded by \((d-1)\)-spheres.

This is -- nearly\fn{A slight complication that can be cured, is explained shortly after Eq.~\eqn{noncommTR} on page~\pageref{localrot}.} -- the best we can do in the Brillouin zone, being the space of the Fourier vectors \(\vec\k\). The reason why we split the Brillouin zones into perfectly spherical regions, rather than other shapes, becomes clear if we inspect the action of this operator in \(\vec x\)-space: how does this operator work in the original space of the lattice sites \(\vec x\)?

Let us first consider the action of a single circle, while the data on the rest of the Brillouin zone are replaced by zero. First take a circle (if \(d=2\)) or sphere (if \(d=3\)) whose centre is at the origin, and its radius is \(r\). Projecting out this circle means that, in \(\vec x\)-space, a wave function \(\j(\vec x)\) is smeared as follows:
	\be \j'(\vec x)&=& (2\pi)^{-d}\sum_{\vec x\,'}\int_{|\vec\k|<r}\dd^d\vec\k\,e^{i\vec\k\cdot(\vec x-\vec x\,')}\,\j(\vec x\,')\ =\nm \\ [3pt]
	&=& \sum_{\vec x\,'}\Big({r\over\pi}\Big)^dK_d\,({ \textstyle {r\over\pi}}|\vec x-\vec x\,'|)\,\j(\vec x\,')\ .\eel{circprojection}
	
The kernel of this rotationally symmetric expression turns out to be a Bessel function:	
	\be K_d\,(y)\iss{\pi^{d-1\over 2}\over 2^d\G({d+1\over 2})}\int_{-1}^{1}\dd k(1-k^2)^{d-1\over 2}e^{i\pi ky}\iss (2y)^{-d/2}J_{d/2}(\pi y)\ . \eel{Kernelbessel}
It is a smooth function, dropping off at infinity as a power of \(y\) (see Figure~\ref{rotkernel.fig}):
	\be K_d\,(y) \longrightarrow {2\sin\pi(y+{\textstyle{1-d\over 4}})\over \pi(2y)^{d+1\over 2}}\qquad\hbox{as}\qquad y\ra\infty\ . \ee

	\begin{figure}[h] \begin{quotation}
\begin{center}\hskip-5mm \lowerwidthfig{0pt}{120mm}{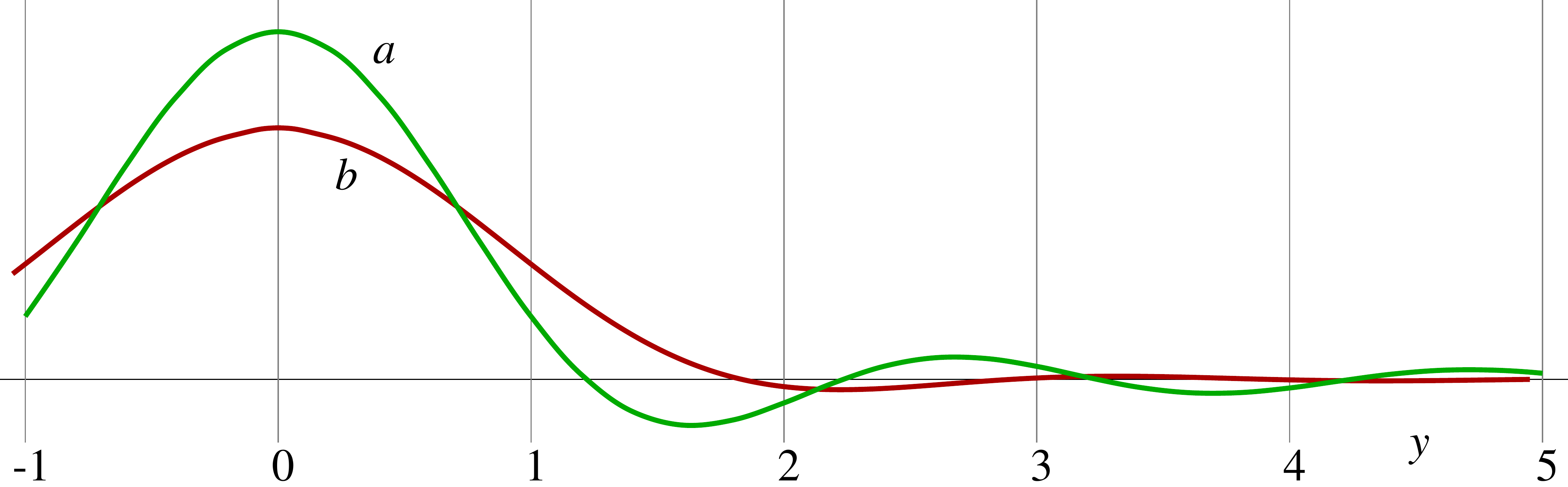}\qquad 
 \caption{\small The function $K_d\,(y)$, $a)$ for $d=2$, and $b)$ for $d=5$. \labell{rotkernel.fig}}
\end{center}\end{quotation}
		\end{figure}
We infer from Eq,~\eqn{circprojection} that, projecting out the inside of a circle with radius \(r\) in the Brillouin zone, implies smearing the data on the lattice over a few lattice sites in all directions, using the kernel \(K_d(y)\). The smaller the radius \(r\), the further out the smearing, which is why we should try to keep our circles (spheres) as large as possible.

Next, we notice that most of the circles in Fig.~\ref{circles.fig}$b$ are off-centre. A displacement by a vector \(\vec\k_1\) in the Brillouin zone corresponds to a multiplication in configuration space by the exponent \(e^{i\vec\k_1\cdot\vec x}\). Projecting out a circle with radius \(r\) and its origin on the spot \(\vec\k_1\) in the Brillouin zone, means dividing the wave function \(\j(\vec x)\) by the exponent \(e^{i\vec\k_1\cdot\vec x}\), smearing it with the kernel \(K_d (y)\), then multiplying with the exponent again (thus, we bring the circle to the origin, project out the centralised circle, then move it back to where it was). This amounts to smearing the original wave function with the modified kernel
	\be K_{d}\,(y,\vec\k_1)=K_d\,(y)e^{i\vec\k_1\cdot(\vec x-\vec x\,')}\ ,\qquad y=\textstyle{r\over\pi}|\vec x-\vec x\,'|\ . \eel{offcenterkernel}
	
If we add the projections of all circles with which we covered the Brillouin zone, the total effect should be that we recover the original wave function on the lattice.

And now we can rotate. Rotating a circle \((r,\,\k_1)\) in the Brillouin zone over any angle \(\vv\) has exactly the same effect as 1) finding the smeared wave function using the kernel \(K_d\,(y)\,e^{-i\vec\k_1\cdot\vec x\,'}\),  rotating the resulting continuous function over the angle \(\vv\) in \(\vec x\)-space, and then multiplying with the exponential \(e^{i\vec\k_1\cdot\vec x}\). If we add together the effects of all circles, we get the rotation operator. If we want the effect of an orthogonal rotation \(\W\) in \(\vec x\)-space, then this results in
	\be \j'(\vec x)=\sum_{\vec x\,'}\sum_{i}K_d\,({\textstyle{r_{i\vphantom{[}}\over\pi}}|\W\vec x-\vec x\,'|)\,e^{i\vec\k_i\cdot(\vec x-\vec x\,')}\,\j(\vec x\,')\ , \eel{contlatticerot}
where the index \(i\) counts the circles covering the Brillouin zone.
	
The transformations described in this subsection form a perfectly acceptable rotation group, converging to the usual rotations in the continuum limit. This can easily be seen by noting that the continuum case is dominated by the small values of \(\vec\k\), which are all in the primary circle. The other circles also rotate the wave functions to the desired location, but they only move along the rapidly oscillating parts, while the vectors \(\vec\k_1\) stay oriented in the original direction. 

The desired group properties of this operator follow from the fact that the circles cover the Brillouin zone exactly once. 
	\be\W_3=\W_1\W_2\ .\eel{rotationgroup}
Of course, the operation \eqn{contlatticerot}, to be referred to as \(R(\W)\), is not quite an ordinary rotation. If \(T(\vec a)\) is the translation over a non-lattice vector \(\vec a\) as described in section~\ref{conttransl.sec}, then
	\be R(\W)\,T(\vec a)\ne T(\W\vec a)\,R(\W)\ , \eel{noncommTR}
and furthermore, if \(\W\) is chosen to be one of the elements of the crystal group of the lattice, \(R(\W)\) does still not coincide with \(\W\) itself.\label{localrot} This latter defect can be cured, but we won't go into these details.

The best feature of this rotation operator is that it appears to act really locally in \(\vec x\)-space, spreading the lattice points only slightly with the Bessel function kernels \eqn{Kernelbessel}, but it also has disadvantages: it will be extremely difficult to construct some deterministic evolution law that respects this transformation as a symmetry. For this reason, we now consider other continuous transformation prescriptions that yield rotations.

\subsubsection{Continuous rotations 2: Using Noether charges and a discrete sub group\labell{Noetherdiscrete}}

In a deterministic theory then, we wish to identify an evolution law that respects our symmetries. This requires a different choice for the definitions of the symmetries involved.
To this end, we enter  Noether's theorem, as it was introduced at the beginning of this chapter. For example, symmetry under time translations is associated to the conservation of energy, translation symmetry is associated to momentum conservation, and rotation symmetry leads to the conservation of angular momentum. We refer to these conserved quantities as \emph{Noether charges}. All these conserved charges are observable quantities, and therefore, if we wish to investigate them in a quantum theory that we relate to a deterministic system, then this deterministic system should also exhibit observable quantities that can be directly related to the Noether charges.

In the \(PQ\) formalism, the Noether charges for translation symmetry are built in, in a sense. Translations in the \(Q_i\) variables are associated to quantities \(p_i\), of which the integer parts \(P_i\) are ontological observables. Only the fractional parts generate translations, which then must be integer steps on the \(Q\) lattice. We need both components of the momentum. If the lattice length is small, the quanta of the integer parts of the momenta are large. Planets have very large momenta, cold atoms have very small momenta. Large momenta also are sources of gravitational fields, and as such directly observable. What about the transition region? It happens to be in a very familiar domain -- the Planck unit of momentum is \(\sim 6.5\  \mathrm{kg\,m/sec}\). Momentum in that domain must be a mixture of the \(P\) observables and the \(Q\) displacement operators, whereas in ordinary physics we notice nothing special in that domain.

In the case of angular momentum, we may note that angular momentum is quantised anyway. Can we associate an ontological observable (beable) to angular momentum? Not so easily, because angular momentum consists of non-commuting components. At best we will have ontological quantised variables playing the role of ``the classical parts" of angular momentum, supplemented by quantum degrees of freedom (changeables) that restore the commutation rules. We observe that, for small particles, angular momentum is only partly observable; sometimes it is a beable, sometimes a changeable. For large systems, angular momentum is observable, with some margin of error.

This leads us to consider the following structure -- and indeed we will have to use similar methods \emph{whenever} a symmetry group becomes large, meaning that it has very many elements. Since angular momenta are non commutative, they cannot be quite ontological, but their `classical parts' must be. Therefore, we assume that the total angular momenta operators \(J_i\) can be written as follows:
	\be J_i=L_i+\l_i\ ,& [J_i,\,J_j]=i\e_{ijk}J_k\ ,	\eel{angmomsplit}
where \(L_i\) represents the expectation values of \(J_i\) in all ontological states, so that \(L_i\) are beables. 
The \(\l_i\) represent the remainder, and their  expectation values in ontological states vanish:
	\be\bra\ont|\l_i|\ont\ket=0\quad\hbox{ for each ontological state } |\ont\ket\ . \eel{ontolexpt}
The following subsection will show an explicit procedure to obtain \(L_i\) and \(\l_i\).

 \subsubsection{Continuous rotations 3: Using the real number operators $p$ and $q$ constructed out of $P$ and $Q$\labell{PQrotations}}

If our theory is defined on a lattice, there is another great way to recover many of the symmetries of the continuum case, by using the \(PQ\) trick as it was exposed in section~\ref{PQ}. We saw that string theory, section~\ref{sustr}, was re-written in such a way that the string moves on a lattice in target space, where the lattice basically describes the integer parts of the coordinates, while the space in between the lattice sites actually correspond to the eigenstates of the displacement operators for the momentum variables \(P\). Together, they form a continuum, and since the entire system is equivalent to the continuum string theory, it also shares all continuous translation and rotation symmetries with that theory.

By allowing the application of this mechanism, string theory appears to be more powerful than theories of point particles; the commutation rules for the operators in target space are fundamentally different, and string theory allows target space to be in a high number of dimensions. 

Thus, in the \(PQ\) formalism, we now use the continuum definition of angular momentum. Consider  the wave function of a single particle in three space dimensions, so that it lives on the product of three \(P,\,Q\) lattices. These lattices generate the three quantum coordinates \(q_i\). Its Hilbert space \(\HH\) is the product space of three Hilbert spaces \(\HH_1,\ \HH_2\) and \(\HH_3\). 

Write, as in Eqs.~\eqn{qopmatrix} and \eqn{popmatrix},
	\be q_i^{\op\,}= Q_i+a_i^{\op\,}\ , \qquad p_i^{\op\,}=2\pi P_i+b_i^{\op\,}\ , \eel{pqsplit1}
so that the angular momentum operator is (in the 3-dimensional case)
	\be J_i=\e_{ijk}q^{\op\,}_jp^{\op\,}_k=\e_{ijk}(2\pi Q_jP_k+2\pi a^{\op\,}_jP_k+Q_jb^{\op\,}_k+a^{\op\,}_jb^{\op\,}_k)\  . \eel{PQL}
Since the expectation values of \(a^\op_i\) and \(b^\op_i\) vanish in the ontological states,  \(|\ont\ket=|\vec P,\vec Q\ket\),  and since the last term will be \(\le\OO(2\pi)\), we can identify \(L_i\) with the first term:
	\be L_i\approx 2\pi\,\e_{ijk}Q_jP_k\ . \eel{Lontapprox}

Note, that the \(L_i\) are quantised in multiples of \(2\pi\) rather than one, as one might have expected, so Eq.~\eqn{Lontapprox} cannot hold exactly.

Let us now inspect the modifications on the commutation rules of these angular momentum operators caused by the edge states. In each of the three Hilbert spaces \(\HH_i\), \(i=1,\,2,\,3,\), we have Eq.~\eqn{PQedge}, while the operators of one of these Hilbert spaces commute with those of the others. Writing the indices explicitly:
	\be [q_1,\,p_1]=i\,{\mathbb I}_2{\mathbb I}_3\,(\,{\mathbb I}_1-|\j_e^1\ket\bra\j_e^1|\,)\ ,\quad [q_1,\,p_2]=0\ ,\quad\hbox{and\ cyclic\ permutations,} \eel{PQcomm3d}		
where \({\mathbb I}_i\) are the identity operators in the \(i^\th\) Hilbert space, and \(|\j_e^i\ket\) are the edge states on the \(i^\th\) \(P,\,Q\) lattice.
One then easily derives that the three angular momentum operators \(J_i\) defined in the usual way, Eq.~\eqn{PQL}, obey the commutation rules
	\be[J_1,\,J_2]=iJ_3\,{\mathbb I}_1{\mathbb I}_2\,(\,{\mathbb I}_3-|\j_e^3\ket\bra\j_e^3|\,)\ ,\quad\hbox{and cyclic permutations.}\eel{LLcommwithedge}
	The importance of this result is that now we observe that the operator \(J_3\) only acts in Hilbert spaces 1 and 2, but is proportional to the identity in \(\HH_3\) (since \(J_3\) contains only \(q_1,\,q_2,\,p_1,\) and \(p_2\)). So the projection operator for the edge state \(|\j_e^3\ket\) commutes with \(J_3\). This implies that, if we limit ourselves to states that are orthogonal to the edge states, they will also rotate to states orthogonal to the edge states. In this subspace of Hilbert space the rotations act normally. And we think that this is remarkable, because certainly the ``ontological" basis defined on the six-dimensional \(\vec P,\,\vec Q\) lattice has no built-in continuous rotation invariance at all.
	
\subsubsection{Quantum symmetries and classical evolution\labell{classqusymm} }
In previous subsections it was observed that, when we project classical models on Hilbert spaces, new symmetries may emerge. These are symmetry transformations that map classical states onto superpositions of states. A few examples were shown. 

None of our procedures are fool proof. In the special case to be discussed next, we study time translation invariance. As stated earlier, we might split the energy \(E\) into a classical part (\(\d E\)) and a quantum part (the generator of discrete time translations, the Hamiltonian \(H\) that lies in the interval \([0,\,2\pi/\d t)\). However, this would suggest that we can only measure energies with \(2\pi/\d t\) as our margin of error. That cannot be right: if \(\d t\) is the Planck time, then the energy quantum is the Planck energy, \(E_{\mathrm{Planck}}\), which is about 543 kiloWatt-hours; yet we pay our electricity bills per kiloWatt-hour, and those bills are certainly ontological. Mutations in our DNA profiles might require only a couple of electronVolts to take place, and these might be crucial for our genetically inherited identities; an electronVolt is about \(10^{-28}\) times the Planck energy. Even that may have to be (mostly) ontological.

Of course we are primarily interested in symmetries that are symmetries of the evolution operator. The cogwheel model, section~\ref{cogwheelN}, for instance has the classical symmetry of rotations over \(N\) steps, if \(N\) is the number of cogwheel position states. But if we go to the energy eigenstates \(|k\ket_H, \ k=0,\cdots N-1\) (Eqs.~\ref{Hstates} and \ref{ontstates}), we see that, there, a translation over \(n\) teeth corresponds to multiplication of these states as follows:
	\be |k\ket_H\ra e^{2\pi i kn/N}|k\ket_H\ . \eel{translinenergybasis}
Since these are eigenstates of the Hamiltonian, this multiplication commutes with \(H\) and hence the symmetry is preserved by the evolution law. 

We now found out that we can enlarge the symmetry group by choosing the multiplication factors in frequency space
	\be |k\ket_H\ra e^{2\pi i k\a/N}|k\ket_H\ . \eel{conttransl}
where \(\a\) now may be any real number, and this also corresponds to a translation in time over the real number \(\a\). This enhances the symmetry group from the group of the cyclic permutations of \(N\) elements to the group of the continuous rotations of a circle.

\subsubsection{Quantum symmetries and classical evolution 2\labell{classqusymm2} }

An other rather trivial yet interesting example of a symmetry that is enlarged if we apply our quantum constructions, occurs in a simple cellular automaton in any number \(d\) of space dimensions. Consider the Boolean variables 
\(\s(\vec x,\,t)=\pm 1\) distributed over all \emph{even} sites in  a lattice space-time, that is, over all points \((\vec x,\,t)=(x_1,\cdots ,\,x_d,\,t)\) with \(x_i\) and \(t\) all integers, and \(x_1+\cdots+x_d+t\ =\) even.

Let the evolution law be
	\be\s(\vec x,\,t+1)=\Big(\prod_{i=1}^d\s(\vec x+\vec e_i\,,\,t)\ \s(\vec x-\vec e_i\,,\, t)\Big)\,\s(\vec x\,,\,t-1)\ , \eel{productCA}
where \(\vec e_i\) are the unit vectors in the \(i^\th\) direction in \(d\) dimensional space. Or: the product of the data on all direct space-time neighbours of 
any odd site \((\vec x,\,t)\) is \(+1\). This law is manifestly invariant under time reversal, and we see that it fixes all variables if the data are given on a Cauchy surface consisting of two consecutive layers in time \(t,\ t-1\). The classical  model has the manifest translation symmetry over vectors \(\d x=(a_1,\cdots,\,a_d,\,\t)\) with \(\sum_ia_i+\t\) even.

Now let us introduce Hilbert space, and consider the \emph{odd} lattice sites. On these odd sites, we define the action of changeables \(\s_1(\vec x_1,\,t_1)\) as follows:
	\begin{quotation}\noindent The data on the time frame \(t=t_1\), are kept unchanged; \\ on the time frame \(t=t_1-1\), only \(\s(\vec x_1,\,t_1-1)\) changes sign, and all others remain unchanged; \\ consequently, according to the evolution law, also on the time frame \(t=t_1+1\), only \(\s(\vec x_1,\,t_1+1)\) changes sign, all others stay the same. \end{quotation}
\noindent The reason for the notation \(\s_1\) is that in a basis of Hilbert space where \(\s(\vec x_1,\,t_1-1)=\s_3=({1\ \ 0\atop 0\ -1})\), our new operator is \(\s_1(\vec x,\,t_1)=\s_1=({0\ \ 1\atop 1\ \ 0})\), as in the Pauli matrices.

Now, checking how the action of \(\s_1(\vec x,\,t)\) propagates through the lattice, we observe that 
	\be \s_1(\vec x,\,t+1)=\Big(\prod_{i=1}^d\s_1(\vec x+\vec e_i\,,\,t)\ \s_1(\vec x-\vec e_i\,,\, t)\Big)\,\s_1(\vec x\,,\,t-1)\ , \eel{productoddsites}
where now the vector \((\vec x,\,t)\) is even, while in Eq.~\eqn{productCA} they were odd. Thus the product of the changeables \(\s_1(\vec x',\,t')\) that are direct space-time neighbours of an even site \((\vec x,\,t)\) is also one.

\begin{figure}[h!]
\begin{center} \lowerwidthfig{0pt}{35mm}{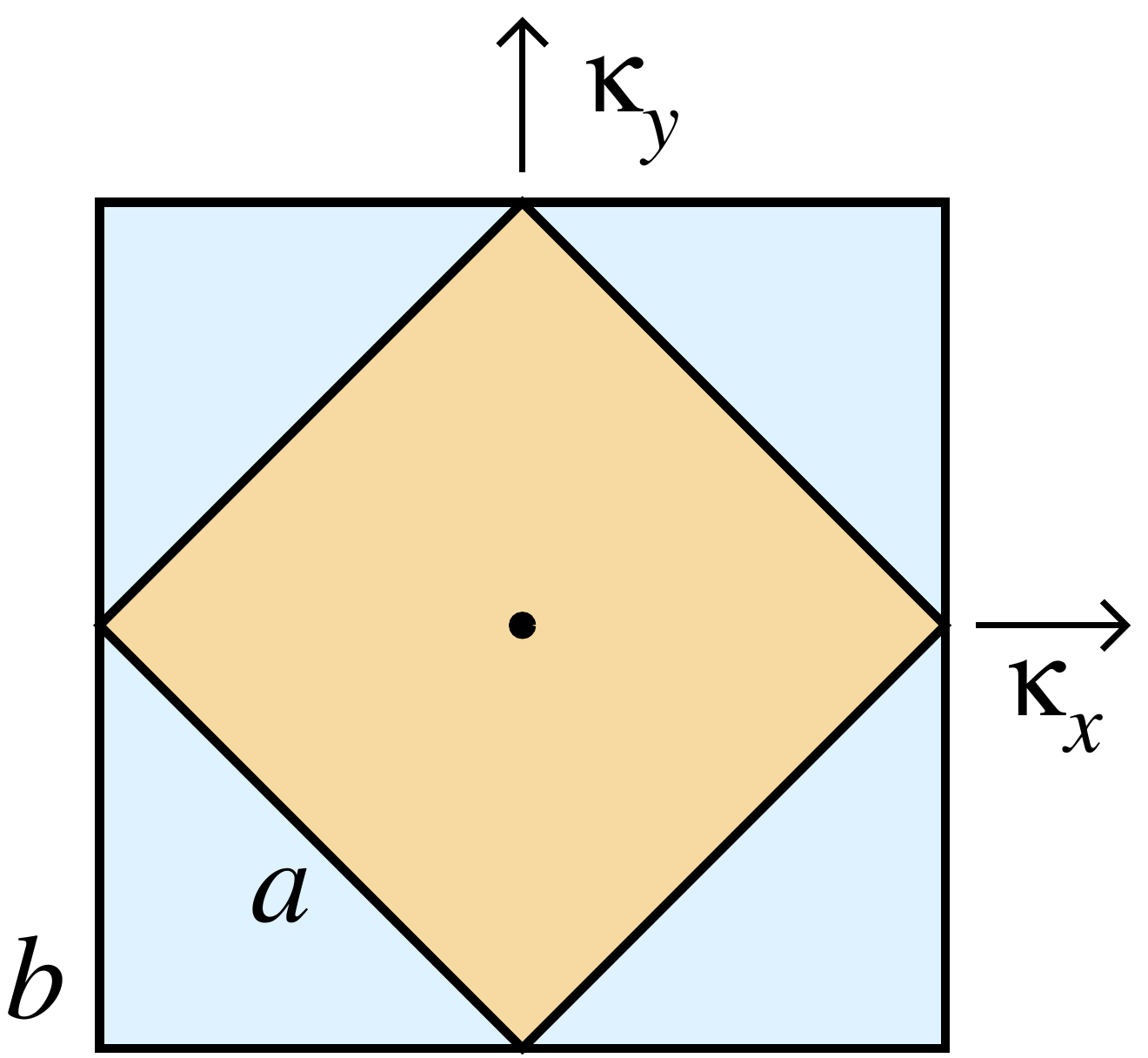} \begin{quote} 
 \caption[\small The Brillouin zones for the lattice momentum $\vec\k$ of the ontological model described by Eq.~\eqn{productCA} in two dimensions. $a)$ the ontological model, $b)$ its Hilbert space description.]{\small $a)$ The Brillouin zone for the lattice momentum \(\vec\k\) of the ontological model described by Eq.~\eqn{productCA}, in two space dimensions, with data only on the even lattice sites (smaller square, tilted by \(45^\circ\)), and $b)$ the Brillouin zone for the Hilbert space description of this model (larger square) \label{brillouin.fig}}
 \end{quote}
\end{center}
\end{figure}	

Since we recovered the same evolution law but now on the sites that before were empty, our translation symmetry group now has twice as many elements. Now, we can perform a translation over a vector, whose sum of components is odd, but the states in Hilbert space then have to undergo a transformation; at every site: 
	\be|\j(\vec x,\,t)\ket\ra U_\op |\j(\vec x,\,t)\ket\ ,\quad U_\op\,\s_1U_\op^{-1}=\s_3\ ;\quad U_\op=\fract 1{\sqrt{2}}(\textstyle{1\ \ 1\atop 1\ \,-1})\ . \eel{evenoaddtrf}
Since \(U^2=1\), this is actually a reflection. This means that the succession of two odd translations gives an even translation without further phase changes.

This simple model shows how the introduction of Hilbert space may enhance the symmetry properties of a theory. In this case it also implies that the Brillouin zone for momentum space becomes twice as large (see Fig.~\ref{brillouin.fig}). A quantum physicist living in this world will not be able to distinguish the even sites from the odd ones.

\subsecl{Large symmetry groups in the CAI}{large}

We end this chapter with a general view of large symmetry groups, such as translations in space and in time, and the Lorentz group. They have infinite numbers of group elements. Now we imagine our automaton models to have discretised amounts of information spread over space and time. How can we have infinite and/or continuous symmetry groups act on them? 

Our impression from the previous results is that the conventional symmetry generators, as used in quantum theories, will be operators that always consist of \emph{combinations} of beables and changeables: the Noether charges, such as angular momentum, energy and momentum, will have classical limits that are perfectly observable, hence they are beables; yet quantum mechanically, the operators do not commute, and so there must also be changeable parts. 

The beable parts will be conjugated to the tiniest symmetry operations such as very tiny translations and rotations. These are unlikely to be useful as genuine transformations among the ontological data -- of course they are not, since they must commute with the beables. 

The changeable parts of these operators are not ontological observables as they do not commute. The \(P\,Q\) formalism, elaborated in section~\ref{PQ}, is a realisation of this concept of splitting the operators: here, both in position space and in momentum space,  the integer parts of the translation operators are beables, the fractional parts are changeables. The continuous translation operators \(p^\op\) consist of both ingredients. We suspect that this will have to become a general feature of all large symmetry groups, in particular the Hamiltonian itself, and this is what we shall attempt to implement in the next chapter, \ref{Hamiltonform}.

\newsecl{The  discretised Hamiltonian formalism in $PQ$ theory}{Hamiltonform}

\subsecl{The vacuum state, and the double role of the Hamiltonian (cont'd)}{doubleHam}
The energy conservation law is usually regarded as an interesting and important feature of both classical and quantum mechanics, but it is often not fully realised how important the role of this law really is. The importance of energy is that it is conserved, it is defined locally, and that it cannot be negative\fn{Often, the \emph{Casimir effect} is brought forward as a counter example. Of course, it is important to realise that this effect can produce small regions of negative energy, but those regions are always accompanied by domains of much larger amounts of positive energy nearby, so that this effect has little impact on the fundamental issues of stability raised here.}. This allows us to define the \emph{vacuum} as the single quantum state of the universe that has the lowest possible energy (or energy per unit of volume).

Consider a small perturbation of this vacuum: a light particle, or a grain of dust. It carries only a small amount of energy. In our world, this energy cannot increase spontaneously, because the surrounding vacuum cannot deliver it, and its own energy cannot increase. All transitions, all processes inside the grain of dust, can only transform the object into other states with exactly the same energy. If the object decays, the decay products must have even lower amounts of energy. Since the number of distinct states with the same or less energy is very limited, not much can happen; the object represents a very \emph{stable} situation.

But now imagine an alien world where the concept of a conserved, positive energy would not exist.   Perhaps our alien world would nevertheless have something like a vacuum state, but it would have to be defined differently. In this alien world, our tiny object could grow spontaneously, since we postulated that there is no conserved quantity such as energy to stop it from doing so. What this means is that the tiniest perturbations around the vacuum state will destabilise this vacuum. Similarly, any other initial state may turn out to be unstable\fn{The absence of a stabiliser does not imply that a dynamical system \emph{has} to destabilise; the solar system is a classical case in point, it stayed in roughly the same state for billions of years, without any conspicuous reason for not converting into a more ``probable" state. Therefore, the argument presented here must be handled with care.}.

We can state this differently: solutions of the equations of motion are stationary if they are in thermal equilibrium (possibly with one or more  chemical potentials added). In a thermal equilibrium, we have the Boltzmann distribution: 
	\be W_i=C\,e^{-\b E_i+\sum_j\m_jR_{ji}}\ , \eel{Boltzmann}
where \(\b=1/kT\) is the inverse of the temperature \(T\), with Boltzmann constant \(k\), and \(i\) labels the states; \(\m_j\) are chemical potentials, and \(R_{ji}\) the corresponding conserved quantities.

If the energies \(E_i\) were not properly bounded from below, the lowest energies would cause this expression to diverge, particularly at low temperatures.

What is needed is a lower bound of the energies \(E_i\) so as to ensure stability of our world. Furthermore, having a ground state is very important to construct systematic approximations to solutions of the time-independent Schr\"odinger equation, using extremum principles. This is not just a technical problem, it would raise doubt on the mere existence of correct solutions to Schr\"odinger's equation, if no procedure could be described that allows one to construct such solutions systematically.

In our world we do have a Hamiltonian function, equal to the total energy, that is locally conserved and bounded from below. Note that ``locally conserved" means that a locally defined tensor \(T_{\m\n}(\vec x,t)\) exists that obeys a local conservation law, \(\pa_\m T_{\m\n}=0\), and this feature is connected in important ways not only to the theory of {special relativity}, but also to \emph{general} relativity. 

Thus, the first role played by the Hamiltonian is that it brings \emph{law and order} in the universe, by being 1) conserved in time, 2) bounded from below, and 3) local (that is, it is the sum of completely localised contributions).

Deriving an equation of motion that permits the existence of such a function, is not easy, but was made possible by the Hamiltonian procedure, first worked out for continuum theories (see section \ref{doubleham} in Part I). 

Hamilton's equations are the most natural ones that guarantee this mechanism to work: first make a judicious choice of kinetic variables \(x_i\) and \(p_i\), then start with any function \(H(\{x_i,\,p_j\})\) that is bounded and local as desired, and subsequently write down the equations for \(\dd x_i/\dd t\) and \(\dd p_j/\dd t\) that guarantee that \(\dd H/\dd t=0\). The principle is then carried over to quantum mechanics in the standard way. 

Thus, in standard physics, we have a function or operator called Hamiltonian that represents the conserved energy on the one hand, and it generates the equations of motion on the other. 

And now, we argue that, being such a fundamental notion, the Hamiltonian principle should also exist for discrete systems. 

\subsecl{The Hamilton problem for discrete deterministic systems}{discHamproblem}
	
Consider now a discrete, deterministic system. Inevitably, time will also be discrete. Time steps must be controlled by a deterministic evolution operator, which implies that there must be a smallest time unit, call it \(\d t\). When we write the evolution operator \(U(\d t)\) as \(U(\d t)=e^{-i E^{\mathrm{\,quant}}\,\d t}\) then \(E^{\mathrm{\,quant}}\) is defined \emph{modulo} \(2\pi/\d t\), which means that we can always choose \(E^{\mathrm{\,quant}}\) to lie in the segment
	\be 0\le E^{\mathrm{\,quant}}<2\pi/\d t\ , \eel{energyperiod}
	
Instead, in the real world, energy is an additively conserved quantity without any periodicity. In the \(PQ\) formalism, we have seen what the best way is to cure such a situation, and it is natural to try the same trick for time and energy: we must add a conserved, discrete, integer quantum to the Hamiltonian operator: \(E^{\mathrm{\,class}}=2\pi N/\d t\), so that we have an absolutely conserved energy,
	\be E\qu E^{\mathrm{\,quant}}+E^{\mathrm{\,class}}\ .\eel{totalenergy}
In the classical theory, we can only use \(E^\mathrm{\,class}\) to ensure that our system is stable, as described in the previous section.

In principle, it may seem to be easy to formulate a deterministic classical system where such a quantity \(E^{\mathrm{\,class}}\) can be defined, but, as we will see, there will be some obstacles of a practical nature. Note that, if Eq.~\eqn{totalenergy} is used to define the total energy, and if \(E^{\mathrm{\,class}}\) reaches to infinity, then time can be redefined to be a continuous variable, since now we can substitute any value \(t\) in the evolution operator \(U(t)=e^{-iEt}\).

One difficulty can be spotted right away: usually, we shall demand that energy be an \emph{extensive} quantity, that is, for two widely separated systems we expect
	\be E^{\mathrm{\,tot}}=E_1+E_2+E^{\mathrm{\,int}}\ , \eel{energysum}
where \(E^{\mathrm{\,int}}\) can be expected to be small, or even negligible. But then, if both \(E_1\) and \(E_2\) are split into a classical part and a quantum part, then either the quantum part of \(E^{\mathrm{\,tot}}\)will exceed its bounds \eqn{energyperiod}, or \(E^{\,\mathrm{class}}\) will \emph{not} be extensive, that is, it will not even approximately be the sum of the classical parts of \(E_1\) and \(E_2\).	

An other way of phrasing the problem is that one might wish to write the total energy \(E^{\mathrm{\,tot}}\) as
	\be E^{\mathrm{\,tot}}=\sum_{\mathrm{lattice\ sites\,} i}E_i\ra\int\dd^d\vec x\ \HH(\vec x)\ , \eel{energyintegral}
where \(E_i\) or \(\HH(\vec x)\) is the energy density. It may be possible to spread \(E^{\mathrm{\,tot}}_{\mathrm{\,class}}\) over the lattice, and it may be possible to rewrite \(E^{\mathrm{\,quant}}\) as a sum over lattice sites, but then it remains hard to see that the total quantum part stays confined to the interval \([0,\,2\pi/\d t)\) while it is treated as an extensive variable at the same time. Can the excesses be stowed in \(E^{\mathrm{\,int}}\)? 

This question will be investigated further in our treatment of the technical details of the cellular automaton, chapter~\ref{quloc}.
	
\subsection{Conserved classical energy in $PQ$ theory\labell{classenergy}}

	If there is a conserved classical energy \( E^{\mathrm{class}}(\vec P,\,\vec Q)\), then the set of \(\vec P,\,\vec Q\) values with the same total energy \(E\) forms closed surfaces \(\SS_E\). All we need to demand for a theory in \((\vec P,\,\vec Q)\) space	is that the finite-time evolution operator \(U(\d t)\) generates motion along these surfaces\,\cite{GtH-2013}. That does not sound hard, but in practice, to generate evolution laws with this property is not so easy. This is because we often also demand that our evolution operator \(U(\d t)\) be time-reversible: there must exist an inverse, \(U^{-1}(\d t)\).
	
	In classical mechanics of continuous systems, the problem of characterising some evolution law that keeps the energy conserved was solved: let the continuous degrees of freedom be some classical real numbers \(\{q_i(t),\,p_i(t)\}\), and take energy \(E\) to be some function
	\be E=H(\vec p,\,\vec q\,)=T(\vec p\,)+V(\vec q\,)+\vec p\cdot\vec A(\vec q\,)\ , \eel{HTV}
although more general functions that are bounded below are also admitted. The last term, describing typically magnetic forces, often occurs in practical examples, but may be omitted for simplicity to follow the general argument.

Then take as our evolution law:
	\be {\dd q_i\over\dd t}=\dot q_i={\pa H(\vec p,\vec q\,)\over\pa p_i}\ , \qquad \dot p_i=-{\pa H(\vec p,\vec q\,)\over\pa q_i}\ . \eel{classhameqs} 
One then derives
	\be{\dd H(\vec p,\vec q\,)\over\dd t}=\dot H=	{\pa H\over\pa q_i}\dot q_i+{\pa H\over\pa p_i}\dot p_i=\dot p_i\dot q_i-\dot q_i\dot p_i=0\ . \eel{hamconservation}
	
This looks so easy in the continuous case that it may seem surprising that this principle is hard to generalise to the discrete systems. Yet \emph{formally} it should be easy to derive some energy-conserving evolution law:
\begin{quote} Take a lattice of integers \(P_i\) and \(Q_i\), and some bounded, integer energy function \(H(\vec P,\,\vec Q)\). Consider some number \(E\) for the total energy. 
Consider all points of the surface \(\SS_E\) on our lattice defined by \(H(\vec P,\,\vec Q)=E\). The number of points on such a surface could be infinite, but let us take the case that it is finite. Then simply consider a path \(P_i(t),\,Q_i(t)\) on \(\SS_E\), where \(t\) enumerates the  integers. 
The path must eventually close onto itself. This way we get a closed path on \(\SS_E\). If there are points on our surface that are not yet on the closed path that we just constructed, then we repeat the procedure starting with one of those points. Repeat until \(\SS_E\) is completely covered by closed paths. 
These closed paths then define our evolution law. 
\end{quote}
At first sight, however, generalising the standard Hamiltonian procedure now seems to fail. Whereas the standard Hamiltonian formalism \eqn{hamconservation} for the continuous case involves just infinitesimal time steps and infinitesimal changes in coordinates and momenta, we now need finite time steps and finite changes. One could think of making finite-size corrections in the lattice equations, but that will not automatically work, since odds are that, after some given time step,  integer-valued points in the surface \(\SS_E\) may be difficult to find. Now with a little more patience, a systematic approach can be formulated, but we postpone it to section~\ref{genintHamilton}.

\subsubsection{Multi-dimensional harmonic oscillator\labell{multiddiscreteham}}

A superior procedure will be discussed in the next subsections, but first let us consider the simpler case of the multi-dimensional harmonic oscillator of section~\ref{highd}, subsection~\ref{oscilform}: take two symmetric integer-valued tensors \(T_{ij}=T_{ji}\), and \(V_{ij}=V_{ji}\). The evolution law alternates between integer and half-odd integer values of the time variable \(t\). See Eqs,~\eqn{Qgeneral} and  \eqn{Pgeneral}:
	\be Q_i(t+1)&=&Q_i(t)+T_{ij}P_j(t+\half)\ ; \crl{Qgeneral1}
		P_i(t+\half)&=&P_i(t-\half)-V_{ij}Q_j(t)\ . \eel{Pgeneral1}
According to Eqs.~\eqn{ham1int}, \eqn{ham2odd}, \eqn{ham2squares} and \eqn{ham2squaresalt}, the conserved classical Hamiltonian is
		\be H&=&\half T_{ij}\,P_i(t+\half)\,P_j(t-\half)+\half V_{ij}\,Q_i(t)\,Q_j(t)\ =\nm\\
		&&\half T_{ij}\,P_i(t+\half)\,P_j(t+\half)+\half V_{ij}\,Q_i(t)\,Q_j(t+1)\ =\nm\\
		&&\half\vec P^+T\vec P^++\half\vec P^+TV\vec Q+\half \vec QV\vec Q\ =\nm\\
		&&\half (\vec P^++\half\vec Q\,V)\,T\,(\vec P^++\half V\,\vec Q)+\vec Q(\half V-\fract18 V\,T\,V)\vec Q\ =\nm\\
		&&\vec P^+(\half T-\fract18 T\,V\,T)\vec P^++\half (\vec Q+\half\vec P^+\,T)\,V\,(\vec Q+\half T\,\vec P^+)\ , \eel{hamoschighd}
where in the last three expressions, \(\vec Q=\vec Q(t)\) and \(\vec P^+=\vec P(t+\half)\).
The equations \eqn{hamoschighd} follow from the evolution equations \eqn{Qgeneral1} and \eqn{Pgeneral1} provided that \(T\) and \(V\) are symmetric.

One reads off that this Hamiltonian is time-independent. It is bounded from below if not only \(V\) and \(T\) but also either \(V-\quart VTV\) or \(T-\quart TVT\) are bounded from below (usually, one implies the other).

Unfortunately, this requirement is very stringent; the only solution where this energy is properly bounded is a linear or periodic chain of coupled oscillators, as in our one-dimensional model of massless bosons. On top of that, this formalism only allows for strictly harmonic forces, which means that, unlike the continuum case, no non-linear interactions can be accommodated for. A much larger class of models will be exhibited in the next section.

Returning first to our model of massless bosons in \(1+1\) dimensions, section~\ref{2dnoint}, we note that the classical evolution operator was defined over time steps \(\d t=1\), and this means that, knowing the evolution operator specifies the Hamiltonian eigenvalue op to multiples of \(2\pi\). This is exactly the range of a single creation or annihilation operator \(a^{L,R}\) and \(a^{L,R\,\dag}\). But these operators can act many times, and therefore the total energy should be allowed to stretch much further. This is where we need the exactly conserved discrete energy function \eqn{hamoschighd}. The fractional part of \(H\), which we could call \(E^{\mathrm{\,quant}}\), follows uniquely from the evolution operator \(U(\d t)\). Then we can add multiples of \(2\pi\) times the energy \eqn{hamoschighd} at will. This is how the entire range of energy values of our 2 dimensional boson model results from our mapping. It cannot be a coincidence that the angular energy function \(E^{\mathrm{\,quant}}\) together with the conserved integer valued energy function \(E^{\mathrm{\,class}}\) taken together exactly represent the spectrum of real energy values for the quantum theory. This is how our mappings work.

\def\class{{\,\mathrm{class}}} \def\quant{{\,\mathrm{quant}}}

\subsection{More general, integer-valued Hamiltonian models with interactions\labell{genintHamilton}}  \def\cl{{\,\mathrm{class}}}
According to the previous section, we recuperate quantum models with a continuous time variable from a discrete classical system if not only the evolution operator over a time step \(\d t\) is time-reversible, but in addition a conserved discrete energy beable \(E^\cl\) exists, taking values \(2\pi\,N/\d t\) where \(N\) is integer. 
Again, let us take \(\d t=1\). If the eigenvalues of \(U^\op(\d t)\) are called \(e^{-iE^\quant}\), with \(0\le E^\quant<2\pi\) then we can define the complete Hamiltonian \(H\) to be
	\be H=E^{\,\mathrm{quant}}+ E^\class\iss 2\pi(\n+N)\ , \eel{EquantplusEclass}		
where \(0\le\n<1\) (or alternatively, \(-\halff<\n\le \halff\)) and \(N\) is integer.
The quantity conjugated to that is a continuous time variable. If we furthermore demand that \(E^{\,\mathrm{class}}\) is bounded from below then Eq.~\eqn{EquantplusEclass}  defines a genuine quantum system with a conserved Hamiltonian that is bounded from below.

As stated earlier, it appears to be difficult to construct explicit, non-trivial examples of such models. If we try to continue along the line of harmonic oscillators, perhaps with some non-harmonic forces added, it seems that the standard Hamiltonian  formalism fails when the time steps are finite, and if we find a Hamiltonian that is conserved, it is usually  not bounded from below. Such models then are unstable; they will not lead to a quantum description of a model that is stable. 

In this section, we shall show how to cure this situation, in principle.  We concentrate on the construction of a Hamiltonian principle that keeps a classical energy function \(E^{\,\mathrm{class}}\) exactly conserved in time. 

In the multidimensional models, we had adopted the principle that we in turn update all variables \(Q_i\), then all \(P_i\). That has to be done differently.  
To obtain better models, let us phrase our assignment as follows: \\[5pt]
\emph{Formulate a discrete, classical time evolution law for some model with the following properties:}
	\bi{$i$} The time evolution operation must be a law that is reversible in time\fn{When information loss is allowed, as in section~\ref{infoloss} of part I, we shall have to relax this condition.}. Only then will we have an operator \(U(\d t)\) that is unitary and as such can be re-written as the exponent of \(-i\) times a hermitean Hamiltonian.
	\itm{$ii$} There must exist a discrete function \(E^{\,\mathrm{class}}\) depending on the dynamical variables of the theory, that is exactly conserved in time.
	\itm{$iii$} This quantity \(E^{\,\mathrm{class}}\) must be bounded from below. \ei
When these first three requirements are met we will be able to map this system on a quantum mechanical model that may be physically acceptable. But we want more:
	\bi{$iv$} Our model should be sufficiently generic, that is, we wish that it features interactions.
	\itm{$v$} Ideally, it should be possible to identify variables such as our \(P_i\) and \(Q_i\) so that we can compare our model with systems that are known in physics, where we have the familiar Hamiltonian canonical variables \(\vec p\) and \(\vec q\).\
	\itm{$vi$} We would like to have some form of \emph{locality}; as in the continuum system, our Hamiltonian should be described as the integral (or sum) of a local Hamiltonian density, \(\HH(\vec x)\), and there should exist a small parameter \(\e>0\) such that at fixed time \(t\), \(\HH(\vec x)\) only depends on variables located at  \(\vec x\,'\) with \(|\vec x\,'-\vec x|<\e\).
	\ei
	
\noindent The last condition turns our system in some discretised version of a field theory (\(\vec P\) and \(\vec Q\) are then fields depending on a space coordinate \(\vec x\) and of course on time \(t\)).
One might think that it would be hopeless to fulfil all these requirements. Yet there exist beautiful solutions which we now construct.  Let us show how our reasoning goes. 

Since we desire an integer-valued energy function that looks like the Hamiltonian of a continuum theory, we start with a Hamiltonian that we like, being a continuous function \(H_{\mathrm{cont}}(\vec q,\,\vec p\,)\) and take its integer part, when also \(\vec p\) and \(\vec q\) are integer. More precisely (with the appropriate factors \(2\pi\), as in Eqs.~\eqn{pqsplit} and \eqn{pqsplit1} in previous chapters): take \(P_i\) and \(Q_i\) integer and write\fn{Later, in order to maintain some form of locality, we will prefer to take our `classical' Hamiltonian to be the sum of many integer parts, as in Eq.~\eqn{Hdensitydiscr}, rather than the floor of the sum of local parts, as in Eq.~\eqn{discrham}.} \def\intt{{\,{\mathrm{int}}}}
		\be E^{\,\mathrm{class}}(\vec Q,\,\vec P)=2\pi H^\class(\vec Q,\,\vec P)\ ,\quad H^\class(\vec Q,\,\vec P)=
	\intt (\fract 1{2\pi} H_{\mathrm{cont}}(\vec Q,\,2\pi\vec P))\ , \eel{discrham}
where `int' stands for the integer part, and 
	\be Q_i=\intt(q_i)\ ,\qquad P_i=\intt(p_i/2\pi)\ ,\quad\hbox{for all }\,i\ . \eel{intparts}	

This gives us a discrete, classical `Hamiltonian function' of the integer degrees of freedom \(P_i\) and \(Q_i\). The index \(i\) may take a finite or an infinite number of values (\(i\) is finite if we discuss a finite number of particles, infinite if we consider some version of a field theory).

Soon, we shall discover that not all classical models are suitable for our construction: first of all: \emph{the oscillatory solutions must oscillate sufficiently slowly to stay visible in our discrete time variable}, but, as we shall see, our restrictions will be somewhat more severe than this.

It will be easy to choose a Hamiltonian obeying these (mild) constraints, but what are the Hamilton equations? Since we wish to consider discrete time steps (\(\d t=1\)), the equations have to be rephrased with some care. As is the case in the standard Hamiltonian formalism, the primary objective that our equations of motion have to satisfy is that the function \(H(\vec Q,\vec P)=E^{\,\mathrm{class}}\) must be conserved. Unlike the standard formalism, however, the changes in the values \(\vec Q\) and \(\vec P\) at the smallest possible time steps cannot be kept infinitesimal because both time \(t\) and the variables \(\vec Q\) and \(\vec P\) contain integer numbers only. 

The evolution equations will take the shape of a computer program. At integer time steps with intervals \(\d t\), the evolution law will ``update" the values of the integer variables \(Q_i\) and \(P_i\). Henceforth, we shall use the word ``update" in this sense. The entire program for the updating procedure is our evolution law.

As stated at the beginning of this section, it should be easy to establish such a program: compute the total energy \(E\) of the initial state, \(H(\,\vec Q(0),\,\vec P(0)\,)=E\). Subsequently, search for all other values of \((\vec Q,\,\vec P)\) for which the total energy is the same number. Together, they form a subspace \(\SS_E\) of the \(\vec Q,\,\vec P\) lattice, which in general may look like a surface. Just consider the set of points in \(\SS_E\), make a mapping \((\vec Q,\,\vec P)\mapsto(\vec Q',\,\vec P')\) that is one-to-one, inside \(\SS_E\). This law will be time-reversible and it will conserve the energy. Just one problem then remains: how do we choose a unique one-to-one mapping? 

To achieve this, we need a strategy. Our strategy now will be that we \emph{order} the values of the index \(i\) in some given way (actually, we will only need a cyclic ordering), and update the \((Q,P)\) pairs sequentially:
first the pair \((Q_1,\,P_1)\), then the pair \((Q_2,\,P_2)\), and so on, until we arrive at the last value of the index. This sequence of updating every pair \((Q_i,\,P_i)\) exactly once will be called a \emph{cycle}. One cycle will define the smallest step \(U^{\,\op}(t,\,t+\d t)\) for the evolution law.

This reduces our problem to that of updating a single \(Q,\,P\) pair, such that the energy is conserved. This should be doable. Therefore, let us first consider a single \(Q,\,P\) pair.

\subsubsection{One-dimensional system: a single $Q,\,P$ pair\labell{oned}}

While concentrating on a single pair, we can drop the index \(i\). The Hamiltonian will be a function of two integers, \(Q\) and \(P\).  For demonstration purposes, we restrict ourselves to the case
	\be H( Q,P)=T( P)+V( Q)+ A(Q)\,B(P)\ , \eel{hamTVdiscr}
which can be handled for fairly generic choices for the functions \(T(P ),\  V(Q),\ A(Q )\) and \(B(P )\). The last term here, the product \(A\,B\), is the lattice generalisation of the magnetic term \(\vec p\cdot \vec A(\vec q\,)\)  in Eq.~\eqn{HTV}.  Many interesting physical systems, such as most many body systems, will  be covered by Eq.~\eqn{hamTVdiscr}. It is possible to choose \(T( P)=P^2\), or better: \(\half P(P-1)\), but \(V(Q)\) must be chosen to vary more slowly with \(Q\), otherwise the system might tend to oscillate too quickly (remember that time is discrete). Often, for sake of simplicity, we shall disregard the \(A\,B\)  term.

The variables \(Q\) and \(P\) form a two-dimensional lattice. Given the energy \(E\), the points on this lattice where the energy \(H(Q,P) =E\) form a subspace \(\SS_E\). We need to define a one-to-one mapping of \(\SS_E\) onto itself. However, since we have just a two-dimensional lattice of points \((Q,P)\), we encounter a risk: if the integer \(H\) tends to be too large, it will often happen that there are no other values of \(Q\) and \(P\) at all that have the same energy. Then, our system cannot evolve. So, we will find out that some choices of the function \(H\) are better than others. In fact, it is not so difficult to see under what conditions this problem will occur, and how we can avoid it: the integer-valued Hamiltonian should not vary too wildly with \(Q\) and \(P\). What does ``too wildly" mean? If, on a small subset of lattice points, a \((Q,\,P)\) pair does not move, this may not be so terrible: when embedded in a larger system, it will move again after the other values changed. But if there are too many values for the initial conditions where the system will remain static, we will run into difficulties that we wish to avoid. Thus, we demand that most of the surfaces \(\SS_E\) contain more than one point on them -- preferably more than two. This means that the functions \(V(Q)\), \(T( P), A(Q)\) and \(B( P)\) should not be allowed to be too steep.  

	\begin{figure}[h!] \begin{quote} \begin{center}
 \lowerwidthfig{0pt}  {110mm}{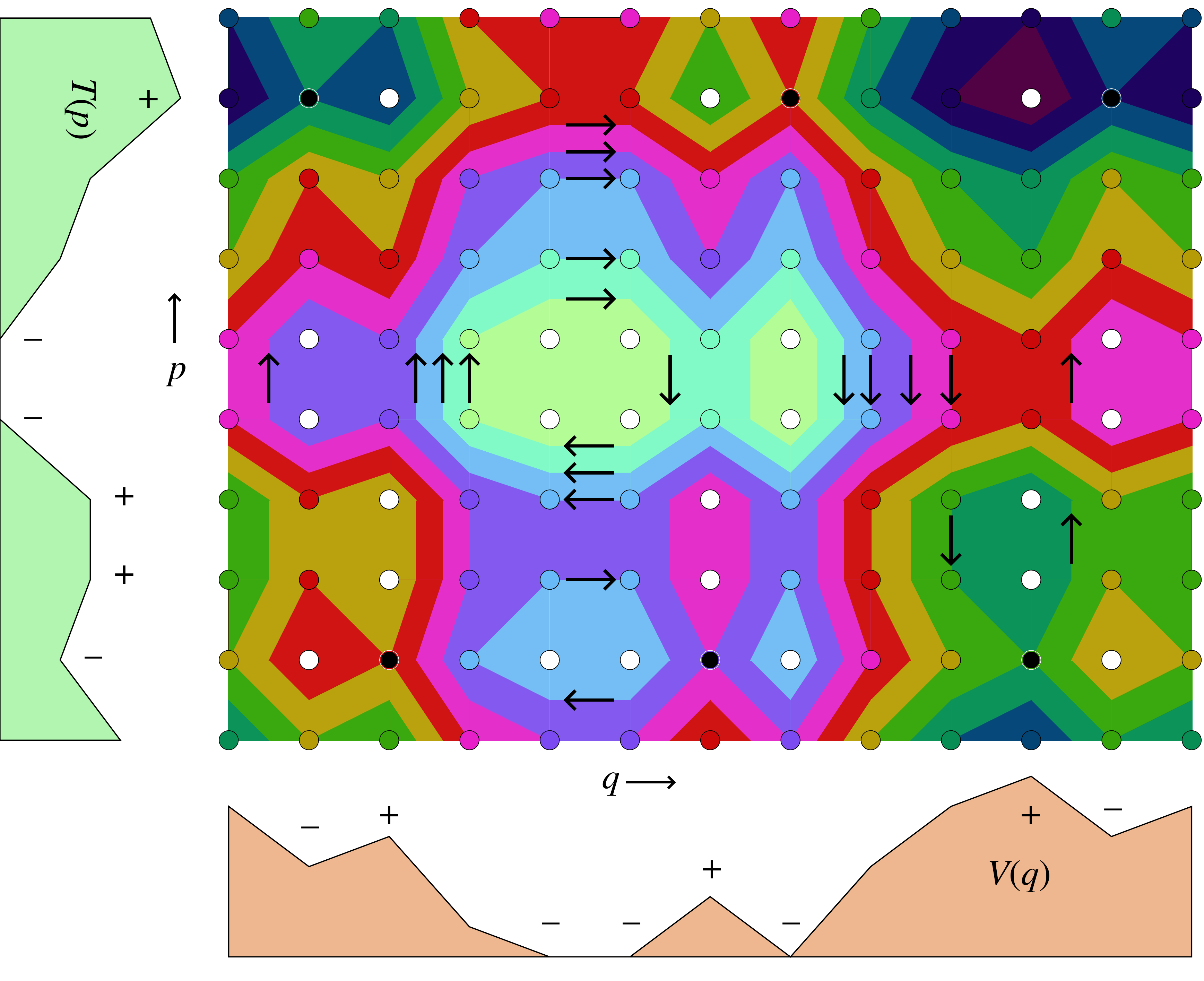} \end{center}
\caption[\small The $Q\,P$ lattice in the 1+1 dimensional case.]{\small The \(Q\,P\) lattice in the 1+1 dimensional case. Constant energy contours are here the boundaries of the differently coloured regions. Points shown in white are local extrema; they are not on a contour and therefore these are stable rest points. Black points are saddle points, where two contours are seen to cross one another. Here, some unique evolution prescriptions must be phrased, such as: ``stick to your right", and it must be specified which of the two contours contains the black dot. All these exceptional points are related to local minima (\(-\)) and maxima (\(+\)) of the functions \(T\) and \(V\).
\labell{PQdomains.fig}}\end{quote}
		\end{figure}	

We then find the desired invertible mapping as follows. First, extrapolate the functions \(T,\ P,\ A\) and \(B\) to all real values of their variables. Write  real numbers \(q\) and \(p\) as
	\be q=Q+\a\ , \quad p=P+\b\ ,\qquad Q\hbox{ and } P\hbox{ integer,}\quad 0\le\a\le 1\ ,\quad 0\le\b\le 1\ . \eel{realintfract}
Then define the continuous functions 
	\be V(q)=(1-\a)V(Q)+\a V(Q+1)\ ,\qquad T(p )=(1-\b)T(P )+\b T(P+1)\ , \eel{VTcont}
and similarly \(A(q)\) and \(B(p )\). Now, the spaces \(\SS_E\) are given by the lines  \(H(q,p)=T(p )+V(q)+A( q)B(p )=E\), which are now sets of oriented, closed contours, see Fig.~\ref{PQdomains.fig}. They are of course the same closed contours as in the standard, continuum Hamiltonian formalism.

The standard Hamiltonian formalism would now dictate how fast our system runs along one of these contours. We cannot quite follow that prescription here, because at \(t\,=\) integer we wish \(P\) and \(Q\) to take integer values, that is, they have to be at one of the lattice sites. But the speed of the evolution does not affect the fact that energy is conserved. Therefore we modify this speed, by now postulating that
\begin{quote} \emph {at every time step \(t\ra t+\d t\), the system moves to the next lattice site that is on its contour \(\SS_E\).} \end{quote}
If there is only one point on the contour, which would be the state at time \(t\), then nothing moves. If there are two points, the system flip-flops, and the orientation of the contour is immaterial. If there are more than two points, the system is postulated to move in the same direction along the contour as in the standard Hamiltonian formalism. In Fig.~\ref{PQdomains.fig}, we see examples of contours with just one point, and contours with two or more points on them. Only if there is more than one point, the evolution will be non-trivial.

In some cases, there will be some ambiguity. Precisely at the lattice sites, our curves will be non-differentiable because the functions \(T,\ C,\ A,\) and \(B\) are non-differentiable there. This gives sone slight complications in particular when we reach extreme values for both \(T( P)\) and \(V(q)\). If both reach a maximum or both a minimum, the contour shrinks to a point and the system cannot move. If one reaches a minimum and the other a maximum, we have a saddle point, and some extra rules must be added. We could demand that the contours ``have to be followed to the right", but we also have to state which of the two contours will have to be followed if we land on such a point; also, regarding time reversal, we have to state which of the two contours has the lattice point on it, and which just passes by. Thus, we can make the evolution law unique and reversible. See Fig.~\ref{PQdomains.fig}. The fact that there are a few (but not too many) stationary points is not problematic if this description is applied to formulate the law for multi-dimensional systems, see section~\ref{highdim}.

Clearly, this gives us the classical orbit in the correct temporal order, but the reader might be concerned about two things: one, what if there is only one point on our contour, the point where we started from, and two, we have the right time ordering, but do we have the correct speed? Does this updating procedure not go too fast or too slowly, when compared to the continuum limit?

As for the first question, we will have no choice but postulating that, if there is only one point on a contour, that point will be at rest, our system does not evolve.
Later, we shall find estimates on how many of such points one might expect.

Let us first concentrate on the second question. How fast will this updating procedure go? how long will it take, on average, to circle one contour? Well, clearly, the discrete period \(T\) of a contour will be equal to the number of points on a contour (with the exception of a single point, where things do not move\fn{But we can also say that, in that case, the period is \(\d t\), the time between two updates.}). How many points do we expect to find on one contour?

	\begin{figure}[h] \begin{quotation}
\begin{center}\hskip-5mm \lowerwidthfig{0pt}{90mm}{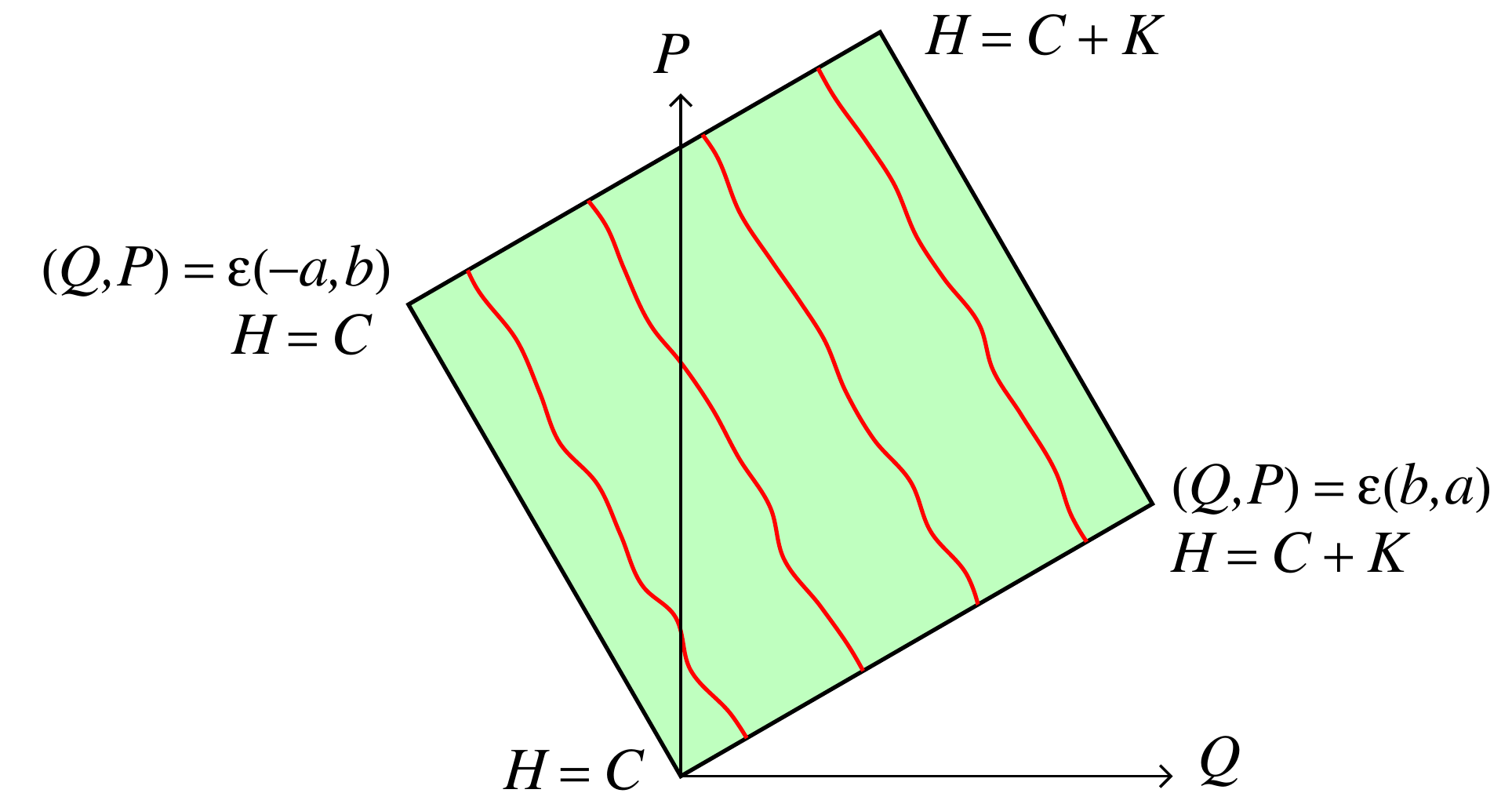}\qquad 
 \caption[\small A small region in the $QP$ lattice where the (integer valued) Hamiltonian is reasonably smooth]{\small A small region in the \(QP\) lattice where the (integer valued) Hamiltonian is reasonably smooth. See Eq.~\eqn{Happroxlinear}. The sides of the tilted square are \(\e\sqrt{a^2+b^2}\). Contours of approximately constant \(H\) values are indicated. \labell{rectangle.fig}}
\end{center}\end{quotation}
		\end{figure}

Consider now a small region on the \((Q,P)\) lattice, where the Hamiltonian \(H^\class\) approximately linearises:
	\be H^\class\approx a\,P+b\,Q+C\ , \eel{Happroxlinear}
with small corrections that ensure that \(H^\class\) is an integer on all lattice points. With a little bit of geometry, one finds a tilted square with sides of length
\(\e\sqrt{a^{2}+b^{2}}\), where the values of \(H^\class\) vary between values \(C\) and \(C+K\), with \(K=\e(a^2+b^2)\). Assuming that all these integers occur at about the same rate, we find that the total number of lattice sites inside the square is \(\e^2(a^2+b^2)\), and since there are \(K\) contours, every contour has, on average, 
	\be \e^2(a^2+b^2)/K=\e\ee
points on it. The lengths of the contours in Figure~\ref{rectangle.fig} is \(\e\sqrt{a^2+b^2}\),  so that, on average, the distance between two points on a contour is \(\sqrt{a^2+b^2}\).

This little calculation shows that, in the continuum limit,  the propagation speed of our updating procedure will be 
		\be\sqrt{\bigg({\d q\over\d t}\bigg)^2+\bigg({\d p\over\d t}\bigg)^2}=\sqrt{\bigg({\pa H\over\pa p}\bigg)^2+\bigg({\pa H\over\pa q}\bigg)^2}\ , \eel{Hamiltonspeed}
		completely in accordance with the standard Hamilton equations! (Note that the factors \(2\pi\) in Eqs.~\eqn{discrham} and \eqn{intparts} cancel out)
		
A deeper mathematical reason why our discrete lattice Hamiltonian formalism generates the same evolution speed as the continuum theory may be traced to the \emph{Liouville theorem}: a co-moving infinitesimal volume element in \((p,q)\)-space stays constant in the continuum theory; in the discrete lattice case, time reversibility ensures that the number of lattice points inside a small volume on the lattice stays fixed, so that we have the same Liouville theorem on the lattice. When increasing values for the partial derivatives of the Hamiltonian cause a squeezing of the infinitesimal volume elements, both the continuum theory and the lattice theory require the same increase in the velocities to keep the volume elements constant.

One concludes that our updating procedure exactly leads to the correct continuum limit. However, the Hamiltonian must be sufficiently smooth so as to have more than one point on a contour. We now know that this must mean that the continuous motion in the continuum limit cannot be allowed to be too rapid. We expect that, on the discrete lattice, the distance between consecutive lattice points on a contour may vary erratically, so that the motion will continue with a variable speed. In the continuum limit, this must average out to a smooth motion, completely in accordance with the standard Hamilton equations.

Returning to the question of the contours with only one point on them, we expect their total lengths, on average, to be such that their classical periods would correspond to a single time unit \(\d t\). These periods will be too fast to monitor on our discrete time scale.

This completes our brief analysis of the 1+1 dimensional case. We found an evolution law that exactly preserves the discrete energy function chosen. The procedure is unique as soon as the energy function can be extended naturally to a continuous function between the lattice sites, as was realised in the case \(H=T+V+AB\) in Eq.~\eqn{VTcont}. Furthermore we must require that the energy function does not vary too steeply, so that most of the closed contours contain more than one lattice point. 

An interesting test case is the choice
	\be T( P)=\half P(P-1)\ ;\qquad V(Q)=\half Q(Q-1)\ , \eel{discrharmosc}
This is a discretised harmonic oscillator whose period is not exactly constant, but this one is easier to generalise to higher dimensions
than the oscillator described in section~\ref{highd} and subsection~\ref{multiddiscreteham}.

\subsubsection{The multi dimensional case\labell{highdim}}

\def\fr{{\mathrm{fract}}}
A single particle in 1 space- and 1 time dimension, as described in the previous section, is rather boring, since the motion occurs on contours that all have rather short periods (indeed, in the harmonic oscillator, where both \(T\) and \(V\) are quadratic functions of their variables, such as in Eq.~\eqn{discrharmosc}, the period will stay close to the fundamental time step \(\d t\) itself).  In higher dimensions (and in multi component oscillators, particularly when they have non-linear interactions), this will be quite different. So now, we consider the variables \(Q_i,\ P_i,\ i=1,\cdots,n\). Again, we postulate a Hamiltonian \(H(\vec Q,\vec P\,)\) that, when \(P_i\) and \(Q_i\) are integer, takes integer values only. Again, let us take the case that
\be H(\vec Q,\,\vec P\,)=T(\vec P\,)+V(\vec Q\,)+A(\vec Q\,)\,B(\vec P\,)\ . \eel{HTVABvec}
To describe an energy conserving evolution law, we simply can apply the procedure described in the previous section \(n\) times for each cycle. For a unique description, it is now mandatory that we introduce a \emph{cyclic ordering} for the values \(1,\cdots,n\) that the index \(i\) can take. Naturally, we adopt the notation of the values for the index \(i\) to whatever ordering might have been chosen: 
	\be 1<2<\,\cdots\,<n<1 \,\cdots\ . \eel{cyclic}
We do emphasise that the procedure described next depends on this ordering.

Let \(U_i^\op\) be our notation for the operation in one dimension, acting on the variables \(Q_i,P_i\) at one given value for the index \(i\). Thus, \(U_i^\op\) maps \((P_i,Q_i)\mapsto(P'_i,Q'_i)\) using the procedure of section~\ref{oned} with the Hamiltonian  \eqn{HTVABvec}, simply keeping all other variables \(Q_j,P_j,\ j\ne i\) fixed. By construction, \(U_i^\op\) has an inverse \(U_i^{\op\,-1}\). Now, it is simple to produce a prescription for the evolution \(U^\op\) for the entire system, for a single time step \(\d t=1\):
	\be U^\op(\d t)=U^\op_n\,U^\op_{n-1}\,\cdots\,U^\op_1\ . \eel{totalU}
where we intend to use the physical notation:   \(U^\op_1\) acts first, then \(U^\op_2\), etc., although the opposite order can also be taken. Note, that we have some parity violation: the operators \(U^\op_i\) and \(U^\op_j\) will not commute if \(i\ne j\), and therefore, if \(n\ge 3\),  the resulting operator \(U^\op\) is not quite the same as the one obtained when the order is reversed. 

Time inversion gives:
	\be U^\op(-\d t)=U^{\op\,-1}(\d t)=U_1^{\op\,-1}\,U_2^{\op\,-1}\cdots\, U_n^{\op\,-1}\ . \eel{timereversal}
Finally, if the exchange \(U^\op_i\leftrightarrow U_i^{\op\,-1}\) might be associated with ``particle-antiparticle conjugation", \(C\), then
the product \(P\)(parity)\,\(T\)(time inversion)\,\(C\) (conjugation) may still be a good symmetry. In the real world, this might lead to a natural explanation of CPT symmetry, while \(P\), \(T\), or \(CP\) are not respected. 

\subsubsection{The Lagrangian\labell{L}} It was emphasised by Elze\,\cite{Elze-2014} that systems with a discrete Hamiltonian should also have an action principle. If both time as well as the variables \(P\) and \(Q\) are discrete, one could consider Lagrangians such as
	\be L(t)\qu\half P(t)(Q(t+1)-Q(t-1))-H\big(P(t),Q(t)\big) \ , \qquad S=\sum_{t\in\Bbb Z}L(t)\ .\ee
This, however, would lead to Lagrange equations that are finite difference equations, at best, while they would no longer guarantee conservation of energy. Some Lagrangians may exist that are purely quadratic in the integers \(P\) and \(Q\), but, as we saw, this would be too strong a restriction that excludes any non-trivial theory. At this moment we have no proposal for a Lagrange principle that works as well as our discrete Hamilton formalism.

\subsubsection{Discrete field theories\labell{discrfields}}
	An important example of an infinite-dimensional \((Q_i,\,P_i)\) system is a local field theory. Suppose that the index \(i\) is replaced by a lattice coordinate \(\vec x\), plus possibly other indices \(j\) labelling species of fields. Let us rename the variables \((\F_j(\vec x\,),\,P_j(\vec x\,))\), where \(\F_j\) are canonical fields and \(P_j\) are their momentum variables (often, in the continuum theory, \(\fract\dd{\dd t}\F_j\)). Now assume that the Hamiltonian of the entire system is the sum of local terms:
		\be H_\intt=\sum_{\vec x}\HH_\intt(\vec x\,)\ , \qquad \HH_\intt(\vec x\,)=V(\vec\F(\vec x\,),\vec\F(\vec x\,'))+T(\vec P(\vec x\,))\ , \eel{Hdensitydiscr}
where the coordinates \(\vec x\,'\) are limited to neighbours of \(\vec x\) only, and all functions \(V\) and \(T\) are integers. This would be a typical discretisation of a (classical or quantum) field theory (ignoring, for simplicity, magnetic terms).

We can apply our multi-dimensional, discrete Hamiltonian equations to this case, but there is one important thing to remember: where in the previous subsections we stated that the indices \(i\) must be cyclically ordered, this now  means that, in the field theory of Eq.~\eqn{Hdensitydiscr},  not only the indices \(i\) but also the coordinates \(\vec x\) must be (cyclically) ordered. The danger of this is that the functions \(V_i(\vec x\,)\) also refer to neighbours, and, consequently, the evolution step defined at point \(\vec x\) affects the evolution at its neighbouring points \(\vec x\,'\), or: \([U^\op(\vec x\,),\,U^\op(\vec x\,'\,)]\ne 0\). Performing the updates in the order of the values of the coordinates \(\vec x\), might therefore produce signals that move much faster than light, possibly generating instantaneous non local effects across the entire system over a single time step \(t\ra t+\d t\). This we need to avoid, and there happens to be an easy way to do this:
	\begin{quote} First make sure that the interaction terms in the Hamiltonian only involve nearest neighbours, The evolution equations (e.o.m.) of the entire system over one time step \(\d t\), are then obtained by ordering the coordinates and other indices as follows: \emph{first} update all even lattice sites, \emph{then} update all odd lattice sites. \end{quote}
	Since the \(U^\op\) operators generated by \(H_i(\vec x\,)\) do commute with the evolution operators \(U^\op(\vec x\,'\,)\) when \(\vec x\) and \(\vec x\,'\) are both on an even site or both on an odd site of the lattice (so that they are \emph{not} nearest neighbours), this ordering does not pass on signals beyond two lattice links. Moreover, there is another huge advantage of this law: the order in which the individual even sites of the lattice are updated is now immaterial, and the same for the set of all odd sites.
	
Thus, we obtained a cellular automaton whose evolution law is of the type 
	\be U^\op=A^\op\,B^\op\ ,\qquad A^\op=\prod_{\vec x\,=\,\mathrm{ even}}A^\op(\vec x)\ ,\qquad B^\op=\prod_{\vec y\,=\,\mathrm{ odd}}B^\op(\vec y)\ ,\eel{ABautomaton}
where the order inside the products over the sites \(\vec x\) and \(\vec y\) is immaterial, except that \(A^\op(\vec x)\) and \(B^\op(\vec y)\) do not commute when \(\vec x\) and \(\vec y\) are direct neighbours. Such automata are interesting objects to be studied, see chapter~\ref{CAdetail}.

\subsubsection{From the integer valued to the quantum Hamiltonian\labell{intquanthamWEG?}}

A deterministic system obeying a discrete Hamiltonian formalism as described in the previous sections is of particular interest when we map it onto a quantum system following the program discussed in this book. This is because we here have two different operators that both play the role of energy: we have the integer valued, discrete Hamiltonian \(H_\cl\) that generates the classical equations of motion, and we have the angular, or fractional valued Hamiltonian \(H_\quant\), defined from the eigenstates and eigenvalues of the one-time step evolution operator \(U^\op(\d t)\):
	\be U^\op(\d t)=e^{-iH^\op_\quant}\ ,\qquad 0\le H_\quant<2\pi\ \qquad (\d t=1)\ , \eel{UH}
where \(H_\quant\) refers to the eigenvalues of the operator  \(H_\quant^\op\).

As anticipated in section~\ref{genintHamilton}, we can now uniquely define a total Hamiltonian that is a real number operator, by
	\be H=H_\cl+H^\op_\quant \  . \eel{Htot}
The bounds imposed in Eq.~\eqn{UH} are important to keep in mind, since \(H_\quant\), as defined, is strictly periodic. \(H_\cl\) is assumed to take only integer values, times \(2\pi/\d t\). In this section we study the quantum theory defined by the Hamiltonian \eqn{Htot}. 

We have seen,  for instance in chapter~\ref{determm}, subsection \ref{cogwheelN}, Eq.~\ref{NHam} in part I, and in chapter~\ref{morecog}, section \ref{infinitediscr}, Eq.~\ref{HfrUn} in part II, how the operator \(H^\op_\quant\) can be calculated from the eigenvalues \(U(\d t)\) of the operator \(U^\op(\d t)\): for instance by Fourier transformations, one derives that, if the eigenvalues of \(H_\quant\) are assumed to lie between 0 and \(2\pi\), then
	\be  H^\op_\quant&=&\pi-\sum_{n-1}^\infty{i\over n}\,(\,U^\op(n\,\d t)-U^\op(-n\,\d t)\,)\  . \eel{HfrUn1}
This sum converges nearly everywhere, but the vacuum is the edge state where the equation does not hold, and it is not quite local, since the evolution operator over \(n\) steps in time, also acts over \(n\) steps in space. 

But both \(H_\cl\) and \(H_\quant\) are uniquely defined, and since \(H_\quant\) is bound to an interval while \(H_\cl\) is bounded from below, also \(H\) is bounded from below.

Note that demanding a large number of low energy states near the vacuum (the absence of a large mass gap) implies that \(U^\op(n\,\d t)\) be non-trivial in the \(H_\cl=0\) sector. This is often not the case in the models described in subsection~\ref{highdim}, but in principle there is no reason why such models should not exist also. In fact, some of the cellular automaton models discussed later in chapter~\ref{CAdetail} have no manifestly conserved \(H_\cl\), so that all their states can be regarded as sitting in the \(H_\cl=0\) sector of the theory.

Because of the non-locality of Eq.~\eqn{HfrUn1}, the Hamiltonian \eqn{HfrUn1} does not obey the rule $vi$,  but if \(U^\op(\d t)\) is the product of local evolution operators, the evolution over integer time steps \(n\,\d t\) \emph{is} local, so the theory can be claimed to obey locality, as long as we refrain from defining its states at time \(t\) when \(t\) is not an integer.\fn{Some have tried to shoot down our theories by objecting that our classical/quantum equivalence only holds for integer times. Of course we simply point out then that, if we restrict ourselves to sufficiently low energies, the time-variability is sufficiently slow that having an equation that only holds rigorously at integer multiples of \(\d t\) is all we need.}

As we have seen in section~\ref{locality}, the sum \eqn{HfrUn1} does not converge rapidly everywhere in Hilbert space. We are particularly interested in the Hamiltonian as it acts on states very close to the vacuum, in our notation: \(H_\cl=0,\quad H_\quant=\w\), where \(0<\w\ll 2\pi\). Suppose then that we introduce a cut-off in the sum  \eqn{HfrUn1} (or \ref{omegasumsin}) by multiplying the summand with \(e^{-n/R}\), where \(R\) is also the range of non-locality of the last significant terms of the sum. As we have seen in section~\ref{locality}, breaking off the expansion at the point \(R\) modifies the Hamiltonian as follows:
	\be H_\quant\ra H_\quant+{2\over R\,H_\quant}\ , \eel{Happrox1}
and this is only acceptable if
	\be R\gg M_{\mathrm{Pl}}/\bra H_\quant\ket ^2\ . \eel{convradius1}	
Here, \(M_{\mathrm{Pl}}\) is the ``Planck mass", or whatever the inverse is of the elementary time scale in the model. This cut-off radius \(R\) must therefore be chosen to be  very large, so that, indeed, the exact quantum description of our local model generates non-locality in the Hamiltonian.

We conclude that the Hamiltonian can be expressed in terms of local terms, but we need to include the operators \(U^\op(\pm\D t)\) where \(\D t\) is large compared to the inverse of the Hamiltonian we wish to calculate. These will develop non localities that are still serious. This is still an obstacle against the construction of a local \emph{quantum} Hamiltonian density (the classical component, \(H^\class\) obeys condition \emph{vi}). As yet, therefore, more has to be done to obtain locality: second quantisation.

The apparent locality clash between the quantum Hamiltonian and the classical theory may well be looked upon as a possible additional explanation of the apparent non-localities expected in `hidden variable' theories: neither the pure quantum system that we usually employ in quantum field theories, nor the associated classical system exhibit any non-locality, but the mapping between them does. This non-locality is spurious, it has no physical consequence whatsoever, but mathematically it may imply that the quantum system should not be split up into local wave functions that do not communicate with each other -- perhaps that is the route along which apparent non-locality arises in classical mechanical models. There is no non-locality in the classical theory, but it is in the representation of the quantum variables, or: the classical - quantum mapping.

\newsecl{Quantum Field Theory}{QFT}
	In this chapter, we give a brief summary of the features of quantum field theory that we shall need to understand in this book. 

	We have seen that producing quantum Schr\"odinger equations starting from non quantum mechanical systems is essentially straightforward. However, to employ this observation as a viable ontological interpretation of quantum mechanics, more is needed.  The main objection against these concepts has always been that quantum theories obeying \emph{locality in the quantum sense},   are much more difficult to reproduce with classical systems obeying locality in the \emph{classical} sense, and impossible according to many. 
	
Locality means that interaction at a distance can only occur with signals that undergo some delay. Locality in the classical sense here means that the classical evolution laws are based on interactions with neighbouring sites only, in such a way that information cannot spread faster than the speed of light. Locality in the quantum sense means the same thing, except that we allow for any kind of quantum mechanical interactions between neighbouring sites. If \(\OO_i(x)\) is an operator only depending on fundamental variables in the immediate vicinity of a space-time point \(x\), enumerated by an index \(i\), then quantum locality means that the commutation property
	\be [\OO_i(x),\,\OO_j(x')]=0\ , \eel{localcomm}
holds as soon as the two space-time points \(x\) and \(x'\) are space-like separated: 
	\be (x-x')^2\equiv(\vec x-\vec x\,')^2-c^2(t-t')^2>0\ . \eel{spacelikesep}
	
	The relativistic quantised field theories employed in the Standard Model are indeed strictly local in the quantum mechanical sense, obeying Eqs.~\eqn{localcomm}, \eqn{spacelikesep}. It is important to recall here the essential features of these systems. A point to be made right-away is, that quantum field theories are quite complicated. This is partly due to the fact that we usually want special relativity to be valid, which is a difficult -- while highly interesting -- demand. But even without special relativity, there are some fairly intricate issues such as second quantisation, perturbation theory, infinities, renormalisation, symmetries and anomalies. This is why the topic of this book is actually quite difficult: not only are we attempting to derive quantum mechanics from scratch, but also (fully renormalised) quantum field theory.
	
	Relativistic quantum field theories with a proper continuum limit can only incorporate elementary fields with spin \(0,\  \half\) and 1. As is well-known, gravity would be propagated by gravitons with spin 2, and supergravity would add one or more gravitino species with spin \(\ffract 3{\!2}\), but then, if we want these fields to interact,  we would have to be close to the Planck scale, and this would require discretisation due to micro states. Since ordinary quantum field theories assume strict continuity, they only apply to the continuum limit, which implies that we can safely omit spin 2 and spin \(\ffract3{\!2}\) fields in those theories. The way this works in quantum field theory is that, at the Standard Model scales, interactions with gravitons and gravitinos are extremely weak.
	
	On the other hand, one could argue that also \emph{special} relativity is not our first priority, and ignoring special relativity would imply no rigorous constraint on spin. If we ignore special as well as general relativity, we could just as well ignore rotation invariance\fn{Ignoring such important symmetries in considering certain models does not mean that we believe these symmetries to be violated, but rather that we wish to focus on simple models where these symmetries do not, or not yet, play a role. In more sophisticated theories of Nature, of course one has to obey all known symmetry requirements.}.  What is left then is a theory of quantised fields enumerated by an index that may or may not represent spin. Later we may wish to reinstate Poincar\'e invariance, at least at the quantum side of the equation,  but this will have to be left as an important exercise for the (hopefully near) future.
	
	What we want to \emph{keep}  is a speed limit for signals that describe interactions, so that the notion of locality can be addressed. In practice, an elegant criterion can be given that guarantees this kind of locality. Consider the quantum system in its Heisenberg notation. We have operators \(\OO_i(\vec x,\,t)\) where both the space coordinates \(\vec x\) and the time coordinate \(t\) may be either continuous or discrete. The discrete index \(i\) enumerates different types of operators. 
	
When our operator fields obey quantum locality, Eqs.~\eqn{localcomm} and \eqn{spacelikesep}, in the continuous case, this means that the Hamiltonian must be the integral of a Hamiltonian density:
	\be H=\int\dd^3\vec x\ \HH(\vec x,\,t)\ , \qquad [\,\HH(\vec x,\,t)\,,\ \HH(\vec x\,',\,t)]=0\quad \hbox{if}\quad \vec x\ne\vec x\,'\ ,\eel{Hamiltondens}
while, when \(x\ra x'\), the commutator \([\,\HH(\vec x,\,t)\,,\ \HH(\vec x\,',\,t)\,]\), may contain derivatives of Dirac delta distributions. 
Note that, here,  we kept equal times \(t\) so that these space-time points are space-like separated unless they coincide.

On a discrete space-like lattice, it is tempting to replace Eqs.~\eqn{Hamiltondens}  by the lattice expressions
	\be H=\sum_{\vec x}\HH(\vec x,\,t)\ ,\qquad [\HH(\vec x,',t),\,\HH(\vec x\,',\,t)]=0\quad\hbox{if}\quad |\vec x-\vec x\,'|>a , \eel{hdensdiscrete}
where \(a\) is the link size of the lattice. The Hamiltonian densities at neighbouring sites, \hbox{\(|\vec x-\vec x\,'|=a\), } will, in general, \emph{not} commute.
	Although Eqs.~\eqn{hdensdiscrete} may well serve as a good definition of locality, they do not guarantee that signals are subject to a speed limit. Only in the continuum limit, one may recover commutation at space-like separations, \eqn{spacelikesep}, if special relativity holds (in that limit).
	
	Cellular automata are typically lattice theories. In general, these theories are difficult to reconcile with Lorentz invariance. This does not mean that we plan to give up Lorentz invariance; quite possibly, this important symmetry will be recovered at some stage. But since we want to understand quantum mechanics as a reflection of discreteness at a scale comparable to the Planck scale, we are unable at present to keep Lorentz invariance in our models, so this price is paid, hopefully temporarily.
	
	For simplicity, let us now return our attention to continuum quantum field theories, which we can either force to be Lorentz invariant, or replace by lattice versions at some later stage. The present chapter is included here just to emphasise some important features.
	
\subsecl{General continuum theories -- the bosonic case}{continuousbosons}
	Let the field variables be real number operators \(\F_i(\vec x,\,t)\) and their canonical conjugates \(P_i(\vec x,\,t)\). Here, \(i\) is a discrete index counting independent fields. The commutation rules are postulated to be
	\be[\F_i(\vec x,t),\,\F_j(\vec x\,',t)] &=& [P_i(\vec x,t),\,P_j(\vec x\,',t)]\iss 0\ , \nm\\[1pt]
		[\F_i (\vec x,t),\,P_j(\vec x\,',t)] &=&i\,\d_{ij}\,\d^3(\vec x-\vec x\,') \eel{phicommrules}
(for simplicity, space was taken to be 3-dimensional).

In bosonic theories, when writing the Hamiltonian as
	\be H=\int\dd\vec x\,\HH(\vec x)\ ,\eel{HHdens}
the Hamiltonian density  \(\HH(\vec x)\) typically takes a form such as
		 \be\HH(\vec x)=\sum_i\bigg(\half P_i^2(\vec x)+\half(\vec\pa\F_i(\vec x))^2\ +\,V(\vec\F(\vec x))\bigg)\ .\eel{scalarHdens}
If we are in 3+1 dimensions, and we want the theory to be renormalizable, \(V(\vec\F(\vec x))\) must be a polynomial function of \(\vec\F\), of degree 4 or lower.
Typically, one starts with \be V(\vec\F)=\half\sum_im_i^2\vec\F_i^2+V_4(\vec\F)\ ,\ee
where \(m_i\) are the (unrenormalized) masses of the particles of type \(i\), and
\(V_4\) is a \emph{homogeneous} quartic expression in the fields \(\vec\F_i(\vec x)\) such as a self interaction \(\fract 1{4!}\l\F^4\).
	
However, when \(\vec\F(\vec x)\) contains components that form a vector in 3-space, Lorentz-invariance dictates a deviation from Eq.~\eqn{scalarHdens}. We then have \emph{local gauge invariance}, which implies that a constraint has to be imposed. Writing these vector fields as \(\vec A(\vec x)\), and their associated momentum fields as \(\vec E(\vec x)\), one is forced to include a time component \(A_0(\vec x)\).

Gauge-invariance must then be invoked to ensure that locality and unitarity of the theory are not lost, but the resulting Hamiltonian deviates a bit from Eq.~\eqn{scalarHdens}. This deviation is minimal if we choose the space-like radiation gauge 
	\be \sum_{i=1}^3\pa_iA_i(\vec x)=0\ , \eel{radiationgauge}
since then the Hamiltonian will have the quadratic terms
	\be\HH_2(\vec x)=\half \vec E^2(\vec x)+\half(\vec\pa A_i(\vec x))^2-\half (\vec\pa A_0(\vec x))^2\ . \eel{gaugeHamiltonian}
In addition, one might have linear terms,
	\be\HH_J(\vec x)=\vec J(\vec x)\cdot\vec A(\vec x)+\r(\vec x)A_0(\vec x)\ ,\qquad \HH\equiv \HH_2+\HH_J+\HH_\intt\ , \eel{sorceterms}
where \(\vec J\) and \(\r\) are some given background functions. \(\vec J(\vec x)\) is a \emph{current density} and \(\r(\vec x)\) a \emph{charge density}. In a relativistic theory, \(\vec J\) and \(\r\) form a 4-vector. All remaining terms in the Hamiltonian, typically higher powers of the fields, which may cause interactions among particles to occur, are collected in \(\HH_\intt\).

Notably, the field \(A_0(\vec x) \) does not have a canonical partner that would have been called \(E_0(\vec x)\), and therefore, the field \(A_0\) can be eliminated classically, by extremising the Hamiltonian \(H=\int\dd{\vec x}\,\HH(\vec x)\), which leads to the Coulomb force between the sources \(\r\). This Coulomb force is instantaneous in time, and would have destroyed locality (and hence Lorentz invariance) if we did not have local gauge invariance.

This, of course, is a description of quantised field theories in a nut shell, as yet only for bosonic particles. How then the Schr\"odinger equation is solved by perturbation expansion in powers of the coupling constant(s) \(\l\) and/or gauge coupling parameters \(g\), is well-known and discussed in the standard text books.\,\cite{ItzyksonHuber-1980}\cite{GtH-V-DIAGR-1973}\cite{GtH-ConcBasis-2007}

The most important point we need to emphasise is that the above formulation of quantised field theories is easy to replace by discretised versions. All we need to do is replace the partial derivatives \(\pa_i=\pa/\pa x_i\) by lattice derivatives:
	\be\pa_i\F(\vec x)\ra {1\over a}\bigg(\F(\vec x+\vec e_ia)-\F(\vec x)\bigg)\ , \eel{discrder}
where \(a\) is the lattice link size, and \(\vec e_i\) is the unit vector along a lattice link in the \(i\) direction. The continuum limit, \(a\downarrow 0\), seems to be deceptively easy to take; in particular, renormalization will now only lead to finite correction terms. Note however, that symmetries such as rotation symmetry and Lorentz invariance will be lost. Recovering such symmetries in more sophisticated models (without taking a continuum limit) is beyond our abilities just now -- but notice that our treatment of string theory, section~\ref{sustr}, appears to be heading in the right direction.

\subsecl{Fermionic field theories}{fermionfields}
	Fermionic field systems are also an essential element in the Standard Model. The fundamental variables are the Dirac fields \(\j_i(\vec x)\) and their canonical associates \(\j^\dag_i(\vec x)\). They are spinor fields, so that \(i\) contains a spinor index. The fields \emph{anti}-commute. The anti-commutation rules are
	\be\{\j^\dag_i(\vec x),\,\j_j(\vec x\,')\}=\d_{ij}\d(\vec x-\vec x\,')\ ,\qquad \{\j_i,\j_j\}=\{\j_i^\dag,\j_j^\dag\}=0\ ,\eel{psianticommute}
where \(\{a,b\}\equiv ab+ba\). Note, that these rules are typical for operators of the form \(({0\ 1\atop 0\ 0})\) and \(({0\ 0\atop 1\ 0})\), so these rules mean that \(\j_i(\vec x)\) is to be regarded as an operator that annihilates an object \(i\) at position \(\vec x\), and \(\j^\dag_i(\vec x)\) creates one. The rules imply that \(\j_i^2(\vec x)=0\) and  \((\j^\dag_i(\vec x))^2=0\), so that we cannot create or annihilate two objects \(i\) at the same spot \(\vec x\).  A state containing two (or more) particles of different type, and/or at different positions \(\vec x\), will always be antisymmetric under interchange of two such fermions, which is Pauli's principle.

In conventional quantum field theory, one now proceeds to the Lagrange formalism, which works magnificently for doing fast calculations of all sorts. For our purpose, however, we need the Hamiltonian. The quantum Hamiltonian density for a fermionic field theory is (compare section~\ref{nu}): 	\def\ol{\overline}
	\be \HH_F(\vec x)=\ol{\j}\,(m+W(\vec\F)+\vec\g\cdot\vec\pa)\,\j\ ,	\eel{fermifieldham}
where \(W(\vec\F)\) stands short for the Yukawa interaction terms that we may expect, and \(\ol\j=\j^\dag\g_4\).

The matrices \(\g_\m,\ \m=1,2,3,4\), need to obey the usual anti-commutation rule
	\be\{\g_\m,\,\g_\n\}=2\d_{\m\n}\ , \eel{gammaanticom}
which requires them to be at least \(4\times 4\) matrices, so that the spinors are 4 dimensional. One can, however, reduce these to 2 component spinors, called Majorana spinors,  by using a constraint such as
	\be \j=C\,\tl{\ol\j}=C\tl\g_4\j^\dag\ , \qquad \j^\dag=C^*\g_4\,\j\ , \eel{majorana}
where \~{ } stands for transposition, and \(C\) is a spinor matrix obeying\,\fn{In section~\ref{nu}, we also used 2 dimensional spinors. The two-component spinor field used there is obtained by using one of the projection operators \(P_\pm=\half(1\pm\g_5)\). The mass term can be made compatible with that.}\(\!^,\)\fn{The reason why Dirac needed a four-dimensional representation is that the constraint \eqn{majorana} would not allow coupling to an electromagnetic field since this would violate gauge-invariance (in particular the mass term).}
	\be \g_\m C=-C\g_\m^*\ , \quad C^\dag C=1\ ,\qquad C=C^\dag\ . \eel{Cmatrix}

Just as in the bosonic case, we may consider replacing the continuum in space by a space-like lattice, using expressions such as Eq.~\eqn{discrder}, at the price of (hopefully temporarily) giving up Lorentz invariance.

The Yukawa term \(W(\F)\) in Eq.~\eqn{fermifieldham} may include interactions with gauge fields in the usual way. The question addressed in this work is to what extent Hamiltonians such as the sum of Eqs.~\eqn{scalarHdens} and \eqn{fermifieldham} can be obtained from deterministic theories.

\subsecl{Standard second quantisation}{2ndquantisation}
	Accurate calculations in field theories for interacting particles are practically impossible without a systematic approximation procedure of some sort. The most efficient approximation scheme used is that of the perturbation expansion in terms of powers of all interaction parameters. This works because, when the interaction terms vanish, the fields will obey linear field equations, which are trivial to solve. 
	
	These linear equations happen to coincide with the linear Schr\"odinger equations obeyed by single particle states. It is as if the wave functions  \ \(|\phi_i(\vec x,t)\ket,\ \ |\j_i(\vec x,t)\ket\)  \ 
and their associated bra states are replaced by classical ontological fields \(\Phi_i(\vec x,t),\ \j_i(\vec x,t)\) and their canonical  conjugates, after which the quantisation procedure is applied to these fields yet again, replacing Poisson brackets by commutators or anticommutators. This explains the term ``second quantisation" by which this procedure is known.
	
	In fact it is not hard to show that the complete Hilbert space of all quantum states of the quantised field system \eqn{phicommrules} and \eqn{psianticommute} can be described as the product space of all sets of multi particle states that can be formed out of the `single-quantised' particles.
	
	Then, however, one has to insert the interaction terms of the Hamiltonian. We write
		\be\HH=\HH_0+\D\HH_0+\HH^\intt+\D\HH^\intt\ , \eel{pertham1}
where \(\HH_0=\HH_2+\HH_J\) is the bilinear part of \(\HH\), and \(\HH^\intt\) contains the higher powers of the fields, causing interactions.  \(\D\HH_0\) and \(\D\HH^\intt\) are extra terms that are of the same form as \(\HH_0\) and \(\HH^\intt\) themselves, but they are taken care of at later stages of the perturbation expansion, just for technical reasons (renormalization). This is how one begins to set up perturbation theory.

	Now, in relativistic quantum theories, the single-quantised free particles have energy spectra that take the form \(E=\pm\sqrt{p^2+m^2}\) (for bosons), or \(E=\vec\a\cdot\vec p+\b\, m\) (for fermions; \(\a_i\) and \(\b\) are the Dirac matrices). This implies that the energies of single particles appear to be unbounded from below. The beauty of the second-quantised theory is that we can replace negative-energy particles by holes of positive energy antiparticles. This automatically ensures a lower bound for the total Hamiltonian.  
	
	For the case of fermions, it is easy to accept the idea that negative energy particles have to be regarded as holes in the sea of antiparticles, because Pauli's exclusion principle forbids the presence of more than one particle in any energy level. In the case of bosons, the situation becomes clear if we regard every mode of the energy spectrum as a harmonic oscillator, controlled by creation and annihilation operators. Its energy is also bounded from below. Note that, in terms of the quantum field variables \(\F\) and \(\j\), the Hamiltonian was non-negative by construction -- if one disregards the complications due to renormalization.
	
	Thus,  the second quantisation procedure restores a lower bound to the Hamiltonian, simply by allowing indefinite numbers of particles. We can allow the same mechanism to work for a cellular automaton, if the automaton also can be described in terms of particles. A particle hops over a grid of points in 3-space, and its evolution operator generates a Hamiltonian that may be unbounded from below. Second quantisation now means that we allow for the presence of indefinite numbers of these particles, which may either behave as fermions or as bosons. The particle - antiparticle procedure then ensures positivity of the total Hamiltonian.

\subsecl{Perturbation theory}{perturb}
	How to compute the effects of these Hamiltonians in perturbation theory, such as mass spectra, cross sections and lifetimes of the quantised particles that it contains, is standard text material, and not the subject of this treatise, but we do need to know about some essential features for our further discussion.
	
	Split up the Hamiltonian into a ``free" part \(\HH_0\) and the various interaction parts,  as in Eq.~\eqn{pertham1}, where the free part only contains terms that are bilinear in the field variables \(\F_i(\vec x),\ P_i(\vec x),\ \j_i(\vec x)\) and \(\ol\j_i(\vec x)\) (possibly after having shifted some of the fields by a vacuum value, such as in the Brout-Englert-Higgs mechanism). The interaction Hamiltonian \(\HH^\intt\) may also contain bilinear terms, here written as \(\D\HH_0\), needed to renormalize divergent effective interactions. There is some freedom as to whether we put parts of these so-called counter terms in \(\HH_0\) or in \(\D\HH_0\), and how to split the interaction terms in \(\HH^\intt\) and \(\D\HH^\intt\), which in fact is a choice concerning the book keeping process of the perturbative expansion. The fact that final results of the calculation should be independent of these choices is an important ingredient of what is called the renormalization group of the theory (see section~\ref{RG}). 

\(\HH^\intt\) is assumed to depend on the coupling parameters \(\l_i,\ g_i,\) etc., of the theory, such that \(\HH^\intt\) vanishes if these parameters are set to zero.

As already mentioned in section~\ref{continuousbosons}, the analysis is facilitated by the introduction of auxiliary terms in the Hamiltonian, called source terms, which are linear in the fields:
	\be \HH_J(\vec x,t)=\sum_iJ_i(\vec x,t)\F_i(\vec x)+\sum_i\ol\eta_i(\vec x,t)\j_i(\vec x)+\sum_i\ol\j_i(\vec x)\eta_i(\vec x,t)\ , \eel{HJ}
where the ``source functions" \(J_i(\vec x,t),\ \ol\eta_i(\vec x,t)\) and \(\eta_i(\vec x,t)\) are freely chosen functions of space and time, to be replaced by zero at the end of the computation (the time dependence is not explicitly mentioned here in the field variables, but, in a Heisenberg representation, of course also the fields are time dependent).

These source terms could serve as simple models for the creation of the initial particles in a scattering experiment, as well as the detection process for the particles in the final state, but they can also simply be regarded as useful devices for a mathematical analysis of the physical properties of the system.
One can then find all amplitudes one needs to know, by computing at any desired order in perturbation theory, that is, up to certain powers of the coupling parameters and the source functions, the \emph{vacuum-to-vacuum amplitude}:
	\be &{}_{t=\infty}\bra\,\emptyset\,|\,\emptyset\,\ket_{t=-\infty}\iss1-\fract i2\int\int\dd^4x\dd^4x'\,J_i(x)J_j(x')\,P_{ij}(x-x')\,+&\nm\\
		&\fract 16\int\int\int\dd^4x\dd^4x'\dd^4x''\,W_{ijk}(x,x',x'')J_i(x)J_j(x')J_k(x'') +
		\cdots\ ,&\eel{pertexp}
where \(x\) stands short for the space-time coordinates \((\vec x,t)\), and the \emph{correlation functions} \(P_{ij}(x-x'),\ W_{ijk}(x,x',x'')\) and many more terms of the sequence are to be calculated. Physically, this means that we compute expectation values of the products of operators \(\F_i(\vec x,t),\ \ol\j_i(\vec x,t)\) and \(\j_i(\vec x,t)\) of Eq.~\eqn{HJ} in a Heisenberg representation. Local products of these operators also follow from the expressions \eqn{pertexp}, if we take space-time points \(x\) and \(x'\) to coincide.

The algorithm for the calculation of the two-point functions \(P\), the three-point functions \( W\), etc., is conveniently summarised in the so-called \emph{Feynman rules}, for which we refer to the standard text books.\,\cite{BdWSmith-1986}\cite{GtH-V-DIAGR-1973}\cite{GtH-ConcBasis-2007}

\subsubsection{Non-convergence of the coupling constant expansion\labell{nonconv}}
	There are some conditions where particles interact strongly. \emph{Quarks} are fermionic particles that interact so strongly that the forces between them keep them permanently bound in hadrons, the so-called quark confinement mechanism. This, however, only happens at the distance scale of the Standard Model, When extrapolated to the Planck scale, these strong interactions have been calculated to be about as weak as the other forces, notably electromagnetism and the weak force. This means that, in a conveniently large domain near the Planck scale, all perturbation expansions may be rapidly convergent: there, one \emph{never} needs to know the very high-order perturbative correction terms, since these are many times smaller than the usual margins of error in our description of the dynamics.
	
	It now so happens that, when we apply second quantisation in our cellular automaton models, something very similar may happen. If we choose our interactions to originate from rare coincidences in the cellular configurations, then most of the interaction events may be far separated at the Planck scale. This may imply that we have freely moving particles interacting only weakly by means of an interaction Hamiltonian. Since this Hamiltonian starts out to be local, only a few higher order calculations may suffice to obtain an accurate description of the dynamics. 
	
	We can now consider combining the quantum field theoretic perturbation expansion with the expansion needed to generate the interaction Hamiltonian itself. The resulting theory will still be accurate in the domain close to the Planck scale. Our proposal is to start from this theory, and to apply the usual renormalization group procedures (section~\ref{RG}) to transform everything to the Standard Model scale.

\subsecl{The algebraic structure of the general, renormalizable, relativistic quantum field theory}{algebra}

The reasons for limiting ourselves to \emph{renormalizable} quantum field theories are not completely obvious. When coupling strengths become large, renormalizable field theories may generate poles where the perturbation expansion diverges. We call these Landau poles. Renormalization is then of little help. The renormalization group (section~\ref{RG}) explains how the Landau poles can arise. If a Landau pole emerges in the small-distance domain, one has to conclude that the renormalization procedure fails, and here is little one can do about this. If however a Landau pole is related to a large distance divergence, it can be attributed to non-canonical behaviour of the force fields at large distances, which can be investigated and understood.

Landau poles do also occur when the couplings are weak, but since they are non-perturbative effects, these poles retreat to very distant domains of extremely high energies, so that they quickly turn harmless. This is the case where, by demanding renormalizability, we can select out a precisely defined class of models that are mathematically accurate, and most useful for comparison with experiments. They are not \emph{infinitely} accurate, but, as we shall see in section~\ref{2ndquantCA}, also the procedure that we can use to derive a field theory out of a cellular automaton, will have an accuracy that appears to be limited by the interaction strength.

Finally, we note that, indeed, in the Standard Model itself, the interaction parameters are remarkably small. This was not known or expected to be the case, a few decades ago.

Relativistically invariant, renormalizable quantum field theories have a remarkably rich mathematical structure. There are \emph{vector fields} (for elementary particles with spin 1), spinor fields (fermions with spin \(\halff\)), and scalars (spin 0).

The vector fields have to be associated with a local gauge theory, usually of the Yang-Mills type. The number of distinct vector particle species equals the number of dimensions, also called the rank, of the local gauge group. Electromagnetism has \(U(1)\) as its local gauge group; the dimension is 1, so there is one photon species.\\
The electro-weak interactions have this local gauge theory enlarged to \(U(1)\otimes SU(2)\), with group dimension 4, while the strong force adds to this \(SU(3)\), with dimension 8.

The fermionic and the scalar fields must all come as \emph{representations} of the gauge group. They each transform trivially or non-trivially under local gauge transformations. This determines how these particles couple to the vectorial gauge fields.

The fermions are based on Dirac's field equation, and the scalars start off with the Klein-Gordon equation. The interactions between these fields are written as Yukawa terms for the fermions, and quartic, sometimes also cubic, self interactions for the scalars.  The allowed couplings are severely constrained by the condition that the system has to be renormalizable and gauge-invariant. 

After all algebraic equations have been written down, it must be checked explicitly whether there are \emph{chiral anomalies}. These are clashes between current-conservation laws in the chiral symmetries one might expect in the theory. Anomalies that would be harmful for the self consistency of the theory only occur when right-handed fermions couple differently to the gauge fields than left handed ones. One then has to see to it that these anomalies cancel out. They indeed do in the Standard Model.

In the Standard Model, the algebra turns out to be arranged in such a way that the fermions come as three identical copies (``generations") of quarks and leptons. Quarks come as triplet representations of the gauge \(SU(3)\) group, while the leptons are \(SU(3)\) singlets. All fermions couple, at least to some extent, to the \(SU(2)\otimes U(1)\) gauge fields, with the exceptions of the right-handed components of the neutrinos, which do not couple to the gauge fields at all.

In principle, however, the mathematical rules known today would have allowed just any compact Lie group as the gauge group, and any kinds of representations for the fermions and the scalars, as long as there are not too many of those.

This summary here illustrates that the mathematical structure of the generic quantum field theory, and the Standard Model in particular, is fairly complex. It would have to be reproduced in a deterministic theory of nature. Further details are to be found in numerous text books. See for instance \,\cite{ItzyksonHuber-1980},
\cite{BdWSmith-1986}.

\subsecl{Vacuum fluctuations, correlations and commutators}{vacfluct}		
		
Because all contributions to our Hamiltonian are translation invariant, one expects the correlation functions to be translation invariant as well, and this is a good reason to consider their Fourier transforms, so, instead of \(x\) space, one considers \(k\) space:
	\be P_{ij}(x^{(1)}-x^{(2)})=(2\pi)^{-4}\int\dd^4k \^P_{ij}(k)\,e^{ik(x^{(1)}-x^{(2)})}\ , \eel{2point}
where we will often omit the caret (\( \^{ } \)).		

Disregarding factors \(2\pi\) for the moment, one finds that the two-point functions are built from elementary expressions such as the \emph{Feynman propagator},
	\be &\displaystyle{ \D^F_m(x)\equiv-i\int\dd^4k{e^{ikx}\over k^2+m^2-i\e}\ ,}&\nm\\[3pt]
	& x=x^{(1)}-x^{(2)}\ ,\qquad k^2=\vec k^2-k_0^2\ ,& \eel{Feynmanprop}
where \(\e\) is an infinitesimal positive number, indicating how one is allowed to arrange the complex contour when \(k_0\) is allowed to be complex. This propagator describes the contribution of a single, non interacting particle to the two-point correlation function. If there are interactions, one finds that, quite generally, the two-point correlation functions take the form of a \emph{dressed propagator}:
	\be \D_F(x)=\int_0^\infty\dd m\,\r(m)\,\D^F_m(x)\ , \eel{dressedprop}
where \(\r(m)\) is only defined for \(m\ge 0\) and it is always non-negative. This property is dictated by unitarity and positivity of the energy, and always holds exactly in a relativistic quantum field theory\,\cite{GtH-ConcBasis-2007}. The function \(\r(m)\) can be regarded as the probability that an intermediate state emerges whose centre-of-mass energy is given by the number \(m\). In turn, \(\r(m)\) can be computed in terms of Feynman diagrams with two external legs; it describes what may happen to a virtual particle as it travels from \(x^{(2)}\) to \(x^{(1)}\).  Diagrams with more external legs (which are usually the contributions to the scattering matrix with given numbers of free particles asymptotically far away in the in-state and the out-state), can be computed with these elementary functions as building blocks.

The two-point function physically corresponds to the vacuum expectation value of a time-ordered product of operators:
	\be P_{ij}(x^{(1)}-x^{(2)})=\bra\,\emptyset\,|T(\F_i(x^{(1)}),\F_j(x^{(2)})|\,\emptyset\,\ket\ , \eel{Tprod}
where 
	\be T(A(t_1),B(t_2))&=&A(t_1)B(t_2)\ ,\qquad\hbox{if }\ t_1>t_2\ ,\nm\\
	&=&B(t_2)A(t_1)\ ,\qquad\hbox{if }\ t_2>t_1\   \eel{Tproddef}
(for fermions, this is to be replaced by the \(P\) product: a minus sign is added if two fermions are interchanged).	

We shall now show how, in explicit calculations, it is always found that two operators \(\OO_1(x^{(1)})\) and \(\OO_2(x^{(2)})\) commute when they both are local functions of the fields \(F_i(\vec x,t)\), and when their space-time points are space-like separated: 
	\be(\vec x\,)^2-(x^0)^2>0\ ,\quad x=x^{(1)}-x^{(2)}\ . \eel{spacelikesep2}
To this end, one introduces the \emph{on-shell propagators}:
	\be\D^\pm_m(x)=2\pi\int\dd^4 k\,e^{ikx}\,\d(k^2+m^2)\theta(\pm k^0)\ ;\quad k\,x=\vec k\cdot\vec x-k^0x^0\ . \eel{Deltapm}	
By contour integration, one easily derives:
	\be \D_m^F(x)&=&\D_m^+(x)\qquad\hbox{if }\ x^0>0\ ; \nm\\
		&=&\D_m^-(x)=\D_m^+(-x)\quad\hbox{if }\ x^0<0\ ;\nm\\
		\D_m^{F*}(x)&=&\D_m^-(x)\qquad\hbox{if }\ x^0>0\ ;\nm\\
		&=&\D_m^+(x)\qquad\hbox{if }\ x^0<0\ . \eel{propproperties}
Here, \(\D_m^{F*} \) is obtained from the Feynman propagator \(D_m^F\) in Eq.~\eqn{Feynmanprop} by replacing \(i\) with \(-i\).	

Now we can use the fact that the expressions for \(\D_m^F(x)\) and \(\D_m^\pm(x)\) are Lorentz-invariant. Therefore, if \(x\) is space-like, one can always go to a Lorentz frame where \(x^0>0\) or a Lorentz frame where \(x^0<0\), so then,
	\be \D_m^F(x)=\D_m^+(x)=\D_m^-(x)=\D_m^{F*}(x)=\D_m^F(-x)\ . \eel{equaltime}
This implies that, in Eq.~\eqn{Tprod}, we can always change the order of the two operators \(\OO(x^{(1)})\) and \(\OO(x^{(2)})\) if \(x^{(1)}\) and \(x^{(2)}\)  are space-like separated. Indeed, for all two-point functions, one can derive from unitarity that they can be described by a dressed propagator	 of the form \eqn{dressedprop}, where, due to Lorentz invariance, \(\r(m)\) cannot depend on the sign of \(x^0\). The only condition needed in this argument is that the operator \(\OO_1(x^{(1)})\) is a local function of the fields \(\F_i(x^{(1)})\), and the same for  \(\OO_2(x^{(2)})\).
To prove that composite fields have two-point functions of the form \eqn{dressedprop}, using unitarity and positivity of the Hamiltonian, we refer to the literature\,\cite{GtH-V-DIAGR-1973}. To see that Eqs.~\eqn{equaltime} indeed imply that commutators between space-like separated operators vanish, and that this implies the non existence of information carrying signals between such points, we refer to section~\ref{commsignals}.
	
Now it is crucial to notice that the Feynman propagator \(\D^F_m(x)\) itself does \emph{not} vanish at space-like separations. In general, one finds for free fields with mass \(m\), at vanishing \(x^{(1)0}-x^{(2)0}\), and writing \(\vec r=\vec x^{(1)}-\vec x^{(2)}\), 
	\be \bra\,\emptyset\,|T(\F(x^{(1)}),\,\F(x^{(2)})\,|\,\emptyset\,\ket=(2\pi)^{-4}\D^F_m(0,\vec r\,)=\int\dd^3\vec k{1\over 2(2\pi)^3\sqrt{\vec k^2+m^2}}\,e^{i\vec k\cdot\vec r}\nm\\
	=\ {1\over (2\pi)^2}\int_0^\infty{k^2\over \sqrt{k^2+m^2}}\,e^{ik|\vec r\,|}\ ,\qquad{ } \eel{spacelikeprop}
but, since the fields here commute, we can omit the \(T\) symbol.
When the product \(m\,|\vec r|\) becomes large, this vanishes rapidly. But when \(m\) vanishes, we have long-range correlations:
	\be\bra\,\emptyset\,|\,\F(0,\vec r)\,\F(0,\vec 0\,)\,|\,\emptyset\,\ket = {1\over (2\pi)^2\,\vec r\,^2}\ . \eel{masslessprop}
For instance, for the photon field, the vacuum correlation function for the two-point function is, in the Feynman gauge,
	\be\bra\,\emptyset\,|\,A_\m(0,\vec r)\,A_\n(0,\vec 0\,)\,|\,\emptyset\,\ket ={g_{\m\n}\over (2\pi)^2\,\vec r\,^2}\ .		\eel{photonprop}

This means that we \emph{do} have correlations over space-like distances. We attribute this to the fact that we always do physics with states that are very close to the vacuum state. The correlations are non-vanishing in the vacuum, and in all states close to the vacuum (such as all \(n\)-particle states, with \(n\) finite). One may imagine that, at very high or infinite temperature, \emph{all} quantum states will contribute with equal probabilities to the intermediate states, and this may wipe out the correlations, but today's physics always stays restricted to temperatures that are very low compared to the Planck scale, most of the time, at most places in the Universe.

There is even more one can say. Due to the special analytic structure of the propagators \(\D^F_m(x)\), the \(n\) point functions can be analytically continued from Minkowski space-time to Euclidean space-time and back. This means that, if the Euclidean correlation functions are known, also the scattering matrix elements in Minkowski space-time follow, so that the entire evolution process at a given initial state can be derived if the space like correlation functions are known. Therefore, if someone thinks there is ``conspiracy" in the space-like correlations that leads to peculiar phenomena later or earlier in time, then this might be explained in terms of the fundamental mathematical structure of a quantum field theory. The author suspects that this explains why ``conspiracy" in ``unlikely" space-like correlations seems to invalidate the Bell and CHSH inequalities, while in fact this may be seen as a natural phenomenon. In any case, it should be obvious from the observations above, that the correlations in quantised field theories do not require any conspiracy, but are totally natural.
	
\subsecl{Commutators and signals}{commsignals}

We shall now show that,  just because  all space-like separated sets of operators commute, no signal can be exchanged that goes faster than light, no matter how entangled the particles are that one looks at. This holds for all relativistic quantum field theories, and in particular for the Standard Model.  This fact is sometimes overlooked in studies of peculiar quantum phenomena.	
	
Of course, if we replace the space-time continuum by a lattice in space, while time stays continuous, we lose Lorentz invariance, so that signals can go much faster, in principle (they still cannot go backwards in time).

 Consider a field \(\f(x)\), where \(x\) is a point in space-time. Let the field be self-adjoint:
	\be \f(x)=\f^\dag(x)\ . \eel{selfadjointfield}
	In conventional quantum field theories, fields are operators in the sense that they measure things and at the same time modify the state, all at one space-time point \(x\). Usually, the field averages in vacuum are zero:
	\be\bra\emptyset|\f(x)|\emptyset\ket=0\ . \ee	
Can a signal arrive at a point \(x^{(1)}\) when  transmitted from a point \(x^{(2)}\)? To find out, take the field operators \(\f(x^{(1)})\) and \(\f(x^{(2)})\). Let us take the case \(t^{(1)}\ge t^{(2)}\). In this case, consider the propagator
	\be&\bra\emptyset|T(\f(x^{(1)}),\, \f(x^{(2)}))|\emptyset\ket = \bra\emptyset|\f(x^{(1)})\,\f(x^{(2)})|\emptyset		\ket=&\nn
	&=\D^F_m(x^{(1)}-x^{(2)})=\D_m^+(x^{(1)}-x^{(2)})\ .&\eel{propagatortpos}
It tells us what the correlations are between the field values at \(x^{(1)}\) and at \(x^{(2)}\). This quantity does not vanish, as is typical for correlation functions, even when points are space-like separated.

The question is now whether the operation of the field at \(x^{(2)}\) can affect the state at \(x^{(1)}\). This would be the case if the result of the product of the actions of the fields depends on their order, and so we ask: to what extent does the expression \eqn{propagatortpos} differ from
	\be &\bra\emptyset|\f(x^{(2)})\,\f(x^{(1)})|\emptyset\ket=(\bra\emptyset|\f^\dag(x^{(1)})\,\f^\dag(x^{(2)})|\emptyset\ket)^*=(\bra\emptyset|T(\f(x^{(1)}),\,\f(x^{(2)}))|\emptyset\ket)^*=&\nn 
	&= \D_m^{F\,*}(x^{(1)}-x^{(2)})=\D^-_m(x^{(1)}-x^{(2)})\ . &\ee
In stead of \(\D_m(x)\) we could have considered the dressed propagators of the interacting fields, which, from general principles, can be shown to take the form of Eq.~\eqn{dressedprop}. We always end up with the identity \eqn{equaltime}, which means that the commutator vanishes:
	\be\bra\emptyset|\,[\f(x^{(1)}),\, \f(x^{(2)})]\,|\emptyset\ket=0\ , \ee
if \(x^{(1)}\) and \(x^{(2)}\) are space-like separated. Thus, it makes no difference whether we act with \(\f(x^{(1)})\) before or after we let \(\f(x^{(2)})\) act on the vacuum. This means that no signal can be sent from \(x^{(2)}\) to \(x^{(1)}\) if it would have to go faster than light.

Since Eqs.~\eqn{equaltime} can be proved to hold exactly in all orders of the perturbation expansion in quantum field theory, just by using the general properties \eqn{propproperties} of the propagators in the theory, one concludes that conventional quantum field theories never allow signals to be passed on faster than light. This is very important since less rigorous reasoning starting from the possible production of entangled particles, sometimes make investigators believe that there are `spooky signals' going faster than light in quantum systems. Whatever propagates faster than light, however, can never carry information. This holds for quantum field theories and it holds for cellular automata.

\subsecl{The renormalization group}{RG}

A feature of quantum field theories that plays a special role in our considerations is the renormalization group. This group consists of symmetry transformations that in their earliest form were assumed to be associated to the procedure of adding renormalization counter terms to masses and interaction coefficients of the theory. These counter terms are necessary to assure that higher order corrections do not become infinitely large when systematic (perturbative) calculations are performed. The ambiguity in separating interaction parameters from the counter terms can be regarded as a symmetry of the theory.\,\cite{PS-1953}

In practice, this kind of symmetry becomes important when one applies scale transformations in a theory: at large distances, the counter terms should be chosen differently from what they are at a small distance scale, if in both cases we require that higher order corrections are kept small. In practice, this has an important consequence for most quantum field theories: a scale transformation must be accompanied by small, calculable corrections to all mass terms and interaction coefficients. This then adds `anomalous dimensions' to the mass and coupling parameters\,\cite{Callan-1970}\cite{Symanzik-1970}\cite{GtH-ConcBasis-2007}. In lowest order, these anomalies are easy to calculate, and the outcome is typically:
	\be{\dd\over\dd\m}\l(\m)=\b_\l\,\l(\m)^2\ ,\qquad {\dd\over\dd\m}m(\m)=\b_m\, \l(\m)\, m(\m)\ , \eel{rgexpl}
with dimensionless coefficients \(\b_\l\) and \(\b_m\). Here, \(\m\) represents the mass scale at which the coupling and mass parameters are being considered. 

In gauge theories such as quantum electrodynamics, it is the charge squared, \(e(\m)^2\), or equivalently,  the fine structure constant \(\a(\m)\), that plays the role of the running coupling parameter \(\l(\m)\). A special  feature for non-Abelian gauge theories is that, there, the coefficient \(\b_{g^2}\) receives a large negative contribution from the gauge self couplings, so, unless there are many charged fields present, this renormalization group coefficient is negative.

It is important to observe, that the consideration of the renormalization group would have been quite insignificant had there not been large scale differences that are relevant for the theory. These differences originate from the fact that we have very large and very tiny masses in the system. In the effective Hamiltonians that we might be able to obtain from a cellular automaton, it is not quite clear how such large scale differences could arise. Presumably, we have to work with different symmetry features, each symmetry being broken at a different scale. Here we just note that this is not self-evident. The problem that we encounter here is the \emph{hierarchy problem}, the fact that enormously different \hbox{length-,} mass- and time scales govern our world, see section~\ref{hierarchy}. This is not only a problem for our theory here, it is a problem that will have to be confronted by any theory addressing physics at the Planck scale.

The mass and coupling parameters of a theory are not the only quantities that are transformed in a non-trivial way under a scale transformation. All local operators \(\OO(\vec x,t)\) will receive finite renormalizations when scale transformations are performed. When composite operators are formed by locally multiplying different kinds of fields, the \emph{operator product expansion} requires scale dependent counter terms. What this means is that operator expressions obtained by multiplying fields together undergo thorough changes and mixtures upon large scale transformations. The transformation that leads us from the Planck scale to the Standard Model scale is probably such a large scale transformation\fn{Unless several extra space-time dimensions show up just beyond the TeV domain, which would bring the Planck scale closer to the Standard Model scale.}, so that not only the masses and couplings that we observe today, but also the fields and operator combinations that we use in the Standard Model today, will be quite different from what they may look like at the Planck scale.

Note that, when a renormalization group transformation is performed, couplings, fields and operators re-arrange themselves according to their canonical dimensions. When going from high mass scales to low mass scales, coefficients with highest mass dimensions, and operators with lowest mass dimensions, become most significant. This implies that, seen from a large distance scale, the most complicated theories simplify since, complicated, composite fields, as well as the coefficients they are associated with, will rapidly become insignificant. This is generally assumed to be the technical reason why all our `effective' theories at the present mass scale are renormalizable field theories. Non-renormalizable coefficients have become insignificant. Even if our local Hamiltonian density may be quite ugly at the Planck scale, it will come out as a clean, renormalizable theory at scales such as the Standard Model scale, exactly as the Standard Model itself, which was arrived at by fitting the phenomena observed today.

The features of the renormalization group briefly discussed here, are strongly linked to Lorentz invariance. Without this invariance group, scaling would be a lot more complex, as we can see in condensed matter physics. This is the reason why we do not plan to give up Lorentz invariance without a fight.

\newsecl{The cellular automaton}{CAdetail}
The fundamental notion of a cellular automaton was briefly introduced in Part I, section~\ref{CAgeneral}. We here resume the discussion of constructing a quantum Hamiltonian for these classical systems, with the intention to arrive at some expression that may be compared with the Hamiltonian of a quantum field theory\,\cite{GtHCA-2010}, resembling Eq.~\eqn{HHdens}, with Hamiltonian density \eqn{scalarHdens}, and/or \eqn{fermifieldham}. In this chapter, we show that one can come very close, although, not surprisingly, we do hit upon difficulties that have not been completely resolved.

\subsecl{Local time reversibility by switching from even to odd sites and back}{evenodd}
	Time reversibility is important for allowing us to perform simple mathematical manipulations. Without time reversibility, one would not be allowed to identify single states of an automaton with basis elements of a Hilbert space. Now this does not invalidate our ideas if time reversibility is not manifest; in that case one has to identify basis states in Hilbert space with information equivalence classes, as was explained in section~\ref{infoloss}. The author does suspect that this more complicated situation might well be inevitable in our ultimate theories of the world, but we have to investigate the simpler models first. They are time reversible. Fortunately, there are rich classes of time reversible models that allow us to sharpen our analytical tools, before making our lives more complicated.
	
Useful models are obtained from systems where the evolution law \(U\) consists of two parts:  \(U_A\) prescribes how to update all even lattice sites, and \(U_B\)  gives the updates of the odd lattice sites. So we have \(U=U_A\cdot U_B\).

\subsubsection{The time reversible cellular automaton\labell{margolus}}

In section~\ref{CAgeneral}, a very simple rule was introduced. The way it was phrased there, the data on two consecutive time layers were required to define the time evolution in the future direction as well as back towards the past -- these automata are time reversible.  Since, quite generally, most of our models work with single time layers that evolve towards the future or the past, we shrink the time variable by a factor 2. Then, one complete time step for this automaton consists of two procedures: one that updates all even sites only, in a simple, reversible manner, leaving the odd sites unchanged, while the procedure does \emph{depend} on the data on the odd sites, and one that updates only the odd sites, while depending on the data at the even sites. The first of these operators is called \(U_A\). It is the operator product of all operations \(U_A(\vec x)\), where \(\vec x\) are all even sites, and we take all these operations to commute:
	\be U_A=\prod_{\vec x\ {\mathrm even}}U_A(\vec x)\ ;\qquad[U_A(\vec x),\,U_A(\vec x\,')]\ =0\ ,\quad \forall\ \vec x,\,\vec x\,'\ . \eel{evensitesop}
The commutation is assured if \(U_A(\vec x)\) depends only on its neighbours, which are odd, but not on the next-to-nearest neighbours, which are even again. Similarly, we have the operation at the odd sites:
	\be U_B=\prod_{\vec y\ {\mathrm odd}}U_B(\vec y)\ ;\qquad[U_B(\vec y),\,U_B(\vec y\,')]\ =0\ ,\quad \forall\ \vec y,\,\vec y\,'\ ,\eel{oddsitesop}
	while \([U_A(\vec x),\,U_B(\vec y)]\ne 0\) only if \(\vec x\) and \(\vec y\) are direct neighbours.

In general, \(U_A(\vec x)\) and \(U_B(\vec y)\) at any single site are sufficiently simple (often they are finite-dimensional, orthogonal matrices) that they are easy to write as exponentials:
	\be U_A(\vec x)&=&e^{-iA(\vec x)}\ ,\qquad [A(\vec x),A(\vec x\,')]=0\ ;\nm\\
	\qquad U_B(\vec y)&=&e^{-iB(\vec y)}\ ,\qquad [B(\vec y),B(\vec y\,')]=0\ . \eel{ABexpab}
\(A(\vec x)\) and \(B(\vec y)\) are defined to lie in the domain \([0,2\pi)\), or sometimes in the domain \((-\pi,\pi]\).
	
The advantage of this notation is that we can now write\fn{The sign in the exponents is chosen such that the operators \(A\) and \(B\) act as Hamiltonians themselves.}
	\be U_A=e^{-iA}\ ,\quad A=\sum_{\vec x\ {\mathrm even}}A(\vec x)\ ;\qquad U_B=e^{-iB}\ ,\quad B=\sum_{\vec y\ {\mathrm odd}}B(\vec y)\ , \eel{ABsums}  
and the complete evolution operator for one time step \(\d t=1\) can be written as
	\be U(\d t)=e^{-iH}=e^{-iA}\,e^{-iB}\ . \eel{Uproduct}
	
Let the data in a cell \(\vec x\) be called \(Q(\vec x)\). In the case that the operation \(U_A(\vec x)\) consists of a simple addition (either a plane addition or an addition \emph{modulo} some integer \(\Bbb{N}\)) by a quantity \(\d Q(Q(\vec y_i))\), where \(\vec y_i\) are the direct neighbours of \(\vec x\), then it is easy to write down explicitly the operators \(A(\vec x)\) and \(B(\vec y)\). Just introduce the translation operator 
	\be U_\eta (\vec x)=e^{i\eta (\vec x)}\ ,\qquad U_\eta |\,Q(\vec x)\,\ket\equiv|\,Q(\vec x)-1\ \mathrm{modulo}\ \Bbb N\,\ket\ , \eel{Qtranslate}
to find	
	\be &U_A(\vec x)=e^{-i\eta (\vec x)\,\d Q(Q(\vec y_i))}\ ,&\nm\\[4pt]
	& A(\vec x)=\eta (\vec x)\,\d Q(Q(\vec y_i))\ ;\qquad 
	B(\vec y)=\eta (\vec y)\,\d Q(Q(\vec x_i))\ . &\eel{AvecxBvecy}

The operator \(\eta (\vec x)\) is not hard to analyse. Assume that we are in a field of additions \emph{modulo}  \(\Bbb N\), as in Eq.~\eqn{Qtranslate}. Go the basis of states \(|k\ket_{\hbox{\tiny{U}}}\), with \(k=0,\,1,\cdots,\, \Bbb N-1\), where the subscript \(\scriptstyle{\mathrm U}\) indicates that they are eigenstates of \(U_\eta \) and \(\eta \) (at the point \(\vec x\)):
	\be \bra Q|k\ket_{\hbox{\tiny{U}}}\equiv {1\over\sqrt{\Bbb N}}\,e^{2\pi ikQ/\Bbb N}\ . \eel{kQmatrix}
We have
	\be \bra Q|U_\eta |k\ket_{\hbox{\tiny{U}}} =\bra Q+1|k\ket_{\hbox{\tiny{U}}}=e^{2\pi  i k/\Bbb N}\bra Q|k\ket_{\hbox{\tiny{U}}}\ ;\qquad U_\eta |k\ket=e^{2\pi i k/\Bbb N}|k\ket_{\hbox{\tiny{U}}} \eel{UPk}
(if \(-\half\Bbb N<k\le\half\Bbb N\)), so we can define \(\eta \) by
	\be   \eta |k\ket_{\hbox{\tiny{U}}}={2\pi\over\Bbb N}k|k\ket_{\hbox{\tiny{U}}}\ , &&
	\bra Q_1|\eta |Q_2\ket=\sum_k\bra Q_1|k\ket_{\hbox{\tiny{U}}}(\fract {2\pi}{\Bbb N }k)\,_{\hbox{\tiny{U}}}
	\;\!\!   \bra k|Q_2\ket \nm\\
	=\  {2\pi\over\Bbb N^2}\sum_{|k|<\halfje\Bbb N}k\,e^{2\pi i k(Q_1-Q_2)/\Bbb N}
	&=&{4\pi i\over\Bbb N^2}\sum_{k=1}^{\halfje\Bbb N}k\sin(2\pi k(Q_1-Q_2)/\Bbb N)\ , \eel{matrixPQ}
mathematical manipulations that must look familiar now, see Eqs.~\eqn{hammatrixelements} and \eqn{NHam} in section~\ref{cogwheelN}.

Now \(\d Q(\vec y_i)\) does not commute with \(\eta (\vec y_i)\), and in Eq.~\eqn{AvecxBvecy} our model assumes the sites \(\vec y_i\) to be only direct neighbours of \(\vec x\) and \(\vec x_i\) are only the direct neighbours of \(\vec y\). Therefore, all \(A(\vec x)\) also commute with \(B(\vec y)\) unless \(|\vec x-\vec y\,|=1\). This simplifies our discussion of the Hamiltonian \(H\) in Eq.~\eqn{Uproduct}.

\subsubsection{The discrete classical Hamiltonian model\labell{discrclassham.sub}}

In subsection~\ref{discrfields}, we have seen how to generate a local discrete evolution law from a classical, discrete Hamiltonian formalism. Starting from a discrete, non negative Hamiltonian function \(H\), typically taking values \(N=0,1,2,\cdots\),one searches for an evolution law that keeps this number invariant. This classical \(H\) may well be defined as a sum of local terms, so that we have a non negative discrete Hamiltonian density. It was decided that a local evolution law \(U(\vec x)\) with the desired properties can be defined, after which one only has to decide in which order this local operation has to be applied to define a unique theory. In order to avoid spurious non-local behaviour, the following rule was proposed:
	\begin{quote} The evolution equations (e.o.m.) of the entire system over one time step \(\d t\), are obtained by ordering the coordinates as follows: \emph{first} update all even lattice sites, \emph{then} update all odd lattice sites  \end{quote}
	(how exactly to choose the order within a given site is immaterial for our discussion). 
	The advantage of this rule is that the \(U(\vec x)\) over all even sites \(\vec x\) can be chosen all to commute, and the operators on all odd sites \(\vec y\) will also all commute with one another; the only non-commutativity then occurs between an evolution operator \(U(\vec x)\) at an even site, and the  operator \(U(\vec y)\) at an \emph{adjacent} site \(\vec y\).
	
	Thus, this model ends up with exactly the same fundamental properties as the time reversible automaton introduced in subsection~\ref{margolus}: we have \(U_A\) as defined in Eq.~\eqn{evensitesop} and \(U_B\) as in \eqn{oddsitesop}, followed by Eqs.~\eqn{ABexpab}--\eqn{Uproduct}.

We conclude that, for model building, splitting a space-time lattice into the even and the odd sub lattices is a trick with wide applications. It does not mean that we should believe that the real world is also organised in  a lattice system, where such a fundamental role is to be attributed to the even and odd sub lattices; it is merely a convenient tool for model building. We shall now discuss why this splitting does seem to lead us \emph{very close} to a quantum field theory.

\subsecl{The Baker Campbell Hausdorff expansion}{BCH.sec}	

The two models of the previous two subsections, the arbitrary cellular automaton and the discrete Hamiltonian model, are very closely related. They are both described by an evolution operator that consists of two steps, \(U_A\) and \(U_B\), or, \(U_\mathrm{even}\) and \(U_\mathrm{odd}\).  The same general principles apply.  We define \(A, \  A(\vec x), \ B\) and \(B(\vec x)\) as in Eq.~\eqn{ABsums}. 

To compute the Hamiltonian \(H\), we may consider using the Baker Campbell Hausdorff expansion\,\cite{BCH-1982}:	
	 \be &e^P\,e^Q=e^R\ ,\qquad\qquad\qquad&\nn
& {\ }\hskip-30pt R=P+Q+\half[P,Q]+\fract{1}{12}[P,[P,Q]]+\fract 1{12}[[P,Q],Q]+\fract{1}{24}[[P,[P,Q]],Q]+\cdots\ ,\quad &\eel{BCH.eq}
a series that continues exclusively with commutators. Replacing \(P\) by \(-iA\), \(Q\) by
\(-iB\) and \(R\) by \(-iH\), we find a series for  \(H\) in the form of an
infinite sequence of commutators. Now we noted at the end of the previous subsection that the commutators between the local operators		\def\rvec{\vec r\;\!}
\(A(\vec x)\) and \(B(\vec x\,')\) are non-vanishing only if \(\vec x\) and \(\vec x\,'\) are
neighbours, \(|\vec x-\vec x\,'|=1\). Therefore, if we insert the sums \eqn{ABsums} into
Eq.~\eqn{BCH.eq}, we obtain again a sum. Writing 
	\be K(\rvec)&=&A(\rvec)\ \hbox{ if \(\rvec\) is even, and }\ B(\rvec)\ \hbox{ if \(\rvec\) is odd,} \nm\\
	L(\rvec)&=&A(\rvec)\ \hbox{ if \(\rvec\) is even, and }-\!B(\rvec)\ \hbox{ if \(\rvec\) is odd,}\eel{ABevenodd}
so that
	\be A(\vec r)=\half\bigg(K(\vec r)+L(\vec r)\bigg) \ \hbox{ and }\ B(\vec r)=\half\bigg(K(\vec r)-L(\vec r)\bigg)\ , \ee
we find
 	\be && H=\sum_{\rvec}\HH(\rvec)\ ,\nn
 &&\HH(\rvec)= \HH_1(\rvec)+\HH_2(\rvec)+\HH_3(\rvec)+\cdots\ ,
 \eel{Hsum}   where
	\be\HH_1(\rvec)&=&K(\rvec)\ ,\nn
 \HH_2(\rvec)&=&\quart i \sum_{\vec s}[K(\rvec),\,L(\vec s\;\!)]\ , \nn
 \HH_3(\rvec)&=&\fract 1{24} \sum_{\vec s_1,\,\vec s_2} [L(\rvec)\,,\,[K(\vec
 s_1),L(\vec s_2)]] \ , \quad\hbox{etc.} \eel{Hdensitysum}
 The sums here are only over close neighbours, so that each term here can be regarded as a local Hamiltonian density term.
 
Note however, that as we proceed to collect higher terms of the expansion, more and more distant sites will eventually contribute; \(\HH_n(\rvec)\) will receive contributions from sites at distance \(n-1\) from the original point \(\vec r\).
 
Note furthermore that the expansion \eqn{Hsum} is infinite, and convergence is not guaranteed; in fact, one may suspect it not to be valid at all, as soon as energies larger than the inverse of the time unit \(\d t\) come into play. We will have to discuss that problem. But first an important observation that improves the expansion.

\subsecl{Conjugacy classes}{conjugacy.sec}
One might wonder what happens if we change the order of the even and the odd sites. We would get
	\be U(\d t)=e^{-iH}\qu e^{-iB}\,e^{-iA}\ , \eel{HBA}
in stead of Eq.~\eqn{Uproduct}. Of course this expression could have been used just as well. In fact, it results from a very simple basis transformation: we went from the states \(|\j\ket\) to the states \(U_B|\j\ket\). As we stated repeatedly, we note that such basis transformations do not affect the physics.

This implies that we do not need to know exactly the operator \(U(\d t)\) as defined in Eqs.~\eqn{Uproduct} or \eqn{HBA}, we just need any one element of its conjugacy class. The conjugacy class is defined by the set of operators of the form
	\be G\,U(\d t)\,G^{-1}\ , \eel{conjugacy.eq}
where \(G\) can be any unitary operator.  Writing \(G=e^{F}\), where \(F\) is anti-hermitean, we replace Eq.~\eqn{BCH.eq} by
	\be e^{\tl{R}}=e^F\,e^P\,e^Q\,e^{-F}\ , \eel{conjugacyBCH}
so that
	\be\tl R=R+[F,R]+\half[F,[F,R]]+\fract 1{3!}[F,[F,[F,R]]]+\cdots . \eel{Rtilde}
We can now decide to choose \(F\) in such a way that the resulting expression becomes as simple as possible. For one, interchanging \(P\) and \(Q\) should not matter. Secondly, replacing \(P\) and \(Q\) by \(-P\) and \(-Q\) should give \(-\tl R\), because then we are looking at the inverse of Eq.~\eqn{Rtilde}. 

The calculations simplify if we write	
	\be S=\half(P+Q)\ ,\quad D=\half(P-Q)\ ;\qquad P=S+D\ ,\quad Q=S-D \eel{BCHvariables}
(or, in the previous section, \(S=-\half i K\,,\ D=-\half i L\)). With 
	\be e^{\tl R}=e^F\,e^{S+D}\,e^{S-D}\,e^{-F}\ , \eel{conjBCH}
we can now demand \(F\) to be such that:
	\be\tl R(S,D)=\tl R(S,-D)=-\tl R(-S,-D)\ , \eel{conjconstraints}
which means that \(\tl R\) contains only even powers of \(D\) and odd powers of \(S\). We can furthermore  demand that \(\tl R\) only contains terms that are commutators of \(D\) with something; contributions that are commutators of \(S\) with something can be removed iteratively by judicious choices of \(F\).

Using these constraints, one finds a function \(F(S,D)\) and \(\tl R(S,D)\). First let us introduce a short hand notation. All our expressions will consist of repeated commutators. Therefore we introduce the notation
	\be\{X_1,X_2,\cdots,X_n\}\equiv [X_1,\,[X_2,\,[\ \cdots,\ X_n\,]]\cdots]\ . \eel{commnotation}
Subsequently, we even drop the accolades \(\{\ \}\). So when we write
	\[ X_1\,X_2\,X_3^2\,X_4\ ,\ \hbox{ we actually mean: }\ [X_1,\,[X_2,\,[X_3,\,[X_3, X_4]]]]\ . \]	

Then, with \(F=-\half D+\fract 1{24}S^2D+\cdots\), one finds
	\be &\tl R(S,D)=2S-\fract 1 {12} DSD+\fract 1{960}D(8S^2-D^2)SD\,+&\nn
	&\fract 1{60480}D(-51\, S^4-76\,DSDS+33\,D^2S^2+44\,SD^2S-\fract 38 D^4)SD\,+\ \OO(S,D)^9\ .\qquad &\eel{conjexp}

There are three remarks to be added to this result:
\bi{(1)} It is much more compact than the original BCH expansion; already the first two terms of the expansion \eqn{conjexp} correspond to all terms shown in Eq.~\eqn{BCH.eq}.
\itm{(2)} The coefficients appear to be much smaller than the ones in \eqn{BCH.eq}, considering the factors \(\halff\) in Eqs.~\eqn{BCHvariables} that have to be included. We found that, in general, sizeable cancellations occur between the coefficients in \eqn{BCH.eq}.
\itm{(3)} However, there is no reason to suspect that the series will actually converge any better. The definitions of \(F\) and \(\tl R\) may be expected to generate singularities when \(P\) and \(Q\), or \(S\) and \(D\) reach values where \(e^{\tl R}\) obtains eigenvalues that return to one, so, for finite matrices, the radius of convergence is expected to be of the order \(2\pi\). \ei

In this representation, all terms \(\HH_n(\rvec)\) in Eq.~\eqn{Hsum} with \(n\) even, vanish. Using 
	\be S=-\half i K\ ,\qquad D=-\half i L\ ,\qquad \tl R=-i\tl H\ , \eel{SKdef}
one now arrives at the Hamiltonian in the new basis:
		\be &\tl\HH_1(\rvec)=K(\rvec)\ , &\nn
		&\tl\HH_3(\rvec)=\fract 1{96}\sum_{\vec s_1,\vec s_2}[L(\rvec),[K(\vec s_1),L(\vec s_2)]]\ ,&\crl{conjHamexp}
		& {\ }\hskip-10pt
		 \tl\HH_5(\rvec)=\fract 1{30720}\sum_{\vec s_1,\cdots,\vec s_4}
		[L(\rvec),\bigg(8[K(\vec s_1),[K(\vec s_2)-
		[L(\vec s_1),[L(\vec s_2)\bigg),\,[K(\vec s_3),L(\vec s_4)]]]]\ ,\qquad		\nm	\ee
and \(\tl\HH_7\) follows from the last line of Eq.~\eqn{conjexp}.	
 
All these commutators are only non-vanishing if the coordinates \(\vec s_1\), \(\vec s_2\),  etc., are all neighbours of the coordinate \(\vec r\). It is true that, in the higher order
terms, next-to-nearest neighbours may enter, but still, one may observe that these operators all are 
local functions of the `fields' \(Q(\vec x,\,t)\), and thus we arrive at a Hamiltonian \(H\) that
can be regarded as the sum over \(d\)-dimensional space of a Hamiltonian density \(\HH(\vec x)\),
which has the property that
 \be[\HH(\vec x),\,\HH(\vec x\,')]=0\ ,\quad\hbox{if}\quad|\vec x,\,\vec x\,'|\gg 1\ .
\eel{Hdensitycomm} At every finite order of the series, the Hamiltonian density \(\HH(\vec x)\) is a
finite-dimensional hermitean matrix, and therefore, it will have a lowest eigenvalue \(h\). In a
large but finite volume \(V\), the total Hamiltonian \(H\) will therefore also have a lowest
eigenvalue, obeying
 \be E_0>h\,V\ . \eel{eigenH}
 
However, it was tacitly assumed that the Baker-Campbell-Hausdorff formula converges. This is usually not the case. One can argue that the
series may perhaps converge  if sandwiched between two eigenstates \(|E_1\ket\) and \(|E_2\ket\)
of \(H\), where \(E_1\) and \(E_2\) are the eigenvalues, that obey
 \be |E_1-E_2|\ll 2\pi\ , \eel{eigenconstraint}
We cannot exclude that the resulting
effective quantum field theory will depend quite non-trivially on the number of Baker-Campbell-Hausdorff terms that
are kept in the expansion.

The Hamiltonian density \eqn{conjHamexp} may appear to be quite complex and unsuitable to serve as a quantum field theory model, but it is here that we actually expect that the renormalization group will thoroughly cleanse our Hamiltonian, by invoking the mechanism described at the end of subsection~\ref{RG}.
		
\newsecl{The problem of quantum locality}{quloc}
	When we have a classical cellular automaton, the condition of locality is easy to formulate and to impose. All one needs to require is that the contents of the cells are being updated at the beat of a clock: once every unit of time, \(\d t\). If we assume the updates to take place in such a way that every cell is only affected by the contents of its direct neighbours, then it will be clear that signals can only be passed on with a limited velocity, \(c\), usually obeying
		\be |c|\le |\d x/\d t|\ , \eel{celvelocity}
where \(|\d x|\) is the distance between neighbouring cells. One could argue that this is a desirable property, which at some point might be tied in with special relativity, a theory that also demands that no signals go faster than a limiting speed \(c\).

	The notion of locality in quantum physics is a bit more subtle, but in quantum field theories one can also demand that no signals go faster than a limiting speed \(c\). If a signal from a space-time point \(x^{(1)}\) can reach an other space-time point \(x^{(2)}\), we say that \(x^{(2)}\) lies in the forward light cone of \(x^{(1)}\). If \(x^{(2)}\) can send a signal to \(x^{(1)}\), then \(x^{(2)}\) is in the backward light cone of \(x^{(1)}\); if neither \(x^{(1)}\) can affect \(x^{(2)}\) nor \(x^{(2)}\) can affect \(x^{(1)}\), we say that \(x^{(1)}\) and \(x^{(2)}\) are space-like separated.
		
The way to implement this in quantum field theories is by constructing a Hamiltonian in such a way that, for space-like separated space-time points, all quantum operators defined at \(x^{(1)}\) \emph{commute} with all operators defined at \(x^{(2)}\). The quantum field theories used to describe the Standard Model obey this constraint. Physically, one can easily see that, under this condition, performing any operation at \(x^{(1)}\) and any measurement at \(x^{(2)}\), give the same result regardless the order of these two operations.

However, the existence of light cones, due to a fixed light velocity \(c\), would not have been easy to deduce from the Hamiltonian unless the theory happens to obey the restrictions of special relativity. If the model is self-consistent in different inertial frames, and space-like operators commute at equal times, then relativity theory tells us they must commute everywhere outside the light cone. Now, most of our cellular automaton models fail to obey special relativity -- not because we might doubt on the validity of the theory of special relativity, but because relativistically invariant cellular automaton models are extremely difficult to construct.		Consequently, our effective Hamiltonians for these models tend to be non-commutative also outside the light cone, in spite of the fact that the automaton cannot send signals faster than light.

This is one of the reasons why our effective Hamiltonians do not even approximately resemble the Standard Model. This does seem to be a mere technical problem, but it is a very important one, and the question we now wish to pose is whether any systematic approach can be found to cure this apparent disease.

This important question may well be one of the principal reasons why as of the present only very few physicists are inclined to take the Cellular Automaton Interpretation seriously. It is as if there is a fundamental obstacle standing in the way of reconstructing existing physical models of the world using cellular automata.

Note, that the Baker-Campbell-Hausdorff expansion does seem to imply a weaker form of quantum locality: if we terminate the expansion at any finite order \(N\), then the effective Hamiltonian density \(\HH(\vec x^{(1)})\) commutes with \(\HH(\vec x^{(2)})\)  at equal times, if \(|\vec x^{(1)}-\vec x^{(2)}|\ge N\,|\d \vec x|\). As we stated earlier, however, this is not good enough, because the BCH expansion is not expected to converge at all. We have to search for better constructions.

\subsection{Second quantisation in Cellular Automata\labell{2ndquantCA}}

A promising approach for dealing with the danger of non-locality and unboundedness of the Hamiltonian may also be to stick more closely to quantum field theories. As it turns out, this requires that we first set up automata that describe freely moving particles; subsequently, one follows the procedure of second quantisation, described in section~\ref{fermions}, and further elaborated in subsection~\ref{secondneutrino}, and section~\ref{2ndquantisation}.

In our theory, this means that we first have to describe deterministic motion of a single particle. We have already examples: the massless ``neutrino" of section~\ref{nu}, and the superstring, section~\ref{sustr}, but if we wish to reproduce anything resembling the Standard Model, we need the complete set, as described in section~\ref{algebra}: fermions, scalar bosons, gauge bosons, and perhaps gravitons. This will be difficult, because the fields we have today are mirroring the wave functions of standard quantum particles, which propagate non-deterministically. 

We now have to replace these by the wave functions of deterministic objects, in line with what has been discussed before, and rely on the expectation that, if we do this right, renormalization group effects may turn these into the more familiar quantised fields we see in the Standard Model. The advantage of this approach should be that now we can start with first-quantised states where the energy needs not be bounded below; second quantisation will take care of that: we get particles and antiparticles, see section~\ref{2ndquantisation}.

A conceivable approach towards deterministic first-quantised particles is to start with discrete \(PQ\) variables: discrete positions of a particle are labelled by three dimensional vectors \(\vec Q\), and their momenta by discrete variables \(\vec P\). The conjugated variables are fractional momenta \(\vec\k\) and fractional positions \(\vec \xi\), and we start with
	\be e^{-i\k_i}|\vec Q,\,\vec P\ket=|\vec Q+e_i,\,\vec P\ket\ ,\qquad e^{i\xi_i}|\vec Q,\,\vec P\ket=|\vec Q,\,\vec P+e_i\ket\ , \eel{PQkappaeta}
where \(e_i\) are unit vectors spanning over one lattice unit in the \(i^{\,\th}\) direction.
Subsequently, we add phases \(\vv(\vec\k,\vec\xi)\) as in chapter~\ref{PQ}. We then choose deterministic evolution equations for our `primordial' particle in terms of its \((\vec P,\,\vec Q)\) coordinates. Our first attempt will be to describe a fermion. It should resemble the ``neutrino" from chapter~\ref{nu}, but we may have to replace the sheets by point particles. This means that the primordial particle cannot obey Dirac's equation. The important point is that we give the primordial particle a Hamiltonian \(h^\op_0\) having a spectrum ranging from \(-\pi/\d t\) to \(\pi/\d t\) in natural units, using the systematic procedure described in chapter~\ref{locality}. \(h_0^\op\) does not yet describe interactions. It is a free Hamiltonian and therefore it allows for a detailed calculation of the particle's properties, which of course will be  trivial, in a sense, until we add interactions.
	
	Upon second quantisation then, if the negative energy levels are filled, and he positive energy levels are kept empty, we have the vacuum state, the state with lowest energy. 
	
	This gives us a local Hamilton density \(H_0^\op\), and the evolution operator over one unit \(\d t\) in time will be generated by the operator (see section~\ref{fermions}; we set \(\d t=1\) again)
		\be U_A^\op=e^{-iH_0^\op}\ ,\qquad H^\op=\ol\j\,h^\op_0\,\j\ , \eel{Hpsipsi}
	This operator obeys locality \emph{and} positivity by construction: locality follows from the observation described in chapter~\ref{locality}, which is that the expansion of arc sines converge rapidly when we limit ourselves to the middle of the energy spectrum, and positivity follows from second quantisation.
	
	Now we carefully insert interactions. These will be described by an evolution operator \(U_B^\op=e^{-iB^\op}\). Of course, this operator must also be deterministic. Our strategy is now that \(U^\op_B\) will only generate \emph{rare, local} interactions; for instance, we can postulate that two particles affect each other's motion only under fairly special circumstances of the surrounding vacuum. Note, that the vacuum is filled with particles, and these degrees of freedom may all play a role. We ensure that the interaction described by \(U^\op_B\) is still local, although perhaps \emph{next-to-next-to} nearest neighbours could interact.
	
	The total evolution operator is then
	\be e^{-iH^\op}=U^\op(\d t)=(U_A^\op)^\half\,U_B^\op\,(U_A^\op)^\half=e^{-\half iH_0^\op}\,e^{-iB^\op}\,e^{-\half iH_0^\op}\ , \eel{2ndquantint}
	which we symmetrised with the powers \(\halff\) for later convenience.
This, we subject to the Baker-Campbell-Hausdorff expansion, section~\ref{BCH.sec}.	

Now, it is important to make use of the fact that \(B^\op\) is small. Expanding with respect to \(B^\op\), we may start with just the first non-trivial term. Using the notation defined in Eq.~\eqn{commnotation} of section~\ref{conjugacy.sec}, we can find the complete set of terms in the expansion of \(H^\op\) to all powers of \(H_0^\op\) but up to terms linear in \(B^\op\) only.
This goes as follows: let \(A,\,B,\) and \(C(t)\) be operators, and \(t\) an arbitrary, small parameter. We use the fact that
	\be e^F\,R\,e^{-F}\iss\{e^F,\,R\}\iss R+[F,R]+\fract 1{2!}[F,[F,R]]+\fract 1{3!}[F,[F,[F,R]]]+\cdots\ ,\eel{expbracket}
and an expression for differentiating the exponent of an operator \(C(t)\), to derive:
	\be e^{C(t)}&=&e^{\half A}\,e^{tB}\,e^{\half A}\ , \nn
	\fract{\dd}{\dd t}e^{C(t)}&=&\int_0^1\dd x\, e^{xC(t)}\,\fract{\dd C}{\dd t}\,e^{(1-x)C(t)}\iss e^{\half A}Be^{tB}e^{\half A}\iss
		e^{\half A}\,B\,e^{-\half A}\,e^{C(t)}\nn
	\ra && \int_0^1\dd x\,\{e^{xC(t)},\,\fract{\dd C}{\dd t}\}\iss\{\fract{e^{C(t)}-1}{C(t)},\,\fract{\dd C}{\dd t}\}
		\iss	\{e^{\half A},\,B\}\ , \eel{BCHdiffvgl}
where, in the last line, we multiplied with \(e^{-C(t)}\) at the right.
Expanding in \(t\), writing \(C(t)=A+tC_1+\OO(t^2)\), we find
	\be \fract{\dd C(t)}{\dd t}=\{\fract{C(t)}{e^{C(t)}-1}\,e^{\half A}\,,\,B\}\ ;\quad C(t)=A+\{\fract{A}{2\sinh(\half A)},\,tB\}+
		\OO(t^2B^2)\ . \eel{linearB}
From this we deduce that the operator \(H^\op\) in Eq.~\eqn{2ndquantint} expands as
	\be H^\op=H_0^\op+\{\fract{H_{0\vphantom{|}}^\op}{2\sin(\half H_0^\op)},\,B^\op\}+\OO(B^\op)^2\ . \eel{HopexpandB}
Expanding the inverse sine, the accolades give
	\be \fract x{2\sin(x/2)}&=&1+\fract 1{24} x^2+\fract 7{5760}x^4+\cdots\quad \ra\nn 
	\{\fract{H_{0\vphantom{|}}^\op}{2\sin(\half H_0^\op)},\,B^\op\}&=&B^\op+\fract1{24}[H_0^\op,[H_0^\op,B^\op]]+
	\fract 7{5760}[H_0^\op,[H_0^\op,[H_0^\op,[H_0^\op,B^\op]]]]+\cdots\nm\\[-5pt]
	&&+\ \OO(B^\op)^2\ .\eel{HexplexpandB}
	
We can now make several remarks:
	\bi{--} This is again a BCH expansion, and again, one can object that it does not converge, neither in powers of \(H_0^\op\) nor in powers of \(B^\op\).
		\itm{--} However, now \(B^\op\) may be assumed to be small, so we do not have to go to high powers of \(B^\op\), when we wish to compute its effect on the Hamiltonian.
		\itm{--} But the expansion in \(H_0^\op\) at first sight looks worrying. However:
		\itm{--} We have the entire expression \eqn{HopexpandB}, \eqn{HexplexpandB} to our disposal. The term linear in \(B^\op\) can be rewritten as follows:\\
			Let \(B_{k\ell}\) be the matrix elements of the operator \(B^\op\) in the basis of eigenstates \(|E_0\ket\) of \(H_0^\op\). 
			Then, the expression in accolades can be seen to generate the matrix elements of \(H^\op\):
		\be H_{k\ell}=\fract {\D E^o_{k\ell\vphantom{|}}}{2\sin(\half\D E_{k\ell}^o)}\,B_{k\ell}\ , \eel{Hopmatrixelem}
			where \(\D E^o_{k\ell}\) is the energy difference between the two basis elements considered.
		\itm{--} Since \(B^\op\) is considered to be small, the energies \(E^0\) of the states considered are expected to stay very close to the total energies of these states.
		\itm{--} It is perhaps not unreasonable now to assume that we may limit ourselves to `soft templates', where only low values for the total energies are involved. This may mean that we might never have to worry about energies as large as \(2\pi/\d t\), where we see the first singularity in the expansion \eqn{HexplexpandB}. \ei
Thus, in this approach, we see good hopes that only the first few terms of the BCH expansion suffice to get a good picture of our interacting Hamiltonian. These terms all obey locality, and the energy will still be bounded from below.

There will still be a long way to go before we can make contact with the Standard Model describing the world as we know it. What our procedure may have given us is a decent, local as well as bounded Hamiltonian at the Planck scale. We know from quantum field theories that to relate such a Hamiltonian to physics that can be experimentally investigated, we have to make a renormalization group transformation covering some 20 orders of magnitude. It is expected that this transformation may wipe out most of the effective non-renormalizable interactions in our primordial Hamiltonian, but all these things still have to be proven.

An interesting twist to the second-quantisation approach advocated here is that we have a small parameter for setting up a perturbation expansion. The sequence of higher order corrections starts out to converge very well, but then, at very high orders, divergence will set in. What this means is that, in practice, our quantum Hamiltonian is defined with a built-in margin of error that is extremely tiny but non-vanishing, just as what we have in quantum field theory. This might lead to a formal non-locality that is far too small to be noticed in our quantum calculations, while it could suffice to take away some of the apparent paradoxes that are still bothering many of us.

What we did in this section is that we produced a credible scenario of how a theory not unlike the Standard Model may emerge from further studies of the approach proposed here. It was an argument, not yet a proof, in favour of the existence of cellular automaton models with this capacity.


	\subsecl{More about edge states}{edge} 
	
Edge states were encountered frequently in this investigation. They arise when we attempt to reproduce canonical commutation rules auch as 
	\be [q,\,p]=i\Bbb I\ , \eel{qpcomm1}
in a finite dimensional vector space. To prove that this is fundamentally impossible is easy: just note that \def\Trce{{\mathrm{Tr}}}
	\be\Trce(p\,q)=\Trce(q\,p)\ ,\qquad \Trce(\Bbb I)=N\ , \ee
where \(N\) is the dimensionality of the vector space. So, we see that in a finite-dimensional vector space, we need at least one state that violates Eq.~\eqn{qpcomm1}. This is then our edge state. When we limit ourselves to states orthogonal to that, we recover  Eq.~\eqn{qpcomm1}, but one cannot avoid that necessarily the operators \(q\) and \(p\) will connect the other states to the edge state eventually. 

In the continuum, this is also true; the operators \(q\) and \(p\) map some states with finite \(L^2\) norm onto states with infinite norm.
	
Often, our edge states are completely delocalised in space-time, or in the space of field variables. To require that we limit ourselves to quantum states that are orthogonal to edge states means that we are making certain restrictions on our boundary conditions. What happens at the boundary of the universe? What happens at the boundary of the Hamiltonian (that is, at infinite energies)? This seems to be hardly of relevance when questions are asked about the local laws of our physical world. In chapter~\ref{PQ}, we identified one edge state to a single point on a two-dimensional  torus. There, we were motivated by the desire to obtain more convergent expressions. Edge states generate effective non-locality, which we would like to see reduced to a minimum.

We also had to confront edge states in our treatment of the constraints for the longitudinal modes of string theory (subsection~\ref{detstr}). Intuitively, these edge state effects seem to be more dangerous there.

Note furthermore, that in our first attempts to identify the vacuum state, 
section~\ref{locality}, it is found that the vacuum state may turn out to be an edge state. This is definitely a situation we need to avoid, for which we now propose to use the procedure of second quantisation.  Edge states are not always as innocent as they look.

	\subsecl{Invisible hidden variables}{invisible}
In the simplest examples of models that we discussed, for instance those in chapters~\ref{harmosc} and \ref{fermions}, the relation between the deterministic states and the quantum basis states is mostly straightforward and unambiguous. However, when we reach more advanced constructions, we find that, given the quantum Hamiltonian and the description of the Hilbert space in which it acts, there is a multitude of ways in which one can define the ontological states. This happens when the quantum model possesses symmetries that are broken in the ontological description. Look at our treatment of string theory in section~\ref{sustr}: the quantum theory has the entire continuous, \(D\) dimensional Poincar\'e group as a symmetry, whereas, in the deterministic description,  this is broken down to the discrete lattice translations and rotations in the \(D-2\) dimensional transverse space. \def\CC{\mathcal{C}}

	Since most of our deterministic models necessarily consist of discretised variables, they will only, at best, have discrete symmetries, which means the all continuous symmetries \(\CC\) of the quantum world that we attempt to account for, must be broken down to discrete subgroups \(\DD\subset\CC\). This means that there is a group
	\(\CC/\DD\) of non-trivial transformations of the set  of ontological variables onto \emph{another} set of  variables that, amongst themselves, also completely commute, so that these could also serve as `the' ontological variables.  We can never know which of these sets are the `true' ontological variables, and this means that the `true' ontological variables are hidden from us. Thus, which operators exactly are to be called true beables, which are changeables and which are superimposables, will be hidden from our view. This is why we are happy to adopt the phrase `hidden variables' to describe our ontological variables. Whether or not we can call them `invisible' depends on the question whether any quantum states can be invisible. That phrase might be misleading.
	
	For our analysis of Bell's theorem, this is an important issue. If the true ontological variables could have been identified, it would have been possible to deduce, in advance, how Alice and Bob will choose their settings. The fact that this is now impossible removes the `conspiracy' aspect ot the CAI.	
	
\subsecl{How essential is the role of gravity?}{gravityrole}
Quantum gravity is not sufficiently well understood to allow us to include the gravitational force in our quantum theories. This may well be the reason why some aspects of this work are leaving holes and question marks. Gravity is active at the smallest conceivable scale of physics, which is exactly the scale where we think our theories are most relevant. So no-one should be surprised that we do not completely succeed in our technical procedures. As stated, what we would wish to be able to do is to find a class of deterministic models that are locally discrete and classical, but that can be cast in a form that can be described by a quantum field theory.

As emphasised before (section~\ref{doubleHam}), our quantum field theories are described by a Hamiltonian that is both extensive and bounded from below. It means that the Hamiltonian can be written as
	\be H=\int\dd^3\vec x\,T^{00}(\vec x)\ . \eel{Hamloc}
The operator \(T^{00}\) is the Hamilton density, and locality and causality in quantum field theory require that, at equal times, \(t=t_0\),
	\be[T^{00}(\vec x_1,t_0),\,T^{00}(\vec x_2,t_0)]=0\qquad\hbox{if}\qquad \vec x_1\ne\vec x_2\ . \eel{emtenscomm}
Having difficulties recuperating Eq.~\eqn{emtenscomm} from our cellular automaton, it may be worth while to observe that the operator \(T^{00}\) pays the important role of generator of \emph{space-dependent time translations}: if we change the metric tensor \(g_{00}\) in the time direction by an amount 
\(\d g_{00}(\vec x,t)\),
then the change in the total action of matter is
	\be\d S=\half\int\dd^3\vec x\,\dd t\,\sqrt{-g}\,T^{00}\d g_{00}(\vec x,t)\ , \eel{dSdg}
(which can actually be seen as a definition of the stress-energy-momentum tensor \(T^{\m\n}\)).
	Indeed, if we take \(\d g_{00}\) to be independent of the space coordinate \(\vec x\), then the amount of time that went by is modified by \(\half\int\d g_{00}\dd t\), which therefore yields a reaction proportional to the total Hamiltonian \(H\). 
	
	The operator \(\int\dd^3\vec x\,\sqrt{-g}\,T^{00}(\vec x)\,f(\vec x)\) is the generator of a space dependent time translation: \(\d t=f(\vec x)\). One finds that \(T^{\m\n}(\vec x,t)\) is the generator of general coordinate transformations. This is the domain of gravity. This gives us reasons to believe that quantum theories in which the Hamiltonian is extrinsic, that is, the integral of local terms \(T^{00}(\vec x,t)\), are intimately connected to quantum gravity. We still have problems formulating completely self-consistent, unambiguous theories of quantum gravity, while this seems to be a necessary ingredient for a theory of quantum mechanics with locality.
	
	Apart from the reasons just mentioned, we suspect an essential role for gravity also in connection with our problem concerning the positivity of the Hamiltonian. In gravity, the energy density of the gravitational field is well-known to be negative. Indeed, Einstein's equation,
	\be T_{\m\n}-\fract 1{8\pi G}G_{\m\n}=0\ , \eel{Einsteineq}
can be interpreted as saying that the negative energy momentum density of gravity itself, the second term in Eq.~\eqn{Einsteineq}, when added to the energy momentum tensor of matter, \(T_{\m\n}\), leads to a total energy-momentum tensor that vanishes. The reason why the total energy-momentum tensor vanishes is that it generates local coordinate transformations, under which all amplitudes should be invariant.	
	
	In section~\ref{infoloss}, and in subsection~\ref{timeinv}, it was indicated that gravity might be associated with local information loss. This would then mean that information loss should become an essential ingredient in theories that explain the emergence of quantum mechanics with locality from a cellular automaton model with locality built in.

\newsecl{Conclusions of part~II}{conc2}

We had three reasons for working out many of our technical exercises in the second part of this book. First, we wanted to show what techniques can be used to support our assertions made earlier, which state that there are many ways to link quantum mechanical models with completely deterministic ones. Most of these deterministic models were chosen just to demonstrate some points; these were too simple to show  delicate structures of direct physical interest. Some of our models are ``computationally universal", which means that they contain sufficient amounts of complexity to investigate their physical interest\,\cite{Fredkin-1982}. 

Secondly, we wish to demonstrate that conventional quantum mechanics contains extensive mathematical tools that can be employed here as well. Fourier expansions, Taylor expansions, unitary transformations in Hilbert space, perturbation expansions, the Noether theorem, and other well-known procedures, are all extremely useful here. We wish to sketch the picture that quantum mechanics, as we know it, should be looked upon as a powerful mathematical tool to handle statistical features of our theories. If the dynamical equations are too complex to allow us to solve them, the quantum statistical approach may be the only option we have.  No other systematic mathematical machinery would allow us to examine statistical features of any non-trivial cellular automaton when stretched over scales a billion times a billion times as large as the elementary scale of the individual cells. In quantised field theories, the tool that makes such jumps over scales is called the renormalization group.

Thirdly, we do not want to belittle the difficulties that are still there. A completely systematic strategy for constructing models as complex as the Standard Model, has not yet been found; instead, we found several procedures that could be considered as useful ingredients for such a strategy, even if still quite incomplete. We emphasise that ``no-go" theorems, such as Bell's theorem and the CHSH inequalities, do contain the loopholes that have been pointed out repeatedly. ``Super determinism", abhorred by a majority of researchers, becomes less fearsome if one realises that it comes with its own conservation law, the conservation of the ontic nature of a quantum state. To describe our universe, we have to limit ourselves to the ontic states. These form a very small subset of the ``template states" that are normally used in quantum mechanics. Ontic states can only evolve into ontic states. 

The issue of `conspiracy' may still be worrisome to the reader, even if it is clear that our theory will not allow us to predict anything about the settings to be used by Alice and Bob. The notion of `free will' can be addressed without religious or emotional overtones; it is simply a statement about correlation functions in the initial state. Complete clarifications of issues such as these may have to wait until more is known about how to handle quantum field theories, such as the Standard Model, in the CAI.

Quite generally, symmetries, symmetry conservation laws and symmetry transformations, are central not only in conventional quantum mechanics but also in its CA interpretation. A special point was raised in connection with the non-compact, or infinite symmetry groups, such as the various components of the Poincar\'e group. Their Noether charges are observable in the classical limit, but in the quantum domain these charges do not commute. This means that these charges must be conglomerates of beables and changeables, and this causes complications in reproducing such symmetries as features in our cellular automata.

We found a conspicuous property of the quantised strings and superstrings. Without modifying any of their physics, it was noted that their ontological degrees of freedom appear to live on a space-time lattice, with a lattice length parameter \(a_{\mathrm{spacetime}}=2\pi\sqrt{\a\,'}\). This not only reflects the fact that string theory is finite, but gives it a clear physical interpretation.

The Cellular Automaton Interpretation has to deal with more mysteries.  We do reproduce quantum mechanics exactly, so also the numerous peculiarities and counter intuitive characteristics of quantum mechanics are duly reproduced.
There are however other reasons why the most explicit model constructions sketched here are not, or not yet, sufficiently refined to serve as models where we can explain all typically quantum mechanical consequences. A critical reader may rightly point to these obstacles; we bring forward as our defence that the Cellular Automaton Interpretation of quantum mechanics yields a considerable amount of clarification of features of quantum mechanics that have been shrouded by mysteries over the years: the collapse question, the measurement problem, Schr\"odinger's cat, the links between quantum mechanical descriptions and classical descriptions of our world, as well as clear indications as to how to avoid the ``many worlds" as well as the "pilot waves". All these came as bonuses, while our real motivation has always been the question of reconciling the gravitational force with quantum mechanics. We suspect that the work presented here will be very helpful in achieving such aims.

Not only the interpretation of quantum mechanics, but also some issues in the foundations of quantum field theories may be at stake in this approach. We observed that the procedure of second quantisation may have to be applied not only in the use of quantum field theories, but also for the construction of our Hamiltonian, and this raises issues of convergence of the perturbation expansion. Our conclusion is now that this expansion may not converge, just as what we have in quantum field theory. Here, however, this leads us to suggest a novel implication:  there is an unavoidable margin of error not only in our use of quantum field theories to calculate particle properties, but a similar margin of error may also exist in our use of quantum mechanics altogether: quantum calculations cannot be done with infinite accuracy.

\vskip40pt \pagebreak[3]
\appendix 
{\Large\textbf{Appendices}}
\newsecl{Some remarks on gravity in 2+1 dimensions}{grav2+1}
	Gravity in 2+1 dimensions is a special example of a classical theory that is difficult to quantise properly, at least if we wish to admit the presence of matter\fn{In 2+1 dimensions, gravity \emph{without} particles present can be quantised\,\cite{Witten-1988}\cite{Carlip-1989}\cite{GtH-1993-1}\cite{GtH-1993-2}, 
but that is a rather esoteric topological theory.}. One can think of scalar (spin zero) particles whose only interactions consist of the exchange of a gravitational force. The classical theory\,\cite{GtH-D-J-1984}\cite{GtH-1993-1}\cite{GtH-1993-2}  
suggests that it can be quantised, but something very special happens\,\cite{GtH2+1discr96}, as we shall illustrate now.

The Einstein equations for regions without matter particles read
	\be R_{\m\n}=0\ , \eel{Einstein2+1}
but in 2+1 dimensions, we can write
	\be &S^{\a\b}=S^{\b\a}=\quart\e^{\a\m\n}\e^{\b\k\l}R_{\m\n\k\l}\ ,\qquad R_{\m\n\k\l}=\e_{\m\n\a}\e_{\k\l\b}S^{\a\b}\ ,& \nn
	&R_{\m\n}=g_{\m\n}S^{\a\a}-S_{\m\n}\ ,\qquad S_{\m\n}=-R_{\m\n}+\half Rg_{\m\n}\ . &\eel{Riemann0}
Consequently, Eq.~\eqn{Einstein2+1} also implies that the Riemann tensor \(R_{\m\n\k\l}\) vanishes. Therefore, matter-free regions are flat pieces of space-time (which implies that, in 2 + 1 dimensions, there are no \emph{tidal forces}). 

When a particle is present, however, \(R_{\m\n}\) does not vanish, and therefore a particle is a local, topological defect.  One finds that a particle, when at rest, cuts out a wedge from the 2-dimensional space surrounding it, turning that 2-space into a cone, and the deficit angle of the excised region is proportional to the mass: In convenient choices of the units, the total wedge angle is exactly twice the mass \(\m\) of the particle.

\begin{figure}[h!]
\begin{center} \lowerwidthfig{0pt}{60mm}{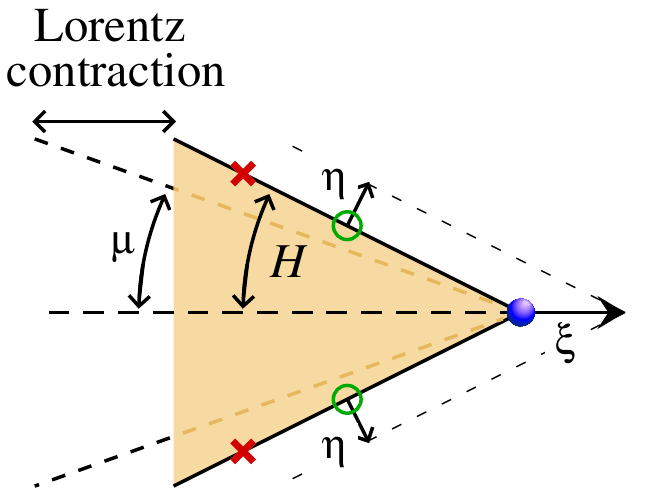}
 \caption[\small The angle cut out of space when a particle moves with velocity $\xi$]{\small The angle cut out of space when a particle moves with velocity $\xi$. See text.
 \labell{wedge.fig}}
\end{center}
\end{figure}

When the particle moves, we choose to orient the wedge with its deficit angle such that the particle moves in the direction of the bisector of the angle. Then, if we ask for the effect of the associated Lorentz transformation, we see that the wedge is Lorentz contracted. This is illustrated in Figure~\ref{wedge.fig}, where the crosses indicate which points are identified when we follow a loop around the particle. We see that, because we chose the particle to move along the bisector, there is no time shift at this identification, otherwise, there would have been. This way we achieve that the surrounding space can be handled as a Cauchy surface for other particles that move around.

Some arithmetic shows that, if the particle's velocity is defined as \(\tanh\xi\), the Lorentz contraction factor is \(\cosh\xi\), and the opening angle \(H\) of the moving particle is given by
	\be \tan H\iss\cosh\xi\,\tan\m\ . \eel{hammass}
If \(\eta\) is the velocity of the seam between the two spaces (arrow in Fig.~\ref{wedge.fig}), then we find
	\be\tanh\eta&=&\sin H\, \tanh\xi\ , \crl{seamvelocity}
	\cos\m&=&\cos H\,\cosh\eta\ ,\crl{cosmuHeta}
	\sinh\eta&=&\sin\m\,\sinh\xi\ . \eel{hypermomentum}
Eq.~\eqn{hammass} gives a relation between \(H\) and \(\m\) that turns into the usual relation between mass in motion and energy in the weak gravity limit, while deficit angles such as \(H\) are additive and conserved. Therefore, we interpret \(H\) as the \emph{energy} of the moving particle, in the presence of gravity.	
		
In Ref.~\,\cite{GtH2+1discr96}, we argued that, taking \(\m\) to be constant, \(\dd(\cos H\,\cosh\eta)=0\), so that 
	\be -\sin H\,\cosh\eta\ \dd H+\cos H\,\sinh\eta\ \dd\eta=0\ ,\qquad {\pa H\over\pa\eta}={\tanh\eta\over\tan H}\ . \eel{dHdeta}
Furthermore, it was derived that we can make tessellations of Cauchy surfaces using configurations such as in Fig.~\ref{wedge.fig} in combination with vertices where no particles are residing, so that the Cauchy surface is built from polygons. The edges \(L_i\) where one polygon is connected to another either end in one of these auxiliary vertices or one of the physical particles. We can then calculate how the lengths of the edges \(L_i\) grow or shrink. 

Both end points of a boundary line make the line grow or shrink with independent velocities, but the orthogonal components are the same. To define these unambiguously, we take a point such as the small circle in Figure \ref{wedge.fig}, which moves in a direction orthogonal to the seam. We now see that our particle gives a contribution to \(\dd L/\dd t\) equal to
	\be {\dd L\over\dd t}\bigg|_\xi\ =\ \tanh\xi\,\cos H\ . \ee
Combining this with Eqs.~\eqn{seamvelocity} and \eqn{dHdeta}, one finds
	\be {\dd L\over\dd t}\bigg|_\xi\ =\ {\tanh\eta\over\tan H}\ =\ {\pa H\over\pa\eta}\ . \eel{hameqL}
Furthermore, the Hamiltonian does not depend on \(L\), while \(\eta\) does not depend on time, so that
	\be{\dd \eta\over\dd t}\ =\ -{\pa H\over\pa L}\ =\ 0\ . \eel{hameqeta}
These last two equations can be seen as the Hamilton equations for \(L\) and \(\eta\).
This means that \(\eta\) and \(L\) are canonically associated with one another. If there are many polygons connected together with seams \(L_i\), moving with transverse velocities \(\tanh\eta_i\), then we obtain Hamiltonian equations for their time dependence, with Poisson brackets
	\be\{L_i,\,\eta_j\}=\d_{ij}\ . \eel{poissonLeta}

Thus, the lengths \(L_i\) are like positions and the \(\eta_i\) are like the associated momenta. This clearly suggests that all one needs to do to obtain the quantum theory is to postulate that these Poisson brackets are replaced by commutators. 

\subsecl{Discreteness of time}{timediscr}
There is, however, a serious complication, lying in the nature of our Hamiltonian. If we have many particles, all adding deficit angles to the shape of our Cauchy surface, one can easily see what might happen:
\begin{quote}\emph{If the total energy due to matter particles exceeds the value \(2\pi\), the universe will close into itself, allowing only the value \(4\pi\) for its total Hamiltonian,}\end{quote}
assuming that the universe is simply connected.
Thus, the question arises what it means to vary the Hamiltonian with respect to $\eta$'s, as in Eq.~\eqn{hameqL}.

There is a way to handle this question: consider some region \(X\) in the universe \(\W\), and ask how it evolves with respect to data in the rest  of the universe, \(\W\backslash X\). The problem is then to define where the boundary between the two regions \(X\) and \(\W\backslash X\) is.

An other peculiar feature of this Hamiltonian is that it is defined as an angle (even if it might exceed the value \(2\pi\); it cannot exceed \(4\pi\)). In the present work, we became quite familiar with Hamiltonians that are actually simple angles: this means that their conjugate variable, time, is \emph{discrete}. The well-defined object is not the Hamiltonian itself, but the evolution operator over one unit of time: \(U=e^{-iH}\). Apparently, what we are dealing with here, is a world where the evolution goes in discretised steps in time. 

The most remarkable thing however, is that we cannot say that the time for the entire universe is discrete. Global time is a meaningless concept, because gravity is a diffeomorphism invariant theory. Time is just a coordinate, and physical states are invariant under coordinate transformations, such as a global time translation. It is \emph{in regions where matter is absent} where we have local flatness, and only in those regions, \emph{relative} time is well-defined, and as we know now, discrete.
Because of the absence of a global time concept, we have no Schr\"odinger equation, or even a discrete time-step equation, that tells us how the entire universe evolves.

Suppose we split the universe \(\W\) into two parts, \(X\) and \(\W\backslash X\). Then the edges \(L_i\) in \(X\) obey a Schr\"odinger equation regarding their dependence on a relative time variable \(t\)  (it is relative to time in \(\W\backslash X\)). The Schr\"odinger equation is derived from Eq.~\eqn{cosmuHeta}, where now \(H\) and \(\eta\) are operators:
	\be \eta=-i{\pa\over\pa L}\ ,\qquad H=i{\pa\over\pa t}\ . \ee
If the wave function is \(\j(L,t)\), then 
	\be(\cos H)\,\j(L,t)=\half\bigg(\j(L,t+1)+\j(L,t-1)\bigg)\ , \ee
and the action of \(1/\cosh\eta\) on \(\j(L,t)\) can be found by Fourier transforming this operator:
	\be(\cosh\eta)^{-1}\j(L,t)=\int_{-\infty}^{\infty}\dd y\,{1\over2\cosh(\pi y/2)}\,\j(L+y,t)\ . \ee 
So, the particle in Fig.~\ref{wedge.fig} obeys the Schr\"odinger equation following from Eq.~\eqn{cosmuHeta}:
	\be\j(L,t+1)+\j(L,t-1)=\int_{-\infty}^{\infty}\dd y\,{\cos\m\over\cosh (\pi y/2)}\,\j(L+y,t)\ . \ee

The problem with this equation is that it involves all \(L\) values, while the polygons forming the tessellation of the Cauchy surface, whose edge lengths are given by the \(L_i\), will have to obey inequalities, and therefore it is not clear to us how to proceed from here. In Ref.~\,\cite{GtH2+1discr96} we tried to replace the edge lengths \(L_i\) by the particle coordinates themselves. It turns out that they indeed have conjugated momenta that form a compact space, so that these coordinates span some sort of lattice, but this is not a rectangular lattice, and again the topological constraints were too difficult to handle.

The author now suspects that,  in a meaningful theory for a system of this sort, we must require all dynamical variables to be sharply defined, so as to be able to define their topological winding properties. Now \emph{that} would force us to search for deterministic, classical models for 2+1 dimensional gravity. In fact, the difficulty of formulating a meaningful `Schr\"odinger equation' for a 2+1 dimensional universe, and the insight that this equation would (probably) have to be deterministic, was one of the first incentives for this author to re-investigate deterministic quantum mechanics as was done in the work reported about here: if we would consider any \emph{classical} model for 2+1 dimensional gravity with matter (which certainly can be formulated in a neat way), declaring its classical states to span a Hilbert space in the sense described in our work, then \emph{that} could become a meaningful, unambiguous quantum system\,\cite{GtH-D-J-1984}\cite{GtH-1993-1}\cite{GtH-1993-2}.

Our treatment of gravity in 2+1 dimensions suggests that the space-time metric and the gravitational fields should be handled as being a set of \emph{beables}. Could we do the same thing in 3+1 dimensions? Remember that the source of gravitational fields, notably the energy density, is not a beable, unless we decide that the gravitational fields generated by energies less than the Planck energy, are negligible anyway (in practice, these fields are too feeble to detect), while our considerations regarding the discretised Hamiltonian, sections~\ref{discHamproblem} and \ref{classenergy}, suggest the one can define large, discretised, amounts of energy that indeed behave as beables.

Consider a 3+1 dimensional gravitating system, where one of the space dimensions is compactified. It will then turn into a 2+1 dimensional world, which we just argued should be subject to the CAI. Should our universe not be regarded as just such a world in the limit where the compactification length of the third spacial dimension tends to infinity? We believe the answer is yes.

\newsecl{A summary of our views on Conformal Gravity}{confgrav}
Whenever a fundamental difficulty is encountered in handling deterministic versions of quantum mechanics, we have to realise that the theory is intended in particular to apply to the Planck scale, and that is exactly where the gravitational force cannot be ignored. Gravity causes several complications when one tries to discretise space and time. One is the obvious fact that any regular lattice will be fundamentally flat, so we have to address the question where the Riemann curvature terms can come from. Clearly, we must have something more complicated that a regular lattice. A sensible suspicion is that we have a discretisation that resembles a glassy lattice.

But this is not all. We commented earlier on the complications caused by having non-compact symmetry groups. The Lorentz group generates unlimited contractions both in the space- and in the time direction. This is also difficult to square with any lattice structure\fn{An important comment was delivered by F.~Dowker: \emph{There is only one type of lattice that reflects perfect Lorentz invariance} (and other non-compact symmetries), \emph{and this is the completely random lattice.}\,\cite{Dowker-2011}}

Furthermore, gravity generates black hole states. The occurrence of stellar-sized black holes is an unavoidable consequence of the theory of General Relativity. They must be interpreted as exotic states of matter, whose mere existence will have to be accommodated for in any ``complete" theory of Nature. It is conceivable that black holes are just large-size limits of more regular field configurations at much smaller scales, but this is also far from being a settled fact. Many theories regard black holes as fundamentally topologically distinct from other forms of matter such as the ones that occur in stars that are highly compressed but did not, or not yet, collapse.
	\def\tot{\mathrm{total}} \def\mat{\mathrm{matter}}\def\EH{\mathrm{EH}}

To make a link to any kind of cellular automaton (thinking of the glassy types, for instance), it seems reasonable first to construct a theory of gravity where space-time, and the fields defined on it, are topologically regular. Consider the standard Einstein-Hilbert action with a standard, renormalizable field theory action for matter added to it: \\[5pt] \(\LL^\tot=\sqrt{-g}(\LL^\EH+\LL^\mat)\), 
	\be\LL^\EH &=& {\textstyle 1\over 16\pi G_N}( R-2\L)\ ,\nm\\
	\LL^\mat&=& -\quart G^a_{\m\n}G^a_{\m\n}-\bar\j\,\g^\m  D_\m\,\j-\half (D\vv)^2  -\half m_\vv^2\vv^2
		 -\fract1{12}R\vv^2\nn
		 &&-V_4(\vv)- V_3(\vv)-\bar\j(y_i\vv_i+iy_i^5\g^5\vv_i+ m_d)\j\ . \eel{EHmat}
Here, \(\L\) is the cosmological constant, \(\vv\) stands for, possibly more than one, scalar matter fields, \(V_4\) is a quartic interaction, \(V_3\) a cubic one, \(y_i\) and \(u_i^5\) are scalar and pseudo-scalar Yukawa couplings, \(m_\vv\) and \(m_d\) are mass terms, and the term \(-\fract 1{12}R\vv^2\) is an interaction between the scalar fields and the Ricci scalar \(R\) that is necessary to keep the kinetic terms for the \(\vv\) field conformally covariant.

Subsequently, one rewrites
	\be g_{\m\n}=\w^2(\vec x,t)\,\^g_{\m\n}\ , \quad\vv=\w^{-1}\^\vv\ ,\quad \j=w^{-3/2}\^\j\ ,\quad\sqrt {-g}=\w^4\sqrt{-\^ g} \ ,\eel{confomega}
and substitutes this everywhere in the total Lagrangian \eqn{EHmat}. 

This leaves a manifest, exact local Weyl invariance in the system:
	\be\^g_{\m\n}\ra\W^2(\vec x,t)\^g_{\m\n}&,\qquad&\w\ra\W(\vec x,t)^{-1}\w\ ,\nn
	\^\vv\ra\W(\vec x,t)^{-1}\^\vv&,\qquad& \^\j\ra\W(\vec x,t)^{-3/2}\,\^\j\ . \eel{Weylinv}

The substitution~\eqn{confomega} turns the Einstein-Hilbert Lagrangian into
	\be {1\over 16\pi G_N}\big(\w^2\hat R-2\,\w^4\L+6\,\hat g^{\,\m\n} \pa_\m\w\,\pa_\n\w\big) \ . \eel{confEH}
Rescaling the \(\w\) field: \(\w=\tl\k\chi\ ,\ \tl\k^2=\fract 43\pi G_N\), turns this into
	\be \half\^g^{\m\n}\pa_\m\chi\pa_\n\chi+\fract 1{12}\^R\chi^2-\fract 16\tl\k^2\L\chi^4\ . \eel{scaledLEH}
	
The resemblance between this Lagrangian for the \(\chi\) field and the kinetic term of the scalar fields \(\vv\) in Eq.~\eqn{EHmat}, suggests that no singularity should occur when \(\chi\ra0\), but we can also conclude directly from the requirement of exact conformal invariance that the coupling constants should not run, but keep constant values under (global or local) scale transformations.  Note, that \(\chi\ra0\) describes the small-distance limit of the theory.

The theory was originally conceived as an attempt to mitigate the black hole information paradox\,\cite{GtHconf-2010}, then it was found that it could serve as a theory that determines the values of physical parameters that up to the present have been theoretically non calculable (this should follow from the requirement that all renormalization group functions \(\b_i\) should cancel out to be zero). 

For this book, however, a third feature may be important: with judiciously chosen conformal gauge-fixing procedures, one may end up with models that feature upper limits on the amount of information that can be stowed in a given volume, or 4-volume, or surface area.

\section*{Acknowledgements}
The author discussed these topics with many colleagues; I often forget who said what, but it is clear that many critical remarks later turned out to be relevant and were picked up. Among them were
A.~Aspect, T.~Banks, N.~Berkovitz, M.~Blasone, M.~Duff, G.~Dvali, Th.~Elze, E.~Fredkin, S.~Giddings, S.~Hawking, M.~Holman, H.~Kleinert,  R.~Maimon, Th.~Nieuwenhuizen, M.~Porter, P.~Shor, L.~Susskind, R.~Werner, E.~Witten, W.~Zurek.

\vskip40pt \pagebreak[4]

 \end{document}